\documentclass[english]{report}
\usepackage{amsmath}
\usepackage{amssymb}
\usepackage[a4paper]{geometry}
\usepackage{array}
\usepackage{longtable}
\usepackage{float}
\usepackage{bm}
\usepackage{multirow}
\usepackage{graphicx}
\usepackage{esint}
\usepackage[unicode=true,
 bookmarks=true,bookmarksnumbered=false,bookmarksopen=false,
 breaklinks=false,pdfborder={0 0 0},pdfborderstyle={},backref=false,colorlinks=false]
 {hyperref}
\hypersetup{pdftitle={Enhancing the Formation of Wigner Negativity in a Kerr Oscillator via Quadrature Squeezing},
 pdfauthor={Christian Anker Rosiek}}

\makeatletter

\providecommand{\tabularnewline}{\\}
\newcommand{\lyxdot}{.}

\usepackage{ifluatex}

\ifluatex
\usepackage{luatex85}

\fi

\makeatletter
\newcommand\footnoteref[1]{\proteted@xdef\@thefnmark{\ref{#1}}\@footnotemark}

\usepackage{graphicx}
\usepackage{svg}
\usepackage{tikz}
\usepackage{etoolbox}
\usepackage{pdflscape}
\usepackage{afterpage}
\usepackage{placeins}
\usepackage{lipsum}

\usepackage{siunitx}

\usepackage{lmodern}
\usepackage{textcase}
\usepackage{amsmath}
\usepackage{amsfonts}
\usepackage{stackengine}
\usepackage{mathtools}
\usepackage{slashed}
\usepackage{bbm}
\usepackage{bm}
\usepackage{tocloft}
\usepackage{mathrsfs}
\usepackage{gensymb}
\usepackage{siunitx}
\usepackage{nccmath}
\usepackage{xcolor}
\usepackage{tikz}
\usepackage{graphicx}

\usepackage[export]{adjustbox}

\usepackage{changepage}
\usepackage{wallpaper}
\usepackage{comment}
\usepackage{microtype}
\usepackage{textcase}

\DeclareMathAlphabet{\mathcal}{OMS}{cmsy}{m}{n}
\let\mathbb\relax 
\DeclareMathAlphabet{\mathbb}{U}{msb}{m}{n}

\usepackage{caption}
\@ifundefined{showcaptionsetup}{}{%
 \PassOptionsToPackage{caption=false}{subfig}}
\usepackage{subfig}
\makeatother

\usepackage[english]{babel}
\usepackage[style=phys,sortcites,biblabel=brackets,eprint=true]{biblatex}
\begin{document}

\makeatletter
\long\def\title#1{\gdef\@title{#1}\gdef\actualtitle{#1}}
\long\def\subtitle#1{\gdef\@subtitle{#1}}
\long\def\supertitle#1{\gdef\@supertitle{#1}}
\long\def\titletext#1{\gdef\@titletext{#1}}
\def\maketitle{%
\begin{titlepage}%
  \renewcommand{\thefootnote}{\fnsymbol{footnote}}
  \setcounter{page}{0}
  \let\footnotesize\small
  \let\footnoterule\relax
  \let \footnote \thanks
  \null\vfil
  \vskip 60\p@
  \begin{center}%
    \ifdefined\@supertitle{\large\sffamily\@supertitle\par}\vskip0.7em\fi
    {\Large\bfseries \@title \par}%
    \ifdefined\@subtitle{\Large\@subtitle\par}\fi
    \vskip 3em%
    {\large
     \lineskip .75em%
      \begin{tabular}[t]{c}%
        \sffamily \@author
      \end{tabular}\par}%
      \vskip 1.5em%
\ifdefined\@supertitle{{\large\sffamily\@titletext\par}\vskip0.7em\par}\fi
\vskip 1.5em%
      {\large \sffamily\@date \par}
  \end{center}\par
  \@thanks
  \vfil\null
  \end{titlepage}%
  \setcounter{footnote}{0}%
}

\thispagestyle{empty}\pagenumbering{arabic}\setcounter{page}{-1}

\supertitle{Master Thesis}

\titletext{\vskip3em Supervised by Prof.~Albert~Schliesser, Prof.~Anders~S.~Sørensen, and~Massimiliano~Rossi}
\title{Enhancing the Formation of Wigner Negativity in a Kerr Oscillator
via Quadrature Squeezing}
\author{Christian Anker Rosiek\thanks{\sffamily\href{mailto:xvm706@alumni.ku.dk}{xvm706@alumni.ku.dk}}}
\date{Submitted at the University of Copenhagen on 1 October 2019\\
\vskip2mm\small Includes minor corrections of February
2022}\maketitle

\cleardoublepage
\phantomsection\addcontentsline{toc}{chapter}{Abstract}
\pagenumbering{roman}
\setcounter{page}{1}

\chapter*{Abstract}

Motivated by quantum experiments with nanomechanical systems, the
evolution of a Kerr oscillator with focus on creation of states with
a negative Wigner function is investigated. Using the phase space
formalism, results are presented that demonstrate an asymptotic behavior
in the large squeezing regime for the negativity of a squeezed vacuum
state under unitary evolution. The analysis and model are extended
to squeezed vacuum states of open systems, adding the decoherence
effects of damping and dephasing. To increase experimental relevance,
the regime of strong damping is considered. These effects are investigated,
yielding similar asymptotic results for the behavior of these effects
in the large squeezing regime. Combining these results, it is shown
that a weak nonlinearity as compared to damping may be improved by
increasing the squeezing of the initial state. It is also shown that
this may be done without exacerbating the effects of dephasing.

\cleardoublepage\phantomsection\addcontentsline{toc}{chapter}{Contents}

\tableofcontents{}

\global\long\def\Tr{\mathrm{Tr}}%
\global\long\def\Im{\mathrm{Im}}%
\global\long\def\Re{\mathrm{Re}}%
\global\long\def\Ai{\mathrm{Ai}}%
\global\long\def\appropto{\approptotex}%
\newcommand\barbelow[1]{\stackunder[1.2pt]{$#1$}{\rule{.8ex}{.075ex}}}
\global\long\def\dunderbar#1{\barbelow{\barbelow{#1}}}%

\global\let\oldsymbf\symbf
\global\long\def\symbf#1{\boldsymbol{#1}}%

\newcommand{\approptoinn}[2]{\mathrel{\vcenter{
  \offinterlineskip\halign{\hfil$##$\cr
    #1\propto\cr\noalign{\kern2pt}#1\sim\cr\noalign{\kern-2pt}}}}}

\newcommand{\approptotex}{\mathpalette\approptoinn\relax}

\cleardoublepage

\cleardoublepage\phantomsection\addcontentsline{toc}{chapter}{Acknowledgments}

\chapter*{Acknowledgments}

{\setlength{\parindent}{0cm}\setlength{\parskip}{0.2cm}

I thank Albert for allowing me to write this thesis with his help
and that of his group. Your trust in my ability has surprised me more
than once and allowed and encouraged me to devote my time to a fascinating
subject that I have thoroughly enjoyed.

I thank Anders his supervision. I am grateful for the time, energy
and ideas you put into the project. I can only assume this to extend
far beyond proposal as originally laid out.

A great thanks goes to Massimiliano, without whom the project's quality
would have been substantially diminished. You provided me with the
rare combination of technical and companionly support and I enjoyed
working with you immensely.

I thank also the other members of SLAB -- in no important order,
David, Eric, Junxin, Letizia, Mads, Sampo, Yannick, Yeghishe. All
of you have contributed to making my work here an enjoyable and learning
experience.

I thank Signe and Timo who kept me company in the office and as part
of The Breakfast Club. Without you, my time in FK-10 would not have
been nearly as compelling or interesting as it was.

I thank my family and friends. Writing a thesis proved a worthy challenge,
but I have found none of you to treat me with anything but patience
and respect. For that I am greatly indebted. I hope to be able to
return this debt should the need ever arise.

In case I forgot to put you here, I did not do so out of spite and
I sincerely apologize.

}

\cleardoublepage\phantomsection\addcontentsline{toc}{chapter}{Introduction}
\setcounter{page}{1}
\pagenumbering{arabic}

\chapter*{Introduction}

Quantum mechanics is a well-established theory to describe the world
at microscopic scales. Yet the classical laws of physics are comfortably
able to explain most macroscopic phenomena that we can observe. Somewhere
in the chain of physical systems of increasing scale, the display
of manifestly quantum behavior discontinues as the system transitions
to the classical regime. The scale at which this happens varies between
systems and an important approach in furthering our understanding
of this transition is thus the observation of quantum phenomena in
systems of ever increasing scale. The observation of quantum behavior
in a macroscopic system could yield fundamental new insights into
the relation between quantum and classical mechanics.

The process by which a quantum system loses its quantum properties
is called decoherence. The Wigner quasiprobability distribution, introduced
\cite{Wigner_QuantumCorrectionThermodynamic_1932} to describe quantum
corrections to a classical theory, provides a way to describe the
state of quantum systems and their coherence. Unlike true probability
distributions, the Wigner function can assume negative values and
this property is of use in assessing the decoherence of a quantum
system. Decoherence effects such as damping lead to the irreversible
loss of coherences causing the negativity of the Wigner function to
decay \cite{Stobinska_WignerFunctionEvolution_2008}. This decay may
explain the reduction of system dynamics to adhere to the classical
description \cite{Milburn_DissipativeQuantumClassical_1986}.

As physical systems grow in size, so does the difficulty in preserving
their coherence. Nanomechanical oscillators can consist of $10^{13}$
to $10^{14}$ atoms, placing them firmly in the category of macroscopic
systems. Thanks to recent advances in nanofabrication technology however,
ultra-coherent nanomechanical resonators have been developed \cite{Tsaturyan_UltracoherentNanomechanicalResonators_2017}
that allow for observation of quantum mechanical effects in such macroscopic
systems. Examples of experiments demonstrating this include the cooling
of a nanomechanical resonator to its ground state \cite{Chan_LaserCoolingNanomechanical_2011}
and squeezing of mechanical fluctuations below those of the ground
state \cite{Wollman_QuantumSqueezingMotion_2015}. These experiments
demonstrate quantum effects though the states involved have still
strictly positive Wigner functions.

To create negativities in the Wigner function, one can consider the
evolution of a nonlinear quantum system. A simple such nonlinearity
is given by the Kerr effect, which finds application in many quantum
experiments and technologies, e.g. the generation of Schrödinger's
cat states \cite{Yurke_GeneratingQuantumMechanical_1986} and continuous
variable quantum computing \cite{Lloyd_QuantumComputationContinuous_1999}.
In a mechanical system, one can exploit the intrinsic Kerr (Duffing)
nonlinearity, present in any mechanical system. For typical mechanical
systems however the strength of the nonlinearity is too small when
compared to decoherence effects such as damping. Table \ref{tab:Nonlinear-systems.-Parameters}
lists a sample of nonlinear nanomechanical systems, showing that the
typical mechanical nonlinearity is several orders of magnitude weaker
than the damping of the same system. In such cases it is nontrivial
to develop of an experiment demonstrating Wigner negativity.

In this thesis we wish to explore the use of a Kerr nonlinearity to
generate states with negative Wigner functions. Guided by the experimental
realities, we expect decoherence effects to be a significant impairment
and we therefore investigate the use of squeezing to counter this
decoherence. We implement in our treatment the decohering effects
of damping and dephasing and apply various degrees of squeezing to
states of the Kerr Oscillator, studying their evolution and decoherence
under damping and dephasing.

The thesis is structured as follows. In Chapters \ref{chap:fundamentals}
and \ref{chap:Numerics}, we review the mathematical fundamentals
and the method used for simulating quantum systems. Chapters \ref{chap:nonlinear-oscillators}
and \ref{chap:coupling-to-the-environment} contains the main results.
Chapter \ref{chap:nonlinear-oscillators} introduces nonlinear oscillators
and considers their unitary evolution for selected initial states.
This yields understanding of some universal qualities of the evolution
of Wigner negativity for a squeezed vacuum state. These will also
be important to the later analysis of open systems. Chapter \ref{chap:coupling-to-the-environment}
then extends the analysis of its prior chapter to include open systems
for the particular case of a squeezed vacuum state. 

\begin{table}

\renewcommand{\arraystretch}{1.5}%
\begin{longtable}[c]{|c|c|c|c|>{\centering}p{5cm}|}
\hline 
$\omega/2\pi$ (Hz) & $g/2\pi$ (Hz) & $\gamma/2\pi$ (Hz) & $g/\gamma$ & System\tabularnewline
\hline 
\hline 
\multicolumn{5}{|c|}{\emph{Experimental systems}}\tabularnewline
\hline 
$10^{6}$ & $10^{-11}$ & $10^{-2}$ & $10^{-9}$ & Silicon nitride membrane \cite{Catalini_MechanicalNonlinearities_2018,Rossi_MeasurementbasedQuantumControl_2018}\tabularnewline
\hline 
$10^{6}$ & $10^{-10}$ & $10^{-2}$ & $10^{-8}$ & Silicon nitride membrane \cite{Catalini_MechanicalNonlinearities_2018,Rossi_MeasurementbasedQuantumControl_2018}\tabularnewline
\hline 
$10^{6}$ & $10^{-4}$ & $10^{5}$ & $10^{-9}$ & Graphene resonator \cite{Singh_GiantTunableMechanical_2019}\tabularnewline
\hline 
$10^{6}$ & $10^{-2}$ & $10^{3}$ & $10^{-5}$ & Graphene/silicon nitride hybrid resonator \cite{Singh_GiantTunableMechanical_2019}\tabularnewline
\hline 
$10^{7}$ & $10^{-4}$ & $10^{4}$ & $10^{-8}$ & Nanomechanical resonator \cite{Woolley_NonlinearQuantumMetrology_2008b}\tabularnewline
\hline 
\hline 
\multicolumn{5}{|c|}{\emph{Theoretical treatments}}\tabularnewline
\hline 
$10^{8}$ & $10^{5}$ & $100$ & $10^{3}$ & Cooper pair box coupled to nanomechanical resonator\cite{Jacobs_EngineeringGiantNonlinearities_2009}\tabularnewline
\hline 
- & $1$ & $10^{5}$ & $10^{-3}$ & Nanomechanical oscillator \cite{Stobinska_WignerFunctionEvolution_2008}\tabularnewline
\hline 
- & 1 & $1$ & $1$ & Rescaling of initial state from above \cite{Stobinska_WignerFunctionEvolution_2008}\tabularnewline
\hline 
\end{longtable}

\setcounter{table}{0}\vskip1em

\caption[Parameters of nonlinear systems]{\label{tab:Nonlinear-systems.-Parameters}\textbf{Parameters of nonlinear
systems.} The table lists the order of magnitude for the base frequency
$\omega$, the frequency $g$ describing the strength of the Kerr
nonlinearity, the damping rate $\gamma$ for some experimental nanomechanical
resonators. The ratio of $g/\gamma$ is important to the viability
of generating mechanical states of motion with negative Wigner functions.
Parameters given for the Duffing oscillator are converted those of
the Kerr oscillator using (\ref{eq:-215}) as derived in Appendix
\ref{chap:Interaction-Picture-and}. Theoretical treatments of mechanical
resonators are listed in the lower part of the table.}
\end{table}

\chapter{Fundamentals \label{chap:fundamentals}}

Before we study nonlinear systems in later chapters, we spend this
chapter reviewing the mathematical techniques used. The first two
sections introduces and motivates the use of the simple quantum harmonic
oscillator as well as its quantum states. We then describe various
unitary operations which may be applied to states and operators of
this system: Section \ref{sec:operator-transformations} defines displacement,
rotation and squeezing while Section \ref{sec:unitary-dynamics} and
\ref{sec:interaction-picture} describe unitary time evolution and
the interaction picture. The introduction to the quantum mechanics
of closed systems finishes with a brief look at expectation values
motivated by projective measurements. Section \ref{sec:Mixed-States}
introduces density matrices and superoperators as required for describing
open quantum systems. With this, open system dynamics can now be described.
This is done in Section \ref{subsec:master-equation} in terms of
the Markovian master equation. The first half of the chapter then
concludes with an example to demonstrate the various techniques and
introduce the concept of squeezing.

Once established, the operator formalism is used as a stepping stone
to introduce the Wigner quasiprobability distribution and the related
phase space formalism. This is a complementary way to describe quantum
systems and their dynamics and a therefore translate a selection of
the initial sections is therefore translated to this new formalism.
The transformation operators are treated in Section \ref{sec:Transformations-in-Phase}.
Section \ref{sec:phase-space-coordinates} introduces two alternate
phase space coordinate systems which may help to simplify discussions.
Phase space dynamics are treated in Section \ref{sec:wigner-pde-derivation}
which describes a procedure for translating the master equation of
Section \ref{subsec:master-equation} to a Fokker-Planck-like equation
for the Wigner function. 

To prepare for the discussion of non-classicality, Section \ref{sec:Gaussian-States-and}
describes states and dynamics that are particularly similar to their
corresponding classical system. Section \ref{sec:wigner-current}
builds on this and introduces the Wigner current as a way to gain
geometrical intuition for the dynamics. Section \ref{sec:measures-of-negativity}
then introduces the two measures which we shall use to gauge non-classicality
in this thesis: negative volume and negative peak. After the phase
space formalism has been treated in Sections \ref{sec:Phase-Space-Representations}--\ref{sec:measures-of-negativity},
its concepts are applied as Section \ref{sec:return-to-parametric-squeezing}
returns to the previous example.

\section{Quantum Harmonic Oscillator \label{sec:the-quantum-harmonic-oscillator}}

The simple harmonic oscillator is a system central to the discussion
in the thesis. We will follow the conventions of \textcite{Gerry_IntroductoryQuantumOptics_2004}
in defining this system and associated mathematical objects. 

The harmonic oscillator can be introduced in terms of the generic
Hamiltonian for a quantum particle
\begin{equation}
\hat{H}=\frac{\hat{p}^{2}}{2m}+V(\hat{q}).\label{eq:-168}
\end{equation}
The Hamiltonian $\hat{H}$ describes the a harmonic oscillator when
the potential is given by
\begin{equation}
V(\hat{q})=\frac{1}{2}m\omega^{2}\hat{q}^{2},\label{eq:-213}
\end{equation}
where $m$ is the mass of the oscillator, $\omega$ is its frequency
and $\hat{q}$ and $\hat{p}$ are, respectively, the canonical position
and momentum operator of the oscillator. These are Hermitian observables,
i.e. $\hat{q}^{\dagger}=\hat{q}$ and $\hat{p}^{\dagger}=\hat{p}.$
They furthermore obey the canonical commutation relation
\begin{equation}
[\hat{q},\hat{p}]=i\hbar.\label{eq:-157}
\end{equation}

We then introduce the annihilation and creation operators $\hat{a}$
and $\hat{a}^{\dagger}$ (collectively known as ladder operators)
by a unitary transformation of the position and momentum operators:
\begin{subequations}
\label{eq:-180}
\begin{equation}
\hat{a}=\sqrt{\frac{1}{2\hbar m\omega}}\left(m\omega\hat{q}+i\hat{p}\right)
\end{equation}
and
\begin{equation}
\hat{a}^{\dagger}=\sqrt{\frac{1}{2\hbar m\omega}}\left(m\omega\hat{q}-i\hat{p}\right).
\end{equation}
\end{subequations}
 The inverse transformations,
\begin{equation}
\hat{q}=\sqrt{\frac{\hbar}{2m\omega}}(\hat{a}+\hat{a}^{\dagger})\qquad\text{and}\qquad\hat{p}=\frac{1}{i}\sqrt{\frac{\hbar m\omega}{2}}(\hat{a}-\hat{a}^{\dagger}).\label{eq:-165}
\end{equation}
allows us to write the Hamiltonian in equation (\ref{eq:-168}) in
the form 
\begin{equation}
\hat{H}=\hbar\omega\left(\hat{a}^{\dagger}\hat{a}+\frac{1}{2}\right).\label{eq:sho-hamiltonian}
\end{equation}
From equation (\ref{eq:-157}) one has for $\hat{a}$ and $\hat{a}^{\dagger}$
that
\begin{equation}
\left[\hat{a},\hat{a}^{\dagger}\right]=1.\label{eq:canonical-commutation}
\end{equation}
The number operator $\hat{n}$ can be defined in terms of the ladder
operators as
\begin{equation}
\hat{n}=\hat{a}^{\dagger}\hat{a}.\label{eq:number-operator}
\end{equation}
We also introduce the quadrature operators $\hat{X}$ and $\hat{Y}$
as the Hermitian and anti-Hermitian parts of $\hat{a}$:
\begin{align}
\hat{X} & =\frac{1}{2}\left(\hat{a}+\hat{a}^{\dagger}\right)\,, & \hat{Y} & =\frac{1}{2i}\left(\hat{a}-\hat{a}^{\dagger}\right)\,.\label{eq:quadrature-operators}
\end{align}
These may be considered dimensionless variants of the position and
momentum operators (compare (\ref{eq:-165}) and (\ref{eq:quadrature-operators})).
From these expressions and (\ref{eq:canonical-commutation}), the
quadrature commutation relation may be derived as

\begin{equation}
[\hat{X},\hat{Y}]=\frac{i}{2}.\label{eq:quadrature-commutation}
\end{equation}

\section{States of the Harmonic Oscillator}

At a given point in time, the system is said to be in a particular
state. The state of the system determines the values of all observables
at that time. We describe the state of a quantum mechanical system
by a ket. For the system in the state denoted by $\Psi$ we write
the state of the system as $|\Psi\rangle$. In this section, we will
discuss some important states for the Harmonic oscillator.

The vacuum state is written $|0\rangle$. It is the lowest energy
state of the system so it is also referred to as the ground state
and it can be found as the solution to the equation
\begin{equation}
\hat{a}|0\rangle=0.
\end{equation}

Central to any quantum mechanical system are the eigenstates of the
system Hamiltonian $\hat{H}$. For the harmonic oscillator, these
states are called number states. Since they are eigenstates of the
system Hamiltonian, they are states of definite energy. We construct
the $n$-th number state from the vacuum state $|0\rangle$ by using
the creation operator:
\begin{equation}
|n\rangle=\frac{(\hat{a}^{\dagger})^{n}}{\sqrt{n!}}|0\rangle.\label{eq:fock-state-def}
\end{equation}
We see that the vacuum state is also the zeroth number state. As the
name suggests, the number states are eigenstates of the number operator:
\begin{equation}
\hat{n}|n\rangle=n|n\rangle.
\end{equation}
Applying this to (\ref{eq:sho-hamiltonian}) shows that $|n\rangle$
is an eigenstate of $\hat{H}$:
\begin{equation}
\hat{H}|n\rangle=\hbar\omega\left(n+\frac{1}{2}\right)|n\rangle.\label{eq:-169}
\end{equation}
Using (\ref{eq:canonical-commutation}) and (\ref{eq:fock-state-def}),
one may also derive the relations 
\begin{equation}
\hat{a}|n\rangle=\sqrt{n}|n-1\rangle\qquad\hat{a}^{\dagger}|n\rangle=\sqrt{n+1}|n+1\rangle.\label{eq:-231}
\end{equation}
The number states form an orthonormal basis for states of the oscillator.
Hence, any state $|\Psi\rangle$ can be written as a linear combination
of the number states
\begin{equation}
|\Psi\rangle=\sum_{n=0}^{\infty}c_{n}|n\rangle.\label{eq:-181}
\end{equation}
Since they are orthogonal and normalized, we have 
\begin{equation}
\langle n|m\rangle=\delta_{nm},
\end{equation}
which can be used to find the coefficient $c_{m}$ as
\begin{equation}
c_{m}=\langle n|\Psi\rangle.\label{eq:-230}
\end{equation}
A state expressed in the form of (\ref{eq:-181}) is said to be expanded
in the number state basis.

Another important class of states are the coherent states $|\alpha\rangle$.
They may be defined in the number state basis as
\begin{equation}
|\alpha\rangle=e^{-|\alpha|^{2}/2}\sum_{n}\frac{\alpha^{n}}{\sqrt{n!}}|n\rangle.\label{eq:-182}
\end{equation}
Applying $\hat{a}$ to (\ref{eq:-182}), $|\alpha\rangle$ is seen
to be an eigenstate of the annihilation operator:
\begin{equation}
\hat{a}|\alpha\rangle=\alpha|\alpha\rangle.
\end{equation}
Coherent states are an important class of states in quantum optics.
They are often considered to be closest analog to states of the classical
harmonic oscillator. 

\section{Transformation Operators \label{sec:operator-transformations}}

To aid in the manipulation of quantum states, it is useful to define
several parameterized unitary transformations.

We introduce first the displacement operator

\begin{equation}
\hat{D}(\lambda)=e^{\lambda\hat{a}^{\dagger}-\lambda^{*}\hat{a}}.\label{eq:displacement-operator}
\end{equation}
The exponential function applied to an operator $\hat{A}$ should
be interpreted using the Taylor expansion of the exponential function:
\begin{equation}
e^{\hat{A}}=\sum_{n=0}^{\infty}\frac{\hat{A}^{n}}{n!}.\label{eq:-200}
\end{equation}
Applied to the ladder operators, the displacement has the effect \cite{Gerry_IntroductoryQuantumOptics_2004}
\begin{subequations}
\label{eq:-116}
\begin{align}
\hat{D}^{\dagger}(\lambda)\hat{a}\hat{D}(\lambda) & =\hat{a}+\lambda,\\
\hat{D}^{\dagger}(\lambda)\hat{a}^{\dagger}\hat{D}(\lambda) & =\hat{a}^{\dagger}+\lambda^{*}.
\end{align}
\end{subequations}
An important theorem for the displacement operator is the disentangling
theorem \cite{Gerry_IntroductoryQuantumOptics_2004}. It states that
\begin{subequations}
\label{eq:disentangling-theorem}
\begin{align}
\hat{D}(\lambda) & =e^{\lambda\hat{a}^{\dagger}-\lambda^{*}\hat{a}}\label{eq:-2}\\
 & =e^{-\lambda\lambda^{*}/2}e^{\lambda\hat{a}^{\dagger}}e^{-\lambda^{*}\hat{a}}\label{eq:-3}\\
 & =e^{\lambda\lambda^{*}/2}e^{-\lambda^{*}\hat{a}}e^{\lambda\hat{a}^{\dagger}}.\label{eq:-4}
\end{align}
\end{subequations}
Applying (\ref{eq:-3}) to the vacuum state, it is seen that the coherent
state can also be written as
\begin{equation}
|\alpha\rangle=\hat{D}(\alpha)|0\rangle.
\end{equation}

We next introduce the rotation or phase shift operator
\begin{equation}
\hat{R}(\phi)=e^{i\hat{n}\phi}.\label{eq:rotation-operator}
\end{equation}
It adds a complex phase to the ladder operator $\hat{a}$:
\begin{subequations}
\label{eq:-115}
\begin{align}
\hat{R}^{\dagger}(\phi)\hat{a}\hat{R}(\phi) & =\hat{a}e^{-i\phi}.\label{eq:-115-1}\\
\hat{R}^{\dagger}(\phi)\hat{a}^{\dagger}\hat{R}(\phi) & =\hat{a}^{\dagger}e^{i\phi}.\label{eq:-149}
\end{align}
\end{subequations}
The specific instance $\hat{R}(\pi)$ is sometimes called the parity
operator \cite{Sakurai_ModernQuantumMechanics_2011a} since the act
of rotating 180° around the origin is the same as the mirroring of
all points through the it. We write
\begin{equation}
\hat{\pi}=\hat{R}(\pi).
\end{equation}

Finally we introduce the squeezing operator
\begin{equation}
\hat{S}(\xi)=e^{\frac{1}{2}\left(\xi^{*}\hat{a}\hat{a}-\xi\hat{a}^{\dagger}\hat{a}^{\dagger}\right)}.\label{eq:squeezing-operator}
\end{equation}
Like the former unitary transformations, $\hat{S}(\xi)$ can be described
in terms of its effect on the ladder operators $\hat{a}$ and $\hat{a}^{\dagger}$:
\begin{subequations}
\label{eq:-70}
\begin{align}
\hat{S}^{\dagger}(\xi)\hat{a}\hat{S}(\xi) & =\hat{a}\cosh r-\hat{a}^{\dagger}e^{i\theta}\sinh r,\label{eq:-71}\\
\hat{S}^{\dagger}(\xi)\hat{a}^{\dagger}\hat{S}(\xi) & =\hat{a}^{\dagger}\cosh r-\hat{a}e^{-i\theta}\sinh r.\label{eq:-72}
\end{align}
\end{subequations}

The transformations $\hat{D}(\lambda)$, $\hat{R}(\phi)$ and $\hat{S}(\xi)$
all share the property that their inverse transformation can be found
by negating their parameter:
\begin{subequations}
\label{eq:-117}
\begin{align}
\hat{D}^{\dagger}(\lambda) & =\hat{D}(-\lambda),\\
\hat{R}^{\dagger}(\phi) & =\hat{R}(-\phi),\\
\hat{S}^{\dagger}(\xi) & =\hat{S}(-\xi).
\end{align}
\end{subequations}

\section{Unitary Dynamics \label{sec:unitary-dynamics}}

The time-evolution of a quantum system is determined by the Hamiltonian
of the system. Given the system Hamiltonian $\hat{H}$, the evolution
of the state $|\Psi(t)\rangle$ obeys the Schrödinger equation 
\begin{equation}
i\hbar\frac{\partial}{\partial t}|\Psi(t)\rangle=\hat{H}|\Psi(t)\rangle.\label{eq:schrodinger-equation}
\end{equation}
If $\hat{H}$ does not vary with time, one can introduce the time-evolution
operator

\begin{equation}
\hat{U}(t)=e^{-i\hat{H}t/\hbar}\label{eq:unitary-propagator}
\end{equation}
 to express the evolved state at any point in time $t$ from the initial
state $|\Psi(0)\rangle$:
\begin{equation}
|\Psi(t)\rangle=\hat{U}(t)|\Psi(0)\rangle.\label{eq:propagated-ket}
\end{equation}
Dynamics which may be described purely in terms of such a unitary
transformation are called unitary dynamics.

A relevant example of unitary dynamics is found in the simple harmonic
oscillator. Inserting its Hamiltonian (\ref{eq:sho-hamiltonian})
into (\ref{eq:unitary-propagator}), the time-evolution operator for
the simple harmonic oscillator is found as
\begin{equation}
\hat{U}(t)=e^{-i\omega(\hat{n}+1/2)t/\hbar}.\label{eq:-186}
\end{equation}
$U(t)$ may be recognized as the product of (\ref{eq:rotation-operator})
and a time dependent complex number of unit magnitude:
\begin{equation}
\hat{U}(t)=e^{i\omega t/2}\hat{R}(-\omega t).
\end{equation}
We may associate the time evolution of the state $|\Psi(0)\rangle$
with a time dependence in the coefficients of the expansion in the
number state basis. Applying $\hat{U}(t)$ to each term of (\ref{eq:-182})
yields
\begin{equation}
|\Psi(t)\rangle=\sum_{n=0}^{\infty}c_{n}e^{-i\omega(n+1/2)t}|n\rangle.\label{eq:-217}
\end{equation}

\section{Interaction Picture \label{sec:interaction-picture} }

In defining the interaction picture, we follow \textcite{Sakurai_ModernQuantumMechanics_2011a}.
Consider a Hamiltonian $\hat{H}$ which assumes the form 
\begin{equation}
\hat{H}=\hat{H}_{0}+\hat{V}
\end{equation}
where the dynamics for $\hat{H}_{0}$ are exactly solvable in the
sense that a basis is known in which $\hat{H}_{0}$ is diagonal (in
which case the time evolution of a state $|\Psi(t)\rangle$ may be
written in that basis in a form similar to (\ref{eq:-217})). One
may now transform the operator $\hat{V}$ to the interaction picture
operator $\hat{V}_{I}$ using the transformation
\begin{equation}
\hat{V}_{I}=e^{i\hat{H}_{0}t/\hbar}\hat{V}e^{-i\hat{H}_{0}t/\hbar}.
\end{equation}
In the context of the interaction picture, we refer to $\hat{H}_{0}$
as the base Hamiltonian. We may furthermore define the interaction
picture state by
\begin{equation}
|\Psi(t)\rangle_{I}=e^{i\hat{H}_{0}t/\hbar}|\Psi(t)\rangle.
\end{equation}
Given that the state $|\Psi(t)\rangle$ is governed by the Schrödinger
equation (\ref{eq:schrodinger-equation}), the equation of motion
for $|\Psi(t)\rangle_{I}$ will then be
\begin{equation}
i\hbar\frac{\partial}{\partial t}|\Psi(t)\rangle_{I}=\hat{V}_{I}|\Psi(t)\rangle_{I}.
\end{equation}
We see that the operation of transforming to the interaction picture
removes $\hat{H}_{0}$ from the dynamics. In quantum optics, the interaction
picture with $\hbar\omega(\hat{n}+\frac{1}{2})$ as the base Hamiltonian
as referred to as the rotating frame.

For demonstrative purposes, this section denotes quantities in the
interaction picture with a subscript $I$. In the following text,
we shall omit the subscript and let the context determine whether
a given quantity is consider in the interaction picture.

\section{Expectation Values and Uncertainties\label{sec:Expectation-Values-and}}

Measurement is a central concept in quantum mechanics. Here, we use
the special case of projective measurements to introduce the statistical
properties of a quantum system. Measurable quantities are called observables.
Mathematically, an observable is represented by a corresponding Hermitian
operator. It is a central axiom in quantum mechanics that the measurement
of an observable $\hat{A}$ will always yield a result that is an
eigenvalue $A_{i}$ of the corresponding operator. The Hermiticity
of $\hat{A}$ ensures that $A_{i}$ is real. After the measurement,
the system is found in the matching eigenstate $|A_{i}\rangle$ of
the operator.\footnote{For the purposes of this thesis, it is amply sufficient to assume
that the eigenstates of the observable are nondegenerate, i.e. that
all the eigenstates have distinct eigenvalues. For a broader understanding
including degeneracy, one may consult material on quantum measurement
\cite{Wiseman_QuantumMeasurementControl_2010}.} In addition, we take it as an axiom that the eigenstates $|A_{i}\rangle$
of any operator form a basis (though not necessarily orthonormal)
in which any state of the system may be expressed exactly.

If the system state before measurement $|\Psi\rangle$ is not an eigenstate
of the observable operator, the system is said to collapse to one
of the eigenstates. The probability $P(A_{i})$ to measure the outcome
$A_{i}$ is given by
\begin{equation}
P(A_{i})=\left|\langle A_{i}|\Psi\rangle\right|^{2}.
\end{equation}
With this, we can determine the expectation value of the operator
$\hat{A}$. This is written
\begin{equation}
\langle\hat{A}\rangle=\sum_{i}P(A_{i})A_{i}.\label{eq:-184}
\end{equation}
For a system in state $|\Psi\rangle$, the expectation value of $\langle\hat{A}\rangle$
can be calculated as
\begin{equation}
\langle\hat{A}\rangle=\langle\Psi|\hat{A}|\Psi\rangle.
\end{equation}
Since the outcome of a measurement is a stochastic quantity, we can
also determine the variance. This is written\footnote{The notation implies that $(\Delta\hat{A})^{2}$ may be viewed as
an operator. One should apply this view cautiously since it implies
that $\Delta\hat{A}$ depends on the particular state under consideration.
Setting e.g. $(\Delta\hat{A})^{2}=(\hat{A}-\langle\Psi|\hat{A}|\Psi\rangle)^{2}$
is consistent with (\ref{eq:-183}). The operator $\hat{A}-\langle\Psi|\hat{A}|\Psi\rangle$
is known as the dispersion of $\hat{A}$ \cite{Sakurai_ModernQuantumMechanics_2011a}.}
\begin{equation}
\left\langle (\Delta\hat{A})^{2}\right\rangle =\left\langle \hat{A}{}^{2}\right\rangle -\langle\hat{A}\rangle^{2}.\label{eq:-183}
\end{equation}
The derived quantity $\sigma_{\hat{A}}=\sqrt{\left\langle (\Delta\hat{A})^{2}\right\rangle }$
is referred to as the uncertainty of the observable. 

One may show that the variances of two observables $\hat{A}$ and
$\hat{B}$ obey the inequality \cite{Sakurai_ModernQuantumMechanics_2011a,Gerry_IntroductoryQuantumOptics_2004}
\begin{equation}
\left\langle (\Delta\hat{A})^{2}\right\rangle \left\langle (\Delta\hat{B})^{2}\right\rangle \geq\frac{1}{4}\left|\langle[\hat{A},\hat{B}]\rangle\right|^{2},\label{eq:-185}
\end{equation}
known as Heisenberg's uncertainty principle. It plays a central role
in quantum mechanics. In particular, it states that the product of
the variances for two non-commuting observables is nonzero. Two such
observables are called incompatible since the system cannot be in
a state where both observables have a definite value. (\ref{eq:-185})
can be used to derive a series of uncertainty relations between incompatible
observables. An important relation of this type is the one between
the two orthogonal quadrature operators $\hat{X}$ and $\hat{Y}$.
Using the commutator (\ref{eq:quadrature-commutation}) we obtain
\begin{equation}
\left\langle (\Delta\hat{X})^{2}\right\rangle \left\langle (\Delta\hat{Y})^{2}\right\rangle \geq\frac{1}{16}.\label{eq:heisenberg-limit}
\end{equation}

\section{Mixed States\label{sec:Mixed-States}}

To increase the relevance of the systems considered, we wish to eventually
allow for coupling to an environment. In preparation for the analysis
of open systems, we introduce in this section density matrices to
represent quantum states.

The density matrix $\hat{\rho}$ of a pure state $|\Psi(t)\rangle$
is formed by taking the outer product between the state ket and bra:
\begin{equation}
\hat{\rho}=|\Psi(t)\rangle\langle\Psi(t)|.\label{eq:-140}
\end{equation}
From the Schrödinger equation (\ref{eq:schrodinger-equation}), it
is now easy to derive an equation which governs the evolution of $\hat{\rho}$
\cite{Schwabl_StatisticalMechanics_2006}
\begin{equation}
\frac{\partial}{\partial t}\hat{\rho}(t)=-\frac{i}{\hbar}\left[\hat{H},\hat{\rho}(t)\right].\label{eq:von-neumann-equation}
\end{equation}
(\ref{eq:von-neumann-equation}) is usually referred to either as
the von Neumann equation. The time-evolution of a state $\hat{\rho}(0)$
explicitly is written using (\ref{eq:unitary-propagator}) as

\begin{equation}
\hat{\rho}(t)=\hat{U}(t)\hat{\rho}(0)\hat{U}^{\dagger}(t).\label{eq:propagated-dm}
\end{equation}
So far, the density matrix simply allows for a description of quantum
systems. Unlike the representations described in Sections \ref{sec:the-quantum-harmonic-oscillator}--\ref{sec:Expectation-Values-and}
however, a density matrix can be constructed which describes an ensemble
of states. 

An important example of a mixed state is a state in thermal equilibrium
with an environment with a finite temperature $T$. The temperature
of the environment determines for each state of a particular energy,
the probability of finding the system in that state. For the harmonic
oscillator, the states of definite energy are the number states $|n\rangle$
(see (\ref{eq:-169})). The probability $p(n)$ of finding the system
in the state $|n\rangle$ is written
\begin{equation}
p(n)=\exp\left(-\frac{\hbar\omega n}{k_{B}T}\right)\left[1-\exp\left(-\frac{\hbar\omega}{k_{B}T}\right)\right]^{-1},\label{eq:-170}
\end{equation}
where $\omega$ is the oscillator frequency and $k_{B}$ the Boltzmann
constant. We construct the density matrix for the thermal state, $\hat{\rho}_{\mathrm{th}}$,
as a linear combination of the number state density matrices $|n\rangle\langle n|$,
each weighted with the matching probability $p(n)$:

\begin{equation}
\hat{\rho}_{\mathrm{th}}=\sum_{n}p(n)|n\rangle\langle n|.\label{eq:-141}
\end{equation}
It is seen that (\ref{eq:-141}) satisfies the requirement that $\langle n|\hat{\rho}_{\mathrm{th}}|n\rangle=p(n)$.
We require for an operator to be a valid density matrix that it is
normalized such that 
\begin{equation}
\Tr\hat{\rho}_{\mathrm{th}}=\sum_{n}p(n)=1.
\end{equation}
$\Tr$ denotes the trace operation which sums the diagonal elements.
It may be carried out in an arbitrary orthonormal basis $|A_{i}\rangle$:
\begin{equation}
\Tr\hat{\rho}=\sum_{i}\langle A_{i}|\hat{\rho}|A_{i}\rangle.
\end{equation}
The mean occupancy of the thermal state is given by
\begin{equation}
\langle\hat{n}\rangle=\frac{1}{e^{\hbar\omega/k_{B}T}+1}.\label{eq:-192}
\end{equation}
Under unitary system dynamics, mixed state density matrices evolve
by (\ref{eq:von-neumann-equation}) and (\ref{eq:propagated-dm})
like pure states. Expectation values of operators with respect to
states described by density matrices are also computed with the trace.
The expectation value of the operator $\hat{A}$ with respect to the
state described by $\hat{\rho}$ is written
\begin{equation}
\langle\hat{A}\rangle=\Tr\left[\hat{\rho}\hat{A}\right].
\end{equation}

Before we end the current section, we note that one can write the
equation of motion for $\hat{\rho}$ in analogy with (\ref{eq:schrodinger-equation}).
To do this, we introduce the superoperator $\mathcal{C}[\hat{O}]$
describing the commutator with the operator $\hat{O}$:
\begin{equation}
\mathcal{C}[\hat{O}]\hat{\rho}=[\hat{O},\hat{\rho}].\label{eq:-212}
\end{equation}
(\ref{eq:von-neumann-equation}) can thus be written 
\begin{equation}
\dot{\hat{\rho}}(t)=-\frac{i}{\hbar}\mathcal{C}[\hat{H}]\hat{\rho}(0).
\end{equation}
With $\hat{H}$ constant in time, we can formally write the explicit
solution as 

\begin{equation}
\hat{\rho}(t)=e^{(-it/\hbar)\mathcal{C}[\hat{H}]}\hat{\rho}(0).
\end{equation}
Additional superoperators will be introduced in the next section.

\section{Open System Dynamics \label{subsec:master-equation}}

So far, the dynamics of the considered quantum systems have been assumed
unitary. This assumption implies that the systems are isolated and
none of the system degrees of freedom interact with their surroundings.
In practice, no physical system is completely isolated from its environment. 

Coupling the quantum system to an environment requires an extension
of the unitary dynamics of Section \ref{sec:unitary-dynamics}. The
evolution of the coupled system is governed by the master equation
\cite{Walls_QuantumOptics_2008}
\begin{equation}
\dot{\hat{\rho}}=-\frac{i}{\hbar}[\hat{H},\hat{\rho}]+\gamma\left(\bar{n}+1\right)\mathcal{D}[\hat{a}]\hat{\rho}+\gamma\bar{n}\mathcal{D}[\hat{a}^{\dagger}]\hat{\rho}+\gamma_{\phi}\mathcal{D}[\hat{n}]\hat{\rho}.\label{eq:general-master-equation}
\end{equation}
The Lindblad superoperator $\mathcal{D}[\hat{O}]$ for an operator
$\hat{O}$ acts on the density matrix as
\begin{equation}
\mathcal{D}[\hat{O}]\hat{\rho}=\hat{O}\hat{\rho}\hat{O}^{\dagger}-\tfrac{1}{2}\hat{O}^{\dagger}\hat{O}\hat{\rho}-\tfrac{1}{2}\hat{\rho}\hat{O}^{\dagger}\hat{O}.\label{eq:lindblad-superoperator}
\end{equation}

The first term of (\ref{eq:general-master-equation}) is inherited
from the von Neumann equation (\ref{eq:von-neumann-equation}). We
refer to this term as the unitary part of the system dynamics.

The terms proportional to $\gamma$ in (\ref{eq:general-master-equation})
produce a damping effect and the frequency $\gamma$ is therefore
called the damping coefficient or damping rate. Physically, (\ref{eq:damped-me})
can be used to describe coupling of the system to a thermal bath of
bosonic oscillators with a temperature $T$. We can relate $\bar{n}$
to the temperature of the bath by considering the steady state solution.
For the harmonic oscillator, this is assumed to correspond to the
thermal equilibrium between system and bath. In thermal equilibrium,
the temperature $T$ is shared between system and bath. The mean occupancy
the harmonic oscillator is then that of the thermal state (\ref{eq:-141}).
For the harmonic oscillator, this is given by (\ref{eq:-192}). The
damping terms of (\ref{eq:general-master-equation}) can be derived
for the harmonic oscillator by considering the combined unitary evolution
the system and the thermal equilibrium bath. Tracing out the bath
degrees of freedom, then making the Markovian assumption and the rotating
wave approximation exactly yields the damping terms \cite{Orszag_QuantumOpticsIncluding_2016,Wiseman_QuantumMeasurementControl_2010}.
In making the rotating wave approximation, it is assumed that the
frequency $\omega$ of the system oscillator is large compared to
the frequency describing the interaction between system and bath.

The term $\gamma_{\phi}\mathcal{D}[\hat{n}]\hat{\rho}$ introduces
a dephasing effect \cite{Wilson_MeasurementbasedControlMechanical_2015}.
In the most narrow sense \cite{Hornberger_IntroductionDecoherenceTheory_2009},
the effect of dephasing is the decay of the off-diagonal elements
of $\hat{\rho}$ expressed in the energy eigenbasis of the system.
These elements are called coherences. We assume that the energy levels
of any considered system are approximately harmonic and thus take
this basis to be the number state basis. The term $\gamma_{\phi}\mathcal{D}[\hat{n}]\hat{\rho}$
is seen to cause decay of the coherences by considering the dephasing-only
equation of motion for the $mn$-matrix element of $\hat{\rho}$.
The right hand side is given by
\begin{equation}
\langle m|\gamma_{\phi}\mathcal{D}[\hat{n}]\hat{\rho}|n\rangle=-\frac{\gamma_{\phi}}{2}(m-n)^{2}\langle m|\hat{\rho}|n\rangle.\label{eq:dephasing-matrix-element}
\end{equation}
It is seen that the dephasing term leads to exponential decay of the
off-diagonal matrix-elements ($n\neq m$) while leaving the diagonal
matrix elements ($n=m$) constant.

The effect of the non-unitary terms are collectively known as decoherence. 

\section{Intermission: Quadrature Squeezing \label{sec:quadrature-squeezing}}

We give now a simple example to better explain the concepts introduced
so far. Thus, consider the following Hamiltonian
\begin{equation}
\hat{H}_{\eta}=i\hbar\left(\eta^{*}\hat{a}^{2}-\eta\left(\hat{a}^{\dagger}\right)^{2}\right),\label{eq:-121}
\end{equation}
yielding the von Neumann equation
\begin{equation}
\dot{\hat{\rho}}=\left[\eta^{*}\hat{a}\hat{a}-\eta\hat{a}^{\dagger}\hat{a}^{\dagger},\hat{\rho}\right].\label{eq:-99}
\end{equation}
The Hamiltonian $\hat{H}_{\eta}$ may be obtained as the effective
Hamiltonian for a parametric amplifier with a strong coherent drive
\cite{Walls_QuantumOptics_2008}. With (\ref{eq:-121}), we see that
the unitary propagator (\ref{eq:unitary-propagator}) can now be written
\begin{equation}
U(t)=e^{\left(\eta^{*}\hat{a}^{2}-\eta\left(\hat{a}^{\dagger}\right)^{2}\right)t}=S(\xi{=}2\eta t).
\end{equation}

To keep the example simple, we consider as initial state the vacuum
state $|0\rangle$. Under the time-evolution of the Hamiltonian (\ref{eq:-121}),
$|0\rangle$ evolves into the squeezed vacuum state
\begin{subequations}
\label{eq:-122}
\begin{equation}
|\xi\rangle=\hat{S}(\xi)|0\rangle\label{eq:-124}
\end{equation}
with a time-dependent squeezing parameter 
\begin{equation}
\xi=2\eta t.
\end{equation}
\end{subequations}
(\ref{eq:-122}) can be regarded as the definition of a squeezed vacuum
state.

In the coming sections, we will introduce the phase space formalism
which provides a more intuitive understanding of the equations above.
Until then though, we can probe the effects of $\hat{H}_{\eta}$ by
considering the time-dependence of the quadrature expectation values
and variances of the state (\ref{eq:-124}). We do this by expressing
the result of applying $\hat{S}$ to the quadrature operators. For
simplicity, we choose $\eta$ to be a real number and write $\xi=r=2\eta t=2\eta^{*}t$.
Combining (\ref{eq:quadrature-operators}) and (\ref{eq:-70}), we
have
\begin{subequations}
\label{eq:-123}
\begin{align}
\hat{S}^{\dagger}(r)\hat{X}\hat{S}(r) & =\hat{X}e^{-r}\,,\\
\hat{S}^{\dagger}(r)\hat{Y}\hat{S}(r) & =\hat{Y}e^{r}.
\end{align}
\end{subequations}

Applying this, we may determine quadrature expectation values as
\begin{subequations}
\label{eq:-118-1}
\begin{align}
\left\langle \hat{X}\right\rangle  & =\langle0|\hat{S}^{\dagger}(r)\hat{X}\hat{S}(r)|0\rangle=0\label{eq:-9-1}\\
\intertext{and}\left\langle \hat{Y}\right\rangle  & =\langle0|\hat{S}^{\dagger}(r)\hat{Y}\hat{S}(r)|0\rangle=0,\label{eq:-10-1}
\end{align}
\end{subequations}
 as well as the quadrature variance as
\begin{subequations}
\label{eq:-118}
\begin{align}
\left\langle \left(\Delta\hat{X}\right)^{2}\right\rangle  & =\langle0|\hat{S}^{\dagger}(r)\hat{X}^{2}\hat{S}(r)|0\rangle=\frac{1}{4}e^{-2r},\label{eq:-9}\\
\intertext{and}\left\langle \left(\Delta\hat{Y}\right)^{2}\right\rangle  & =\langle0|\hat{S}^{\dagger}(r)\hat{Y}^{2}\hat{S}(r)|0\rangle=\frac{1}{4}e^{2r}.\label{eq:-10}
\end{align}
\end{subequations}
In this case, the variance of the quadrature $\hat{X}$ is reduced
or squeezed. The orthogonal quadrature $\hat{Y}$ is called the anti-squeezed
quadrature. The effect of $\hat{S}(r)$ is therefore also referred
to as quadrature squeezing. We also note that even though the variance
of $\hat{X}$ may be made arbitrarily small by choosing a sufficiently
large $t$,\footnote{Of course, this purely a theoretical consideration. Experimental factors
limit the amount of squeezing that can realistically be achived.} the fundamental Heisenberg limit (\ref{eq:heisenberg-limit}) is
still obeyed due to the matching increase in the variance of $\hat{Y}$.
In the case of vanishing squeezing, the quadrature variances are both
$1/4$.

\section{The Wigner Quasiprobability Distribution\label{sec:Phase-Space-Representations}}

In the study of classical systems, the concept of a phase space is
ubiquitous in modern physics. It provides a concise and abstracted
view of system dynamics and allows for the straightforward description
of a stochastic system whose state is described by a probability distribution.
With the introduction of the Wigner function \cite{Wigner_QuantumCorrectionThermodynamic_1932},
an analogous concept was made available in the study of quantum systems.
The Wigner function shares many properties with a probability distribution
in classical phase space, however it can not be viewed as such since
it can assume negative values. For this reason it is known as a quasiprobability
distribution.

To introduce the Wigner function, we define first the symmetrically
ordered characteristic function $\chi(\lambda,\lambda^{*})$ of a
state $\hat{\rho}$ by \cite{Gerry_IntroductoryQuantumOptics_2004}

\begin{equation}
\chi(\lambda,\lambda^{*})=\Tr[\hat{\rho}e^{\lambda\hat{a}^{\dagger}-\lambda^{*}\hat{a}}]=\Tr[\hat{\rho}\hat{D}(\lambda)]=\langle\hat{D}(\lambda)\rangle\label{eq:wigner-characteristic-function}
\end{equation}
where $\hat{D}(\lambda)$ is given by (\ref{eq:displacement-operator}).
We then define the Wigner function\footnote{The Wigner function is an instance of a broader family of phase space
qausiprobability distributions \cite{Gerry_IntroductoryQuantumOptics_2004}.
Other important distributions are the Husimi Q function and the Glauber-Sudarshan
P function. Except for a brief mention of the Q function in Chapter
\ref{chap:coupling-to-the-environment}, we will only consider the
Wigner function.} as the complex Fourier transform of $\chi(\lambda,\lambda^{*})$:\footnote{As exemplified by (\ref{eq:wigner-function}), we shall generally
omit the limits of integrals when they may be derived from the context.
For instance, integrals over the complex phase space coordinates should
be taken over the entire phase space. In case of $\lambda$ and $\lambda^{*}$,
this is the set $\mathbb{C}$.}

\begin{equation}
W(\alpha,\alpha^{*})=\frac{1}{\pi^{2}}\int d\lambda\,d\lambda^{*}\,e^{\lambda^{*}\alpha-\alpha^{*}\lambda}\chi(\lambda,\lambda^{*}).\label{eq:wigner-function}
\end{equation}

The characteristic function $\chi(\lambda,\lambda^{*})$ can be recovered
from a Wigner function by means of the inverse Fourier transform:
\begin{equation}
\chi(\lambda,\lambda^{*})=\int d\alpha\,d\alpha^{*}\,e^{\lambda^{*}\alpha-\alpha^{*}\lambda}W(\alpha,\alpha^{*}).\label{eq:-166}
\end{equation}
There exists a one-to-one mapping between Wigner functions and density
matrices \cite{Stobinska_WignerFunctionEvolution_2008}. The inverse
of (\ref{eq:wigner-characteristic-function}) is given by \cite{Cahill_DensityOperatorsQuasiprobability_1969a}
\begin{equation}
\hat{\rho}=\int d^{2}\lambda\,\chi(\lambda)\hat{D}^{\dagger}(\lambda).\label{eq:-1}
\end{equation}
In combination, (\ref{eq:-166}) and (\ref{eq:-1}) allows one to
express the density matrix given a Wigner function. Using the Wigner
function expressed in terms of $\alpha$ also allows one to express
an expectation value of an operator expression $f(\hat{a},\hat{a}^{\dagger})$
written symmetrically or Weyl ordered \cite{Scully_QuantumOptics_1997}
in terms of creation and annihilation operators as \cite{Gerry_IntroductoryQuantumOptics_2004}
\begin{equation}
\left\langle f(\hat{a},\hat{a}^{\dagger})\right\rangle =\int d\alpha\,d\alpha^{*}\,W(\alpha,\alpha^{*})f(\alpha,\alpha^{*}).
\end{equation}

As an example, we derive the Wigner function for the vacuum state
$|0\rangle$. Write the vacuum state Wigner function $W_{|0\rangle}$
as
\begin{equation}
W_{|0\rangle}(\alpha,\alpha^{*})=\frac{1}{\pi^{2}}\int d\lambda\,d\lambda^{*}\,e^{\lambda^{*}\alpha-\lambda\alpha^{*}}\langle0|e^{\lambda\hat{a}^{\dagger}-\lambda^{*}\hat{a}}|0\rangle.
\end{equation}
Application of the disentangling theorem (\ref{eq:disentangling-theorem})
yields
\begin{equation}
\langle0|e^{\lambda\hat{a}^{\dagger}-\lambda^{*}\hat{a}}|0\rangle=e^{-\lambda\lambda^{*}/2}\langle0|e^{\lambda\hat{a}^{\dagger}}e^{-\lambda^{*}\hat{a}}|0\rangle=e^{-|\lambda|^{2}/2}.
\end{equation}
This is Fourier transformed to find $W_{|0\rangle}$ as 
\begin{equation}
W_{|0\rangle}(\alpha,\alpha^{*})=\frac{1}{\pi^{2}}\int d\lambda\,d\lambda^{*}\,e^{\lambda^{*}\alpha-\lambda\alpha^{*}}e^{-|\lambda|^{2}/2}=\frac{2}{\pi}e^{-2|\alpha|^{2}},\label{eq:vacuum-state-wigner-function}
\end{equation}
showing that the Wigner function of the vacuum state is simply a normalized
Gaussian function centered at the origin and with variance $1/4$
in both quadrature coordinates.

\section{Transformations in Phase Space\label{sec:Transformations-in-Phase}}

We can exercise the formalism introduced in the previous section by
computing the effect on the Wigner function for the various transformations
from Section \ref{sec:operator-transformations}. The resulting identities
will also be useful later to compute the Wigner function for a state
constructed from application of $\hat{D}$, $\hat{S}$ and $\hat{R}$
to some base state with a known Wigner function.

The argument goes in general as follows: For the transformation $\hat{U}(t)$
(which could be $\hat{D}(\alpha_{0})$, $\hat{S}(\xi_{0})$, or $\hat{R}(\phi_{0})$)
with $t$ as a generic parameter, the Wigner function of the transformed
state $\hat{U}(t)\hat{\rho}\hat{U}^{\dagger}(t)$ is written using
(\ref{eq:wigner-function}). This yields an expression containing
a trace of the transformed state with a displacement operator. Using
the cyclic property of the trace, the operators are rearranged to
transform the displacement operator instead:
\begin{equation}
\Tr\left[\hat{U}(t)\hat{\rho}\hat{U}^{\dagger}(t)\hat{D}(\lambda)\right]=\Tr\left[\hat{\rho}\hat{U}^{\dagger}(t)\hat{D}(\lambda)\hat{U}(t)\right].
\end{equation}
Expanding the displacement operator using its Taylor series (\ref{eq:-200}),
\begin{subequations}
\label{eq:-201}
\begin{align}
\hat{U}^{\dagger}(t)\hat{D}(\lambda)\hat{U}(t) & =\sum_{n}\frac{1}{n!}\hat{U}^{\dagger}(t)(-\lambda^{*}\hat{a}+\lambda\hat{a}^{\dagger})^{n}\hat{U}(t)\\
 & =\sum_{n}\frac{1}{n!}\left(-\lambda^{*}\hat{U}^{\dagger}(t)\hat{a}\hat{U}(t)+\lambda\hat{U}^{\dagger}(t)\hat{a}^{\dagger}\hat{U}(t)\right)^{n}\\
 & =e^{-\lambda^{*}\hat{U}^{\dagger}(t)\hat{a}\hat{U}(t)+\lambda\hat{U}^{\dagger}(t)\hat{a}^{\dagger}\hat{U}(t)}
\end{align}
\end{subequations}
allows for the application of the appropriate identities from Section
\ref{sec:operator-transformations} to $\hat{U}^{\dagger}(t)\hat{a}\hat{U}(t)$
and its Hermitian conjugate. The resulting expression is then rewritten
to one containing manifestly the Wigner function before transformation
which then reveals the transformation's effect on $W$.

We first consider the displacement operator $\hat{D}(\alpha_{0})$
from (\ref{eq:displacement-operator}). We have (making the corresponding
density matrix explicit as a subscript of the Wigner function)
\begin{subequations}
\label{eq:-114}
\begin{align}
W_{\hat{D}(\alpha_{0})\hat{\rho}\hat{D}^{\dagger}(\alpha_{0})}(\alpha,\alpha^{*}) & =\frac{1}{\pi^{2}}\int d\lambda\,d\lambda^{*}\,e^{\alpha\lambda^{*}-\alpha^{*}\lambda}\Tr\left[\hat{D}(\alpha_{0})\hat{\rho}\hat{D}^{\dagger}(\alpha_{0})\hat{D}(\lambda)\right]\\
 & =\frac{1}{\pi^{2}}\int d\lambda\,d\lambda^{*}\,e^{\alpha\lambda^{*}-\alpha^{*}\lambda}\Tr\left[\hat{\rho}e^{-\lambda^{*}\hat{a}-\lambda^{*}\alpha_{0}+\lambda\hat{a}^{\dagger}+\lambda\alpha_{0}^{*}}\right]\\
 & =W_{\hat{\rho}}(\alpha-\alpha_{0},\alpha^{*}-\alpha_{0}^{*}).\label{eq:-143}
\end{align}
\end{subequations}
Hence applying the displacement operator $\hat{D}(\alpha_{0})$ to
a state simply causes its Wigner function to move rigidly in phase
space by a distance corresponding to the displacement parameter $\alpha_{0}$.

Next, consider rotation by an angle $\phi_{0}$ represented by the
operator $\hat{R}(\phi_{0})$ of (\ref{eq:-115}). Repeating (\ref{eq:-114}),
we insert the transformed state $\hat{R}(\phi_{0})\hat{\rho}\hat{R}^{\dagger}(\phi_{0})$
into (\ref{eq:wigner-function}) and use (\ref{eq:-115}) to find
\begin{subequations}
\label{eq:-114-1}
\begin{align}
W_{\hat{R}(\phi_{0})\hat{\rho}\hat{R}^{\dagger}(\phi_{0})}(\alpha,\alpha^{*}) & =\frac{1}{\pi^{2}}\int d\lambda\,d\lambda^{*}\,e^{\alpha\lambda^{*}-\alpha^{*}\lambda}\Tr\left[\hat{R}(\phi_{0})\hat{\rho}\hat{R}^{\dagger}(\phi_{0})\hat{D}(\lambda)\right]\\
 & =\frac{1}{\pi^{2}}\int d\lambda\,d\lambda^{*}\,e^{\alpha\lambda^{*}-\alpha^{*}\lambda}\Tr\left[\hat{\rho}e^{-\lambda^{*}\exp(-i\phi_{0})\hat{a}+\lambda\exp(i\phi_{0})\hat{a}^{\dagger}}\right]\\
 & \begin{aligned}[t]=\frac{1}{\pi^{2}}\int d\left(\lambda e^{i\phi_{0}}\right)d\left(\lambda e^{i\phi_{0}}\right)^{*}\, & \exp\left(\alpha e^{i\phi_{0}}\left(\lambda^{*}e^{i\phi_{0}}\right)^{*}-\alpha^{*}e^{-i\phi_{0}}\left(\lambda e^{i\phi_{0}}\right)\right)\\
 & \Tr\left[\hat{\rho}e^{-\lambda^{*}\exp(-i\phi_{0})\hat{a}+\lambda\exp(i\phi_{0})\hat{a}^{\dagger}}\right]
\end{aligned}
\\
 & =W_{\hat{\rho}}(\alpha e^{i\phi_{0}},\alpha^{*}e^{-i\phi_{0}}).
\end{align}
\end{subequations}
We find that the rotation operator applied to a state causes causes
the complex argument of the Wigner function to pick up a corresponding
complex phase. Geometrically, this is simply rigid rotation around
the origin.

Finally, we may introduce squeezing through a coordinate transformation
in the expression for the Wigner function. The transformation of the
displacement operator is found using (\ref{eq:-70}). For the squeezing
parameter $\xi_{0}=r_{0}e^{i\theta_{0}}$, the argument goes as
\begin{subequations}
\label{eq:-114-1-1}
\begin{align}
W_{\hat{S}(\xi_{0})\hat{\rho}\hat{S}^{\dagger}(\xi_{0})}(\alpha,\alpha^{*}) & =\frac{1}{\pi^{2}}\int d\lambda\,d\lambda^{*}\,e^{\alpha\lambda^{*}-\alpha^{*}\lambda}\Tr\left[\hat{S}(\xi_{0})\hat{\rho}\hat{S}^{\dagger}(\xi_{0})\hat{D}(\lambda)\right]\\
 & \begin{aligned}[t]=\frac{1}{\pi^{2}}\int & d\lambda\,d\lambda^{*}\,e^{\alpha\lambda^{*}-\alpha^{*}\lambda}\\
 & \cdot\Tr\left[\hat{\rho}e^{\left(\lambda\cosh r_{0}-\lambda^{*}e^{i\theta_{0}}\sinh r_{0}\right)\hat{a}^{\dagger}-\left(\lambda^{*}\cosh r_{0}-\lambda e^{-i\theta_{0}}\sinh r_{0}\right)\hat{a}}\right]
\end{aligned}
\label{eq:-20}\\
 & \begin{aligned}[t]=\frac{1}{\pi^{2}}\int & d\mu\,d\mu^{*}\,e^{\left(\alpha\cosh r_{0}+\alpha^{*}e^{i\theta_{0}}\sinh r_{0}\right)\mu^{*}-\left(\alpha^{*}\cosh r_{0}+\alpha e^{-i\theta_{0}}\sinh r_{0}\right)\mu}\\
 & \Tr\left[\hat{\rho}\hat{D}(\mu)\right]
\end{aligned}
\label{eq:-119}\\
 & =W_{\hat{\rho}}(\alpha\cosh r_{0}+\alpha^{*}e^{i\theta_{0}}\sinh r_{0},\alpha^{*}\cosh r_{0}+\alpha e^{-i\theta_{0}}\sinh r_{0}).
\end{align}
\end{subequations}
Between (\ref{eq:-20}) and (\ref{eq:-119}), the integral was rewritten
in terms of the new coordinates $(\mu,\mu^{*})=(\lambda\cosh r-\lambda^{*}e^{i\theta}\sinh r,\lambda^{*}\cosh r-\lambda e^{-i\theta}\sinh r)$.

Equations (\ref{eq:-114}), (\ref{eq:-114-1}) and (\ref{eq:-114-1-1})
show that the effect of the operators $\hat{D}(\alpha_{0})$, $\hat{S}(\xi_{0})$,
or $\hat{R}(\phi_{0})$ can all be expressed simply as coordinate
transforms of $W$.

\section{Phase Space Coordinates \label{sec:phase-space-coordinates}}

\begin{figure}
\noindent \begin{centering}
\makebox[0pt][c]{\mbox{\includegraphics{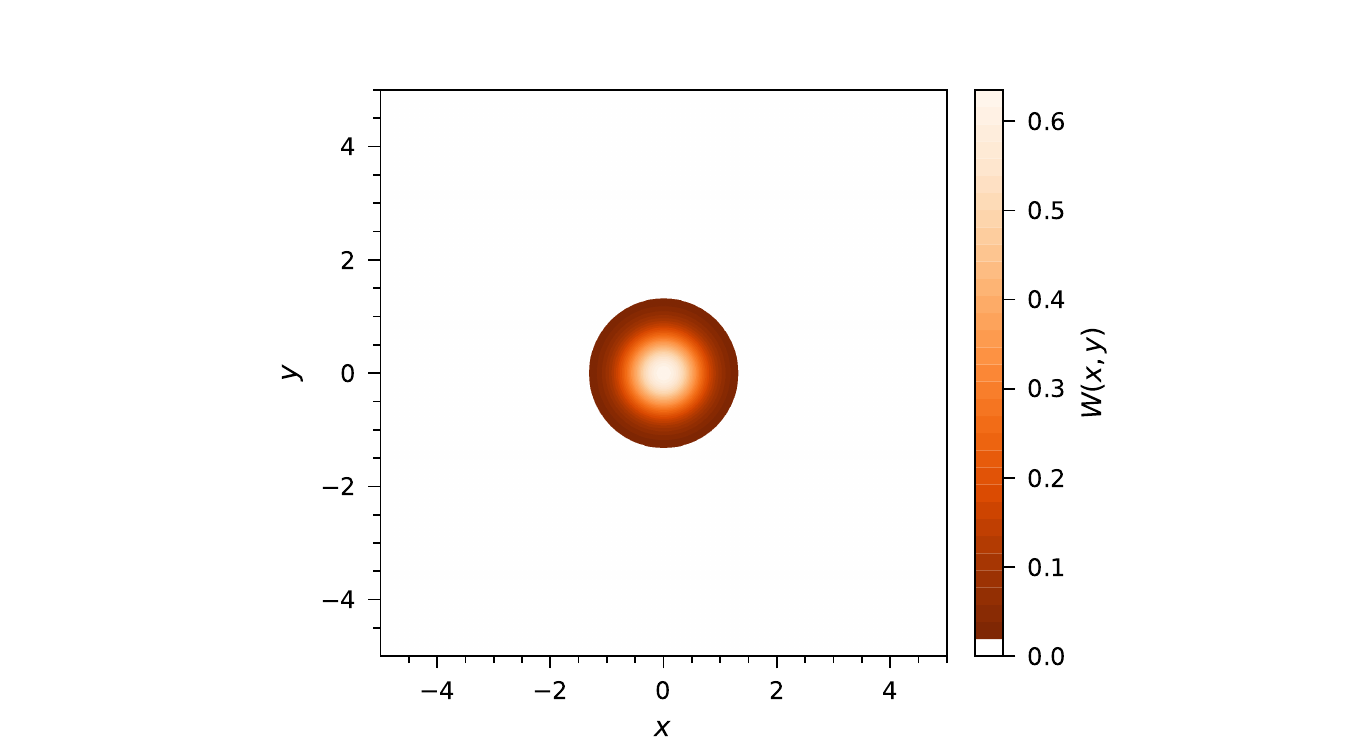}\hskip1em}}
\par\end{centering}
\caption[The Wigner function of the vacuum state]{\label{fig:vacuum-state-plot}\textbf{The Wigner function of the
vacuum state.} The Wigner function for the vacuum state $|0\rangle$
is a Gaussian function centered at the origin and with an isotropic
variance of $1/4$ in the coordinates $(x,y)$. The expression is
shown in (\ref{eq:vacuum-wigner-carteesian}). Note that the used
color map displays any value of $W$ below a certain threshold as
white. }
\end{figure}

We will find it useful to express the Wigner function in various coordinate
systems. This section establishes the different coordinate systems.
Until now, we have considered the Wigner function solely in terms
of the coherent amplitude $\alpha$. For instance, the Wigner function
for the vacuum state $|0\rangle$ was found in (\ref{eq:vacuum-state-wigner-function})
to be
\begin{equation}
W_{|0\rangle}(\alpha,\alpha^{*})=\frac{2}{\pi}e^{-2|\alpha|^{2}}.\label{eq:vacuum-wigner-complex}
\end{equation}
The Wigner function of the vacuum state shall serve to demonstrate
the normalization conventions in this section.

We now define the Cartesian or quadrature coordinates $(x,y)$ by
\begin{equation}
\alpha=x+iy\label{eq:-17}
\end{equation}
with real numbers $x$ and $y$. The relation between $\alpha$ and
$(x,y)$ mirrors that between the annihilation operator $\hat{a}$
and the quadrature operators $(\hat{X},\hat{Y})$ (although definitions
without this property could have been chosen); compare (\ref{eq:quadrature-operators})
with the inverse relations of (\ref{eq:-17})
\begin{subequations}
\label{eq:-15}
\begin{align}
x & =\Re\alpha=\frac{1}{2}(\alpha+\alpha^{*}),\label{eq:-242}\\
y & =\Im\alpha=\frac{1}{2i}(\alpha-\alpha^{*}).\label{eq:-243}
\end{align}
\end{subequations}
With the definition (\ref{eq:-17}), the vacuum state is expressed
in Cartesian coordinates as
\begin{equation}
W_{|0\rangle}(x,y)=\frac{2}{\pi}e^{-2x^{2}-2y^{2}}.\label{eq:vacuum-wigner-carteesian}
\end{equation}
Writing the Wigner function as a function of two reals $x$ and $y$,
we can visualize it using a density plot. This is done for (\ref{eq:vacuum-wigner-carteesian})
in Figure \ref{fig:vacuum-state-plot}. To demonstrate the usefulness
of Cartesian coordinates, apply the squeezing transformation $\hat{S}(\xi_{0})$
with $\xi_{0}=r_{0}e^{i\theta_{0}}$ to the Wigner function of the
vacuum state. This forms the Wigner function for the squeezed vacuum
state (\ref{eq:-124})
\begin{equation}
W_{\hat{S}(\xi_{0})|0\rangle}(x,y)=\frac{2}{\pi}\exp\left[-2e^{r_{0}}\left(x\cos\frac{\theta_{0}}{2}+y\sin\frac{\theta_{0}}{2}\right)^{2}-2e^{-r_{0}}\left(x\sin\frac{\theta_{0}}{2}-y\cos\frac{\theta_{0}}{2}\right)^{2}\right]\label{eq:-190}
\end{equation}

Finally, polar coordinates are defined by the relation
\begin{equation}
\alpha=re^{i\phi}\label{eq:-16}
\end{equation}
with real numbers $r$ and $\phi$. In polar coordinates, the vacuum
state assumes the form
\begin{equation}
W_{|0\rangle}(r,\phi)=\frac{2}{\pi}e^{-2r^{2}}.\label{eq:vacuum-wigner-polar}
\end{equation}
(\ref{eq:-17}) and (\ref{eq:-16}) are both each real-valued functions
of two real arguments. Choosing these coordinate conventions means
that the normalization factor is the same between (\ref{eq:vacuum-wigner-complex}),
(\ref{eq:vacuum-wigner-carteesian}) and (\ref{eq:vacuum-wigner-polar}),
e.g.
\begin{subequations}
\label{eq:-167}
\begin{align}
1 & =\int d\alpha\,d\alpha^{*}\,\frac{2}{\pi}e^{-2|\alpha|^{2}}\\
 & =\int dx\,dy\,\frac{2}{\pi}e^{-2x^{2}-2y^{2}}\\
 & =\int_{0}^{\infty}dr\int_{0}^{2\pi}d\phi\,r\,\frac{2}{\pi}e^{-2r^{2}}.
\end{align}
\end{subequations}
 This allows one to compare the value of the Wigner function without
regard for the coordinates used, e.g.\footnote{These definitions also tie back to those of the quadrature operators,
defined by (\ref{eq:quadrature-operators}). Other conventions for
introducing quadrature operators of dimension $1$, e.g. $\hat{X}=\frac{1}{\sqrt{2}}\left(\hat{a}+\hat{a}^{\dagger}\right)$
\cite{Bowen_QuantumOptomechanics_2015} or $\hat{X}=\left(\hat{a}+\hat{a}^{\dagger}\right)$
\cite{Walls_QuantumOptics_2008}, would cause the peak value of the
Wigner function to vary between coordinate systems.}
\begin{equation}
W_{|0\rangle}(r{=}0,\phi{=}0)=W_{|0\rangle}(x{=}0,y{=}0)=W_{|0\rangle}(\alpha{=}0,\alpha^{*}{=}0)=\frac{2}{\pi}.
\end{equation}

In some cases, we will also consider vector quantities in this coordinates.
To write these, we can introduce the unit vectors for the various
coordinates as well. The Cartesian coordinate unit vectors are written\footnote{Note that the hat “$\hat{\cdot}$” does not here signify operator
quantities. Operator valued vector quantities will not be needed,
so any vector marked in this way can be assumed to be a unit vector.} $\hat{\mathbf{x}}$ and $\hat{\mathbf{y}}$ while the polar unit
vectors are written $\hat{\symbf{\phi}}$ and $\hat{\mathbf{r}}$.

Either one of $\alpha$, $(x,y)$, or $(r,\phi)$ refers to a point
in phase space. For this reason, we shall refer to $\alpha,x,y,r$
and $\phi$ collectively as spatial coordinates. This should be contrasted
with the time coordinate $t$ which naturally enters the discussion
when system dynamics are considered.

\section{Phase Space Dynamics \label{sec:wigner-pde-derivation}}

As noted in Section \ref{sec:Phase-Space-Representations}, there
exists a one-to-one mapping between Wigner functions and density matrices.
Because of this, a description of the Wigner function dynamics can
be used to uniquely determine the evolution of a system instead of
the density matrix dynamics. 

In this section, we outline a procedure for deriving the equation
of motion for the Wigner function corresponding to the equation of
motion for a density matrix. The master equation (\ref{eq:general-master-equation})
is used here as the most general equation of motion for a density
matrix. The master equation may be transformed into a partial differential
equation for the Wigner function as is shown below. The goal is to
write an equation of the form
\begin{equation}
\partial_{t}W(\alpha,\alpha^{*},t)=L(\alpha,\alpha^{*},\partial_{\alpha},\partial_{\alpha^{*}})W(\alpha,\alpha^{*},t)\label{eq:-130}
\end{equation}
where $L(\alpha,\alpha^{*},\partial_{\alpha},\partial_{\alpha^{*}})$
is a differential operator expression. All differential operators
on the right hand side of (\ref{eq:-130}) should be spatial ones.
To achieve the goal, we follow the procedure of \textcite{Walls_QuantumOptics_2008}.

Take from (\ref{eq:wigner-function}) that 
\begin{equation}
W(\alpha,\alpha^{*},t)=\frac{1}{\pi^{2}}\int d\lambda\,d\lambda^{*}\,e^{\lambda^{*}\alpha-\lambda\alpha^{*}}\,\Tr\left[\hat{\rho}(t)e^{\lambda\hat{a}^{\dagger}-\lambda^{*}\hat{a}}\right]\label{eq:-129}
\end{equation}
and consider the equation of motion for $W(\alpha,\alpha^{*},t)$:
\begin{equation}
\partial_{t}W(\alpha,\alpha^{*},t)=\frac{1}{\pi^{2}}\int d\lambda\,d\lambda^{*}\,e^{\lambda^{*}\alpha-\lambda\alpha^{*}}\partial_{t}\chi(\lambda,\lambda^{*},t).\label{eq:-12}
\end{equation}
$\partial_{t}\chi(\lambda,\lambda^{*},t)$ is rewritten by using the
relevant equation of motion for the density matrix:
\begin{equation}
\partial_{t}\chi(\lambda,\lambda^{*},t)=\partial_{t}\Tr\left[\hat{\rho}(t)\hat{D}(\lambda)\right]=\Tr\left[\dot{\hat{\rho}}(t)\hat{D}(\lambda)\right]\label{eq:-128}
\end{equation}
Using the cyclic property of the trace, $\Tr\left[\hat{a}\hat{\rho}\hat{D}(\lambda)\right]=\Tr\left[\hat{\rho}\hat{D}(\lambda)\hat{a}\right]$,
the effect of creation and annihilation operators on the displacement
operator may be written as differential operators with respect to
$\lambda$ and $\lambda^{*}$. For example
\begin{equation}
\hat{a}\hat{D}(\lambda)=\hat{a}e^{\lambda\lambda^{*}/2}e^{-\lambda^{*}\hat{a}}e^{\lambda\hat{a}^{\dagger}}=\left(\frac{\partial}{\partial\lambda^{*}}-\frac{\lambda}{2}\right)e^{\lambda\lambda^{*}/2}e^{-\lambda^{*}\hat{a}}e^{\lambda\hat{a}^{\dagger}}=\left(\frac{\partial}{\partial\lambda^{*}}-\frac{\lambda}{2}\right)\hat{D}(\lambda).
\end{equation}
Using as appropriate (\ref{eq:-3}) or (\ref{eq:-4}), all relevant
combinations of $\hat{D}(\lambda)$ and either $\hat{a}$ and $\hat{a}^{\dagger}$
may be written as \cite{Walls_QuantumOptics_2008}
\begin{subequations}
\label{eq:-5}
\begin{align}
\hat{a}\hat{D}(\lambda) & =\left(-\partial_{\lambda^{*}}+\frac{\lambda}{2}\right)\hat{D}(\lambda),\label{eq:-6}\\
\hat{a}^{\dagger}\hat{D}(\lambda) & =\left(\partial_{\lambda}+\frac{\lambda^{*}}{2}\right)\hat{D}(\lambda),\label{eq:1}\\
\hat{D}(\lambda)\hat{a}^{\dagger} & =\left(\partial_{\lambda}-\frac{\lambda^{*}}{2}\right)\hat{D}(\lambda),\label{eq:-7}\\
\hat{D}(\lambda)\hat{a} & =-\left(\partial_{\lambda^{*}}+\frac{\lambda}{2}\right)\hat{D}(\lambda).\label{eq:-8}
\end{align}
\end{subequations}
With this, we can write (\ref{eq:-128}) with a left hand side of
$\partial_{t}\chi(\lambda,\lambda^{*},t)$ and a right hand side consisting
of a sum of terms of the form $\lambda^{m}\left(\lambda^{*}\right)^{n}\partial_{\lambda}^{p}\partial_{\lambda^{*}}^{q}\chi(\lambda,\lambda^{*},t)$.
The result is a partial differential equation for the characteristic
function $\chi(\lambda,\lambda^{*},t)$. We apply the Fourier transform
(\ref{eq:-129}) on both sides. This yields the correct left hand
side of (\ref{eq:-130}). Each term on the right hand side is rewritten
separately. A general right hand side term is rewritten as
\begin{subequations}
\label{eq:-79-1}
\begin{align}
\frac{1}{\pi^{2}}\int d\lambda\,d\lambda^{*}\, & e^{\alpha\lambda^{*}-\alpha^{*}\lambda}\lambda^{m}\left(\lambda^{*}\right)^{n}\partial_{\lambda}^{p}\partial_{\lambda^{*}}^{q}\left\langle \hat{D}(\lambda)\right\rangle \nonumber \\
 & =\frac{1}{\pi^{2}}(-1)^{m}\partial_{\alpha^{*}}^{m}\partial_{\alpha}^{n}\int d\lambda\,d\lambda^{*}\,e^{\alpha\lambda^{*}-\alpha^{*}\lambda}\partial_{\lambda}^{p}\partial_{\lambda^{*}}^{q}\left\langle \hat{D}(\lambda)\right\rangle \\
 & =\frac{1}{\pi^{2}}(-1)^{m+p+q}\partial_{\alpha^{*}}^{m}\partial_{\alpha}^{n}\left[\left(\alpha^{*}\right)^{q}\alpha^{p}\int d\lambda\,d\lambda^{*}\,\partial_{\lambda}^{p}\partial_{\lambda^{*}}^{q}e^{\alpha\lambda^{*}-\alpha^{*}\lambda}\left\langle \hat{D}(\lambda)\right\rangle \right]\\
 & =(-1)^{m+q}\partial_{\alpha^{*}}^{m}\partial_{\alpha}^{n}\left[\left(\alpha^{*}\right)^{q}\alpha^{p}W(\alpha,\alpha^{*},t)\right].
\end{align}
\end{subequations}
The result is (\ref{eq:-130}) with an explicit right hand side.\footnote{This derivation of the equation of motion for $W$ uses the corresponding
master equation. Alternately, one may disregard the operator picture
completely and instead derive Wigner function equation of motion using
the symmetrically ordered system Hamiltonian. The right hand side
of (\ref{eq:-12}) can then be written as a concise expression using
the Moyal bracket \cite{Moyal_QuantumMechanicsStatistical_1949,Curtright_ConciseTreatiseQuantum_2014}.}

As with the master equation, we will normally express any partial
differential equation for the Wigner function in the form (\ref{eq:-130})
with the left hand side $\partial_{t}W$ and a right hand side containing
only spatial derivatives. Statements referring to the left and right
hand sides of an equation should be interpreted given the equation
in this particular form.

\section{Gaussian States and Quadratic Hamiltonians\label{sec:Gaussian-States-and}}

Gaussian states are central to continuous variable quantum mechanics.
They may be defined as the set of states whose Wigner function is
a Gaussian function \cite{Ferraro_GaussianStatesQuantum_2005a}. From
the inverse Fourier transformation (\ref{eq:-166}) it is seen that
an equivalent statement is that the characteristic function is a Gaussian
function. An important theorem due to Hudson states that only Gaussian
states have completely non-negative Wigner functions \cite{Hudson_WhenWignerQuasiprobability_1974,Kenfack_NegativityWignerFunction_2004}.
Thus any non-Gaussian pure state assumes negative values somewhere
in phase space.

We say that a Hamiltonian is quadratic if it consists of terms that
are at most quadratic in the ladder operators. We see from this definition
that examples of quadratic Hamiltonians include the simple harmonic
oscillator (\ref{eq:sho-hamiltonian}) and parametric squeezing (\ref{eq:-121}). 

In classical mechanics, one may describe the evolution of a phase
space probability density $P$ is described by the Liouville equation.\footnote{The appropriate equation of motion for the phase space probability
when noise is present is the Fokker-Planck equation.} Systems quadratic Hamiltonians have the unique property that the
equation of motion for the Wigner function (\ref{eq:-130}) assumes
a form identical to the classical Liouville equation \cite{Katz_ClassicalQuantumTransition_2008}.
Hence the classical and quantum mechanical systems is the same as
far as the phase space distributions and the expectation values that
may be calculated from them is concerned. Furthermore, any evolution
of a Gaussian state with a quadratic Hamiltonian will always result
in a Gaussian state. The preservation of the Gaussian quality holds
even when adding the damping terms of the master equation (\ref{eq:general-master-equation})
to the evolution \cite{Ferraro_GaussianStatesQuantum_2005a} (although
not for the dephasing term).

One may show that the equation of motion for the Wigner function (\ref{eq:-130})
will contain higher order derivatives if and only if the Hamiltonian
contains terms of higher than quadratic order in the annihilation
and creation operators \cite{Corney_NonGaussianPureStates_2015}.
These higher terms are precisely the terms removed when expressing
the Liouville equation for the classical system \cite{Katz_ClassicalQuantumTransition_2008}.\footnote{Note however, that one can not for unitary evolution connect smoothly
the limit $\hbar\to0$ to the classical case \cite{Habib_QuantumClassicalTransitionNonlinear_2002,Katz_ClassicalQuantumTransition_2008}.} For this reason, we can also identify these higher order derivatives
with the creation of negative regions of the Wigner function. 

We note that the unitary transformations $\hat{D}(\lambda)$, $\hat{R}(\phi)$,
or $\hat{S}(\xi)$ are instances of time evolution operators arising
from a quadratic Hamiltonian. Hence, any application of these transformation
operators to a Gaussian state results in a Gaussian state as well.
In fact, appropriately applying $\hat{D}(\lambda)$ and $\hat{S}(\xi)$
to the vacuum state $|0\rangle$ is sufficient to reach any pure Gaussian
state.

\section{Wigner Current\label{sec:wigner-current}}

The classical phase space dynamics as described by the Liouville equation
can be formulated as a continuity equation. This is done by defining
a current $\mathbf{J}_{P}$ in terms of the probability density $P$
such that the equation of motion equates the time derivative of the
density to the negative divergence of the defined current: $\partial_{t}P=-\nabla\cdot\mathrm{J}_{P}$.

We may define a similar current $\mathbf{J}$ for the Wigner function
\cite{Oliva_QuantumKerrOscillators_2019,Bauke_VisualizingQuantumMechanics_2011,Friedman_WignerFlowOpen_2017}
(also known as the Wigner flow) by choosing $\mathbf{J}$ such that
(\ref{eq:-130}) can be written
\begin{equation}
\partial_{t}W=-\nabla\cdot\mathbf{J}.\label{eq:quantum-continuity-equation}
\end{equation}
In (\ref{eq:quantum-continuity-equation}) $W$ takes the role of
the classical probability density. For this reason, one might refer
to $W$ as the Wigner density.

In the case of quadratic Hamiltonians, the phase space dynamics are
unchanged between the classical and quantum mechanical systems. Hence
the interpretation of $\mathbf{J}$ is the same as in the classical
case. This provides an intuitive geometric view of the evolution of
the Wigner function. In contrast to the classical phase space continuity
equation, the quantum mechanical expression for $-\nabla\cdot\mathbf{J}$
for a non-quadratic Hamiltonian will contain spatial derivatives of
$W$ that are of higher order than $1$. In that case some freedom
(or ambiguity \cite{Oliva_QuantumKerrOscillators_2019}) can arise
in choosing $\mathbf{J}$ for the equation of motion for $W$. Given
the presence of such higher-order terms, the dynamics of the Wigner
function can not generally be described as a flow of Wigner density
along trajectories in phase space \cite{Oliva_QuantumKerrOscillators_2019}.
Nevertheless, the expression for $\mathbf{J}$ can in such cases still
provide insight into the evolution of the Wigner function. 

In fact, these higher order terms will give rise to derivatives of
$W$ in the expression of $\mathbf{J}$. The inclusion of derivatives
of $W$ in the expression of $\mathbf{J}$ causes the $\mathbf{J}$
to depend not only on the value of $W$ at the specific point but
also on adjacent values. For this reason, terms causing these derivatives
to appear in $\mathbf{J}$ can be referred to as non-local terms \cite{Oliva_QuantumKerrOscillators_2019}.
As noted in Section \ref{sec:Gaussian-States-and}, these higher order
derivatives are also a necessary condition for the evolution of a
negative Wigner function from a non-negative one.

\section{Measures of Non-classicality \label{sec:measures-of-negativity}}

We define in this section two measures of non-classicality based the
negativity of the Wigner function. A Wigner function that assumes
negative values in certain parts of phase space is an indicator of
non-classicality \cite{Gerry_IntroductoryQuantumOptics_2004}. Experimentally,
negativity of the Wigner function has been used to demonstrate non-classicality
\cite{Leibfried_ExperimentalDeterminationMotional_1996}. To quantify
the amount of negativity here, we introduce two quantities derived
from the Wigner function: The negative peak and the negative volume.\footnote{The terms “volume” and, later, “peak” are used in place
of, perhaps, more natural terms such as “integral”, “minimum”,
or “maximum”. This is to distinguish from the maximum with respect
to variables other than the phase space coordinates. Most commonly,
the quantities $\max_{t}\left\{ N_{\mathrm{vol}}(t)\right\} $ and
$\max_{t}\left\{ N_{\mathrm{peak}}(t)\right\} $ which might then
be referred to as “maximum negative volume” and “maximum
negative peak,” respectively.}

We define the negative peak as

\begin{equation}
N_{\mathrm{peak}}=-\min_{x,y}\left(\min\{0,W(x,y,t)\}\right).\label{eq:negpeak-definition}
\end{equation}
The Wigner function may also be expressed as the expectation value
of the displaced parity operator \cite{Royer_WignerFunctionExpectation_1977}:
\begin{equation}
W(\alpha,\alpha^{*})=\frac{2}{\pi}\langle\psi|\hat{D}(\alpha)\hat{\pi}\hat{D}^{\dagger}(\alpha)|\psi\rangle.\label{eq:-81}
\end{equation}
Writing $\hat{\pi}=e^{i\pi\hat{n}}$ in (\ref{eq:-81}) and using
the normalization of the displaced state $\hat{D}^{\dagger}(\alpha)|\psi\rangle$
allows one to establish the bounds \cite{Cahill_DensityOperatorsQuasiprobability_1969a}
\begin{equation}
-\frac{2}{\pi}\leq W(\alpha,\alpha^{*})\leq\frac{2}{\pi}\qquad\text{for all \ensuremath{\alpha},}\label{eq:-82}
\end{equation}
which by extension bounds $N_{\mathrm{peak}}\leq2/\pi$. The upper
and lower bounds of (\ref{eq:-81}) are for example reached at $\alpha=0$
for the states $|0\rangle$ and $|1\rangle$ respectively.\footnote{In general, all Wigner functions for pure states with even (odd) wave
functions will reach the upper (lower) bound of (\ref{eq:-82}) at
the origin $\alpha=0$. This can be seen by considering the definition
of $W$ in terms of wave functions \cite{Case_WignerFunctionsWeyl_2008}
or alternately expanding $|\psi\rangle$ in the number state basis
in (\ref{eq:-81}) and setting $\alpha=0$. Even (odd) wave function
states have contain only even (odd) basis elements in the number state
basis. $|0\rangle$ and $|1\rangle$ are trivial examples of such
states.} For $|0\rangle$ one has from (\ref{eq:vacuum-wigner-complex})
\begin{equation}
W_{|0\rangle}(\alpha=0)=\frac{2}{\pi}.
\end{equation}
To evaluate $W(\alpha=0)$ for the state $|1\rangle$, use the general
expression for the number state Wigner function \cite{Gerry_IntroductoryQuantumOptics_2004}
\begin{equation}
W_{|n\rangle}(\alpha,\alpha^{*})=\frac{2}{\pi}(-1)^{n}L_{n}(4|\alpha|^{2})e^{-2|\alpha|^{2}}\label{eq:-83}
\end{equation}
and that \cite{Arfken_MathematicalMethodsPhysicists_2012}
\begin{equation}
L_{1}(x)=1-x.\label{eq:-84}
\end{equation}
Combining (\ref{eq:-83}) and (\ref{eq:-84}), we obtain 
\begin{equation}
W_{|1\rangle}(\alpha=0)=-\frac{2}{\pi}.
\end{equation}
The bounds of (\ref{eq:-82}) tend to plus and minus infinity as $\hbar\rightarrow0$
\cite{Zachos_QuantumMechanicsPhase_2005}. This behavior is necessary
for the Wigner function to agree with the classical phase space probability
density in the classical limit (a delta function is an example of
a valid classical phase space probability density which obviously
violates (\ref{eq:-82}) if the bounds remain finite).

The second measure of non-classicality which will be defined is the
negative volume $N_{\mathrm{vol}}$. The negative volume is defined
as the integral over all regions where $W$ assumes negative values.
We can write this as

\begin{equation}
N_{\mathrm{vol}}=-\int dx\,dy\,\min\{0,W(x,y,t)\}.\label{eq:negvol-definition}
\end{equation}
Note that an equivalent definition (used by \textcite{Kenfack_NegativityWignerFunction_2004})
is given by
\begin{equation}
N_{\mathrm{vol}}=\frac{1}{2}\int dx\,dy\,\left(|W(x,y,t)|-W(x,y,t)\right).
\end{equation}
$N_{\mathrm{vol}}$ can grow much larger than $N_{\mathrm{peak}}$.
For instance, \textcite{Kenfack_NegativityWignerFunction_2004} demonstrate
numerically for $0\leq n\leq250$ that $N_{\mathrm{vol}}$ of the
number state $|n\rangle$ increases monotonically with $n$ and approximately
as $\tfrac{1}{2}\sqrt{n}$.

$N_{\mathrm{vol}}$ and $N_{\mathrm{peak}}$ are both functionals
of the Wigner function $W$. Thus they could be written with $W$
(and optionally $t$) as their argument. As done above, we will however
leave the arguments implicit and interpret them from the context of
the symbol.

For a Wigner function that is everywhere non-negative, it follows
from their definitions that $N_{\mathrm{vol}}=N_{\mathrm{peak}}=0$.
As soon as the Wigner function departs from this, $N_{\mathrm{vol}}\neq0$
and $N_{\mathrm{peak}}\neq0$. It may be shown that the set of all
non-negative Wigner functions and all Wigner functions for Gaussian
states (i.e. all Gaussian functions) are equal \cite{Hudson_WhenWignerQuasiprobability_1974}.
Such a theorem does not exist for mixed states however.

As pointed out by \textcite{Kenfack_NegativityWignerFunction_2004},
the space of all possible states is too large for a single quantity
to characterize all non-classical features of a state. In this thesis,
the quantities $N_{\mathrm{vol}}$ and $N_{\mathrm{peak}}$ have been
chosen to quantify the negativity of a quantum state. Prior use as
indicators of non-classicality exists for both negative volume \cite{Kenfack_NegativityWignerFunction_2004,Arkhipov_NegativityVolumeGeneralized_2018}
and negative peak \cite{Koppenhofer_HeraldedDissipativePreparation_2019}.
However two wildly differing Wigner functions may still share both
$N_{\mathrm{vol}}$ and $N_{\mathrm{peak}}$. For this reason, we
shall also discuss the geometry of the Wigner function and its negative
domains supported by plots such as the one found in Figure \ref{fig:vacuum-state-plot}.

\section{Return to Quadrature Squeezing \label{sec:return-to-parametric-squeezing}}

\begin{figure}
\noindent \begin{centering}
\makebox[0pt][c]{\mbox{\includegraphics{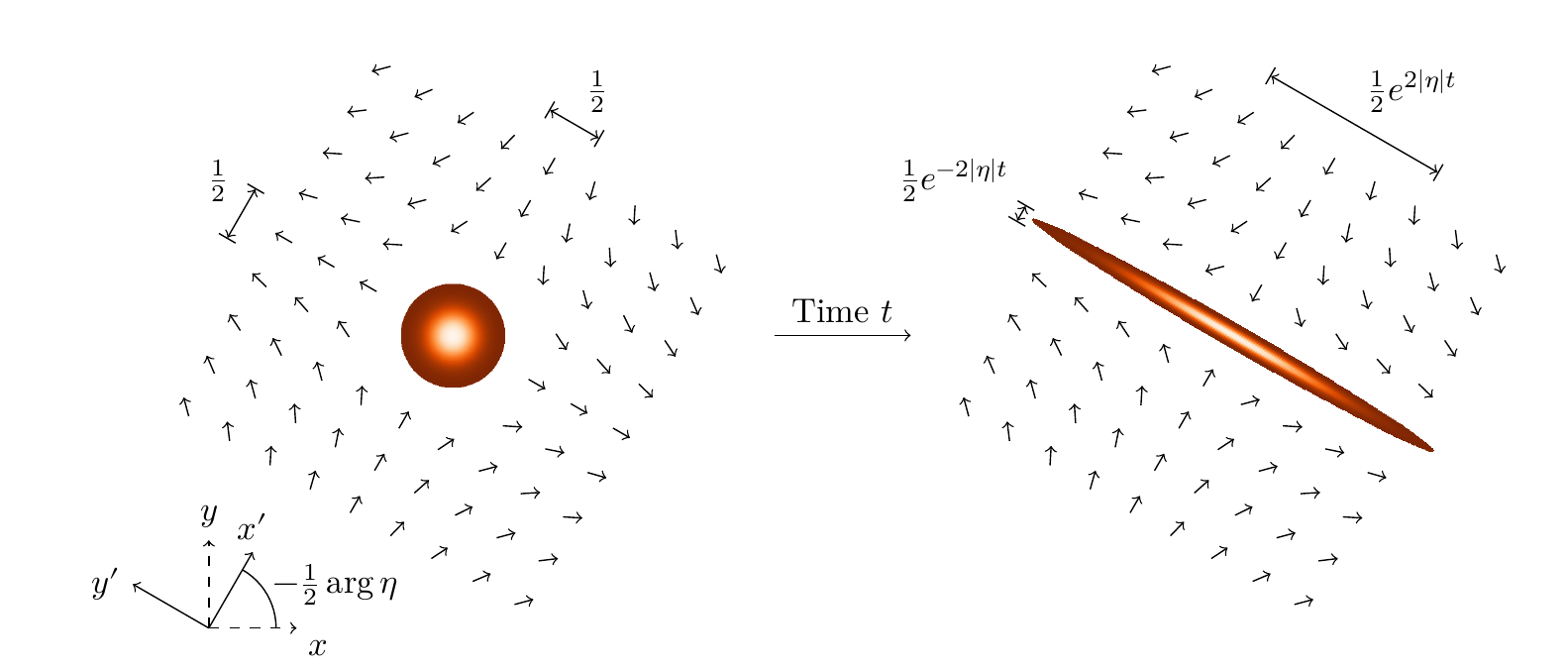}\hskip1em}}
\par\end{centering}
\caption[Illustration of quadrature squeezing dynamics]{\label{fig:parametric-squeezing-sketch}\textbf{Illustration of quadrature
squeezing dynamics.} Sketch of the dynamics of the Hamiltonian $\hat{H}_{\eta}$
as discussed in Section \ref{sec:return-to-parametric-squeezing}.
As time evolves, the isotropic initial state is transformed to an
elliptical Gaussian with its major axis rotated from the $y$-axis
by an angle equal to half of the complex argument of the parameter
$\eta$. The direction of the Wigner current $\mathbf{J}$ as given
in (\ref{eq:-135}) is indicated by arrows (arrow lengths are not
scaled with the magnitude of $\mathbf{J}$).}
\end{figure}

We return now to the problem Section \ref{sec:quadrature-squeezing}
with the tools of the subsequent section. We can gain a better intuition
for the system by considering it using the phase space formalism since
developed.

We first apply Section \ref{sec:wigner-pde-derivation} to derive
a partial differential equation for $W(\alpha,\alpha^{*},t)$. Inserting
(\ref{eq:-99}) into (\ref{eq:-128}), the partial differential equation
for the symmetrically ordered characteristic function takes the form
\begin{subequations}
\label{eq:-126}
\begin{align}
\partial_{t}\langle\hat{D}(\lambda)\rangle & =\eta^{*}\Tr\left(\left[\hat{a}\hat{a},\hat{\rho}\right]\hat{D}(\lambda)\right)-\eta\Tr\left(\left[\hat{a}^{\dagger}\hat{a}^{\dagger},\hat{\rho}\right]\hat{D}(\lambda)\right)\\
 & =\eta^{*}\Tr\left(\hat{\rho}\left[\hat{D}(\lambda),\hat{a}\hat{a}\right]\right)-\eta\Tr\left(\hat{\rho}\left[\hat{D}(\lambda),\hat{a}^{\dagger}\hat{a}^{\dagger}\right]\right)\label{eq:-125}
\end{align}
with the terms of (\ref{eq:-125}) given by
\begin{align}
\eta^{*}\Tr\left(\hat{\rho}\left[\hat{D}(\lambda),\hat{a}\hat{a}\right]\right) & =\eta^{*}\left[\left(\partial_{\lambda^{*}}+\frac{\lambda}{2}\right)^{2}-\left(-\partial_{\lambda^{*}}+\frac{\lambda}{2}\right)^{2}\right]\Tr\left[\hat{\rho}\hat{D}(\lambda)\right]\\
 & =2\eta^{*}\lambda\partial_{\lambda^{*}}\Tr\left[\hat{\rho}\hat{D}(\lambda)\right],\\
-\eta\Tr\left(\hat{\rho}\left[\hat{D}(\lambda),\hat{a}^{\dagger}\hat{a}^{\dagger}\right]\right) & =-\eta\left[\left(\partial_{\lambda}-\frac{\lambda^{*}}{2}\right)^{2}-\left(\partial_{\lambda}+\frac{\lambda^{*}}{2}\right)^{2}\right]\Tr\left[\hat{\rho}\hat{D}(\lambda)\right]\\
 & =2\eta\lambda^{*}\partial_{\lambda}\Tr\left[\hat{\rho}\hat{D}(\lambda)\right].
\end{align}
\end{subequations}
$\Tr\left(\left[\hat{A},\hat{B}\right]\hat{C}\right)=\Tr\left(\left[\hat{B},\hat{C}\right]\hat{A}\right)$
for operators $\hat{A}$, $\hat{B}$ and $\hat{C}$. The result of
the calculations (\ref{eq:-126}) is written
\begin{equation}
\partial_{t}\chi(\lambda,\lambda^{*},t)=2\left(\eta^{*}\lambda\partial_{\lambda^{*}}+\eta\lambda^{*}\partial_{\lambda}\right)\chi(\lambda,\lambda^{*},t).\label{eq:-127}
\end{equation}
The next step is to convert (\ref{eq:-127}) into the equivalent equation
of motion for $W(\alpha,\alpha^{*},t)$. Use (\ref{eq:wigner-function})
on both sides of (\ref{eq:-127}) to obtain
\begin{equation}
\partial_{t}W(\alpha,\alpha^{*},t)=\frac{2}{\pi^{2}}\int d\lambda\,d\lambda^{*}\,e^{\lambda^{*}\alpha-\alpha^{*}\lambda}\left(\eta^{*}\lambda\partial_{\lambda^{*}}+\eta\lambda^{*}\partial_{\lambda}\right)\chi(\lambda,\lambda^{*},t).\label{eq:-131}
\end{equation}
Then (\ref{eq:-79-1}) is applied to each right hand side term to
find
\begin{equation}
\partial_{t}W(\alpha,\alpha^{*},t)=2\eta^{*}\alpha\partial_{\alpha^{*}}W(\alpha,\alpha^{*},t)+2\eta\alpha^{*}\partial_{\alpha}W(\alpha,\alpha^{*},t).\label{eq:-197}
\end{equation}
The effect of this equation becomes more clear, if we write it in
Cartesian coordinates instead. Using (\ref{eq:-17}), we obtain

\begin{equation}
\partial_{t}W(x,y;t)=2\left(\Re\eta\,x+\Im\eta\,y\right)\frac{\partial}{\partial x}W\left(x,y,t\right)+2\left(\Im\eta\,x-\Re\eta\,y\right)\frac{\partial}{\partial y}W\left(x,y,t\right).\label{eq:-120}
\end{equation}

\begin{figure}
\noindent \begin{centering}
\makebox[0pt][c]{\mbox{\includegraphics{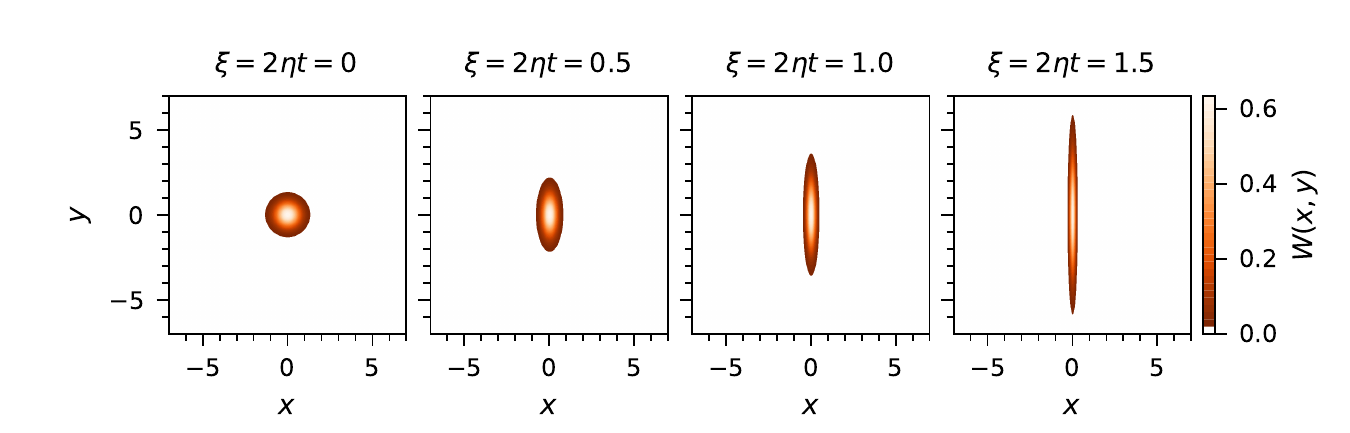}\hskip1em}}
\par\end{centering}
\caption[Evolution of vacuum state undergoing quadrature squeezing]{\label{fig:vacuum-state-plot-1}\textbf{Evolution of vacuum state
undergoing quadrature squeezing.} The evolution of the vacuum state
$|0\rangle$ under dynamics of the  squeezing Hamiltonian $\hat{H}_{\eta}$.
The analytical solution is (\ref{eq:-193}). $\eta$ is real and thus
the major (or anti-squeezed) axis of the evolved Gaussian coincides
with the $y$-axis.}
\end{figure}

Using the tools of Section \ref{sec:wigner-current}, we can define
a Wigner current $\mathbf{J}$ such that (\ref{eq:-120}) takes the
form of the continuity-like equation (\ref{eq:quantum-continuity-equation}):
\begin{equation}
\mathbf{J}=-2\left(\Re\eta\,x+\Im\eta\,y\right)\hat{\mathbf{x}}W\left(x,y,t\right)-2\left(\Im\eta\,x-\Re\eta\,y\right)\hat{\mathbf{y}}W\left(x,y,t\right).\label{eq:-135}
\end{equation}
Since the Hamiltonian (\ref{eq:-121}) is quadratic, the right hand
side of (\ref{eq:-197}) contains only first order derivatives. The
equation is therefore unchanged from the classical case Liouville
equation. Furthermore, the current can be described as flow along
trajectories. This may be done by constructing a velocity field $\mathbf{v}$
from (\ref{eq:-135}) such that
\begin{equation}
\mathbf{J}=W\mathbf{v}.\label{eq:-198}
\end{equation}
Comparing (\ref{eq:-135}) and (\ref{eq:-198}) it is seen that $\mathbf{v}$
is independent of $W$. In other words, $\mathbf{J}$ depends only
on the local value of $W$ and not adjacent values. Such non-local
dependence would have been expressed using spatial derivatives of
$W$ (cf. Section \ref{sec:wigner-current}).

Choosing, as in Section \ref{sec:quadrature-squeezing}, a real $\eta$,
the current becomes 
\begin{equation}
\mathbf{J}=-2\eta x\hat{\mathbf{x}}W\left(x,y,t\right)+2\eta y\hat{\mathbf{y}}W\left(x,y,t\right)\qquad\text{for real \ensuremath{\eta}.}\label{eq:-132}
\end{equation}
(\ref{eq:-132}) describes a flow of density toward the $y$ axis
(the term $-2\eta x\hat{\mathbf{x}}$) and away from the $x$ axis
(the term $2\eta y\hat{\mathbf{y}}$). This is intuitively consistent
with a decrease in the variance of the quadrature $\hat{X}$ and an
increase in the variance of the quadrature $\hat{Y}$. This evolution
in the quadrature variances is exactly the one found in (\ref{eq:-118}).

We extend to the general case of a complex $\eta$ by considering
the equation of motion in polar coordinates. Using $r$ and $\phi$
of (\ref{eq:-16}), we can write
\begin{equation}
\partial_{t}W(r,\phi,t)=2r\Re\left(\eta e^{-2i\phi}\right)\frac{\partial}{\partial r}W\left(r,\phi,t\right)+2\Im\left(\eta e^{-2i\phi}\right)\frac{\partial}{\partial\phi}W\left(r,\phi,t\right).\label{eq:-133}
\end{equation}
To demonstrate the implication of the argument of $\eta$, we introduce
a rotated coordinate system 
\begin{equation}
\phi\rightarrow\phi'=\phi+\frac{1}{2}\arg\eta
\end{equation}
 and a rotated Wigner function 
\[
W'(r,\phi',t)=W'(r,\phi'-\frac{1}{2}\arg\eta,t).
\]
Inserting into (\ref{eq:-133}), we find that $W'$ evolves according
to the equation of motion
\begin{equation}
\partial_{t}W'(r,\phi',t)=2r\Re\left(\eta e^{-i\arg\eta}e^{-2i\phi}\right)\frac{\partial}{\partial r}W\left(r,\phi;t\right)+2\Im\left(\eta e^{-i\arg\eta}e^{-2i\phi}\right)\frac{\partial}{\partial\phi}W\left(r,\phi;t\right)
\end{equation}
or equivalently
\begin{equation}
\partial_{t}W(r,\phi',t)=\begin{aligned}[t]2r\eta'\Re\left(e^{-2i\phi'}\right)\frac{\partial}{\partial r} & W\left(r,\phi',t\right)+2\eta'\Im\left(e^{-2i\phi'}\right)\frac{\partial}{\partial\phi'}W\left(r,\phi',t\right)\\
 & \text{where \ensuremath{\eta'=|\eta|}.}
\end{aligned}
\label{eq:-134}
\end{equation}
We may interpret this as the fact that $W'$ evolves under (\ref{eq:-134})
as $W$ does under (\ref{eq:-133}) for real $\eta$, i.e. with the
current (\ref{eq:-132}) (although with the unit vectors rotated correspondingly).
Hence the conclusions made from (\ref{eq:-132}) for the original
non-rotated axes, e.g. the quadrature variance evolution, may be applied
unchanged to the axes rotated by $\frac{1}{2}\arg\eta$.

Note finally, that we can construct a concise analytical expression
for the evolution of the vacuum state Wigner function under quadrature
squeezing. In Section \ref{sec:quadrature-squeezing}, it was established
that the system evolves to the squeezed state $|\xi{=}2\eta t\rangle$
as given by (\ref{eq:-122}). The Wigner function is then found by
inserting the appropriate parameter into the Wigner function for the
squeezed vacuum state (\ref{eq:-190}):
\begin{equation}
W(x,y,t)=\begin{aligned}[t]\frac{2}{\pi}\exp\bigg[ & -2e^{2|\eta|t}\left(x\cos\frac{\arg\eta}{2}+y\sin\frac{\arg\eta}{2}\right)^{2}\\
 & -2e^{-2|\eta|t}\left(x\sin\frac{\arg\eta}{2}-y\cos\frac{\arg\eta}{2}\right)^{2}\bigg].
\end{aligned}
\label{eq:-193}
\end{equation}
The evolution is sketched in Figure \ref{fig:parametric-squeezing-sketch}
where the current $\mathbf{J}$ has also been overlayed. The solution
is plotted in Figure \ref{fig:vacuum-state-plot-1}. While a trivial
example, we nonetheless note that the Wigner function remains non-negative
in accordance with the statements in Section \ref{sec:Gaussian-States-and}.

This Section illustrates a general method which will be employed later,
namely the introduction of a transformed coordinate system and corresponding
transformed Wigner function. It also demonstrates the value in being
able to freely move between coordinate systems, choosing at any one
time the most appropriate one for the problem.

\chapter{Numerics\label{chap:Numerics}}

Discussions in the following chapters are supported by the numerical
analysis of the discussed physical problems with example parameters.
To avoid weighing the discussion down with persistent description
of the numerical details, we present in this chapter the techniques
used to obtain the numerical results.

The objective of the numerical analysis will often be to compute the
Wigner function of a quantum state or some quantity derived from the
Wigner function. The state is usually obtained as the result of either
unitary (see Section \ref{sec:unitary-dynamics}) or non-unitary (see
Section \ref{subsec:master-equation}) evolution given some initial
state. Section \ref{sec:measures-of-negativity} defines the derived
quantities which will be used. Section \ref{sec:Simulation-of-System}
describes the steps to simulate quantum systems thereby obtaining.
Section \ref{sec:Wigner-Function-and} then details the steps to evaluate
the Wigner function at points in phase space and as well as computing
its negativity.

The Python library QuTiP \cite{Johansson_QuTiPOpensourcePython_2012,Johansson_QuTiPPythonFramework_2013}
(version 4.3) was used to simulate the evolution of the studied quantum
systems. We briefly outline the methods used below. We should note
however, that QuTiP provides functions abstracting away most of the
details in ordinary use.

\section{Simulation of System Dynamics\label{sec:Simulation-of-System}}

Systems are described using the number state basis. To reduce dimensions
to a finite number, the basis is truncated and only the lowest $N$
states are considered. A state $|\Psi\rangle$ is then represented
by a vector $\vec{c}$ with components $c_{n}$ such that
\begin{equation}
|\Psi\rangle=\sum_{n=0}^{N-1}c_{n}|n\rangle\label{eq:-238}
\end{equation}
given by (\ref{eq:-230}) (note the difference to (\ref{eq:-181})).
Likewise density matrix can now be represented by an $N$-by-$N$
matrix $\dunderbar{\rho}$ with components $\rho_{mn}$: 
\begin{equation}
\hat{\rho}=\sum_{\substack{m=0\\
n=0
}
}^{N-1}\rho_{mn}|m\rangle\langle n|\qquad\text{with}\qquad\rho_{mn}=\langle m|\hat{\rho}|n\rangle.\label{eq:-229}
\end{equation}
The states are normalized in the finite basis such that
\begin{equation}
\sum_{n=0}^{N-1}\left|c_{n}\right|^{2}=1\qquad\text{and}\qquad\sum_{n=0}^{N-1}\rho_{nn}=1.\label{eq:-239}
\end{equation}
It is assumed that the magnitude of $c_{n}$ and $\rho_{mn}$ falls
off sufficiently fast (see e.g. (\ref{eq:-182}) or (\ref{eq:-86}))
that the truncation of the basis is a good approximation. An operator
$\hat{O}$ is represented by a matrix $\dunderbar O$ with components
$O_{mn}$ such that
\begin{equation}
\hat{O}=\sum_{\substack{m=0\\
n=0
}
}^{N-1}O_{mn}|m\rangle\langle n|\qquad\text{with}\qquad O_{mn}=\langle m|\hat{O}|n\rangle.
\end{equation}
For an operator expressed in terms of of $\hat{a}$ and $\hat{a}^{\dagger}$,
one may determine $\dunderbar O$ by noting from (\ref{eq:-231})
that $\langle m|\hat{a}|n\rangle=\sqrt{n}\delta_{m(n-1)}$ and $\langle m|\hat{a}^{\dagger}|n\rangle=\sqrt{n+1}\delta_{m(n+1)}$
and combining factors of each term by matrix multiplication.

Systems governed by the Schrödinger equation are solved using the
time evolution operator $\hat{U}$ of (\ref{eq:unitary-propagator}).
Its matrix $\dunderbar U$ is found as the matrix exponential of the
matrix for $\hat{H}$ expressed in terms of $\hat{a}$ and $\hat{a}^{\dagger}$.
The evolved state is then obtained with matrix elements given by $\dunderbar U\vec{c}$
or $\dunderbar U\dunderbar{\rho}\dunderbar U^{\dagger}$.

For open systems, the master equation (\ref{eq:general-master-equation})
is written in element-wise form by applying $\langle m|$ and $|n\rangle$
to both sides:
\begin{equation}
\dot{\rho}_{mn}=-\frac{i}{\hbar}\langle m|[\hat{H},\hat{\rho}]|n\rangle+\gamma\left(\bar{n}+1\right)\langle m|\mathcal{D}[\hat{a}]\hat{\rho}|n\rangle+\gamma\bar{n}\langle m|\mathcal{D}[\hat{a}^{\dagger}]\hat{\rho}|n\rangle+\gamma_{\phi}\langle m|\mathcal{D}[\hat{n}]\hat{\rho}|n\rangle.\label{eq:-232}
\end{equation}
(\ref{eq:-232}) is now a system of ordinary differential equations
for a finite system of variables $\rho_{mn}$ which may be solved
by a standard solve of which QuTiP has a selection to choose from.
Simulations in this thesis were computed with ZVODE \cite{Brown_ZVODEVariablecoefficientOrdinary_}
with the method “BDF”.

Matrices and vectors representing the initial states were obtained
by appropriate use of the matrices derived from $\hat{D}$ and $\hat{S}$
in a way similar to $\hat{U}$. These were used on the vector of the
vacuum state and the matrix of a thermal state. The vacuum state vector
(\ref{eq:-238}) is specified component-wise as $c_{n}=\delta_{0n}$.
In the truncated basis, the thermal state matrix is computed from
\begin{equation}
\hat{\rho}=\frac{1}{Z}\sum_{n=0}^{N-1}e^{-\hbar\omega n/k_{B}T}|n\rangle\langle n|\label{eq:-240}
\end{equation}
where $Z$ is chosen such that (\ref{eq:-239}) holds. (\ref{eq:-240})
and (\ref{eq:-170}--\ref{eq:-141}) agree in the limit where $N\to\infty$.

\section{Wigner Function and Derived Quantities\label{sec:Wigner-Function-and}}

QuTiP includes multiple methods for evaluating the Wigner function
given a density matrix $\dunderbar{\rho}$. The default method which
was also used in this thesis employs the Wigner function transition
probabilities of the number states.

To evaluate the Wigner function, insert (\ref{eq:-229}) into the
definition of the Wigner function as given in (\ref{eq:wigner-characteristic-function}--\ref{eq:wigner-function}).
Rearranging the order of sum, trace and integral yields
\begin{equation}
W_{\hat{\rho}}(\alpha,\alpha^{*})=\sum_{\substack{m=0\\
n=0
}
}^{N-1}\rho_{mn}\frac{1}{\pi^{2}}\int d\lambda\,d\lambda^{*}\,e^{\lambda^{*}\alpha-\alpha^{*}\lambda}\Tr[|m\rangle\langle n|\hat{D}(\lambda)].\label{eq:-233}
\end{equation}
Defining $W_{|k\rangle\langle n|}(\alpha,\alpha^{*})$ by
\begin{equation}
W_{|m\rangle\langle n|}(\alpha,\alpha^{*})=\frac{1}{\pi^{2}}\int d\lambda\,d\lambda^{*}\,e^{\lambda^{*}\alpha-\alpha^{*}\lambda}\Tr[|m\rangle\langle n|\hat{D}(\lambda)]
\end{equation}
allows one to write (\ref{eq:-233}) as
\begin{equation}
W_{\hat{\rho}}(\alpha,\alpha^{*})=\sum_{\substack{m=0\\
n=0
}
}^{N-1}\rho_{mn}W_{|m\rangle\langle n|}(\alpha,\alpha^{*}).\label{eq:-234}
\end{equation}
The quantity $W_{|m\rangle\langle n|}(\alpha,\alpha^{*})$ is referred
to as the transition probability. One may show that \cite{Bartlette_ExactTransitionProbabilities_1948,Curtright_ConciseTreatiseQuantum_2014} 

\begin{equation}
W_{|m\rangle\langle n|}(\alpha,\alpha^{*})=\frac{2}{\pi}(-1)^{m}\sqrt{\frac{m!}{n!}}|\alpha|^{(n-m)}e^{-2|\alpha|^{2}}L_{m}^{n-m}(4|\alpha|^{2})e^{i(m-n)\arg\alpha}\label{eq:-235}
\end{equation}
where $L_{m}^{n-m}$ denotes the associated Laguerre polynomial. The
combination of (\ref{eq:-234}) and (\ref{eq:-235}) allows one to
evaluate the Wigner function.

We will also need to evaluate the quantities $N_{\mathrm{vol}}$ and
$N_{\mathrm{peak}}$ defined in Section \ref{sec:measures-of-negativity}.
$N_{\mathrm{vol}}$ is computed using a Riemann sum in a bounded region
centered on the origin to approximate the definite integral. We define
an $N_{x}$-by-$N_{y}$ grid of points $(x_{i},y_{j})$ and write
the distance between adjacent points as
\begin{subequations}
\label{eq:-78}
\begin{align}
x_{i+1}-x_{i} & =\Delta x,\\
y_{i+1}-y_{i} & =\Delta y.
\end{align}
\end{subequations}
To center the region of integration on the origin, we define the extent
of the grid $(x_{\mathrm{ext}},y_{\mathrm{ext}})$ such that
\begin{subequations}
\label{eq:-78-1}
\begin{align}
x_{N_{x}} & =x_{\mathrm{ext}}=-x_{1},\\
y_{N_{y}} & =y_{\mathrm{ext}}=-y_{1},
\end{align}
\end{subequations}
Together with (\ref{eq:-78}) and (\ref{eq:-78-1}), any two of the
three pairs $(N_{x},N_{y})$, $(x_{\mathrm{ext}},y_{\mathrm{ext}})$
and $(\Delta x,\Delta y)$ specify the grid uniquely. Using the Cartesian
coordinates for the Wigner function, the approximation to the integral
of $N_{\mathrm{vol}}$ of the state $\hat{\rho}$ is then written
as 

\begin{equation}
N_{\mathrm{vol}}=-\sum_{i=1}^{N_{x}}\sum_{j=1}^{N_{y}}\min\left\{ 0,W_{\hat{\rho}}(x_{i},y_{j})\right\} \Delta x\Delta y.\label{eq:-236}
\end{equation}
$W_{\hat{\rho}}(x_{i},y_{j})$ is evaluated using (\ref{eq:-234})
and (\ref{eq:-235}). $N_{\mathrm{peak}}$ is also evaluated using
a grid:

\begin{equation}
N_{\mathrm{peak}}=-\min_{\substack{1\leq i\leq N_{x}\\
1\leq j\leq N_{y}
}
}\left\{ \min\left\{ 0,W_{\hat{\rho}}(x_{i},y_{j})\right\} \right\} .\label{eq:-237}
\end{equation}

To determine a sufficient extent and refinement of the grid, the convergence
of the expressions (\ref{eq:-236}) and (\ref{eq:-237}) were investigated
with respect to $N$, $(\Delta x,\Delta y)$ and $(x_{\mathrm{ext}},y_{\mathrm{ext}})$
separately. In practice, the evaluation of $N_{\mathrm{vol}}$ and
$N_{\mathrm{peak}}$ to similar accuracy was found to require similar
values of $(\Delta x,\Delta y)$. Hence, the evaluations performed
in (\ref{eq:-236}) were reused in the computation of $N_{\mathrm{vol}}$
as given by (\ref{eq:-236}). The convergence was assessed with respect
to Gaussian states of known negativity ($N_{\mathrm{vol}}=N_{\mathrm{peak}}=0$)
and also sample simulated states with nonzero negativity.

Specifically in the case of a squeezed vacuum or thermal state, the
necessary minimum values of the parameters $(x_{\mathrm{ext}},y_{\mathrm{ext}})$
are expected vary with the largest variance of the Gaussian function
(e.g. the variance of the anti-squeezed quadrature). This was verified
for various values of the squeezing parameter $r_{0}$ from the interval
$[0.5,2.5]$.

\chapter{Unitary Oscillator Dynamics \label{chap:nonlinear-oscillators}}

The current chapter is dedicated to the understanding of the ways
in which the negative regions of the Wigner function form in the Kerr
oscillator. We use the phenomenon of nonlinear oscillators in the
field of quantum optomechanics to motivate and derive the system Hamiltonian.
The rest of the chapter considers exclusively the unitary dynamics
of the Kerr oscillator which is general to many quantum systems beside
optomechanical ones.

Dynamics general to all initial states are discussed in Section \ref{sec:Kerr-Oscillator}.
In particular, the periodicity of the Kerr oscillator is shown. From
there, we move on to study the dynamics of specific initial states,
starting with the trivial case of the vacuum state in Section \ref{sec:Kerr-Evolution-of}.

We then consider as initial state the squeezed vacuum state. This
state as well as related states form the basis for most discussion
in this thesis and a large part of the chapter is therefore dedicated
to their treatment. The evolution of the Wigner function and its negativity
throughout the period is discussed. We then consider the evolution
over short times for which a universal scaling behavior for the negativity
is observed and characterized. We also construct a Fourier space solution
of the Wigner function for large squeezing.

The conclusions drawn from the squeezed vacuum for short times are
readily generalized to thermal states, squeezed below the vacuum state
variance. This is done in Section \ref{sec:squeezed-thermal-state}.
The section finishes with a discussion of the applicability to squeezed
thermal states that obey the standard quantum limit. Section \ref{sec:Coherent-Initial-State}
finally considers the coherent state dynamics, serving mainly as perspective
for the results obtained for the other initial states.

\section{Nonlinear Resonators\label{sec:Nonlinear-Oscillators}}

Nonlinear effects are visible in many physical systems. Within the
field of quantum optics they are found systems such as fibers \cite{Levenson_SqueezingClassicalNoise_1985}
and trapped ions \cite{Home_NormalModesTrapped_2011}. We focus here
on nanomechanical oscillators. These take on many forms \cite{Bowen_QuantumOptomechanics_2015,Aspelmeyer_CavityOptomechanics_2014},
including silicon nitride membranes and strings, microtoroidal optomechanical
cavities and photonic-phononic systems and Fabry--Pérot cavities
with a membrane in the middle or as one of its mirrors. It is common
to model each of these systems quantum mechanically as a particle
in an harmonic potential (such as (\ref{eq:-213})). Several phenomena
can however give rise to an anharmonic potential which can not modeled
in this way. In nanonechanical systems, such potential anharmonicities
can for example arise from intrinsic material properties, the deformation
of the oscillator as it vibrates or electrostatic displacement \cite{Schmid_FundamentalsNanomechanicalResonators_2016}
(see also Table \ref{tab:Nonlinear-systems.-Parameters}). For oscillations
that are small in amplitude such as those of quantum fluctuations,
the potential takes the form of 

\begin{equation}
V(\hat{q})=\frac{1}{2}m\omega^{2}\hat{q}^{2}+\frac{\beta}{4}\hat{q}^{4}.
\end{equation}
This potential describes the Duffing oscillator \cite{Babourina-Brooks_QuantumNoiseNanomechanical_2008}.
The quantity $\beta$ is the Duffing parameter and has appropriate
dimensions such that $\beta\hat{q}^{4}$ takes the form of an energy
(see Appendix \ref{chap:Interaction-Picture-and}). For the systems
studied in quantum optics, the effects of the harmonic contribution
to the potential usually happen on a much shorter time scale than
those of the anharmonic contribution (this statement will shortly
be formalized as (\ref{eq:-214})). It is therefore useful to consider
the system in a rotating frame and with the rotating wave approximation
as this removes the harmonic contribution from $\hat{H}$ and simplifies
the remaining expression. Practically, the rotating frame expressions
are found by transforming to the interaction picture with the base
Hamiltonian $\hbar\omega\hat{a}^{\dagger}\hat{a}$. The rotating wave
approximation is then made by removing all terms with an explicit
phase that oscillates with a multiple of the base frequency $\omega$.
Appendix \ref{chap:Interaction-Picture-and} details these steps.
The result is a Hamiltonian of the form
\begin{equation}
\hat{H}=\hbar g\hat{a}^{\dagger}\hat{a}^{\dagger}\hat{a}\hat{a}\label{eq:kerr-hamiltonian}
\end{equation}
where the identification

\begin{equation}
g=\frac{3\hbar\beta}{8m^{2}\omega^{2}}\label{eq:-215}
\end{equation}
has been made. $\hat{H}$ is called the Kerr Hamiltonian.\footnote{Using the commutation relation (\ref{eq:canonical-commutation}) and
disregarding added terms proportional two or less ladder operators
$\hat{a}^{\dagger}$ and $\hat{a}$, any expression consisting of
two creation and two annihilation operators may be written in the
form (\ref{eq:kerr-hamiltonian}). Hence, any such Hamiltonian would
be referred to as the Kerr Hamiltonian. Here, we choose to keep the
Hamiltonian normal ordered as seen in (\ref{eq:kerr-hamiltonian}).} The earlier requirement that the dynamics arising from the harmonic
contribution to the potential have much shorter time scales than those
from the anharmonic contributions (required for the validity of the
rotating wave approximation) can now be expressed as

\begin{equation}
\omega\gg g.\label{eq:-214}
\end{equation}

\section{Kerr Oscillator\label{sec:Kerr-Oscillator}}

We take now as the system under investigation the Hamiltonian derived
in the previous section:
\[
\hat{H}=\hbar g\hat{a}^{\dagger}\hat{a}^{\dagger}\hat{a}\hat{a},\tag{{\ref{eq:kerr-hamiltonian}}}
\]
where $\hat{a}$ is the annihilation operator of a bosonic mode and
$g$ is the frequency describing the strength of the Kerr nonlinearity.
It should be noted that commuting the operators of the first term
using the canonical commutation relation (\ref{eq:canonical-commutation}),
renders (\ref{eq:kerr-hamiltonian}) in the form
\begin{equation}
\hat{H}=\hbar g(\hat{n}^{2}-\hat{n}).\label{eq:-216}
\end{equation}
In the form of (\ref{eq:-216}), $\hat{H}$ is manifestly diagonal
in the basis of number states $|n\rangle$. In other words, $\hat{H}$
shares eigenstates with the harmonic oscillator. It is therefore trivial
to apply the time evolution to a state expanded in the number state
basis to obtain an expression similar in character to (\ref{eq:-217}).

\subsection{Periodic Evolution \label{sec:operator-unitary-evolution}}

In the operator formalism, evolution of the system can be described
with the unitary time-evolution operator. Inserting $\hat{H}$ into
(\ref{eq:unitary-propagator}) yields
\begin{equation}
\hat{U}(t)=e^{-ig(\hat{n}^{2}-\hat{n})t}.\label{eq:-65}
\end{equation}
The system evolution is periodic for any initial state which may be
seen from the formal solution given an arbitrary initial state $|\Psi(0)\rangle$.
We expand the state in the basis of number states as
\begin{equation}
|\Psi(0)\rangle=\sum_{n}c_{n}|n\rangle.\label{eq:-144}
\end{equation}
Using (\ref{eq:propagated-ket}) with the expansion of the initial
state yields
\begin{equation}
|\Psi(t)\rangle=\hat{U}(t)|\Psi(0)\rangle=\sum_{n}e^{-ig(n^{2}-n)t}c_{n}|n\rangle.\label{eq:-87}
\end{equation}
Inserting $t=\pi/g$ and rewriting the exponential as
\begin{equation}
e^{-i\pi n^{2}}e^{i\pi n}=(-1)^{n^{2}}e^{i\pi n}=(-1)^{n}e^{i\pi n}=e^{-i\pi n}e^{i\pi n}=1,\label{eq:-152}
\end{equation}
it is seen that
\begin{equation}
|\Psi(\pi/g)\rangle=\sum_{n}e^{-i\pi n}e^{i\pi n}c_{n}|n\rangle=|\Psi(0)\rangle.\label{eq:-145}
\end{equation}
Hence the system is periodic with a period of $\pi/g$.\footnote{In fact any quantum system with discrete energy levels is at least
approximately periodic \cite{Bocchieri_QuantumRecurrenceTheorem_1957}.
The analysis of (\ref{eq:-152}--\ref{eq:-145}) is trivially extended
to show exact periodicity (as in (\ref{eq:-145})) for any Hamiltonian
which is a polynomial function of $\hat{n}$. Other systems exhibiting
exact periodicity also exists, e.g. the infinite square well.}

\subsection{Generation of Superposition States}

Halfway through the period when $t=\pi/2g$, the system evolves to
form a balanced superposition of two instances of the initial state,
rotated to be out of phase by 180°. This can be shown by adapting
an argument due to \textcite{Yurke_GeneratingQuantumMechanical_1986}.
Consider again the evolution of an initial state $|\Psi(0)\rangle$.
Using (\ref{eq:-65}), $|\Psi(\pi/2g)\rangle$ is written
\begin{equation}
|\Psi(\pi/2g)\rangle=e^{-i(\hat{n}^{2}-\hat{n})\pi/2}|\Psi(0)\rangle.
\end{equation}
We apply again the expansion in the number state basis to write
\begin{subequations}
\label{eq:-146}
\begin{align}
|\Psi(\pi/2g)\rangle & =\sum_{n}e^{-i(n^{2}-n)\pi/2}c_{n}|n\rangle\\
 & =e^{-i\hat{n}\pi/2}\sum_{n}e^{-in^{2}\pi/2}c_{n}|n\rangle.\label{eq:-148}
\end{align}
\end{subequations}
Notice now that
\begin{equation}
e^{-in^{2}\pi/2}=\frac{e^{-i\pi/4}+(-1)^{n}e^{i\pi/4}}{\sqrt{2}}=\begin{cases}
-i & \text{\ensuremath{n} odd,}\\
1 & \text{\ensuremath{n} even,}
\end{cases}\label{eq:-147}
\end{equation}
for integer $n$. Using (\ref{eq:-147}) on (\ref{eq:-148}), we continue\footnote{Alternately, to move from (\ref{eq:-151}) to (\ref{eq:-150}), notice
the effect of rotating a number state 180°. Combining (\ref{eq:fock-state-def})
and (\ref{eq:-149}), we write $\hat{R}(\pi)|n\rangle=(n!)^{-1/2}(\hat{R}(\pi)\hat{a}^{\dagger}\hat{R}^{\dagger}(\pi))^{n}\hat{R}(\pi)|0\rangle=(-1)^{n}|n\rangle$
since $\hat{R}(\pi)\hat{a}^{\dagger}\hat{R}^{\dagger}(\pi)=-\hat{a}^{\dagger}$
and $\hat{R}(\pi)|0\rangle=|0\rangle$ (see e.g. (\ref{eq:rotation-operator})). }
\begin{subequations}
\label{eq:-146-1}
\begin{align}
|\Psi(\pi/2g)\rangle & =e^{-i\hat{n}\pi/2}\sum_{n}\frac{\left(e^{-i\pi/4}+(-1)^{n}e^{i\pi/4}\right)}{\sqrt{2}}c_{n}|n\rangle\label{eq:-151}\\
 & =\frac{e^{-i\hat{n}\pi/2}}{\sqrt{2}}\left(e^{-i\pi/4}\sum_{n}c_{n}|n\rangle+e^{i\pi/4}\sum_{n}e^{-in\pi}c_{n}|n\rangle\right)\label{eq:-148-1}\\
 & =\frac{1}{\sqrt{2}}\left(e^{-i\pi/4}e^{-i\hat{n}\pi/2}|\Psi(0)\rangle+e^{i\pi/4}e^{i\hat{n}\pi/2}|\Psi(0)\rangle\right).\label{eq:-150}
\end{align}
\end{subequations}
(\ref{eq:-150}) expresses $|\Psi(\pi/2g)\rangle$ as a superposition
of the states $e^{-i\hat{n}\pi/2}|\Psi(0)\rangle$ and $e^{i\hat{n}\pi/2}|\Psi(0)\rangle$.
These are, apart from a global phase, exactly a superposition of two
instances of the initial state $|\Psi(0)\rangle$ rotated to be 180°
out of phase. Note however, that $|\Psi(\pi/2g)\rangle$ may not be
a “true superposition”.\footnote{Of course, the term “true superposition” is somewhat ill-defined
since a change of basis allows any ket to be expressed as a linear
combination of more than one basis kets. Here, we shall take the term
to mean that the superposition state cannot be written by simply transforming
the initial state using the operators of Section \ref{sec:operator-transformations}
and addition of a global phase.} For example, with a number state as initial state $|\Psi(0)\rangle=|n\rangle$,
we obtain
\begin{equation}
e^{-i(\hat{n}^{2}-\hat{n})\pi/2}|n\rangle=\frac{1}{\sqrt{2}}\left(e^{-i\pi/4}e^{-in\pi/2}|n\rangle+e^{i\pi/4}e^{in\pi/2}|n\rangle\right)=\sqrt{2}\Re\left(e^{-i\pi/4}e^{-in\pi/2}\right)|n\rangle=|n\rangle,
\end{equation}
which is clearly not a state any more exotic than the initial state
(it \emph{is} the initial state). Other examples include the the vacuum
state and the superposition formed between the vacuum state and any
single other number state.

A less trivial but still nonconforming example is given by the squeezed
vacuum state $|\xi\rangle$ as defined in (\ref{eq:-124}). The squeezed
vacuum state does not possess continuous rotational symmetry like
the number states, but it does exhibit a discrete rotational symmetry
of exactly 180°. When $t=\pi/2$ we therefore recover the original
state with an added global phase. In the case of the squeezed vacuum
state, a nontrivial balanced superposition is however achieved when
$t=\pi/8g$. Evolution of the squeezed vacuum state is examined in
Section \ref{sec:squeezed-vacuum-initial-state}.

For a conforming example, the evolution of a coherent state under
the Kerr Hamiltonian does generate a superposition at time $t=\pi/2g$.
The periodic evolution of the coherent state is discussed in Section
\ref{subsec:coherent-state-periodic-evolution}.

\subsection{Equation of Motion for the Wigner Function \label{subsec:Wigner-Function-Equation}}

To explore the non-classical aspects of the evolution of the Kerr
oscillator, it is useful to study the equation of motion describing
the evolution of the Wigner function directly. Starting from the von
Neumann equation for the system, one can derive a partial differential
equation for the Wigner function which describes the same dynamics.
The von Neumann equation for the system at hand is found as
\begin{equation}
\dot{\hat{\rho}}=-ig\left[\hat{a}^{\dagger}\hat{a}^{\dagger}\hat{a}\hat{a},\hat{\rho}\right]\label{eq:kerr-vonneumann-eq}
\end{equation}
by insertion of the Kerr Hamiltonian (\ref{eq:kerr-hamiltonian})
into (\ref{eq:von-neumann-equation}). Section \ref{sec:wigner-pde-derivation}
describes a procedure for obtaining the equation of motion for $W$
given (\ref{eq:kerr-vonneumann-eq}). The result is most easily expressed
in polar coordinates (cf. Section \ref{sec:phase-space-coordinates})
as
\begin{equation}
\partial_{t}W(r,\phi,t)=2g\left(r^{2}-1\right)\partial_{\phi}W(r,\phi,t)-\frac{g}{8}\nabla^{2}\partial_{\phi}W(r,\phi,t).\label{eq:-18}
\end{equation}
Further details of this derivation may be found in Appendix \ref{app:derivations-of-c-eqs}.
Before we proceed, note that all terms on the right hand side are
linear in $g$. This is expected since $g$ is the only frequency
of the problem as is seen in (\ref{eq:kerr-vonneumann-eq}). This
means that system evolution can be considered as a function of a rescaled
time 
\begin{equation}
(\text{rescaled time})=gt,
\end{equation}
eliminating the parameter $g$ from the system dynamics.

Let us now break down the contents of equation (\ref{eq:-18}). It
is useful to organize the terms by the order of the spatial derivative
of $W$. Look first to the terms in which first-order spatial derivatives
of $W$ appear. These are the terms shared with the classical Liouville
equation. The first of these is $2gr^{2}\partial_{\phi}$. One might
describe this as the introduction of a radially dependent angular
frequency. It contains purely first-order spatial derivatives and
thus causes a flow at every point proportional to the Wigner function
at that point. The second term, $-2g\partial_{\phi}$, is simply an
additional oscillator frequency which  causes the Wigner function
to rotate rigidly in phase space.\footnote{The term $-2g\partial_{\phi}W(r,\phi,t)$ corresponds to the right
hand side of the simple harmonic oscillator equation of motion. This
oscillator has frequency $\omega=-2g$ (see Appendix \ref{app:derivations-of-c-eqs}).
By choosing a frame rotating at the proper frequency, one can change
the subexpression $\left(r^{2}-1\right)$ of (\ref{eq:-18}) to $\left(r^{2}-k\right)$
for any desired real constant $k$. The appearance of $1$ could be
regarded as the consequence of the choice of the normally ordered
Hamiltonian (\ref{eq:kerr-hamiltonian}). Had $\hbar g\hat{a}^{\dagger}\hat{a}\hat{a}^{\dagger}\hat{a}$
been used instead, the $1$ would have vanished. This could also be
considered to be an adjustment of the angular frequency of the rotating
frame.} This term is usually not included in the classical Kerr oscillator,
since it is the result of the choice ordering of ladder operators
in (\ref{eq:kerr-hamiltonian}).

Since (\ref{eq:-18}) describes the unitary evolution, it is expected
to contain no even-order spatial derivatives (cf. Section \ref{sec:Gaussian-States-and}
or \textcite{Corney_NonGaussianPureStates_2015}). This is indeed
found to be the case. Even-order terms arising from non-unitary evolution
will be considered in the next chapter.

The final term $2g\nabla^{2}\partial_{\phi}$, contains third-order
spatial derivatives and is the only term in (\ref{eq:-18}) to do
so. For this reason, one might say that this term is non-local \cite{Oliva_QuantumKerrOscillators_2019}
and indeed, this term describes an effect that cannot be captured
in the evolution of a classical phase space probability distribution.
Without this term, no negative regions could form in the Wigner function.
In case of the Kerr oscillator, this is the only such term.

Equation (\ref{eq:-18}) can also be stated as a continuity equation
as introduced in Section \ref{sec:wigner-current}. This is done by
choosing the Wigner current $\mathbf{J}$ such that (\ref{eq:-18})
assumes the form of (\ref{eq:quantum-continuity-equation}). The term
$(g/8)\nabla^{2}\partial_{\phi}$ contains both $\partial_{\phi}$
and $\partial_{r}$ so one has some freedom \cite{Oliva_QuantumKerrOscillators_2019}
in how to represent it in the expression of $\mathbf{J}$. The form
of (\ref{eq:-18}) suggests placing the entire current in the $\phi$-component
of $\mathbf{J}$ though.\footnote{There are in fact infinitely many valid choices for distributing the
contribution of the term $(g/8)\nabla^{2}\partial_{\phi}$ between
$J_{r}$ and $J_{\phi}$. Reference \cite{Oliva_QuantumKerrOscillators_2019}
expresses these choices using a continuous parameter. For the purposes
of the current considerations, it is sufficient to simply take the
current as being parallel to $\hat{\symbf{\phi}}$. For most quantum
states, a large contribution will anyway come from the right hand
side term $2g\left(r^{2}-1\right)\partial_{\phi}W$ whose current
terms are unambiguously parallel to $\hat{\symbf{\phi}}$.} We thus write
\begin{equation}
\mathbf{J}=\left(-2g\left(r^{2}-1\right)W+\frac{g}{8}\nabla^{2}W\right)r\hat{\symbf{\phi}}+0\cdot\hat{\mathbf{r}}.\label{eq:-101}
\end{equation}
With this choice of $\mathbf{J}$, we can view the dynamics as a circular
flow around the origin. Supporting this view, the Wigner density $W$
is preserved on rings around the origin \cite{Oliva_QuantumKerrOscillators_2019}:

\begin{equation}
\partial_{t}\oint d\phi\,W=-\oint d\phi\,\nabla\cdot\mathbf{J}=0.\label{eq:-100}
\end{equation}
(\ref{eq:-100}) should be treated with caution however, since the
current does not evolve independently on each ring. The appearance
of $\partial_{r}$ and $\partial_{r}^{2}$ in (\ref{eq:-101}) means
that current $\mathbf{J}$ on a ring depends on $W$ on adjacent rings.
For this reason, the flow in quantum phase space has been called “viscous”
\cite{Oliva_DynamicShearSuppression_2019,Oliva_QuantumKerrOscillators_2019}
when compared to the classical phase space flow.

\section{Kerr Evolution of Vacuum State\label{sec:Kerr-Evolution-of}}

As a trivial example demonstrating the phase space dynamics, consider
the vacuum state $|0\rangle$ as initial state. In the operator picture,
this is seen to be constant under the time-evolution described by
$\hat{U}(t)$ (equations (\ref{eq:unitary-propagator}) and (\ref{eq:propagated-ket})):
\begin{equation}
\hat{\rho}(t)=e^{-ig(\hat{n}^{2}-\hat{n})t}|0\rangle=|0\rangle.\label{eq:-77}
\end{equation}
We can draw the same conclusions from the phase space picture. Recall
from Section \ref{sec:phase-space-coordinates}, the corresponding
vacuum state Wigner function
\begin{equation}
W_{|0\rangle}(r,\phi)=\frac{2}{\pi}e^{-2r^{2}}.\label{eq:-199}
\end{equation}
$W_{|0\rangle}(r,\phi)$ is an isotropic Gaussian function centered
at the origin and with a variance $\frac{1}{4}$ (measured in both
phase space coordinates $x$ and $y$). The right hand side of the
equation of motion (\ref{eq:-19}) is seen to vanish when applied
to $W_{|0\rangle}(r,\phi)$ due to the occurrence of the factor 
\begin{equation}
\partial_{\phi}W_{|0\rangle}(r,\phi)=0
\end{equation}
 in every term. For the initial state 
\begin{equation}
W(r,\phi,0)=W_{|0\rangle}(r,\phi),
\end{equation}
one has therefore that
\begin{equation}
\partial_{t}W(r,\phi,0)=0.\label{eq:-88}
\end{equation}
Since the time-evolution of $W(r,\phi,t)$ is governed by a differential
equation that is first-order in $t$, (\ref{eq:-88}) completely fixes
the evolution of the vacuum state (\ref{eq:vacuum-wigner-polar})
to 
\begin{equation}
W(r,\phi,t)=W(r,\phi,0).
\end{equation}
This matches the conclusion drawn from (\ref{eq:-77}). In geometrical
terms it may be concluded by recalling the current $\mathbf{J}$ from
(\ref{eq:-101}). As seen from this equation, $\mathbf{J}$ describes
a flow in the angular direction. Since the vacuum state Wigner function
is rotationally invariant: $W_{|0\rangle}(r,\phi)=W_{|0\rangle}(r,0)$,
$\mathbf{J}$ thus leads to no change in $W$.

\section{Kerr Evolution of Squeezed Vacuum \label{sec:squeezed-vacuum-initial-state}}

\begin{figure}
\noindent \begin{centering}
\includegraphics{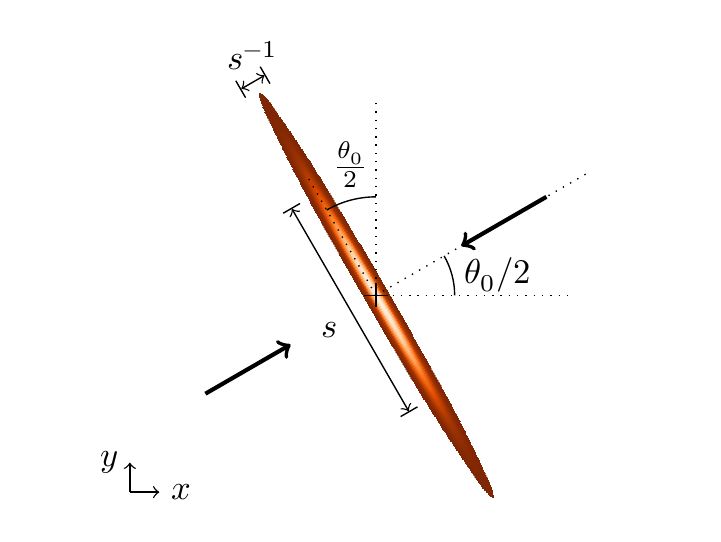}
\par\end{centering}
\caption[Illustration of the squeezed vacuum Wigner function]{\label{fig:evo-gallery-medium-1-1}\textbf{Illustration of the squeezed
vacuum Wigner function.} The Wigner function (\ref{eq:-85}) for the
squeezed vacuum state is shown. The squeezed axis (indicated by two
arrows) is rotated $\theta_{0}/2$ from the $x$-axis. The anti-squeezed
axis is rotated the same angle from the $y$-axis. The shown dimensions
are not to scale. The squeezed vacuum state Wigner function is discussed
in Section \ref{sec:squeezed-vacuum-initial-state}.}
\end{figure}
We continue with a generalization of the previous example and consider
as initial state the squeezed vacuum state of (\ref{eq:-124}): 
\begin{equation}
|\Psi(0)\rangle=|\xi\rangle\label{eq:-73}
\end{equation}
The squeezing parameter $\xi$ is written explicitly in terms of its
modulus and argument $r_{0}$ and $\theta_{0}$. The state $|\xi\rangle$
can be expanded in the number-state basis as \cite{Scully_QuantumOptics_1997}
\begin{equation}
|\xi{=}r_{0}e^{i\theta_{0}}\rangle=\frac{1}{\sqrt{\cosh r_{0}}}\sum_{m=0}^{\infty}(-1)^{m}\frac{\sqrt{(2m)!}}{2^{m}m!}e^{im\theta_{0}}\left(\tanh r_{0}\right)^{m}|2m\rangle.\label{eq:-86}
\end{equation}
When writing the Wigner function for the squeezed vacuum state, it
is convenient to define the parameter $s$ as
\begin{equation}
s=e^{r_{0}}.\label{eq:-228}
\end{equation}
We term $s$ simply “squeezing” to distinguish it from the squeezing
parameter $\xi$. The value of $s$ for the the squeezing parameters
$r_{0}$ that we use later are listed in Table \ref{tab:Scaled-negativity-decay-1}
for reference. The Wigner function for (\ref{eq:-73}) can now be
found by combining (\ref{eq:vacuum-wigner-polar}) and (\ref{eq:-114-1-1}).
In Cartesian coordinates, the resulting initial state Wigner function
is written
\begin{equation}
W_{|\xi\rangle}(x,y)=\frac{2}{\pi}\exp\left(-2s^{2}\left(x\cos\frac{\theta_{0}}{2}+y\sin\frac{\theta_{0}}{2}\right)^{2}-\frac{2}{s^{2}}\left(x\sin\frac{\theta_{0}}{2}-y\cos\frac{\theta_{0}}{2}\right)^{2}\right).\label{eq:-85}
\end{equation}
(\ref{eq:-85}) defines an elliptical Gaussian function in phase space
with its major and minor axes rotated an angle $\theta_{0}/2$ from
the $y$ and $x$ axes respectively. This is illustrated in Figure
\ref{fig:evo-gallery-medium-1-1}. Since the equation of motion (\ref{eq:-18})
lack dependence on the angular coordinate $\phi$, the system dynamics
are concluded to be rotationally invariant. We may therefore disregard
the parameter $\theta_{0}$ and simply set $\theta_{0}=0$ without
loss of generality. The change in parameters $s\to1/s$ and $\theta_{0}\to\theta_{0}+\pi/2$
leaves the Wigner function invariant. Hence the change $s\to1/s$
also corresponds to a rotation of the phase space coordinate system
and we may assume $s\geq1$ without loss of generality. This leaves
us with the initial state
\begin{equation}
|\Psi(0)\rangle=|\xi{=}r_{0}\rangle\label{eq:-225}
\end{equation}
which has the Wigner function
\begin{equation}
W(x,y,0)=\frac{2}{\pi}e^{-2s^{2}x^{2}-2y^{2}/s^{2}}.\label{eq:squeezed-vacuum-initial-state}
\end{equation}

\begin{table}
\renewcommand{\arraystretch}{1.5}
\noindent \centering{}%
\begin{tabular}{|c|c|}
\hline 
$r_{0}$ & $s$\tabularnewline
\hline 
\hline 
$0.5$ & $1.65$\tabularnewline
\hline 
$0.75$ & $2.12$\tabularnewline
\hline 
$1$ & $2.72$\tabularnewline
\hline 
\end{tabular}\qquad%
\begin{tabular}{|c|c|}
\hline 
$r_{0}$ & $s$\tabularnewline
\hline 
\hline 
$1.25$ & $3.49$\tabularnewline
\hline 
$1.5$ & $4.48$\tabularnewline
\hline 
$1.75$ & $5.75$\tabularnewline
\hline 
\end{tabular} \qquad%
\begin{tabular}{|c|c|}
\hline 
$r_{0}$ & $s$\tabularnewline
\hline 
\hline 
$2$ & $7.39$\tabularnewline
\hline 
$2.25$ & $9.49$\tabularnewline
\hline 
$2.5$ & $12.18$\tabularnewline
\hline 
\end{tabular}\caption[Relation between often used squeezing parameters and squeezing]{\label{tab:Scaled-negativity-decay-1}\textbf{Relation between often
used squeezing parameters and squeezing. }The relation between the
squeezing parameter $r_{0}$ and the squeezing $s=\exp r_{0}$ is
defined by (\ref{eq:-228}).}
\end{table}

\subsection{Periodic Evolution}

\begin{figure}
\noindent \begin{centering}
\includegraphics{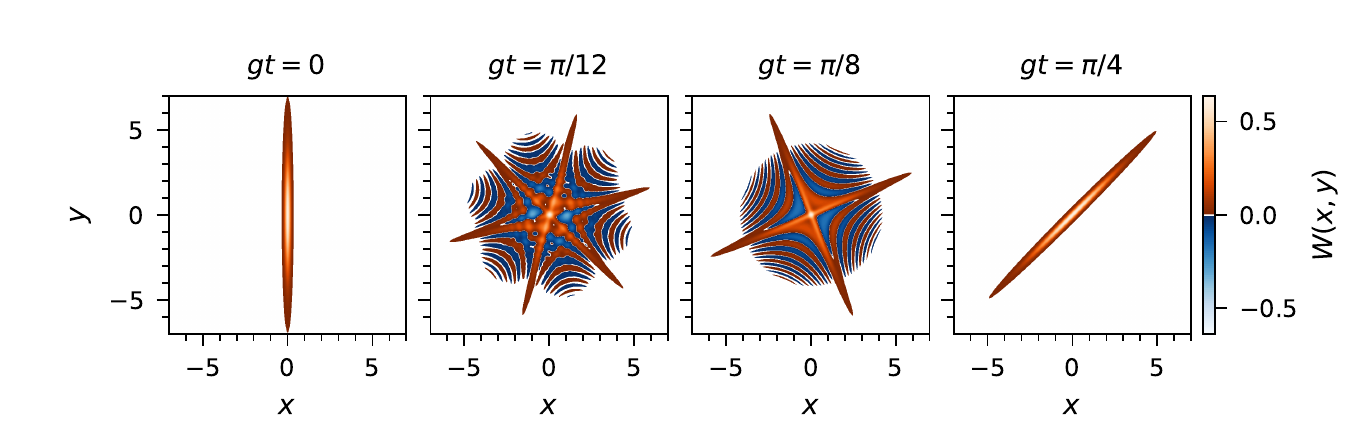}
\par\end{centering}
\caption[Notable states during unitary evolution of squeezed vacuum]{\label{fig:evo-gallery-long}\textbf{Notable states during unitary
evolution of squeezed vacuum. }The initial state is a squeezed vacuum
state (\ref{eq:-124}) with $\xi=1.5$. Contour plots show $W(x,y,t)$
at points of fractional revival, $t=\pi/8g$ (with the state (\ref{eq:-154}))
and $t=\pi/12g$. Also shown are the initial state ($gt=0$) and the
state after one period ($gt=\pi/4$). After one period, $t=\pi/4g$,
the Wigner function has been rotated by $\pi/4$ as described by (\ref{eq:-153}).}
\end{figure}

As stated in Section \ref{sec:operator-unitary-evolution}, the unitary
evolution described by (\ref{eq:-87}) is periodic with period $\pi/g$.
In particular, the periodicity of squeezed vacuum state evolution
is just $\pi/4g$. Re-purposing the arguments in Section \ref{sec:operator-unitary-evolution},
this is shown as follows: Consider the squeezed state in the number
state basis (\ref{eq:-86}). For this derivation, it is sufficient
that the squeezed vacuum state contains only even terms in its expansion
in the number state basis. That this is the case can be seen from
(\ref{eq:-86}). We shall express it here simply as
\begin{equation}
|\Psi(0)\rangle=\sum_{m}c_{2m}|2m\rangle.\label{eq:-144-1}
\end{equation}
The time evolution follows from (\ref{eq:-65}) and (\ref{eq:-144-1})
as
\begin{equation}
|\Psi(t)\rangle=\hat{U}(t)|\Psi(0)\rangle=\sum_{m}e^{-ig(4m^{2}-2m)t}c_{2m}|2m\rangle.
\end{equation}
We evolve the state to a time $t=\pi/4g$ using (\ref{eq:-65}) to
find
\begin{equation}
|\Psi(\pi/4g)\rangle=\sum_{m}e^{-i(m^{2}-m/2)\pi}c_{2m}|2m\rangle=e^{-i\hat{n}\pi/4}\sum_{m}e^{-i(m^{2}-m)}c_{2m}|2m\rangle.\label{eq:squeezed-periodicity}
\end{equation}
Using the identity (\ref{eq:-152}), we now have
\begin{equation}
|\Psi(\pi/4g)\rangle=e^{-i\hat{n}\pi/4}\sum_{m}c_{2m}|2m\rangle=e^{-i\hat{n}\pi/4}|\Psi(0)\rangle=\hat{R}(-\frac{\pi}{4})|\Psi(0)\rangle.\label{eq:-153}
\end{equation}
We see from (\ref{eq:-153}) that the squeezed vacuum state exhibits
a periodicity of only $\pi/4g$ (compared to the general periodicity
of $\pi/g$, see Section \ref{sec:operator-unitary-evolution}). The
periodicity can be seen demonstrated in Figure \ref{fig:evo-gallery-long}.

When time reaches a certain rational multiple of the period, special
states are observed. These are somewhat reminiscent of the fractional
revival states found when considering a coherent state evolving with
the same dynamics (see also Section \ref{sec:Coherent-Initial-State}).
Specifically at time $t=\pi/8$, the system state is a coherent superposition
of two squeezed vacuum states. To show this we re-purpose the arguments
of Section \ref{sec:operator-unitary-evolution}. Applying the time
evolution operator (\ref{eq:-65}) to the expansion in number states
(\ref{eq:-144-1}), we obtain
\begin{equation}
|\Psi(\pi/4g)\rangle=\sum_{m}e^{-i(4m^{2}-2m)\pi/8}c_{2m}|2m\rangle=e^{i\hat{n}\pi/8}\sum_{m}e^{-im^{2}\pi/2}c_{2m}|2m\rangle.\label{eq:-153-1}
\end{equation}
Then apply (\ref{eq:-147}) to obtain 
\begin{subequations}
\label{eq:-154}
\begin{align}
|\Psi(\pi/4g)\rangle & =e^{i\hat{n}\pi/8}\sum_{m}\frac{\left(e^{-i\pi/4}+(-1)^{m}e^{i\pi/4}\right)}{\sqrt{2}}c_{2m}|2m\rangle\label{eq:-153-1-1}\\
 & =\frac{e^{i\hat{n}\pi/8}}{\sqrt{2}}\left(e^{-i\pi/4}\sum_{m}c_{2m}|2m\rangle+e^{i\pi/4}\sum_{m}e^{-im\pi}c_{2m}|2m\rangle\right)\\
 & =\frac{1}{\sqrt{2}}\left(e^{-i\pi/4}e^{i\hat{n}\pi/8}|\Psi(0)\rangle+e^{i\pi/4}e^{-3i\hat{n}\pi/8}|\Psi(0)\rangle\right).
\end{align}
\end{subequations}
We thus see that the state evolves to a coherent superposition of
the states $e^{i\hat{n}\pi/8}|\Psi(0)\rangle$ and $e^{-3i\hat{n}\pi/8}|\Psi(0)\rangle$.
These are both instances of the initial state $|\Psi(0)\rangle=|\xi\rangle$
but rotated 90° out of phase. This state can also be seen in Figure
\ref{fig:evo-gallery-long}.

\subsection{Negativity during a Full Period\label{subsec:Negativity-during-a}}

\begin{figure}
\subfloat[\label{fig:negvol-period}Negative volume]{\noindent \centering{}\captionsetup{position=top}\centerline{\includegraphics{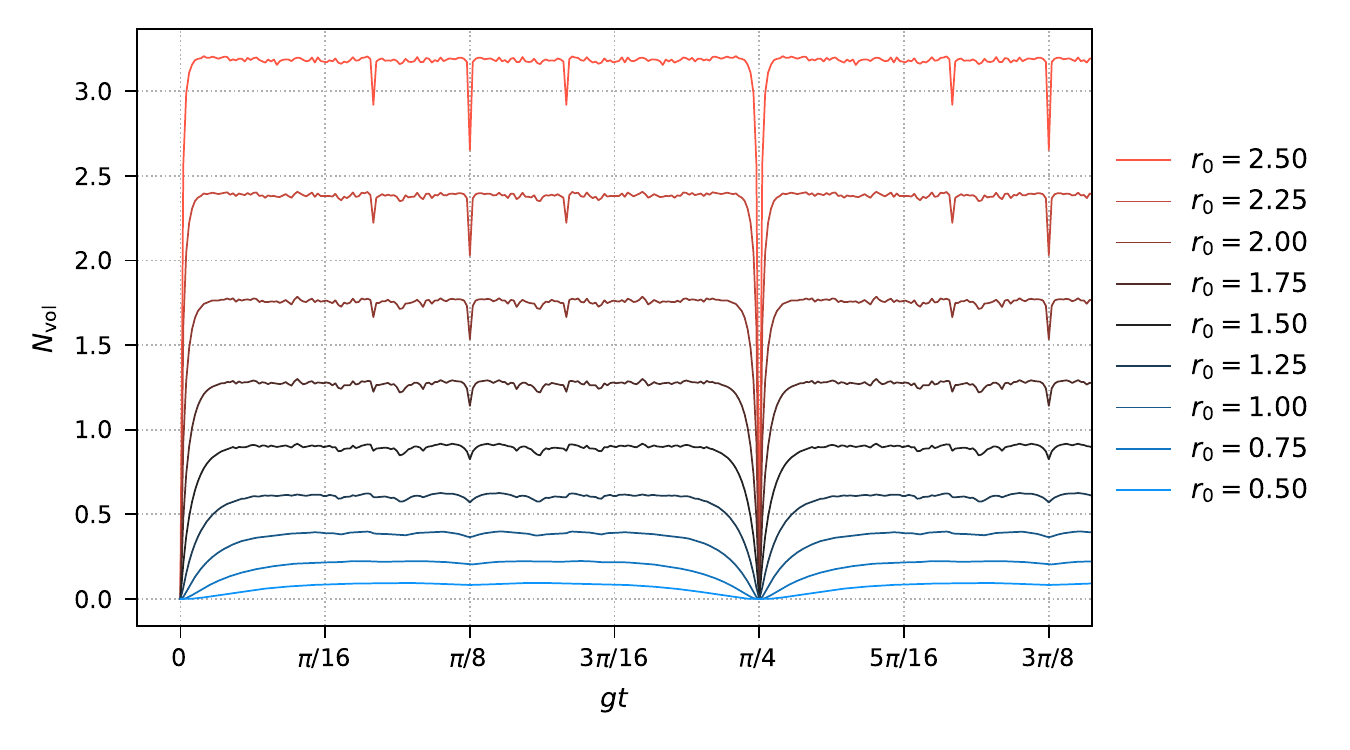}}}

\subfloat[\label{fig:negpeak-period}Negative peak]{\noindent \centering{}\captionsetup{position=top}\centerline{\includegraphics{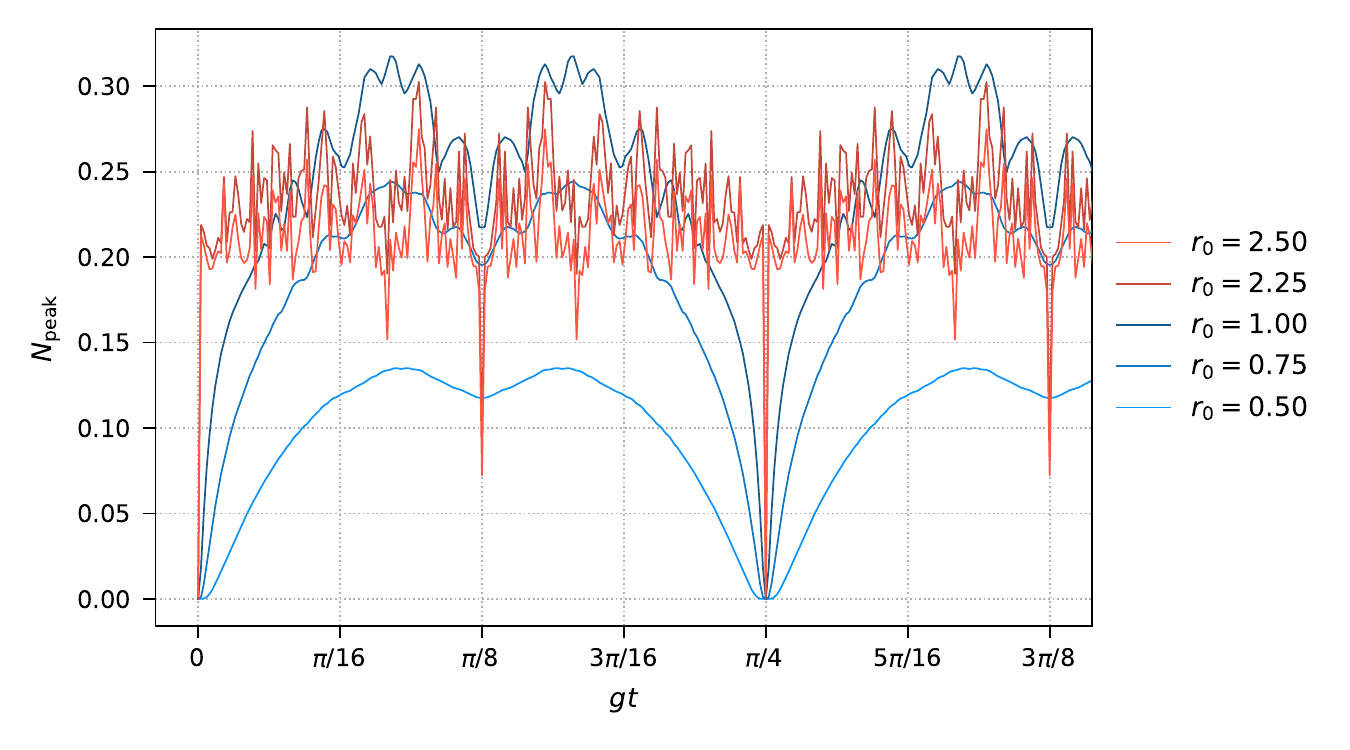}}}

\caption[Negativity during periodic evolution of squeezed vacuum]{\label{fig:}\textbf{Negativity during periodic evolution of squeezed
vacuum. }The initial states are squeezed vacuum states with varying
squeezing parameters $r_{0}$. To reduce clutter, the graph of negative
peak has been limited to a few different choices of $r_{0}$. }
\end{figure}

We now wish to characterize the evolution of negativity for the state.
To this end, $N_{\mathrm{vol}}$ and $N_{\mathrm{peak}}$ have been
plotted for an entire period in Figure \ref{fig:}.

\begin{figure}
\noindent \centering{}\centerline{\includegraphics{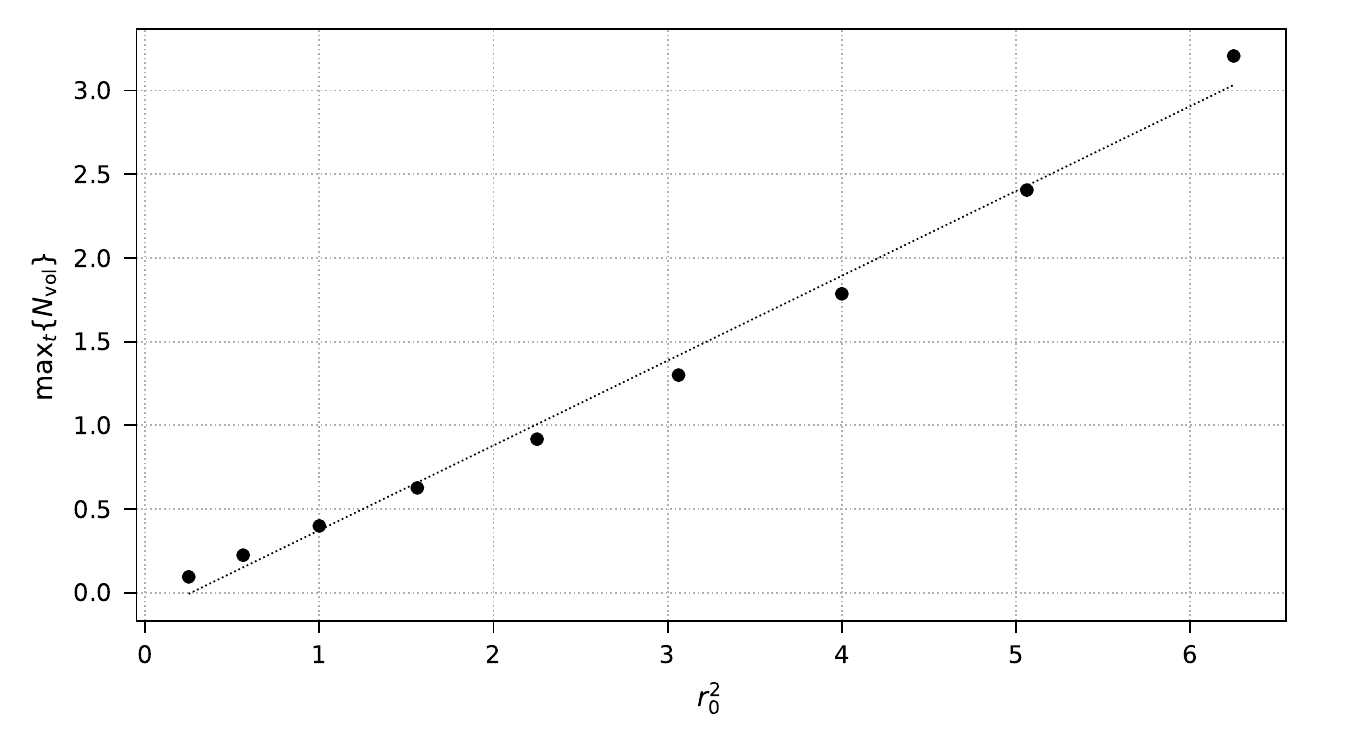}}\caption[Maximum negative volume for varying squeezing]{\label{fig:maxvol-period}\textbf{ Maximum negative volume for varying
squeezing.} The maximum negative volume \textbf{$\max_{t}\{N_{\mathrm{vol}}(t)\}$}
was computed from the simulations in Figure \ref{fig:negvol-period}.
The quantity scales roughly quadratically with $r_{0}$. The dotted
line is a linear function of $r_{0}$ to guide the eye.}
\end{figure}

We consider first the negative volume $N_{\mathrm{vol}}$. For larger
squeezing parameters, the negative volume increases rapidly until
it reaches a plateau. The plateau becomes more clear as $r_{0}$ increases
and the $N_{\mathrm{vol}}$ as a function of time takes on a more
square appearance. As the squeezing increases, $N_{\mathrm{vol}}$
additionally starts to fluctuate strongly and an increasing finer
structure of details appear in the plateau region. At special points
in time even larger features of $N_{\mathrm{vol}}$ become visible.
This is most evident when $gt=\pi/8$ and (to a lesser degree) when
$gt=\pi/12$. Here, a dip in the negative volume can be clearly made
out. The Wigner functions of these special states can be seen in Figure
\ref{fig:evo-gallery-long}. The height of the plateau appears to
scale roughly quadratically with $r_{0}$ (though not exactly). Figure
\ref{fig:maxvol-period} shows the height of the plateau $\max_{t}\left\{ N_{\mathrm{vol}}(t)\right\} $
as a function of the squared squeezing parameter $r_{0}^{2}$ for
$r_{0}\in[0,2.5]$. 

We next consider the negative peak $N_{\mathrm{peak}}$. This is shown
for a full period in Figure \ref{fig:negpeak-period}. $N_{\mathrm{peak}}$
generally increases with squeezing until around $r_{0}=1$. Further
increasing the squeezing from $r_{0}=1$ increases the frequency and
amplitude of the fluctuations but does not apparently increase the
peak negativity. Within the investigated parameter regime, the peak
negativity does not reach the bound of $2/\pi=0.637$ set by (\ref{eq:-82}).
As with $N_{\mathrm{vol}}$, the points $gt=\pi/8$ and $gt=\pi/12$
can be made out as dips in the graph of $N_{\mathrm{peak}}$. 

Some behavior is shared between $N_{\mathrm{vol}}$ and $N_{\mathrm{peak}}$.
Both start with a value of zero at $t=0$. This is expected since
the initial state is a Gaussian state. As the time first evolves,
they both increase rapidly and monotonically for some time. After
that, the evolution changes character and the negativity does not
clearly increase or decrease. We are mainly interested in this initial
evolution of the negativity where the rate at which the negativity
grows increases significantly with squeezing. 

\subsection{Evolution over Short Time\label{subsec:Evolution-over-Short}}

\begin{figure}
\noindent \begin{centering}
\includegraphics{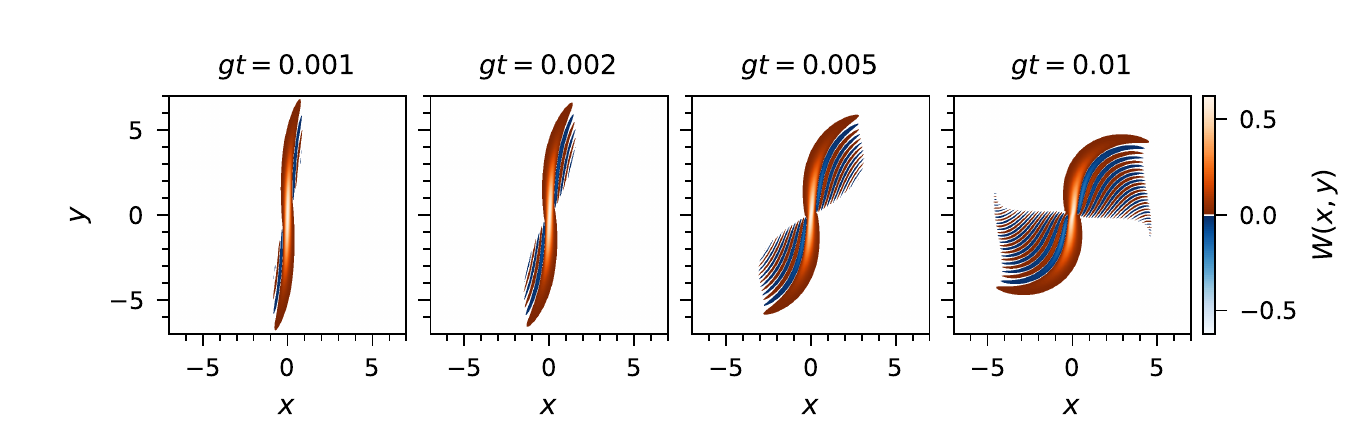}
\par\end{centering}
\caption[Short time unitary evolution of squeezed vacuum]{\label{fig:Squeezed-vacuum-state}\textbf{Short time unitary evolution
of squeezed vacuum.} Demonstration of short-time evolution for a squeezed
vacuum state. The squeezing parameter $r_{0}=1.5$ was used. As the
Wigner function evolves it forms an S-shape in phase space. The short
time evolution is discussed in Section \ref{subsec:Evolution-over-Short}.}
\end{figure}

We anticipate that the decoherence effects introduced in Chapter \ref{chap:coupling-to-the-environment}
will have a diminishing effect on the negativity, in some cases causing
the negativity to completely vanish before the plateau. Hence, we
shall focus on the initial stages of evolution. Figure \ref{fig:}
demonstrates that the rate of growth for the negativity increases
with squeezing and thus indicates that it may be possible to compensate
for strong decoherence effects by using states of stronger squeezing. 

\begin{figure}
\subfloat[\label{fig:-3}Negative volume]{\noindent \begin{centering}
\captionsetup{position=top}\includegraphics{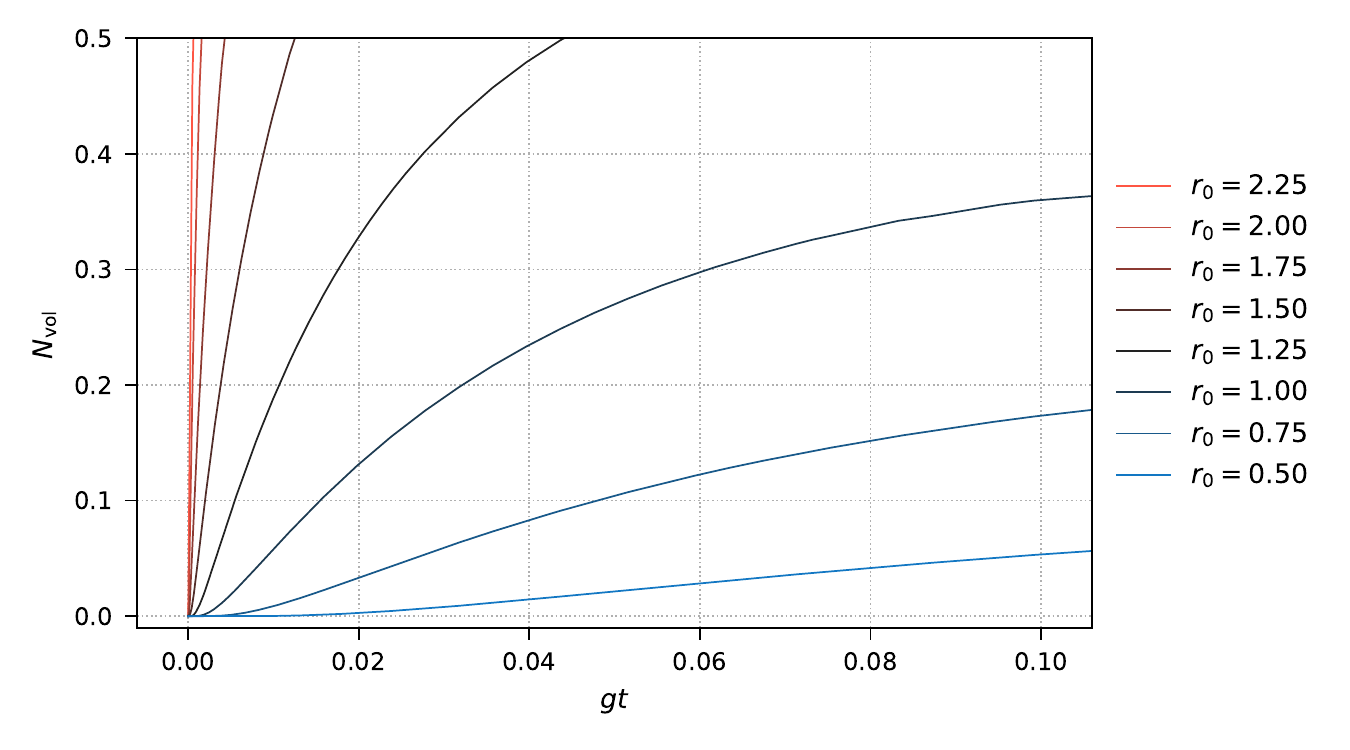}
\par\end{centering}
\noindent \centering{}}

\subfloat[\label{fig:-1-1}Negative peak]{\noindent \begin{centering}
\captionsetup{position=top}\includegraphics{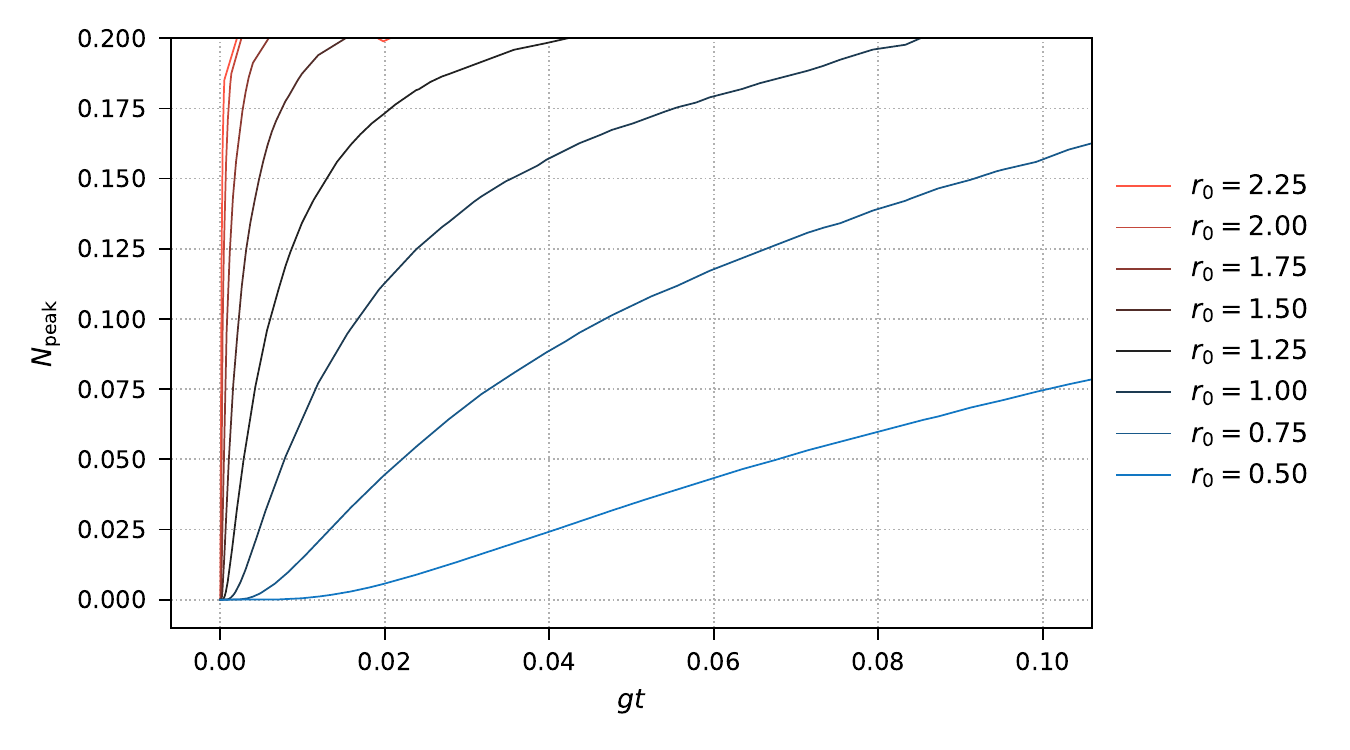}
\par\end{centering}
\noindent \centering{}}

\caption[Negativity during short time unitary evolution of squeezed vacuum]{\label{fig:-1}\textbf{Negativity during short time unitary evolution
of squeezed vacuum. }The initial states are squeezed vacuum states
with varying $r_{0}$. The evolution of $N_{\mathrm{vol}}$ and $N_{\mathrm{peak}}$
appear qualitatively similar in that an increase in $r_{0}$ causes
the negativity to initially grow more rapidly. The short time evolution
is discussed in Section \ref{subsec:Evolution-over-Short}.}
\end{figure}

Figure \ref{fig:evo-gallery-medium} shows the time-evolution of the
squeezed state state $|\xi{=}1.5\rangle$ over short times. Analogously
to the classical evolution of a probability density \cite{Huber_SqueezingThermalFluctuations_2019},
the squeezed state evolves to form an “S”-like shape in phase
space. Unlike the classical evolution however, negative and positive
fringes appear in the concave regions of the S-shape. These fringes
constitute the negative regions of the Wigner function. As the state
evolves, the fringes increase in number and amplitude as the curve
of the S-shape becomes more pronounced. Figure \ref{fig:-1} shows
the negativity in the initial stages of evolution, demonstrating that
this corresponds to a growth in negativity.

\begin{figure}
\noindent \begin{centering}
\includegraphics{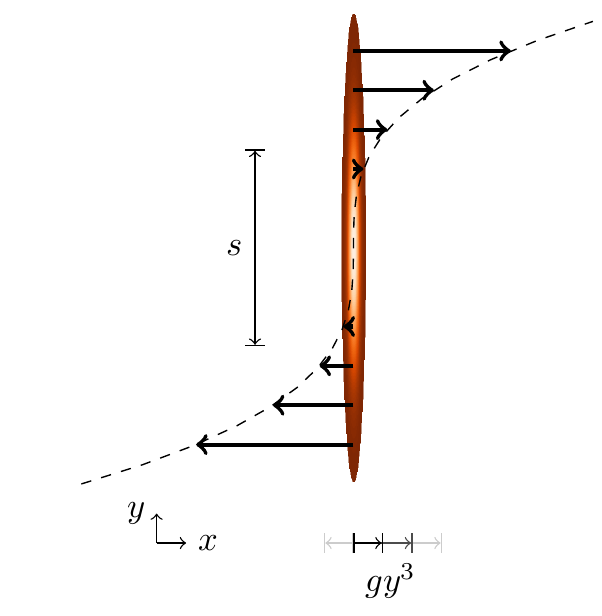}
\par\end{centering}
\caption[Illustration of Wigner current for squeezed vacuum]{\label{fig:evo-gallery-medium-1}\textbf{Illustration of Wigner current
for squeezed vacuum.} The arrow lengths are proportional to $y^{3}$
to illustrate the part of the current $\mathbf{J}$ on the $y$-axis
not linear in $y$. These cause the bending of the Wigner function.
An expression for $\mathbf{J}$ is found in equation (\ref{eq:-221}).
This illustrates that the current increases super-linearly with the
distance to the origin causing a bending of the initial squeezed state.}
\end{figure}

A geometrical understanding of the short time behavior is gained by
considering the Wigner current (\ref{eq:-101}) for the initial state
(\ref{eq:squeezed-vacuum-initial-state}) along the $y$-axis. In
Cartesian coordinates, the current for the initial state on the $y$-axis
($x=0$) reads
\begin{align}
\mathbf{J} & =g\left[s^{2}+s^{-2}+1+2(1-s^{-4})y^{2}\right]yW(0,y,0)\hat{\mathbf{x}}.\label{eq:-221}
\end{align}
The bending of the shape is caused by the terms not linear in $y$,
i.e. $2g(1-s^{-4})y^{3}W(0,y,0)\hat{\mathbf{x}}$. This dependence
on $y^{3}$ is illustrated in Figure \ref{fig:evo-gallery-medium-1}.

\subsection{Preliminary Algebraic View of Negativity \label{subsec:Preliminary-Algebraic-View}}

We shall now try to build up some intuition for the scaling of negativity.
Let us start by considering the evolution of the squeezed vacuum state
$|\Psi(0)\rangle$ where
\begin{equation}
|\Psi_{\xi}(0)\rangle=\hat{S}(r_{0})|0\rangle.
\end{equation}
We can write the time-evolution of the state as
\begin{equation}
|\Psi_{\xi}(0)\rangle=\hat{U}(t)|\Psi_{\xi}(0)\rangle\label{eq:-53}
\end{equation}
with the time evolution operator obtained by combining (\ref{eq:-65})
and (\ref{eq:-144-1}) as
\begin{equation}
\hat{U}(t)=e^{-ig\hat{a}^{\dagger}\hat{a}^{\dagger}\hat{a}\hat{a}t}.
\end{equation}
The squeezing transformation $\hat{S}(\xi')$ for any choice of the
parameter $\xi'$ leaves the negativity unchanged as can be seen by
applying (\ref{eq:-114-1-1}) to the definitions (\ref{eq:negpeak-definition})
and (\ref{eq:negvol-definition}). Computing the negativity of the
state $|\Psi_{\xi}(t)\rangle$ is thus the same as computing the negativity
of the state $\hat{S}^{\dagger}(r_{0})|\Psi_{\xi}(t)\rangle$, i.e.
\begin{equation}
N_{\mathrm{vol}}[|\Psi_{\xi}(t)\rangle]=N_{\mathrm{vol}}[\hat{S}^{\dagger}(\xi)|\Psi_{\xi}(t)\rangle]=N_{\mathrm{vol}}\left[\hat{S}^{\dagger}(r_{0})\hat{U}(t)\hat{S}(r_{0})|0\rangle\right],\label{eq:-203}
\end{equation}
writing the state explicitly as an argument to $N_{\mathrm{vol}}$.
Equation (\ref{eq:-203}) moves the squeezing parameter $r_{0}$ from
the initial state to the equation of motion. The squeezing transformation
applied to $\hat{a}$ may be stated as \cite{Gerry_IntroductoryQuantumOptics_2004} 

\begin{align}
\hat{S}(r_{0})\hat{a}\hat{S}^{\dagger}(r_{0}) & =\hat{X}s+i\hat{Y}s^{-1}
\end{align}
and the operator part of the Kerr Hamiltonian thus transforms as
\begin{align}
\hat{S}(r_{0})\hat{a}^{\dagger}\hat{a}^{\dagger}\hat{a}\hat{a}\hat{S}^{\dagger}(r_{0}) & =\left(\hat{X}s+i\hat{Y}s^{-1}\right)^{2}\left(\hat{X}s-i\hat{Y}s^{-1}\right)^{2}.\label{eq:-24}
\end{align}
Keeping only the highest power of $s$ in (\ref{eq:-24}), we arrive
at
\begin{equation}
\hat{S}(r_{0})\hat{a}^{\dagger}\hat{a}^{\dagger}\hat{a}\hat{a}\hat{S}^{\dagger}(r_{0})\xrightarrow[s\to\infty]{}s^{4}\hat{X}^{4}.\label{eq:-202}
\end{equation}
We expect then that the dynamics are dominated by the term proportional
to $s^{4}$ in the limit of large squeezing. We shall summarize this
statement symbolically by writing (\ref{eq:-203}) as

\begin{equation}
N_{\mathrm{vol}}\left[|\Psi_{\xi}(t)\rangle\right]\approx N_{\mathrm{vol}}\left[e^{-igts^{4}\hat{X}^{4}}|0\rangle\right]\qquad\text{for large \ensuremath{s}.}\label{eq:-54}
\end{equation}
Here, the squeezing transformation (\ref{eq:-202}) was applied to
the Taylor expansion of $\hat{U}(t)$ as in (\ref{eq:-201}). The
derivation (\ref{eq:-203}--\ref{eq:-54}) may be repeated for $N_{\mathrm{peak}}$
to similarly write 
\begin{equation}
N_{\mathrm{peak}}\left[\Psi_{\xi}(t)\right]\approx N_{\mathrm{peak}}\left[e^{-igts^{4}\hat{X}^{4}}|0\rangle\right]\qquad\text{for large \ensuremath{s}.}\label{eq:-204}
\end{equation}

Hence, having disregarded all but the leading order terms in the expressions
for $N_{\mathrm{vol}}$ and $N_{\mathrm{peak}}$, it could be suggested
that the negativity for a highly squeezed initial state is constant
as a function of the quantity $gts^{4}$. With (\ref{eq:-54}) and
(\ref{eq:-204}), we have however no indication of the validity of
(\ref{eq:-202}). Applying an analogous transformation directly to
the Wigner function and phase space dynamics yields greater insight
into the meaning of (\ref{eq:-202}). Before this is done however,
we first view the problem in a geometric setting.

\subsection{Preliminary Geometric View of Negativity\label{subsec:Preliminary-Geometric-View}}

\begin{figure}
\noindent \begin{centering}
\includegraphics{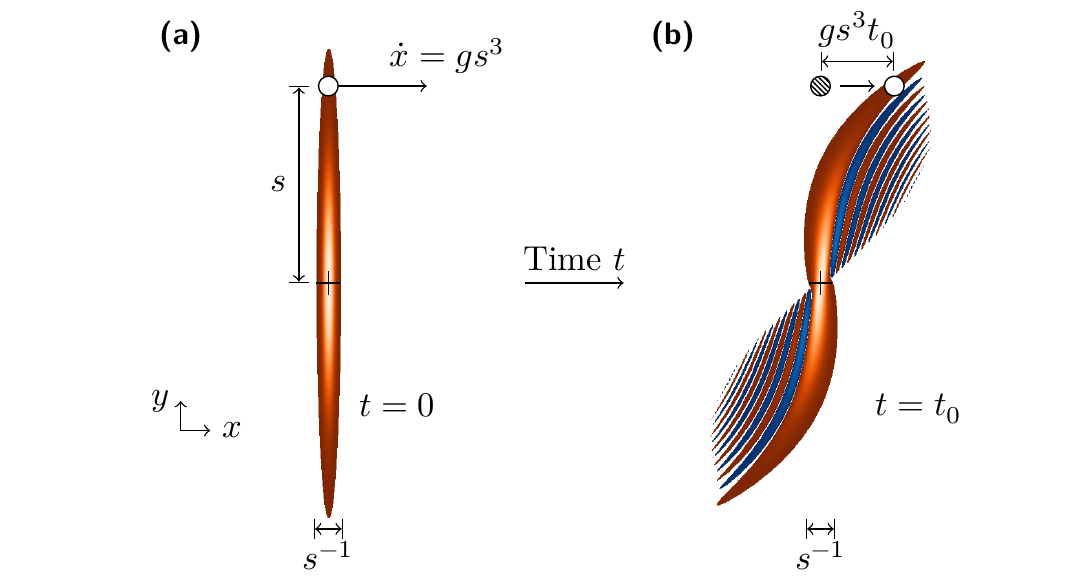}
\par\end{centering}
\noindent \centering{}\caption[Illustration of the short time evolution of a squeezed vacuum state]{\label{fig:negativity-mechanism-sketch}\textbf{Illustration of the
short time evolution of a squeezed vacuum state.} The drawing illustrates
the argument of Section \ref{subsec:Preliminary-Geometric-View}.
(a) shows the initial state. The white circle represents a particle
obeying the classical Liouville equation. Its velocity, shown by the
arrow, is given by (\ref{eq:-220}). After a short time $t_{0}$ has
passed (b), the particle has moved $gs^{3}t_{0}$. This distance is
used to quantify the bending of the Wigner function. Orange denotes
regions of positive $W$ and blue denotes regions where $W$ is negative
(see e.g. Figure \ref{fig:Squeezed-vacuum-state}).}
\end{figure}
 An estimate similar to (\ref{eq:-54}) may be reached by considering
geometrically the time-evolution in the phase space picture. We consider
again as the initial state a highly squeezed vacuum state. The initial
state is displayed in Figure \ref{fig:negativity-mechanism-sketch}a.
Let the squeezed state evolve over a short time so that it forms an
S-shape in phase space as displayed in Figure \ref{fig:negativity-mechanism-sketch}b.
It is known from simulations (recall Figure \ref{fig:Squeezed-vacuum-state})
that the negative parts of the Wigner function first appear as fringes
in the concave region of the S-shape. We might intuitively expect
the time at which the negativity first appears to bear some relation
to the magnitude of the initial state squeezing (e.g. for no squeezing,
no negativity will be observed). To support this, let us apply some
dimensions to Figure \ref{fig:negativity-mechanism-sketch}a. For
the initial state, we may define two characteristic phase space-length
scales from the variances in the anti-squeezed and squeezed direction.\footnote{Of course, these quantities, as displayed in Figure \ref{fig:negativity-mechanism-sketch},
are “lengths” in phase space coordinates and are therefore both
dimensionless. It is possible to use a Wigner function where the arguments
have differing dimensions (indeed this was the case when the function
was first introduced by Eugene Wigner in 1932 \cite{Wigner_QuantumCorrectionThermodynamic_1932})
making it manifestly impossible to compare lengths measured in anything
but parallel directions in phase space. Here, however, we compare
lengths in parallel directions since the length proportional to $s$
is transformed to a length in its orthogonal direction using the equation
of motion of a particle. Thus only lengths measured in parallel directions
are compared as can be clearly seen in Figure \ref{fig:negativity-mechanism-sketch}b.} In terms of the parameter $s$, these characteristic length scales
are\footnote{The aforementioned variances are actually given by $s^{2}/4$ and
$1/4s^{2}$ cf. equations (\ref{eq:-9}) and (\ref{eq:-10}). The
constant factor $\sqrt{1/4}$ may be absorbed into $k_{0}$ when they
are compared in (\ref{eq:-11}).} $s$ and $s^{-1}$.

Consider now the hypothetical motion of a classical particle\footnote{A related concept is the Ehrenfest time $t_{E}$ \cite{Katz_ClassicalQuantumTransition_2008}.
It is a time until which evolution leaves classical and quantum mechanical
phase space distributions in general agreement.} placed a distance $s$ up from the origin in the anti-squeezed direction.
We can find the instantaneous phase space velocity of the particle
from the classical probability current. The classical current is found
by removing all derivative expressions from $\mathbf{J}$ of (\ref{eq:-101}).
The result can be written as
\begin{equation}
\mathbf{J}=W\mathbf{v}
\end{equation}
where $\mathbf{v}$ is a vector quantity independent of the value
of $W$. $\mathbf{v}$ corresponds to the phase space velocity of
a classical particle: $\mathbf{v}=(\dot{x},\dot{y})$. From (\ref{eq:-101})
we obtain
\[
\mathbf{v}=-2g\left(r^{2}-1\right)r\hat{\symbf{\phi}}.
\]
Evaluating $\mathbf{v}$ at the position of the particle $(x,y)=(0,s)$
yields
\begin{align}
(\dot{x},\dot{y}) & \propto(2gs^{3},0)\label{eq:-220}
\end{align}
where the term $2gsr$ has been neglected from $\dot{x}$ due to the
assumption of large squeezing $s$. $(\dot{x},\dot{y})$ is displayed
as an arrow in Figure \ref{fig:negativity-mechanism-sketch}a. We
expect some amount negativity to appear once the particle has moved
some fixed multiple of the squeezed width $l_{0}=k_{0}s^{-1}$.\footnote{An argument for this may be found in the statements of Section \ref{sec:Gaussian-States-and}.
It is clear that the formation of the S-shape in phase space removes
the state from the Gaussian initial state. Since any pure state that
is non-Gaussian exhibits negativity (see Section \ref{sec:Gaussian-States-and}),
we expect the negativity to increase more as state evolves farther
from the Gaussian initial state. } It is the hope that $l_{0}$ encapsulates the geometrical considerations
in such a way that $k_{0}$ is independent of the squeezing $s$.
This being the case, one would find some fixed degree of negativity
to appear once $gs^{3}t_{0}=l_{0}$ or, making all $s$-dependence
explicit, 
\begin{equation}
gt_{0}s^{4}=k_{0}.\label{eq:-11}
\end{equation}
Extending this to several values of $t_{0}$, the expression $gts^{4}$
shows up as in (\ref{eq:-54}).

\subsection{Phase Space View of Negativity \label{subsec:negativity-mechanism}}

Our goal is now to substantiate the relevance of the quantity $gts^{4}$
in the description of the initial growth of negativity. To do this,
it is useful to first gain a more accurate intuition of the relevant
mechanism than the one developed above. Recall from Figure \ref{fig:evo-gallery-medium}
the general features of the first stages of evolution of the Wigner
function for a squeezed state. As the S-shape forms the negative regions
develop as fringes in the concave regions of the S-shape. We wish
to understand the development of the fringes for short timescales
from the viewpoint of the partial differential equation for $W$.
It is clear that no fringes develop with vanishing squeezing $s=0$
(see Section \ref{sec:Kerr-Evolution-of}). More interestingly, we
may consider the case of large squeezing, i.e. the limit $s\to\infty$.
With large squeezing the problem is most easily stated in Cartesian
coordinates. The initial state is given by (\ref{eq:squeezed-vacuum-initial-state}).
\begin{equation}
W(x,y,0)=\frac{2}{\pi}e^{-2x^{2}s^{2}-2y^{2}/s^{2}}.\label{eq:-74}
\end{equation}
The equation of motion is recast in Cartesian coordinates from (\ref{eq:-18}),
yielding the partial differential equation
\begin{equation}
\partial_{t}W(x,y,t)=\begin{aligned}[t] & 2g\left(-x^{2}y\partial_{x}-y^{3}\partial_{x}+x^{3}\partial_{y}+xy^{2}\partial_{y}\right)W(x,y,t)\\
 & -2g\left(-y\partial_{x}+x\partial_{y}\right)W(x,y,t)\\
 & -\frac{g}{8}\left(-y\partial_{x}^{3}+x\partial_{y}^{3}+x\partial_{y}\partial_{x}^{2}-y\partial_{x}\partial_{y}^{2}\right)W(x,y,t)
\end{aligned}
\label{eq:-31-2}
\end{equation}
in variables $x$, $y$ and $t$. 

Consider now for each term in (\ref{eq:-31-2}) its relative magnitude
in the vicinity of the region of negativity. (To simplify the following
discussion, consider only the negativity present above the $x$-axis
(i.e. $y>0$). By rotational symmetry, the evolution in negativity
is the same for negative $y$-coordinates so the following also applies
for negative $y$-coordinates.) Negative values of $W$ are first
seen in the concave regions of the S-shape. To gain intuition for
the short-time effects of the various terms of (\ref{eq:-31-2}),
we can write $W(x,y,t)$ as a power series in $t$ and expand to linear
order:
\begin{equation}
W(x,y,t)=W(x,y,0)+t\partial_{t}W(x,y,0)+\mathcal{O}(t^{2}).\label{eq:-26}
\end{equation}
Insert now the right hand side of (\ref{eq:-31-2}) in the first-order
expansion (\ref{eq:-26}) and consider each term separately. As we
imagine the squeezing $s$ increase towards $\infty$, the terms containing
the highest power of $y$ and $\partial_{x}$ are seen to dominate:
$y$ dominates because the area of negativity for the highly squeezed
state is hypothesized to move towards larger $y$-coordinates and
$\partial_{x}$ dominates since the highly squeezed state varies more
quickly in the $x$-direction. For instance, applying the operators
$\partial_{x}$ and $\partial_{y}$ to the initial state:
\begin{subequations}
\label{eq:-76}

\begin{equation}
\partial_{x}W(x,y,0)=-4s^{2}xW(x,y,0),
\end{equation}

\begin{equation}
\partial_{y}W(x,y,0)=-\frac{4y}{s^{2}}W(x,y,0),
\end{equation}
\end{subequations}
we see that $\partial_{x}W$ scales with a positive power of $s$
whereas $\partial_{y}W$ scales with a negative power, suggesting
that terms containing $\partial_{x}$ are relatively more significant
than terms containing $\partial_{y}$. This relationship between the
components of the equation of motion for $W$ is not apparent in (\ref{eq:-31-2}),
but can be made explicit using an appropriate coordinate transformation
as is done in the following section.

\subsection{Introduction of Rescaled Coordinates \label{subsec:intro-rescaled-coords}}

\begin{figure}
\noindent \begin{centering}
\makebox[0pt][c]{\mbox{\includegraphics{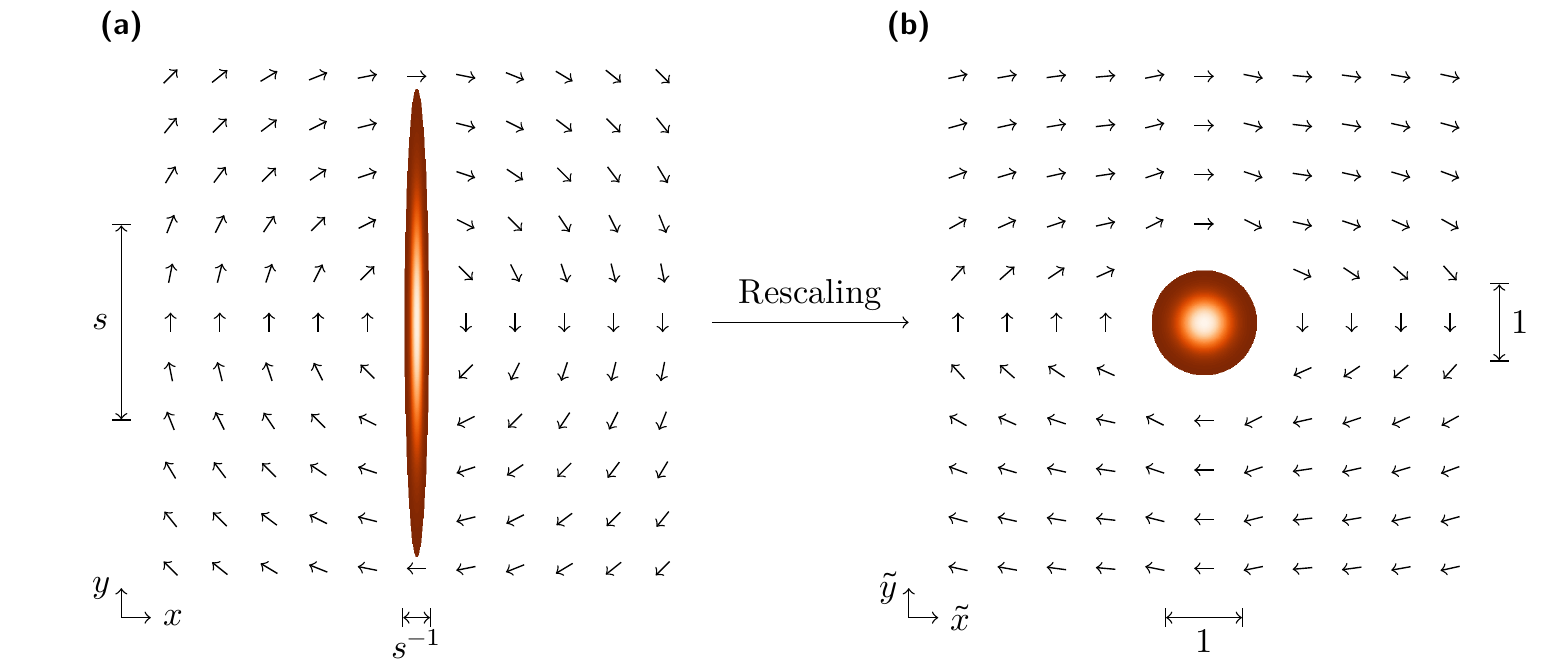}\hskip2em}}
\par\end{centering}
\noindent \centering{}\caption[Illustration of the rescaling of the squeezed vacuum state]{\textbf{Illustration of the rescaling of the squeezed vacuum state.}
Illustration of the rescaling of the initial state as described in
Section \ref{subsec:intro-rescaled-coords}. (a) shows the initial
state (\ref{eq:squeezed-vacuum-initial-state}) in the regular Cartesian
coordinates $(x,y)$. The direction of the Wigner current as given
by (\ref{eq:-101}) is shown with arrows (whose lengths are not scaled
with the magnitude however). (b) shows the initial state in the rescaled
coordinate system where, notably, the parameter $s$ has vanished
from the characteristic lengths (both shown as $1$). The direction
of the rescaled Wigner current as given by (\ref{eq:-222}) is also
shown. The dynamics in (b) have lost their manifest rotational symmetry.
\label{fig:initial-state-rescaling}}
\end{figure}

To formalize the loosely formed scaling arguments from the previous
section, consider again the initial state in Cartesian coordinates.
We can now introduce the coordinates 

\begin{equation}
\tilde{x}=sx\text{\ensuremath{\qquad}and}\qquad\tilde{y}=\frac{y}{s}\label{eq:-30-2}
\end{equation}
in which the initial state (\ref{eq:squeezed-vacuum-initial-state})
takes the simpler form

\begin{equation}
\tilde{W}(\tilde{x},\tilde{y},0)=\frac{2}{\pi}e^{-2\tilde{x}^{2}-2\tilde{y}^{2}}.\label{eq:-30}
\end{equation}
Significantly, (\ref{eq:-30}) contains no reference to $s$. Instead,
the initial state $\tilde{W}(\tilde{x},\tilde{y},0)$ now has the
same form as the Wigner function for a vacuum state (in regular Cartesian
coordinates $(x,y)$, see (\ref{eq:vacuum-wigner-carteesian})). Having
introduced $\tilde{W}(\tilde{x},\tilde{y},t)$ to denote the Wigner
function in rescaled coordinates, we state its relation to the unscaled
Wigner function $W$:
\begin{equation}
\tilde{W}(\tilde{x},\tilde{y},t)=W(\tilde{x}/s,s\tilde{y},t).\label{eq:-19}
\end{equation}
The evolution of $\tilde{W}(\tilde{x},\tilde{y},t)$ is described
by a partial differential equation in the coordinates $(\tilde{x},\tilde{y},t)$.
To derive this equation, we write the relevant differential operators
in the rescaled coordinates. These are
\begin{equation}
\partial_{\tilde{x}}=\frac{1}{s}\partial_{x}\qquad\text{and}\qquad\partial_{\tilde{y}}=s\partial_{y}.\label{eq:-27}
\end{equation}
Using (\ref{eq:-19}) and the chain rule, the corresponding equation
of motion for $\tilde{W}(\tilde{x},\tilde{y},t)$ is found to be

\begin{equation}
\partial_{t}\tilde{W}(\tilde{x},\tilde{y},t)=\begin{aligned}[t] & 2g\left(-\tilde{x}^{2}\tilde{y}\partial_{\tilde{x}}-s^{4}\tilde{y}^{3}\partial_{\tilde{x}}+\frac{1}{s^{4}}\tilde{x}^{3}\partial_{\tilde{y}}+\tilde{x}\tilde{y}^{2}\partial_{\tilde{y}}\right)\tilde{W}(\tilde{x},\tilde{y},t)\\
 & -2g\left(-s^{2}\tilde{y}\partial_{\tilde{x}}+\frac{1}{s^{2}}\tilde{x}\partial_{\tilde{y}}\right)\tilde{W}(\tilde{x},\tilde{y},t)\\
 & -\frac{g}{8}\left(-s^{4}\tilde{y}\partial_{\tilde{x}}^{3}+\frac{1}{s^{4}}\tilde{x}\partial_{\tilde{y}}^{3}+\tilde{x}\partial_{\tilde{y}}\partial_{\tilde{x}}^{2}-\tilde{y}\partial_{\tilde{x}}\partial_{\tilde{y}}^{2}\right)\tilde{W}(\tilde{x},\tilde{y},t).
\end{aligned}
\label{eq:-31-2-1}
\end{equation}
In general, terms containing subexpressions that describe the spatial
variation in the direction of the $\tilde{x}$-axis ($\partial_{\tilde{x}}$)
or the distance to $\tilde{x}$-axis ($\tilde{y}$) are multiplied
by $s$ to some positive power (e.g. the term $(g/8)s^{4}\tilde{y}\partial_{\tilde{x}}^{3}\tilde{W}$).
Correspondingly, terms which describe the spatial variation in the
direction of the $\tilde{y}$-axis ($\partial_{\tilde{y}}$) or the
distance to the $\tilde{y}$-axis ($\tilde{x}$) are divided by $s$
to some positive power (e.g. the term $2gs^{-4}\tilde{x}^{3}\partial_{\tilde{y}}\tilde{W}$).
Terms that contain some balance of the two remain unchanged with respect
to $s$ (e.g. the term $-2g\tilde{x}^{2}\tilde{y}\partial_{\tilde{x}}\tilde{W}$). 

To summarize, we might now say that choosing a new coordinate system
in which to express the Wigner function, allows one to “normalize”
the initial state to (\ref{eq:-30}) regardless of its squeezing $s$.
In return for this, the equation of motion in these new rescaled coordinates
changes to (\ref{eq:-31-2-1}). This rescaled equation of motion takes
on the characteristic features of the initial state, e.g. if the unscaled
initial state varies greatly in the $x$-direction (as is the case
for a squeezed state with $x$ as its squeezed axis) the terms describing
this variation are amplified in the rescaled equation of motion. Conceptually,
this transformation is identical to the one applied in Section \ref{subsec:Preliminary-Algebraic-View}.
Moving from $(x,y)$ to $(\tilde{x},\tilde{y})$, we have obscured
the rotational symmetry allowing for the concise expression of the
equation in polar coordinates (as done in (\ref{eq:-18})) in return
for making the squeezing $s$ explicit in the equation of motion.

\subsection{Rescaled Wigner Current}

To visualize the effect of rescaling, we can compare the Wigner current
in the regular and rescaled coordinates. The regular current is given
by (\ref{eq:-101}). To find the rescaled current $\tilde{\mathbf{J}}$,
we look for a $\tilde{\mathbf{J}}$ such that
\begin{equation}
\partial_{t}\tilde{W}=-\tilde{\nabla}\cdot\tilde{\mathbf{J}}
\end{equation}
where
\begin{equation}
\tilde{\nabla}\cdot\tilde{\mathbf{J}}=\partial_{\tilde{x}}J_{\tilde{x}}+\partial_{\tilde{y}}J_{\tilde{y}}.
\end{equation}
Inserting the transformed operators (\ref{eq:-27}), we can write
\begin{equation}
\tilde{\nabla}\cdot\tilde{\mathbf{J}}=\frac{1}{s}\partial_{x}J_{\tilde{x}}+s\partial_{y}J_{\tilde{y}}.
\end{equation}
We have defined the rescaled current $\tilde{\mathbf{J}}$ such that
\begin{equation}
\tilde{\nabla}\cdot\tilde{\mathbf{J}}=\partial_{t}\tilde{W}=\partial_{t}W=\nabla\cdot\mathbf{J},
\end{equation}
from which it follows that 
\begin{align}
\tilde{J}_{\tilde{x}} & =sJ_{x}, & \tilde{J}_{\tilde{y}} & =\frac{1}{s}J_{y}.
\end{align}
We obtain the regular Cartesian coordinate current from (\ref{eq:-101}):
\begin{equation}
\mathbf{J}=\left(-2g\left(r^{2}-1\right)W+\frac{g}{8}\nabla^{2}W\right)\left(-y\hat{\mathbf{x}}+x\hat{\mathbf{y}}\right).
\end{equation}
The rescaled current $\tilde{\mathbf{J}}$ then takes the form\footnote{Note that the vectors $\tilde{\hat{\mathbf{x}}}$ and $\tilde{\hat{\mathbf{y}}}$
have unit length in the coordinates $(\tilde{x},\tilde{y})$ and hence
vary in length in the coordinates $(x,y)$. }
\begin{equation}
\mathbf{\tilde{J}}=\left(-2g\left(\frac{1}{s^{2}}\tilde{x}^{2}+s^{2}\tilde{y}^{2}-1\right)\tilde{W}+\frac{g}{8}\nabla^{2}\tilde{W}\right)\left(-s^{2}\tilde{y}\tilde{\hat{\mathbf{x}}}+s^{-2}\tilde{x}\tilde{\hat{\mathbf{y}}}\right),\label{eq:-222}
\end{equation}
where
\begin{equation}
\nabla^{2}\tilde{W}=s^{2}\partial_{\tilde{x}}^{2}\tilde{W}+\frac{1}{s^{2}}\partial_{\tilde{y}}^{2}\tilde{W}.
\end{equation}
Most importantly, notice that the current now describes a flow along
a vector different from the angular unit vector, namely $(-s^{2}\tilde{y}\tilde{\hat{\mathbf{x}}}+s^{-2}\tilde{x}\tilde{\hat{\mathbf{y}}}).$
As $s$ increases, the $\tilde{y}$-component of the current becomes
negligible. The scaled and unscaled currents are illustrated in Figure
\ref{fig:initial-state-rescaling}.

\subsection{Large Squeezing Approximation\label{subsec:large-squeezing-approximation}}

\begin{figure}
\noindent \begin{centering}
\includegraphics{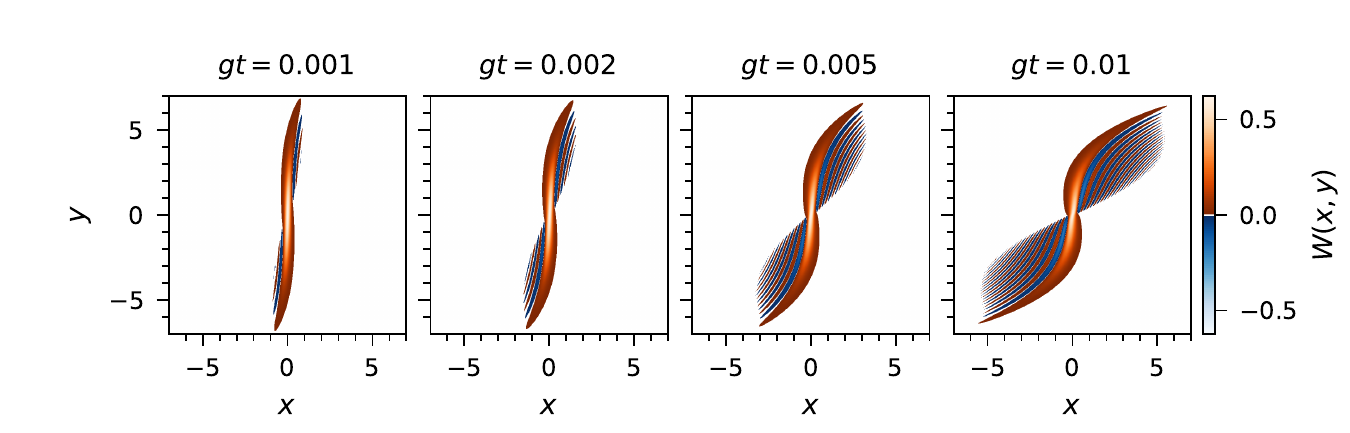}
\par\end{centering}
\caption[Unitary evolution of squeezed vacuum with large squeezing approximation]{\label{fig:asymp-medium-evo}\textbf{Unitary evolution of squeezed
vacuum with large squeezing approximation.} Application of the large
squeezing approximation introduced in Section \ref{subsec:large-squeezing-approximation}
to evolve the squeezed vacuum state with $r_{0}=1.5$. The Wigner
function was computed from (\ref{eq:-52}) with $\tilde{u}$ found
from the Fourier domain solution given by equation (\ref{eq:-28}).
This figure should therefore be compared with Figure \ref{fig:evo-gallery-medium}
which shows the Wigner function computed from solution of the full
master equation, i.e. without the approximation. The solution with
the large squeezing approximation fails to capture the bending of
the Wigner function towards the $x$-axis.}
\end{figure}

Since the initial state in the squeezed coordinates $\tilde{W}(\tilde{x},\tilde{y},0)$
is invariant with respect to the squeezing $s$, all dependence on
squeezing is captured in the equation of motion (\ref{eq:-31-2-1}).
In the limit of large squeezing, we expect only the terms carrying
the highest power of $s$ to bear significance. We therefore disregard
any term of (\ref{eq:-31-2-1}) which is not proportional to $s^{4}$
and look for solutions to the equation
\begin{equation}
\partial_{t}\tilde{W}(\tilde{x},\tilde{y},t)=-2gs^{4}\tilde{y}^{3}\partial_{\tilde{x}}\tilde{W}(\tilde{x},\tilde{y},t)+\frac{gs^{4}}{8}\tilde{y}\partial_{\tilde{x}}^{3}\tilde{W}(\tilde{x},\tilde{y},t).\label{eq:-61}
\end{equation}
We refer to this step as the large squeezing approximation. All derivatives
with respect to $\tilde{y}$ have been discarded, as has any term
dependent on $\tilde{x}$. The disappearance of $\partial_{\tilde{y}}$
means that $\tilde{y}$ can be regarded simply as a parameter. Hence
(\ref{eq:-61}) can be characterized as a linear homogeneous partial
differential equation with constant coefficients in the two variables
$\tilde{x}$ and $t$. The spatial first order term can be eliminated
by looking for a solution $u_{\tilde{y}}(\mu,t)$ such that 
\begin{equation}
\tilde{W}(\tilde{x},\tilde{y},t)=u_{\tilde{y}}(\tilde{x}-2gs^{4}\tilde{y}^{3}t,t).\label{eq:-62}
\end{equation}
In this new function $u_{\tilde{y}}$, $\tilde{y}$ should simply
be considered a parameter (and denoted by subscript). The corresponding
equation of motion for $u_{\tilde{y}}$ is found by insertion of (\ref{eq:-62})
into (\ref{eq:-61}):
\begin{equation}
\partial_{t}u_{\tilde{y}}(\mu,t)=\frac{gs^{4}}{8}\tilde{y}\partial_{\mu}^{3}u_{\tilde{y}}(\mu,t).\label{eq:-75}
\end{equation}
We may now rescale time to $\tilde{\tau}=gs^{4}\tilde{y}t/8$, thus
defining a new function $\tilde{u}_{\tilde{y}}(\mu,\tilde{\tau})$
by
\begin{equation}
u_{\tilde{y}}(\mu,t)=\tilde{u}_{\tilde{y}}(\mu,gs^{4}\tilde{y}t/8).\label{eq:-64}
\end{equation}
Note that $\tilde{\tau}$ has an implicit dependence on $\tilde{y}$.
The evolution of $\tilde{u}_{\tilde{y}}(\mu,\tilde{\tau})$ is then
governed by the equation 
\begin{equation}
\partial_{\tilde{\tau}}\tilde{u}_{\tilde{y}}(\mu,\tilde{\tau})=\partial_{\mu}^{3}\tilde{u}_{\tilde{y}}(\mu,\tilde{\tau}).\label{eq:-63}
\end{equation}
Before we continue, note that (\ref{eq:-63}) contains no reference
to the squeezing $s$ of the initial state. Hence, all information
about $s$ is contained in the transformation from $\tilde{u}_{\tilde{y}}(\mu,\tilde{\tau})$
to $W(x,y,t)$. 

As a small digression it should be mentioned that equation (\ref{eq:-63})
sometimes is referred to as the linearized Korteweg de Vries equation
\cite{Villanueva_LinearizedKortewegdeVries_2012}. Since its solution
may be expressed in terms of the Airy function $\Ai$, it is also
sometimes called the Airy equation \cite{Olver_PointsConstancyPeriodic_2018}.
Namely, for the general initial condition $\tilde{u}_{\tilde{y}}(\mu,0)=f_{\tilde{y}}(\mu)$,
the solution to (\ref{eq:-63}) can be expressed as \cite{Olver_IntroductionPartialDifferential_2013}
\begin{equation}
\tilde{u}_{\tilde{y}}(\mu,\tilde{\tau})=\frac{1}{(3\tilde{\tau})^{3/2}}\int_{-\infty}^{\infty}d\xi\,f_{\tilde{y}}(\xi)\,\Ai\left(\frac{\mu-\xi}{(3\tilde{\tau})^{3/2}}\right).\label{eq:-29}
\end{equation}
Recalling definitions (\ref{eq:-64}) and (\ref{eq:-62}) allows one
to express the solution to the Wigner function within the confines
set by the large squeezing approximation (\ref{eq:-61}):
\begin{equation}
\tilde{W}(\tilde{x},\tilde{y},t)=\frac{1}{(3gs^{4}t\tilde{y}/8)^{3/2}}\int_{-\infty}^{\infty}d\xi\,\tilde{W}(\xi,\tilde{y},0)\,\Ai\left(\frac{\tilde{x}+gs^{4}t\tilde{y}^{3}/8-\xi}{(3gs^{4}t\tilde{y}/8)^{3/2}}\right),\label{eq:-31}
\end{equation}
where $\tilde{W}(\xi,\tilde{y},0)$ is the Wigner function of an arbitrary
initial state.

Of greater interest here,\footnote{Unlike (\ref{eq:-29}), (\ref{eq:-174}) allows for the straightforward
inclusion of terms other than $\partial_{\mu}^{3}\tilde{u}_{\tilde{y}}$
on the right hand side of (\ref{eq:-63}) -- notably $\partial_{\mu}^{2}\tilde{u}_{\tilde{y}}$
which will be needed for the description of decoherence effects in
Chapter \ref{chap:coupling-to-the-environment}.} one can also express the solution in the Fourier domain as
\begin{subequations}
\label{eq:-174}
\begin{equation}
\tilde{u}_{\tilde{y}}(\mu,\tilde{\tau})=\frac{1}{\sqrt{2\pi}}\int_{-\infty}^{\infty}dk\,h_{\tilde{y}}(k)\,e^{i(k\mu-k^{3}\tilde{\tau})}\label{eq:-28}
\end{equation}
with the Fourier transform of the initial state computed as\footnote{The link between the two solutions (\ref{eq:-29}) and (\ref{eq:-174})
can be seen by considering the defining integral for the Airy function:
$\Ai(z)=\int_{0}^{\infty}ds\,\cos\left(sz+\frac{1}{3}s^{3}\right)$.}
\begin{equation}
h_{\tilde{y}}(k)=\frac{1}{\sqrt{2\pi}}\int_{-\infty}^{\infty}d\mu\,\tilde{u}_{\tilde{y}}(\mu,0)\,e^{-ik\mu}.\label{eq:-80}
\end{equation}
\end{subequations}

Generally, upon obtaining a solution for $\tilde{u}_{\tilde{y}}(\mu,\tilde{\tau})$
(using e.g. (\ref{eq:-29}) or (\ref{eq:-28})), one may return to
the rescaled Wigner function with the relation
\begin{equation}
\tilde{W}(\tilde{x},\tilde{y},t)=\tilde{u}_{\tilde{y}}(\tilde{x}-2gts^{4}\tilde{y}^{3},gts^{4}\tilde{y}/8).\label{eq:-51}
\end{equation}
Using the Fourier transformed solution, one has the solution
\begin{subequations}
\label{eq:-160}
\begin{align}
\tilde{W}(\tilde{x},\tilde{y},t) & =\frac{1}{\sqrt{2\pi}}\int_{-\infty}^{\infty}dk\,h_{\tilde{y}}(k)\,e^{i(k\tilde{x}-2kgts^{4}\tilde{y}^{3}-gts^{4}k^{3}\tilde{y}/8)}\label{eq:-102}
\end{align}
where the Fourier transform of the initial state is given by
\begin{equation}
h_{\tilde{y}}(k)=\frac{1}{\sqrt{2\pi}}\int_{-\infty}^{\infty}d\tilde{x}\,\tilde{W}(\tilde{x},\tilde{y},0)\,e^{-ik\tilde{x}}.
\end{equation}
\end{subequations}
For an initial state as given in (\ref{eq:-30}), 
\begin{equation}
h_{\tilde{y}}(k)=\frac{2}{\pi}e^{-2\tilde{y}^{2}}\frac{1}{\sqrt{2\pi}}\int_{-\infty}^{\infty}d\tilde{x}\,e^{-2\tilde{x}^{2}}e^{-ik\tilde{x}}=\frac{1}{\pi}e^{-2\tilde{y}^{2}}e^{-k^{2}/8}.
\end{equation}
Inserting into (\ref{eq:-102}) yields
\begin{equation}
\tilde{W}(\tilde{x},\tilde{y},t)=\frac{1}{\pi}\frac{1}{\sqrt{2\pi}}\int_{-\infty}^{\infty}dk\,e^{-2\tilde{y}^{2}}e^{-k^{2}/8}\,e^{i(k\tilde{x}-2kgts^{4}\tilde{y}^{3}-gts^{4}k^{3}\tilde{y}/8)}.\label{eq:-223}
\end{equation}
Finally for completeness, one can return to the unscaled Wigner function
by applying the inversion of (\ref{eq:-19}) to the above equation
to arrive at
\begin{equation}
W(x,y,t)=\tilde{u}_{\frac{y}{s}}(sx-2gs^{2}y^{3}t,gs^{3}yt/8).\label{eq:-52}
\end{equation}
To demonstrate the effect of the large squeezing approximation the
Wigner function $W(x,y,t)$ from (\ref{eq:-223}) has been plotted
in Figure \ref{fig:asymp-medium-evo} for $r_{0}=1.5$. Comparing
with the solution to the von Neumann equation (\ref{eq:kerr-vonneumann-eq})
as shown in Figure \ref{fig:Squeezed-vacuum-state}, we see that the
approximation agrees well for shorter times. In the non-approximated
solution (Figure \ref{fig:Squeezed-vacuum-state}) the ends of the
Wigner function are seen to bend toward the $x$-axis toward the end
of the displayed time interval (at $gt=0.01$). In the approximate
solution (Figure \ref{fig:asymp-medium-evo}), this bending does not
occur.

\begin{figure}[p]
\noindent \begin{centering}
\centerline{\includegraphics[viewport=0bp 40bp 428bp 500bp,clip]{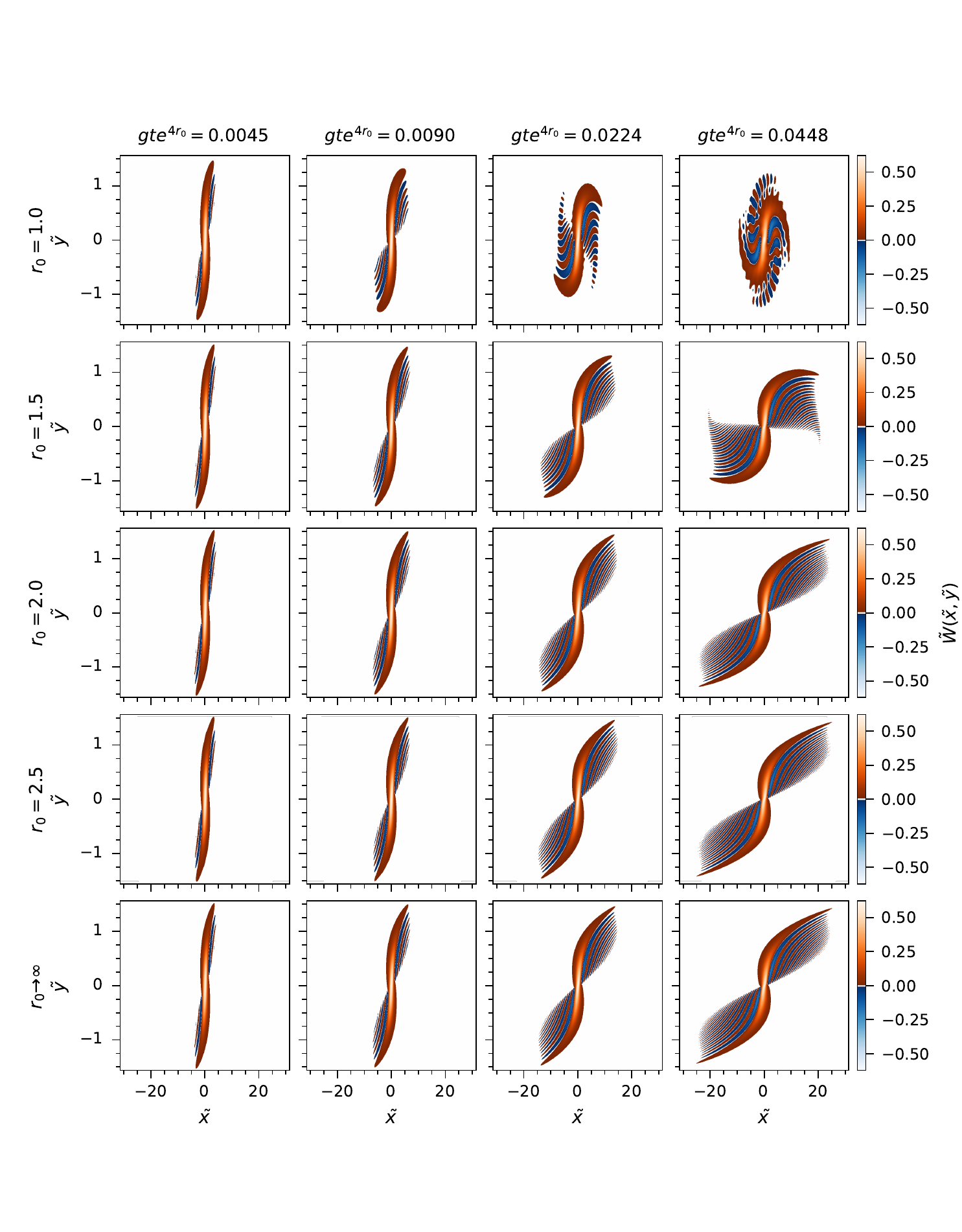}}
\par\end{centering}
\caption[Short time unitary evolution with scaled coordinates  for varying
squeezing]{\label{fig:scaled-squeezed-gallery-1}\textbf{Short time unitary
evolution with scaled coordinates for varying squeezing. }The evolution
of $\tilde{W}(\tilde{x},\tilde{y},t)$ as defined in\textbf{ }(\ref{eq:-19}).
The bottom row (labeled $r_{0}\to\infty$) shows the solution with
the large squeezing approximation, obtained as the Fourier transformed
solution (\ref{eq:-160}). As time progresses (left to right), it
is seen that the universal behavior breaks down earlier for smaller
values of the squeezing parameter $r_{0}$. The evolution times match
Figure \ref{fig:Squeezed-vacuum-state}. Note that the rescaling is
only meaningful in the short time initial evolution (Figure \ref{fig:scaled-squeezed-gallery-1-1}
shows the evolution for longer times).}
\end{figure}

Figure \ref{fig:scaled-squeezed-gallery-1} shows the scaled Wigner
function $\tilde{W}(\tilde{x},\tilde{y},t)$ computed from the solution
of (\ref{eq:kerr-vonneumann-eq}) as well as from (\ref{eq:-223}).
This demonstrates the increasing accuracy of the approximation as
$r_{0}$ is increased. 

\subsection{Validity of Large Squeezing Approximation \label{subsec:Validity-of-Large}}

\begin{figure}
\noindent \begin{centering}
\includegraphics{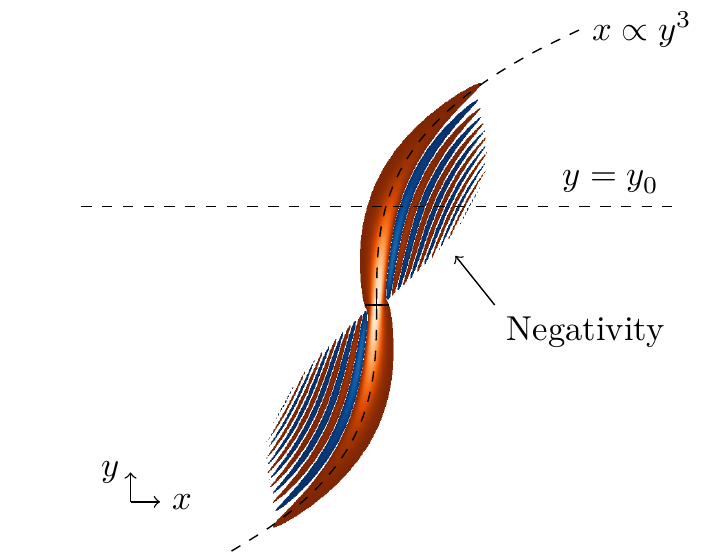}
\par\end{centering}
\noindent \centering{}\caption[Illustration of cut along squeezed axis]{\label{fig:wigner-1dcut}\textbf{Illustration of cut along squeezed
axis.} Looking at the evolution along the cut yields insight into
the mechanism generating the negativity. The Wigner function varies
rapidly along the straight dashed line while varying slowly in the
direction perpendicular to it. This fact is exploited to disregard
the terms describing the slower variation when obtaining equation
(\ref{eq:-61}). A graph where $x\propto y^{3}$ has also been plotted
to illustrate the transformation (\ref{eq:-62}) which takes $x=2gty^{3}$
to $\mu=0$ (see also equation (\ref{eq:-206})).}
\end{figure}

Before concluding on the squeezed vacuum state, we return briefly
to the regular coordinates $(x,y)$ to discuss the validity of the
approximation which led us to (\ref{eq:-61}) and from there (\ref{eq:-52}).
We can use the insight provided by the rescaling transformations (\ref{eq:-30-2})
and (\ref{eq:-27}) to expand the loosely defined scalings of Section
\ref{subsec:negativity-mechanism}. We see from the transformations
(\ref{eq:-30-2}) and (\ref{eq:-27}) that the components of the equation
of motion for the Wigner function roughly scale as 
\begin{subequations}
\label{eq:-76-1}
\begin{align}
\partial_{y} & \appropto\frac{1}{s}, & \partial_{x} & \appropto s,\\
y & \appropto s, & x & \appropto\frac{1}{s}.
\end{align}
\end{subequations}
The symbol $\appropto$ should be read as “approximately proportional
to” since (\ref{eq:-76-1}) describes the exact behavior for the
initial state only. We can only expect the large squeezing approximation
to hold as long as the these scalings are approximately true. Geometrically,
the Wigner function can only assume large values where $x$ is small
and can only vary slowly in the direction of the $y$-axis. Looking
at the evolution qualitatively (recall the initial state of Figure
\ref{fig:evo-gallery-long}), this is seen to be the case for the
initial state and the short time evolution. Recall however Figure
\ref{fig:Squeezed-vacuum-state}: As the evolution progresses, the
regions where the Wigner density is largest are pulled outward from
the $y$-axis and bent toward larger $x$-values. Simultaneously,
the Wigner function bends into the S-shape causing parts of the $x$-component
of the gradient $\partial_{x}$ to shift to the $y$-component ($\partial_{y}$)
. Both mechanisms tend to worsen the large squeezing approximation.

\begin{figure}
\noindent \begin{centering}
\includegraphics{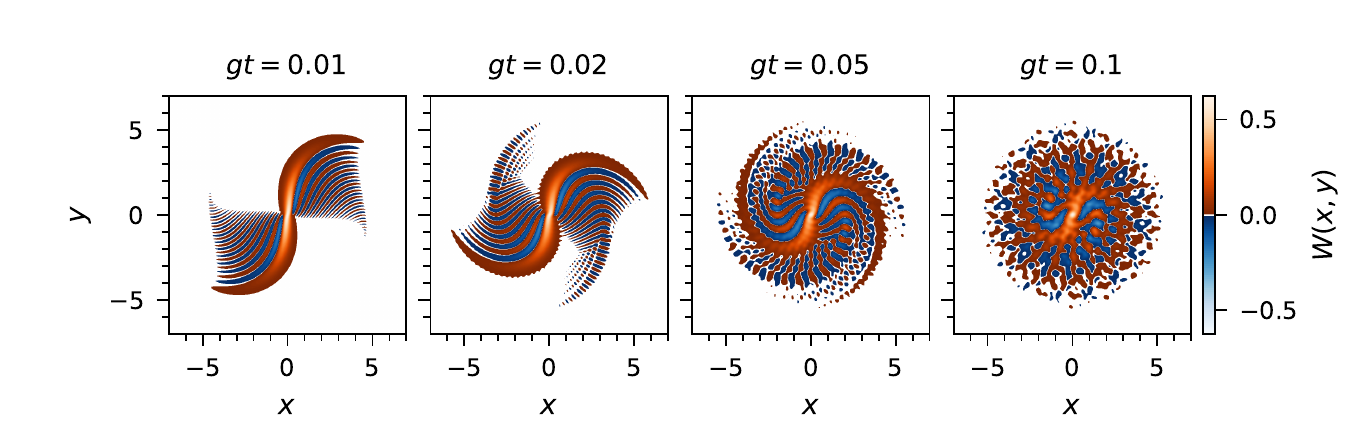}
\par\end{centering}
\caption[Intermediate time unitary evolution of squeezed vacuum]{\label{fig:evo-gallery-medium}\textbf{Intermediate time evolution
of squeezed vacuum.} Demonstration of the transition away from the
initial negativity mechanism as described in Section \ref{subsec:Evolution-over-Short}.
As the Wigner function starts to bend toward the $x$-axis, the fringes
eventually reach the opposite side of the bulk of the Wigner function.
Around this time, the character of the evolution changes significantly
and the initial “cigar” shape of the Wigner function is now
no longer visible. The plots show the Wigner function evolved from
the squeezed vacuum state with $r_{0}=1.5$. Figure \ref{fig:Squeezed-vacuum-state}
shows the same initial state at points earlier in the evolution.}
\end{figure}

Deriving and using the differential equation (\ref{eq:-75}) is effectively
equivalent to considering the evolution of $W$ on a single line parallel
to the squeezed axis.\footnote{One might also conceive of a similar argument but using polar coordinates:
Instead of looking at the evolution of $W$ along a line, one could
instead consider $W$ on a circle centered at the origin. The corresponding
algebraic view would be to disregard terms of (\ref{eq:-18}) containing
$\partial_{r}$, thereby obtaining an equation for $W(r,\phi,t)$
in which $r$ can be regarded as a parameter. This resulting partial
differential equation would be in the variables $\phi$ and $t$ and,
due to the rotational symmetry of the Kerr Hamiltonian, have constant
coefficients (independent of $\phi$ and $t$ though not necessarily
of $r$). For an example, see \textcite{Oliva_QuantumKerrOscillators_2019}
who employ such an argument, albeit only qualitatively, to describe
the appearance of negativity for a coherent initial state. (\ref{eq:-100})
can also be used to support this view.} This situation is illustrated in Figure \ref{fig:wigner-1dcut}.
We note that the first moment of $\tilde{u}_{\tilde{y}}$ with respect
to $\mu$ is unchanged under the evolution of (\ref{eq:-63}):
\begin{align}
\partial_{\tilde{\tau}}\int d\mu\,\mu\tilde{u}_{\tilde{y}}(\mu,\tilde{\tau}) & =\int d\mu\,\mu\partial_{\mu}^{3}\tilde{u}_{\tilde{y}}(\mu,\tilde{\tau})=0.\label{eq:-205}
\end{align}
Hence, any change in the conditional expectation value of $\tilde{x}$
(conditioned on $\tilde{y}$) is described by the rescaling in equation
(\ref{eq:-62}) alone:
\begin{subequations}
\label{eq:-207}
\begin{align}
\partial_{t}\int d\tilde{x}\,\tilde{x}\tilde{W}(\tilde{x},\tilde{y},t) & =\int d\tilde{x}\,\tilde{x}\partial_{t}u_{\tilde{y}}(\tilde{x}-2gs^{4}\tilde{y}^{3}t,t)\\
 & =\int d\tilde{x}\,\tilde{x}\left(\partial_{t}u_{\tilde{y}}(\tilde{x},t)-2gs^{4}\tilde{y}^{3}\partial_{\mu}u_{\tilde{y}}(\mu,t)\right)
\end{align}
\end{subequations}
The first term disappears by (\ref{eq:-205}). The second term is
then rewritten as
\begin{subequations}
\label{eq:-206}
\begin{align}
\partial_{t}\int d\tilde{x}\,\tilde{x}\tilde{W}(\tilde{x},\tilde{y},t) & =-\int d\mu\,\left(\mu+2gs^{4}\tilde{y}^{3}\right)2gs^{4}\tilde{y}^{3}\partial_{\mu}u_{\tilde{y}}(\mu,t)\\
 & =\int d\mu\,gs^{4}\tilde{y}^{3}u_{\tilde{y}}(\mu,t)\\
 & \propto gs^{4}\tilde{y}^{3}.
\end{align}
\end{subequations}
The bulk of the initial Wigner function (what is sometimes described
as a cigar-shape \cite{Kenfack_NegativityWignerFunction_2004} parallel
to the $y$-axis) is therefore expected to evolve such that the major
axis moves to form a cubic monomial proportional to $gs^{4}\tilde{y}^{3}$.
This is also illustrated in Figure (\ref{fig:wigner-1dcut}) where
an appropriate cubic monomial has been superimposed on top of the
Wigner function. This evolution fails to describe the bending toward
the $x$-axis which is observed for intermediate times (compare Figures
\ref{fig:Squeezed-vacuum-state} and \ref{fig:asymp-medium-evo}).
As the state evolves further, the fringes reach the opposing side
of the bulk and cause the appearance of the state to change character
completely, loosing most of its resemblance with the initial state.
This is shown in Figure \ref{fig:evo-gallery-medium}.

\subsection{Evolution of Negativity\label{subsec:Evolution-of-Negativity-1}}

\begin{figure}
\subfloat[\label{fig:sctime-negvol} Negative volume]{\noindent \begin{centering}
\captionsetup{position=top}\includegraphics{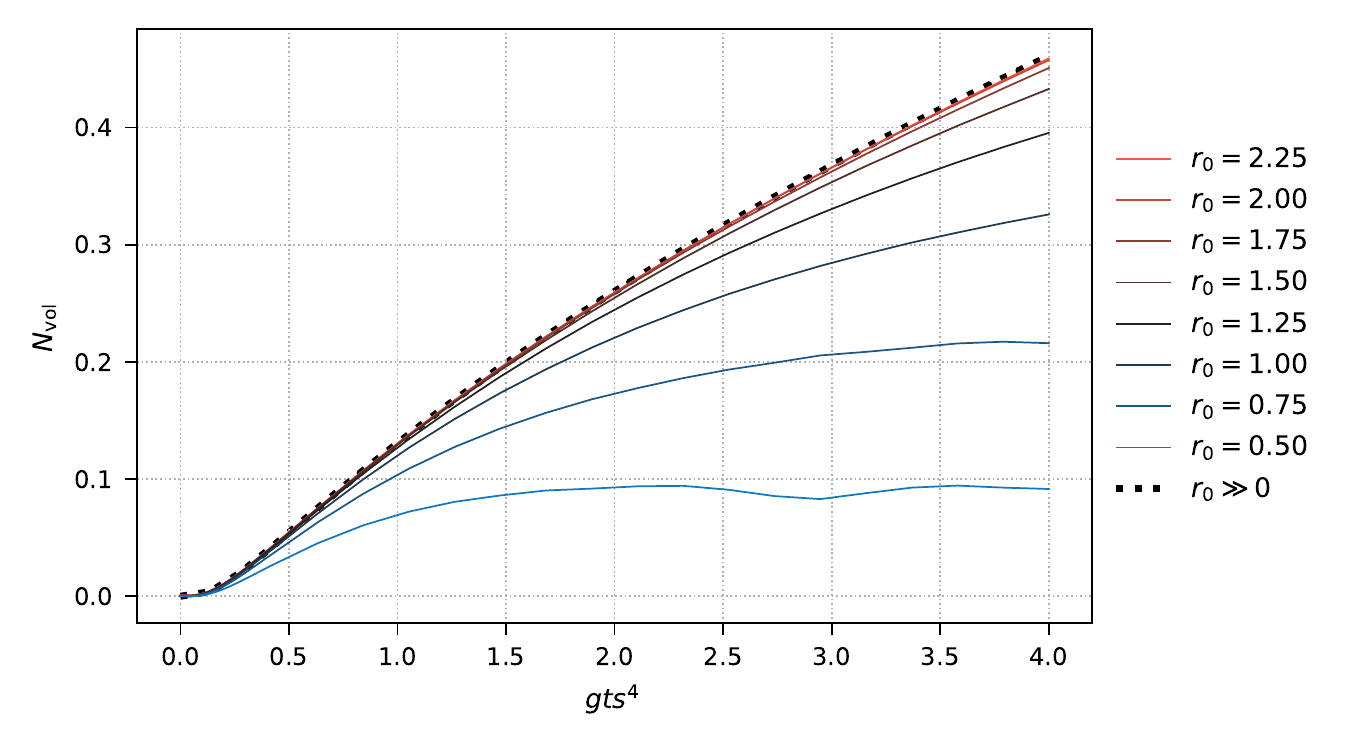}
\par\end{centering}
\noindent \centering{}}

\subfloat[\label{fig:sctime-negpeak}Negative peak]{\noindent \begin{centering}
\captionsetup{position=top}\includegraphics{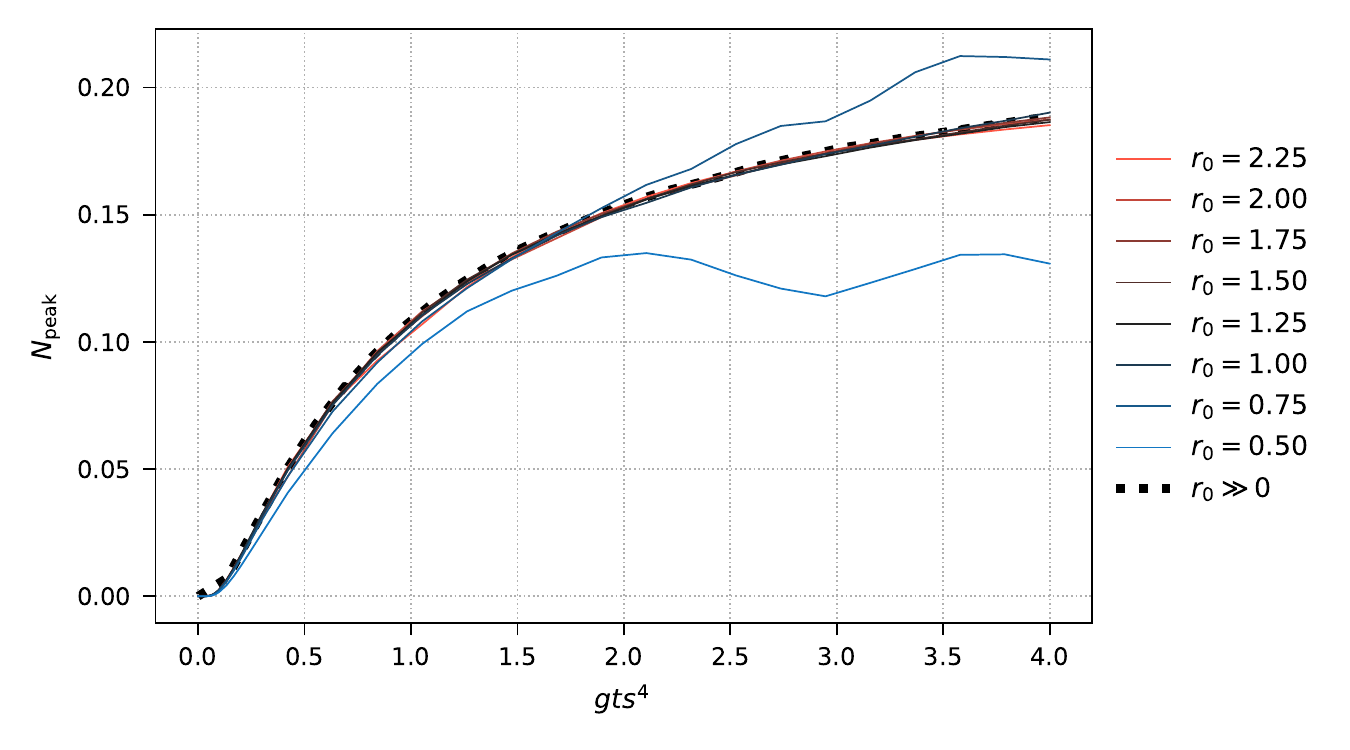}
\par\end{centering}
\noindent \centering{}}

\caption[Negativity versus scaled time for squeezed vacuum]{\label{fig:-2}\textbf{Negativity versus scaled time for squeezed
vacuum. }The graphs of Figure \ref{fig:-1} plotted as a function
of $gts^{4}$. This reveals the universal scaling described in Section
\ref{subsec:Evolution-of-Negativity-1}. The negativity obtained in
the large squeezing approximation from (\ref{eq:-223}) is shown as
the thick dotted line.}
\end{figure}

To conclude on the unitary evolution of the squeezed state, we return
to the consideration of the quantities $N_{\mathrm{vol}}$ and $N_{\mathrm{peak}}$.
With the considerations of Sections \ref{subsec:intro-rescaled-coords}--\ref{subsec:large-squeezing-approximation},
we can define a scaled time that reveals the universal large squeezing
behavior of the graphs shown in Figure \ref{fig:-1}. Consider first
the negative volume $N_{\mathrm{vol}}$. Inserting $\tilde{W}(\tilde{x},\tilde{y},t)$
into the definition of negative volume (\ref{eq:negvol-definition})
and changing the integration variables, one obtains

\begin{equation}
N_{\mathrm{vol}}=-\int d\tilde{x}d\tilde{y}\,\min\{0,\tilde{W}(\tilde{x},\tilde{y},t)\}.\label{eq:negvol-definition-1}
\end{equation}
This expression still depends on squeezing. Using (\ref{eq:-51})
we can however express $N_{\mathrm{vol}}$ in terms of $\tilde{u}_{\tilde{y}}$
which is independent of $s$ (see (\ref{eq:-63}) and remarks below):
\begin{equation}
N_{\mathrm{vol}}=-\int d\tilde{x}d\tilde{y}\,\min\{0,\tilde{u}_{\tilde{y}}(\tilde{x}-2gts^{4}\tilde{y}^{3},gts^{4}\tilde{y}/8)\}.\label{eq:-162}
\end{equation}
The same analysis may be performed for $N_{\mathrm{peak}}$ (defined
in equation (\ref{eq:negpeak-definition})) yielding
\begin{equation}
N_{\mathrm{peak}}=-\min_{\tilde{x},\tilde{y}}\left(\min\{0,\tilde{W}(\tilde{x},\tilde{y},t)\}\right).\label{eq:negvol-definition-1-1}
\end{equation}
The peak of $\tilde{W}$ (denoted by $\min$) is the same as $W$
since no rescaling was performed in the transformation (\ref{eq:-19})
(Section \ref{sec:squeezed-thermal-state} treats an initial state
for which this is not the case). Inserting $\tilde{u}_{\tilde{y}}$
yields
\begin{equation}
N_{\mathrm{peak}}=-\min_{\tilde{x},\tilde{y}}\left(\min\{0,\tilde{u}_{\tilde{y}}(\tilde{x}-2gts^{4}\tilde{y}^{3},gts^{4}\tilde{y}/8)\}\right).\label{eq:-163}
\end{equation}
Since$\tilde{u}_{\tilde{y}}$ is independent of the squeezing $s$,
all information about $s$ is explicit in (\ref{eq:-162}) and (\ref{eq:-163}).
From this we expect that $N_{\mathrm{vol}}$ and $N_{\mathrm{peak}}$
as functions of a rescaled time $gts^{4}$ are both invariant of $s$.
We can use this to rescale the time axes of Figures \ref{fig:-1}.
Doing this yields Figure \ref{fig:-2}. The graphs display asymptotic
behavior for the combination of large squeezing and small time. The
time axes have been extended far enough to clearly show the breakdown
of the approximation for various values of $r_{0}$. Figure \ref{fig:-2}
suggests that an increase in the squeezing $s$ may be used to compensate
for a weak nonlinearity, i.e. small $g$ (and vice versa), which is
at least seen to be possible for the quantities $N_{\mathrm{vol}}$
or $N_{\mathrm{peak}}$.

\section{Kerr Evolution of Squeezed Thermal State\label{sec:squeezed-thermal-state}}

\begin{figure}
\noindent \begin{centering}
\includegraphics{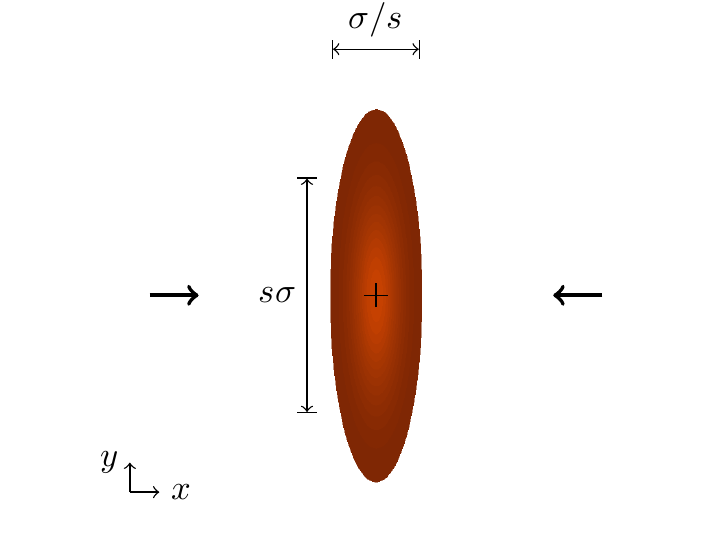}
\par\end{centering}
\noindent \centering{}\caption[Illustration of squeezed thermal state]{\label{fig:squeezed-thermal-initial-state}\textbf{Illustration of
squeezed thermal state.} The Wigner function of a squeezed thermal
state is also a Gaussian function. The expression may be found in
equation (\ref{eq:-95}). The parameter $\sigma$ increases the variance
in both axes while the parameter $s$ decreases and increases the
variance in the squeezed and anti-squeezed axis respectively. Compare
this with the squeezed vacuum state as illustrated in Figure \ref{fig:evo-gallery-medium-1-1}.
The parameter $\theta_{0}$ has been set to zero causing the major
axis of the Wigner function to coincide with the $y$-axis.}
\end{figure}

To broaden the relevance of the results of the previous section, we
consider now squeezed thermal states. Moving from a squeezed vacuum
state to a squeezed thermal state extends the results to a broader
class of Gaussian initial states (in fact all valid Gaussian states
that are centered at the origin) yet the analysis remains largely
unchanged. The Wigner function for the squeezed thermal state is given
by a Gaussian function centered on the origin (see Appendix \ref{app:squeezed-thermal-state-wig}):
\begin{equation}
W(x,y,0)=\frac{2}{\pi\sigma^{2}}\begin{aligned}[t]\exp & \left[-\frac{2x^{2}s^{2}}{\sigma^{2}}-\frac{2y^{2}}{s^{2}\sigma^{2}}\right]\end{aligned}
.\label{eq:-95}
\end{equation}
Since the dynamics are rotationally invariant, we have set $\theta_{0}=0$
without loss of generality. The parameter $s$ is defined by (\ref{eq:-228})
and describes the squeezing as in the previous section. To describe
the temperature of the state, the parameter
\begin{equation}
\sigma=\sqrt{2\bar{n}_{0}+1}
\end{equation}
is introduced. Here $\bar{n}_{0}$ is the mean occupancy of the initial
non-squeezed thermal state.\footnote{$\bar{n}_{0}$ is not equal to the mean occupancy of the state (\ref{eq:-95})
though, which may be calculated from (\ref{eq:-95}) as $\langle\hat{n}\rangle=\int dxdy\frac{1}{2}\left(x^{2}+y^{2}\right)W(x,y)=\frac{1}{4}\left(2\bar{n}_{0}+1\right)\left(s^{2}+s^{-2}\right)$.} The state is illustrated in Figure \ref{fig:squeezed-thermal-initial-state}.
The quadrature variances are found as
\begin{subequations}
\label{eq:-96}
\begin{align}
\left\langle (\Delta\hat{X})^{2}\right\rangle  & =\int dxdy\,(x^{2}-x\langle\hat{X}\rangle)W(x,y)=\frac{\sigma^{2}}{4s^{2}}\label{eq:-107}\\
\intertext{and}\left\langle (\Delta\hat{Y})^{2}\right\rangle  & =\int dxdy\,(y^{2}-y\langle\hat{Y}\rangle)W(x,y)=\frac{s^{2}\sigma^{2}}{4},
\end{align}
\end{subequations}
which is in agreement with (\ref{eq:-9}) and (\ref{eq:-10}) for
the vacuum state ($\sigma=1$). Note that $\langle\hat{X}\rangle=\langle\hat{Y}\rangle=0$
which can be found in the same way. Since the parameter $\bar{n}_{0}$
can be any non-negative real, we see that $\sigma$ can be chosen
as $\sigma\in[1,\infty)$. Comparing (\ref{eq:-96}) with (\ref{eq:heisenberg-limit}),
this is found to be exactly the condition for the state (\ref{eq:-95})
to obey the fundamental quadrature uncertainty relation (\ref{eq:heisenberg-limit}).
Choosing $s\in[1,\infty)$, means that (\ref{eq:-95}) can represent
any Gaussian function with the $x$- and $y$-axes as its minor and
major axes.\footnote{One can generalize to a Gaussian function with its major and minor
axes rotated to any angle by reintroducing the parameter $\theta_{0}$,
however for the dynamics considered here which are rotationally invariant,
this is unnecessary. Further generalization to any Gaussian function
obeying (\ref{eq:heisenberg-limit}) the can be achieved with use
of the displacement operator (\ref{eq:displacement-operator}). Both
are generalizations can be found in Appendix \ref{app:displaced-squeezed-thermal-state-wig}.}

We focus again on the short time evolution of negativity. Figure \ref{fig:-2}
show the negative volume and peak plotted as a function of time for
various values of the parameter $\bar{n}_{0}$. Increasing $\sigma$
generally causes the negativity to decrease. We wish to find a way
to scale the axes in Figure \ref{fig:-2} such that the graphs converge
to a single graph independent of $\bar{n}_{0}$. This is similar to
what was done for the squeezed vacuum state in Figures \ref{fig:sctime-negvol}
and \ref{fig:sctime-negpeak} with respect to the parameter $s$.
With this in mind, we therefore repeat the analysis of Section \ref{subsec:intro-rescaled-coords}.

\subsection{Introduction of Rescaled Coordinates}

In Section \ref{subsec:intro-rescaled-coords}, we introduced a new
set of coordinates which allowed us to express the initial squeezed
vacuum state in a form independent of the squeezing $s$ (equation
(\ref{eq:-30})). To do the same for the thermal state, the new coordinates
$(\tilde{x},\tilde{y})$ should also depend on the parameter $\sigma$.
In this case, they take the form
\begin{equation}
\tilde{x}=\frac{sx}{\sigma}\text{\ensuremath{\qquad}and}\qquad\tilde{y}=\frac{y}{s\sigma}.\label{eq:-97}
\end{equation}
With this choice of coordinates, the initial state Wigner function
is again a two-dimensional isotropic Gaussian with both of its variances
equal to $1/4$ (the same form as (\ref{eq:-30})). In analogy with
(\ref{eq:-19}) and (\ref{eq:-27}), the scaled Wigner function $\tilde{W}(\tilde{x},\tilde{y},t)$
is introduced in terms of the regular Wigner function $W(x,y,t)$
with
\begin{equation}
\tilde{W}(\tilde{x},\tilde{y},t)=\sigma^{2}W(\tilde{x}/s,s\tilde{y},t)\label{eq:-98}
\end{equation}
and the scaled differential operators with
\begin{equation}
\partial_{\tilde{x}}=\frac{\sigma}{s}\partial_{x},\qquad\text{and}\qquad\partial_{\tilde{y}}=s\sigma\partial_{y}.\label{eq:-27-1}
\end{equation}
The additional factor of $\sigma^{2}$ in (\ref{eq:-98}) is required
for $\tilde{W}(\tilde{x},\tilde{y},0)$ to take the exact form of
(\ref{eq:-30}). It also retains the normalization of the Wigner function
with respect to the new coordinates: $\int d\tilde{x}\,d\tilde{y}\,\tilde{W}(\tilde{x},\tilde{y},t)=1$.
The unscaled equation of motion for $W(x,y,t)$ remains (\ref{eq:-31-2}).
Using (\ref{eq:-97}--\ref{eq:-27-1}) to express (\ref{eq:-31-2}),
the scaled coordinate equation of motion therefore becomes
\begin{equation}
\partial_{t}\tilde{W}(\tilde{x},\tilde{y},t)=\begin{aligned}[t] & 2g\sigma^{2}\left(-\tilde{x}^{2}\tilde{y}\partial_{\tilde{x}}-s^{4}\tilde{y}^{3}\partial_{\tilde{x}}+\frac{1}{s^{4}}\tilde{x}^{3}\partial_{\tilde{y}}+\tilde{x}\tilde{y}^{2}\partial_{\tilde{y}}\right)\tilde{W}(\tilde{x},\tilde{y},t)\\
 & -2g\left(-s^{2}\tilde{y}\partial_{\tilde{x}}+\frac{1}{s^{2}}\tilde{x}\partial_{\tilde{y}}\right)\tilde{W}(\tilde{x},\tilde{y},t)\\
 & -\frac{g}{8\sigma^{2}}\left(-s^{4}\tilde{y}\partial_{\tilde{x}}^{3}+\frac{1}{s^{4}}\tilde{x}\partial_{\tilde{y}}^{3}+\tilde{x}\partial_{\tilde{y}}\partial_{\tilde{x}}^{2}-\tilde{y}\partial_{\tilde{x}}\partial_{\tilde{y}}^{2}\right)\tilde{W}(\tilde{x},\tilde{y},t).
\end{aligned}
\label{eq:-106}
\end{equation}
Compare this with the rescaled equation of motion (\ref{eq:-31-2-1})
for a squeezed vacuum initial state. The remarks on the power of $s$
in the different terms made below equation (\ref{eq:-31-2-1}) are
still valid. Additionally, terms describing the distance to the $x$-
or $y$-axes are multiplied with some positive power of $\sigma$
(e.g. $-2g\sigma^{2}\tilde{x}^{2}\tilde{y}\partial_{\tilde{x}}\tilde{W}$),
whereas terms describing the spatial variation are divided by $\sigma$
to some positive power (e.g. $-(g/8)\tilde{x}\partial_{\tilde{y}}\partial_{\tilde{x}}^{2}\tilde{W}$).
Terms that are balanced between the two remain unchanged (e.g. $2gs^{2}\tilde{y}\partial_{\tilde{x}}$).
This follows from the fact that increasing $\sigma$ decreases the
spatial variation in the Wigner function and increases the average
distance to the origin for the Wigner density.

\begin{figure}
\subfloat[\label{fig:-2-2-1-1}Negative volume]{\noindent \begin{centering}
\captionsetup{position=top}\includegraphics{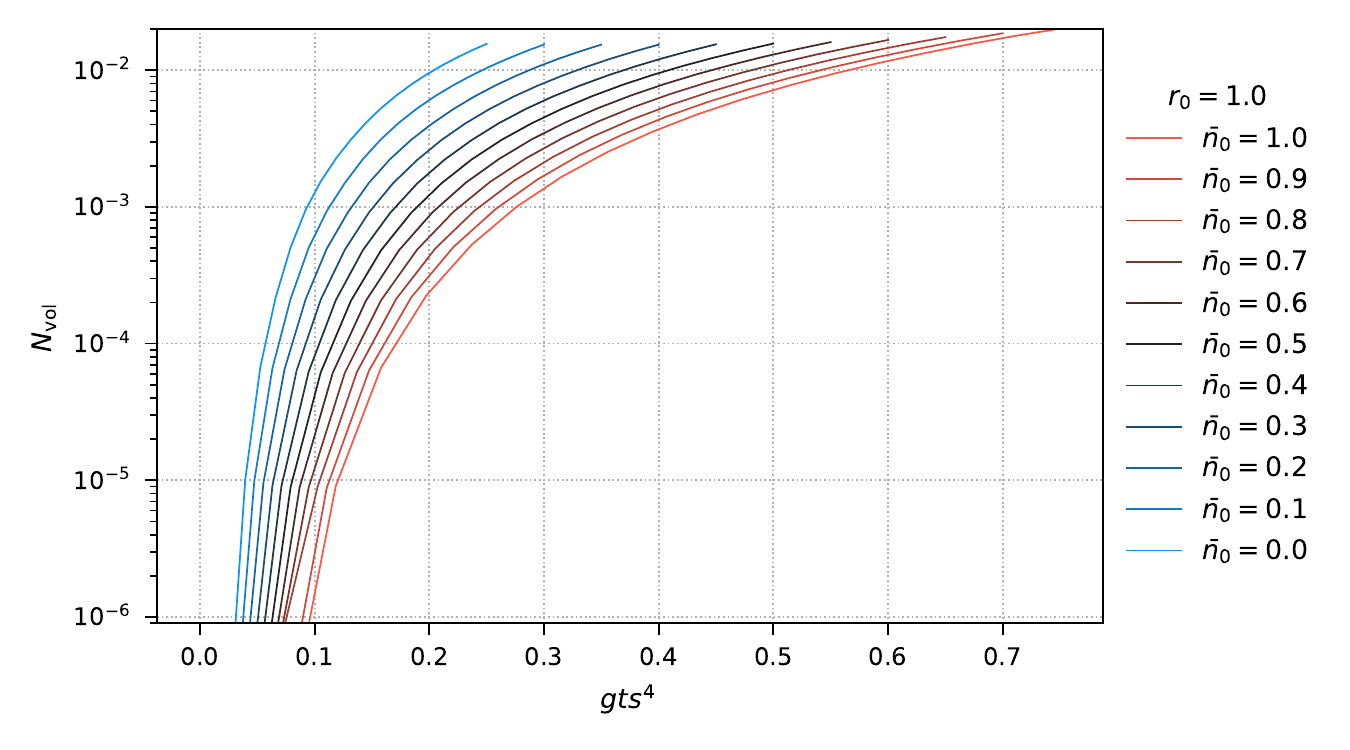}
\par\end{centering}
\noindent \centering{}}

\subfloat[\label{fig:-2-1-1-1-1} Negative peak]{\noindent \begin{centering}
\captionsetup{position=top}\includegraphics{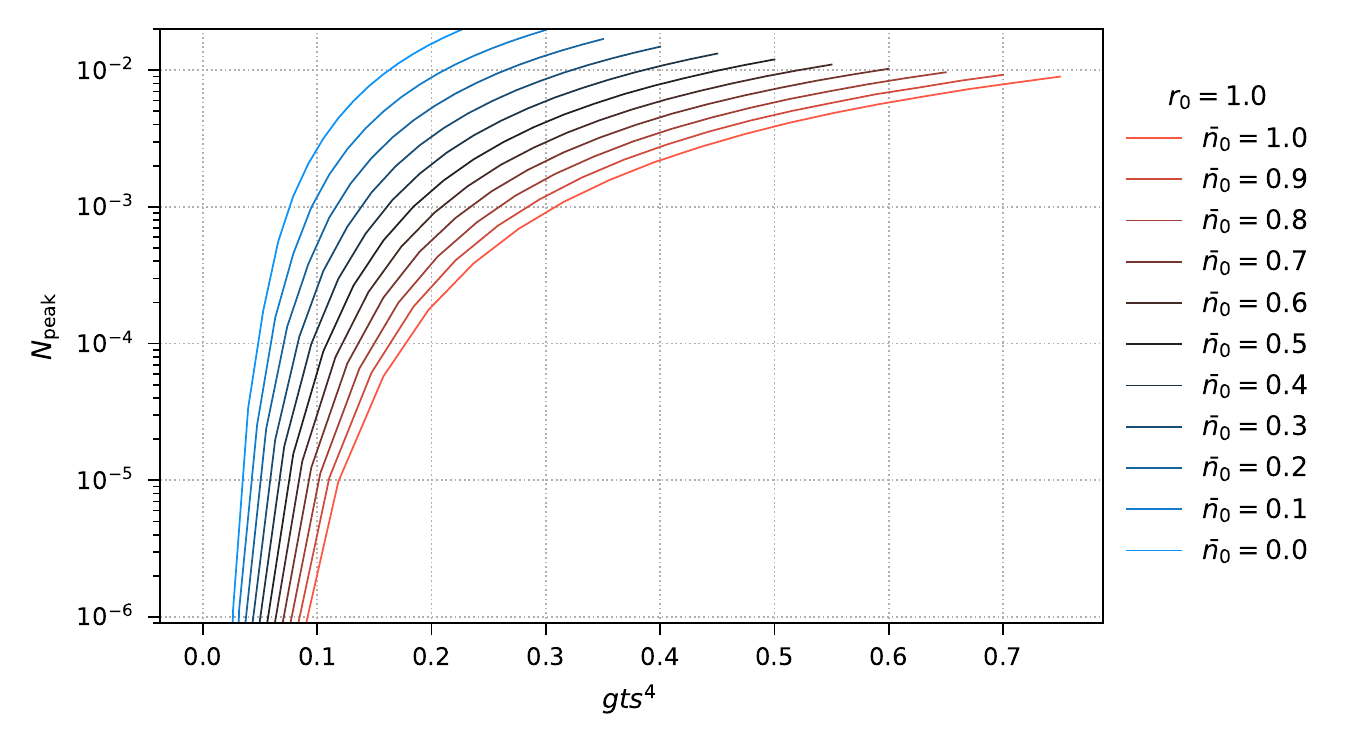}
\par\end{centering}
\noindent \centering{}}

\caption[Negativity during short time unitary evolution of squeezed thermal
state]{\label{fig:-1-2}\textbf{Negativity during short time unitary evolution
of squeezed thermal state. }The evolution in negativity as simulated
for various squeezed thermal states. The squeezing parameter is held
fixed $r_{0}=1$ while the parameter $\bar{n}_{0}$, describing the
mean occupancy of the thermal state (before squeezing), is varied.
Increasing the temperature of the state and thus $\bar{n}_{0}$ causes
the negativity to drop. Evolution of the squeezed thermal state is
described in Section \ref{sec:squeezed-thermal-state}.}
\end{figure}

All terms containing third-order derivatives also contain the factor
$\sigma^{-2}$. These are the terms generating negativity and we therefore
expect the negativity to decrease with increasing $\sigma$. Figure
\ref{fig:-1-2} show the quantities $N_{\mathrm{vol}}$ and $N_{\mathrm{peak}}$
computed for the system evolved under the master equation (\ref{eq:kerr-vonneumann-eq}).
The initial state is a squeezed thermal state with $r_{0}=1$. The
thermal occupancy $\bar{n}_{0}$ was varied between $0$ and $1$
whereby $\sigma$ varies between $1$ and $\sqrt{3}=1.73$. We see
that both $N_{\mathrm{vol}}$ and $N_{\mathrm{peak}}$ decrease monotonically
with increasing $\bar{n}_{0}$.

\subsection{Large Squeezing Approximation}

Having transformed the problem such as to express the squeezing and
temperature from the initial state to the equation of motion (\ref{eq:-106}),
we continue in analogy with Section \ref{subsec:large-squeezing-approximation}.
We wish to construct an approximate equation of motion from (\ref{eq:-106})
by retaining only terms significant in the limit of large squeezing.
We consider here the case where $s\gg\sigma$ for any valid value
of $\sigma$. In this case, we keep from (\ref{eq:-106}) only terms
proportional to $s^{4}$ to obtain
\begin{equation}
\partial_{t}\tilde{W}(\tilde{x},\tilde{y},t)=-2g\sigma^{2}s^{4}\tilde{y}^{3}\partial_{\tilde{x}}\tilde{W}(\tilde{x},\tilde{y},t)+\frac{g}{8\sigma^{2}}s^{4}\tilde{y}\partial_{\tilde{x}}^{3}\tilde{W}(\tilde{x},\tilde{y},t).\label{eq:-113}
\end{equation}
This is similar to what was done to reach (\ref{eq:-61}). We can
treat the approximate equation (\ref{eq:-113}) in the same way as
we did that of a squeezed vacuum state. To solve (\ref{eq:-113}),
introduce a new function $\tilde{u}(\mu,\tilde{\tau})$ by
\begin{equation}
\tilde{W}(\tilde{x},\tilde{y},t)=\tilde{u}_{\tilde{y}}(\tilde{x}-2g\sigma^{2}s^{4}\tilde{y}^{3}t,gs^{4}\tilde{y}t/8\sigma^{2}),\label{eq:-158}
\end{equation}
whose equation of motion will now be given by
\begin{equation}
\partial_{\tilde{\tau}}\tilde{u}_{\tilde{y}}(\mu,\tilde{\tau})=\partial_{\mu}^{3}\tilde{u}_{\tilde{y}}(\mu,\tilde{\tau}).\label{eq:-159}
\end{equation}
Equations (\ref{eq:-158}) and (\ref{eq:-159}) are analogous to the
equations (\ref{eq:-62}--\ref{eq:-63}) for the squeezed vacuum
state. Equation (\ref{eq:-159}) is identical to (\ref{eq:-63}) and
its solutions may thus be obtained using the methods described in
Section \ref{subsec:large-squeezing-approximation}. We simply state
here the Fourier series solution for $\tilde{W}(\tilde{x},\tilde{y},t)$
obtained in a way analogous to (\ref{eq:-160}):
\begin{subequations}
\label{eq:-161}
\begin{equation}
\tilde{W}(\tilde{x},\tilde{y},t)=\frac{1}{\sqrt{2\pi}}\int_{-\infty}^{\infty}dk\,h_{\tilde{y}}(k)\,e^{i(k\tilde{x}-2kgt\sigma^{2}s^{4}\tilde{y}^{3}-gts^{4}k^{3}\tilde{y}/8\sigma^{2})}
\end{equation}
with
\begin{equation}
h_{\tilde{y}}(k)=\frac{1}{\sqrt{2\pi}}\int_{-\infty}^{\infty}d\tilde{x}\,\tilde{W}(\tilde{x},\tilde{y},0)\,e^{-ik\tilde{x}}=\frac{1}{\pi}e^{-2\tilde{y}^{2}}e^{-k^{2}/8}.
\end{equation}
\end{subequations}

\subsection{Evolution of Negativity\label{subsec:Evolution-of-Negativity}}

\begin{figure}
\subfloat[\label{fig:-2-2-1} Negative volume]{\noindent \begin{centering}
\captionsetup{position=top}\includegraphics{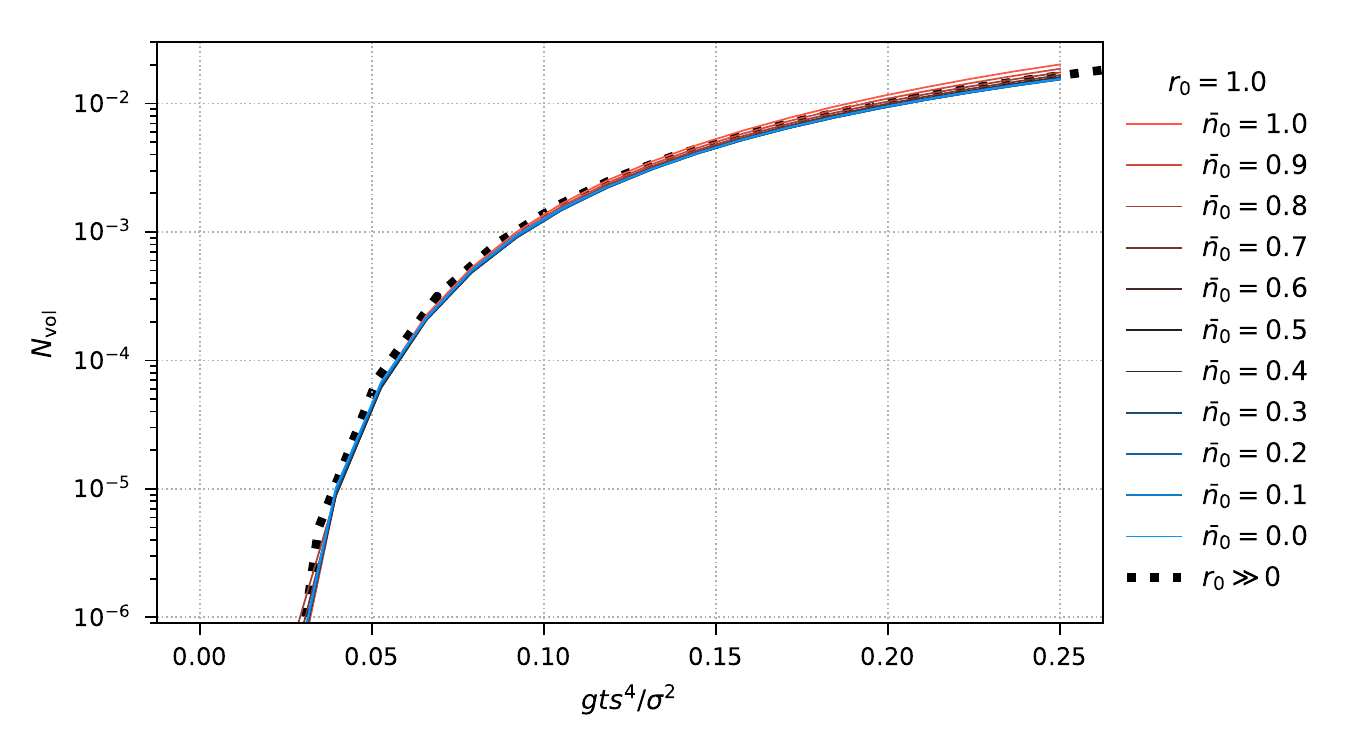}
\par\end{centering}
\noindent \centering{}}

\subfloat[\label{fig:-2-1-1-1}Negative peak]{\noindent \begin{centering}
\captionsetup{position=top}\includegraphics{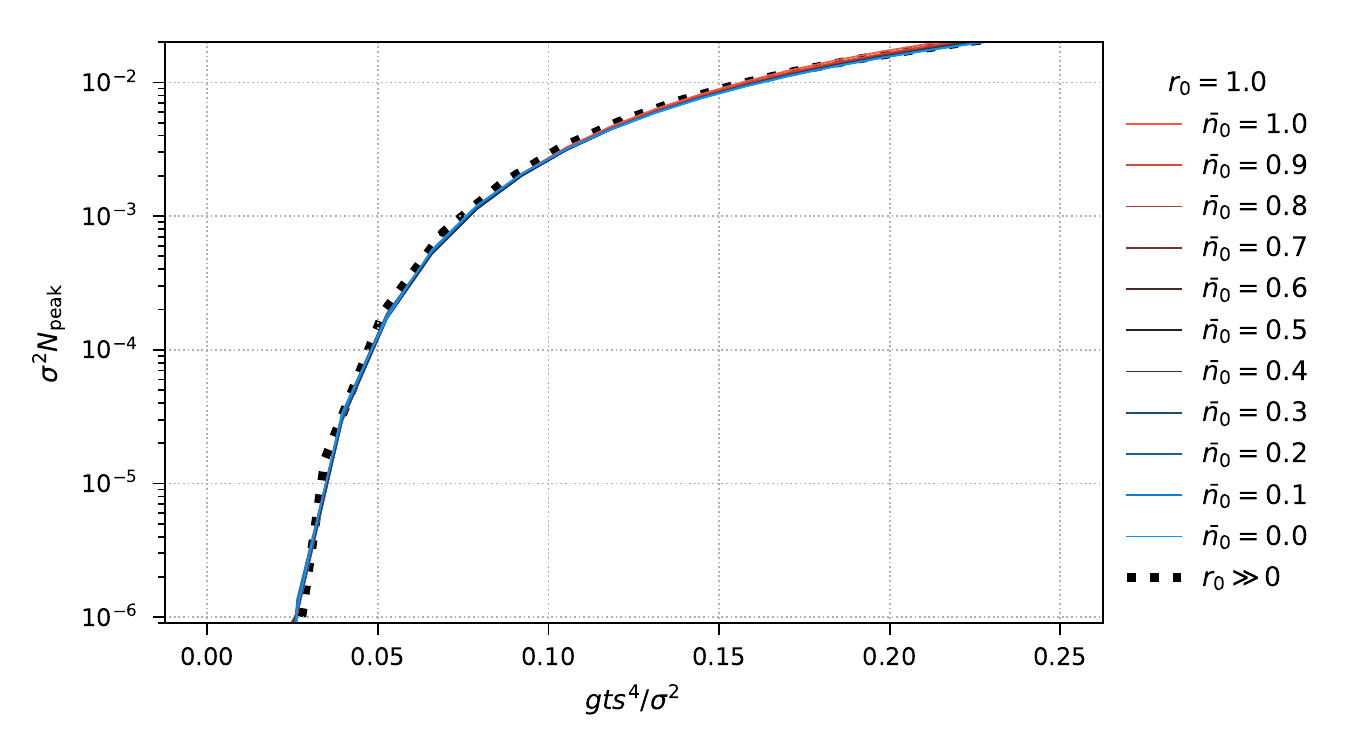}
\par\end{centering}
\noindent \centering{}}

\caption[Negativity versus rescaled time for squeezed thermal state]{\label{fig:Negativity-versus-rescaled}\textbf{Negativity versus
rescaled time for squeezed thermal state. }Rescaling the time axis
to $gts^{4}/\sigma^{2}$ reveals the universal scaling described in
Section \ref{subsec:Evolution-of-Negativity}. The negativity obtained
in the large squeezing approximation from (\ref{eq:-161}) is shown
as the thick dotted line.}
\end{figure}

We now wish to express $N_{\mathrm{vol}}$ and $N_{\mathrm{peak}}$
with all dependence on $s$ and $\sigma$ explicit. Using the definition
(\ref{eq:-98}) of $\tilde{W}(\tilde{x},\tilde{y},t)$ with the definition
of negative volume (\ref{eq:negvol-definition}) and changing the
the integration variables according to (\ref{eq:-97}), one obtains

\begin{equation}
N_{\mathrm{vol}}=-\int d\tilde{x}d\tilde{y}\,\min\{0,\tilde{W}(\tilde{x},\tilde{y},t)\}.\label{eq:negvol-definition-1-2}
\end{equation}
Even though a factor $\sigma^{2}$ is now present in both (\ref{eq:-97})
and (\ref{eq:-98}), they exactly cancel and the resulting form of
$N_{\mathrm{vol}}$ matches (\ref{eq:negvol-definition-1}). (\ref{eq:-158})
is applied to write 
\begin{equation}
N_{\mathrm{vol}}=-\int d\tilde{x}d\tilde{y}\,\min\{0,\tilde{u}_{\tilde{y}}(\tilde{x}-2g\sigma^{2}s^{4}\tilde{y}^{3}t,gs^{4}\tilde{y}t/8\sigma^{2})\}.\label{eq:negvol-definition-1-2-1}
\end{equation}
We perform the integral substitution with $\mu=\tilde{x}-2g\sigma^{2}s^{4}\tilde{y}^{3}t$
to obtain\footnote{Notice that this relation between $\tilde{x}$ and $\mu$ in (\ref{eq:-62})
where $\tilde{u}_{\tilde{y}}(\mu,t)$ is defined. This substitution
is needed here specifically since the subexpressions $gts^{4}\sigma^{2}$
and $gts^{4}\sigma^{-2}$ both appear in (\ref{eq:negvol-definition-1-2-1})
(whereas time only enters into (\ref{eq:-162}) as $gts^{4}$). The
limits of the integral are $\pm\infty$ and thus unchanged by the
substitution.} 
\begin{equation}
N_{\mathrm{vol}}=-\int d\tilde{y}\int d\mu\,\min\{0,\tilde{u}_{\tilde{y}}(\mu,gs^{4}\tilde{y}t/8\sigma^{2})\}.\label{eq:negvol-definition-1-2-1-1}
\end{equation}
Steps for the negative peak are similar. From its definition (\ref{eq:negpeak-definition}),
the negative peak is expressed in terms of $\tilde{W}(\tilde{x},\tilde{y},t)$
as
\begin{equation}
N_{\mathrm{peak}}=-\min_{\tilde{x},\tilde{y}}\left(\min\{0,\sigma^{-2}\tilde{W}(\tilde{x},\tilde{y},t)\}\right).
\end{equation}
Insertion of $\tilde{u}_{\tilde{y}}$ yields 
\begin{subequations}
\label{eq:-164}
\begin{align}
\sigma^{2}N_{\mathrm{peak}} & =-\min_{\tilde{x},\tilde{y}}\left(\min\{0,\tilde{u}_{\tilde{y}}(\tilde{x}-2g\sigma^{2}s^{4}\tilde{y}^{3}t,gts^{4}\tilde{y}/8\sigma^{2})\}\right)\label{eq:-163-1}\\
 & =-\min_{\tilde{x},\tilde{y}}\left(\min\{0,\tilde{u}_{\tilde{y}}(\tilde{x},gts^{4}\tilde{y}/8\sigma^{2})\}\right).
\end{align}
\end{subequations}

Plotting $N_{\mathrm{vol}}$ and $\sigma^{2}N_{\mathrm{peak}}$ as
functions of $gts^{4}/\sigma^{2}$, they are seen to be invariant
of $\bar{n}_{0}$ ($s$ is held constant here so using $gt/\sigma^{2}$
for the $x$-axis would simply scale the axis and lead to the same
conclusion, but we preserve $s$ for consistency with earlier figures).
Figure \ref{fig:Negativity-versus-rescaled} shows $N_{\mathrm{vol}}$
and $N_{\mathrm{peak}}$ with this scaled axis. The chosen axis scaling
is seen to shift the graphs to lie atop each other (compare with Figure
\ref{fig:-1-2}). An increase in $\sigma$ generally leads to broader
and shallower features of the Wigner function which is why $N_{\mathrm{vol}}$
is unchanged whereas $N_{\mathrm{peak}}$ is reduced in size by a
factor of $\sigma^{2}$. 

\subsection{Validity of Approximation for Squeezed Thermal State}

We shall briefly discuss the validity of the approximation made for
the thermal state, as was done for the squeezed vacuum state in Section
\ref{subsec:Validity-of-Large}. Assuming $s\gg\sigma$ to reach (\ref{eq:-113})
carries with it the same assumptions as the case for the squeezed
vacuum state. As seen from (\ref{eq:negvol-definition-1-2-1-1}) and
(\ref{eq:-164}) the term of (\ref{eq:-113}) proportional to $\partial_{\tilde{x}}$
has no influence on either $N_{\mathrm{vol}}$ or $N_{\mathrm{peak}}$
and so neither does the relative magnitude of the terms containing
$\partial_{\tilde{x}}$ and $\partial_{\tilde{x}}^{3}$. So long as
$s\gg\sigma$ holds, we therefore expect (\ref{eq:-113}) to be of
applicable to both the squeezed vacuum state and the squeezed thermal
state.

The relation $s>\sigma$ however requires the squeezing of one quadrature
beyond the vacuum state variance as can been seen by insertion into
(\ref{eq:-107}). We have not considered the case where $\sigma$
is of similar to or greater than $s$ in magnitude. To simplify the
discussion, we consider the case where $\sigma$ and $s$ are similar
in magnitude. Returning again to the terms of the scaled equation
of motion (\ref{eq:-106}) and retaining terms of significance equal
to or greater than the most significant third order term yields
\begin{equation}
\partial_{t}\tilde{W}(\tilde{x},\tilde{y},t)=\begin{aligned}[t] & -2g\sigma^{2}\tilde{x}^{2}\tilde{y}\partial_{\tilde{x}}\tilde{W}(\tilde{x},\tilde{y},t)-2g\sigma^{2}s^{4}\tilde{y}^{3}\partial_{\tilde{x}}\tilde{W}(\tilde{x},\tilde{y},t)+2g\sigma^{2}\tilde{x}\tilde{y}^{2}\partial_{\tilde{y}}\tilde{W}(\tilde{x},\tilde{y},t)\\
 & +2gs^{2}\tilde{y}\partial_{\tilde{x}}\tilde{W}(\tilde{x},\tilde{y},t)\\
 & +\frac{gs^{4}}{8\sigma^{2}}\tilde{y}\partial_{\tilde{x}}^{3}\tilde{W}(\tilde{x},\tilde{y},t).
\end{aligned}
\label{eq:-106-1}
\end{equation}
Several more terms describing the formation of the S-shape are kept
in (\ref{eq:-106-1}). Recalling the discussion of (\ref{subsec:Validity-of-Large}),
these terms cause the large squeezing approximation to lose its validity
sooner in the evolution. We also see that both $\tilde{x}$ and $\partial_{\tilde{y}}$
enter into the equation. The coordinate $\tilde{y}$ can therefore
no longer be considered simply a parameter as was done when introducing
the function $\tilde{u}_{\tilde{y}}$ in (\ref{eq:-158}). Furthermore,
the equation (\ref{eq:-106-1}) no longer has constant coefficients.
In summary, the treatment of the case where $s\gg\sigma$ does not
hold will likely require adjustments to the arguments made here.

\section{Kerr Evolution of Coherent State \label{sec:Coherent-Initial-State}}

\begin{figure}
\noindent \begin{centering}
\includegraphics{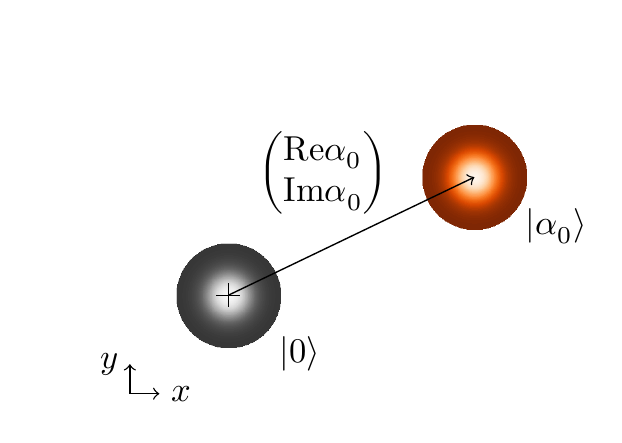}
\par\end{centering}
\noindent \centering{}\caption[Illustration of a coherent state]{\textbf{Illustration of a coherent state.} The coherent state with
parameter $\alpha_{0}$ is shown. Any coherent state is simply a displaced
vacuum state (see (\ref{eq:-142})). The state $|\alpha_{0}\rangle$
can be obtained as the displacement of the vacuum state by the vector
$(\protect\Re\alpha_{0},\protect\Im\alpha_{0})$ in Cartesian coordinates.
Hence the quadrature variances and general shape of the Wigner function
is shared between coherent states and the vacuum state. The vacuum
state is displayed in grayscale at the origin. The Kerr oscillator
with a coherent initial state is treated in Section \ref{sec:Coherent-Initial-State}.
\label{fig:coherent-initial-state}}
\end{figure}
Before progressing to open quantum systems, we superficially treat
the negativity of a coherent initial state to provide some perspective
for the previous results. The system is still defined by (\ref{eq:kerr-hamiltonian})
and the equations of motion are thus shared with Sections \ref{sec:Kerr-Evolution-of}--\ref{sec:squeezed-thermal-state}.
With respect to initial state, we can construct a coherent state by
applying the displacement operator $\hat{D}$ to the vacuum state
$|0\rangle$:
\begin{equation}
|\alpha_{0}\rangle=\hat{D}(\alpha_{0})|0\rangle,\label{eq:-142}
\end{equation}
constructing the coherent state with parameter $\alpha_{0}$. With
(\ref{eq:-142}) in mind, we find the corresponding Wigner function
by applying (\ref{eq:-142}) to (\ref{eq:vacuum-wigner-complex}).
We thus obtain
\begin{equation}
W(\alpha,\alpha^{*})=\frac{2}{\pi}e^{-2|\alpha-\alpha_{0}|^{2}}.
\end{equation}
We see that the coherent state is simply a displacement of the Wigner
function in phase space. The state is illustrated in Figure \ref{fig:coherent-initial-state}.

\subsection{Periodic Evolution\label{subsec:coherent-state-periodic-evolution}}

We can straightforwardly specialize the conclusions of Section \ref{sec:operator-unitary-evolution}
to the case of a coherent initial state. From (\ref{eq:-145}), we
may write
\begin{equation}
\hat{U}(\pi/g)|\alpha_{0}\rangle=|\alpha_{0}\rangle.\label{eq:-224}
\end{equation}

\begin{figure}
\noindent \begin{centering}
\includegraphics{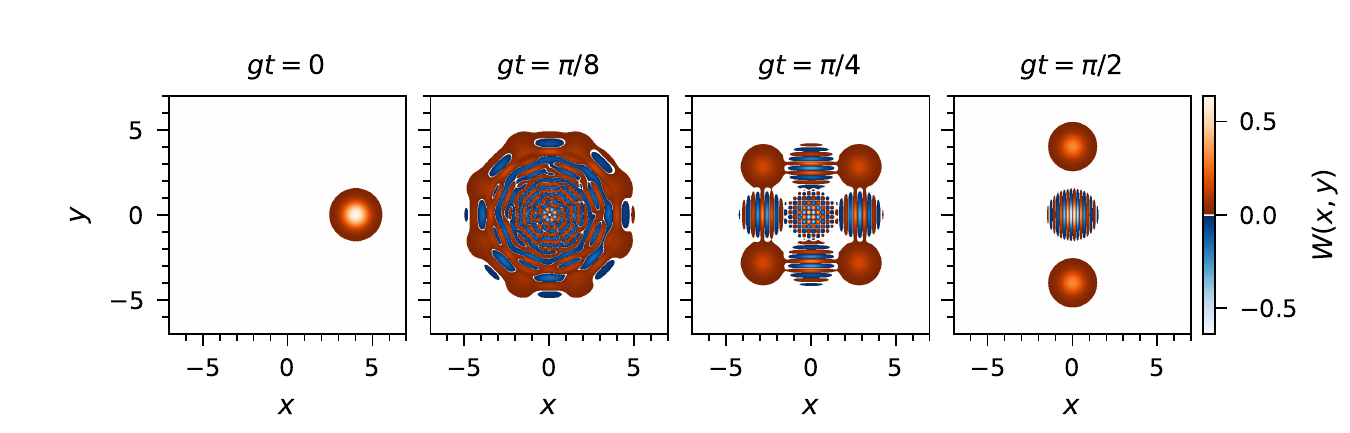}
\par\end{centering}
\caption[Notable states during unitary evolution of coherent state]{\label{fig:evo-gallery-coherent-2}\textbf{Notable states during
unitary evolution of coherent state.} The initial state ($gt=0$)
is a coherent state (\ref{eq:-142}) with $\alpha_{0}=4$. Contour
plots show $W(x,y,t)$ at points of fractional revival $t=\pi/8g$,
$t=\pi/4g$ and $t=\pi/2g$ (the period is $t=\pi/g$, see (\ref{eq:-224})).}
\end{figure}

We also describe the state found halfway through a period. We then
apply (\ref{eq:-146-1}) to the initial state to obtain
\begin{equation}
\hat{U}(\pi/2g)|\alpha_{0}\rangle=\frac{1}{\sqrt{2}}\left(e^{-i\pi/4}e^{-i\hat{n}\pi/2}|\alpha_{0}\rangle+e^{i\pi/4}e^{i\hat{n}\pi/2}|\alpha_{0}\rangle\right).
\end{equation}
Using (\ref{eq:-155}), we can finally express the evolved coherent
state as
\begin{equation}
\hat{U}(\pi/2g)|\alpha_{0}\rangle=\frac{1}{\sqrt{2}}\left(e^{-i\pi/4}|-i\alpha_{0}\rangle+e^{i\pi/4}|i\alpha_{0}\rangle\right).
\end{equation}
The kets $|-i\alpha_{0}\rangle$ and $|i\alpha_{0}\rangle$ are seen
to represent two coherent states with opposite displacements. The
state $\hat{U}(\pi/2g)|\alpha_{0}\rangle$, being a superposition
of two coherent states, is sometimes referred to as a cat state. The
action of the rotation operator on the coherent state is given by
the relation 
\begin{equation}
\hat{R}(\phi)|\alpha_{0}\rangle=|\alpha_{0}e^{i\phi}\rangle,\label{eq:-155}
\end{equation}
and we can thus write
\begin{equation}
\hat{U}(\pi/2g)|\alpha_{0}\rangle=\frac{1}{\sqrt{2}}\left(e^{-i\pi/4}\hat{R}(-\pi/2)|-i\alpha_{0}\rangle+\hat{R}(\pi/2)|i\alpha_{0}\rangle\right).
\end{equation}
The state $\hat{U}(\pi/2g)|\alpha_{0}\rangle$ can be seen in Figure
\ref{fig:evo-gallery-coherent-2}.

\begin{figure}
\subfloat[\label{fig:negvol-period-1}Negative volume]{\noindent \centering{}\captionsetup{position=top}\centerline{\includegraphics{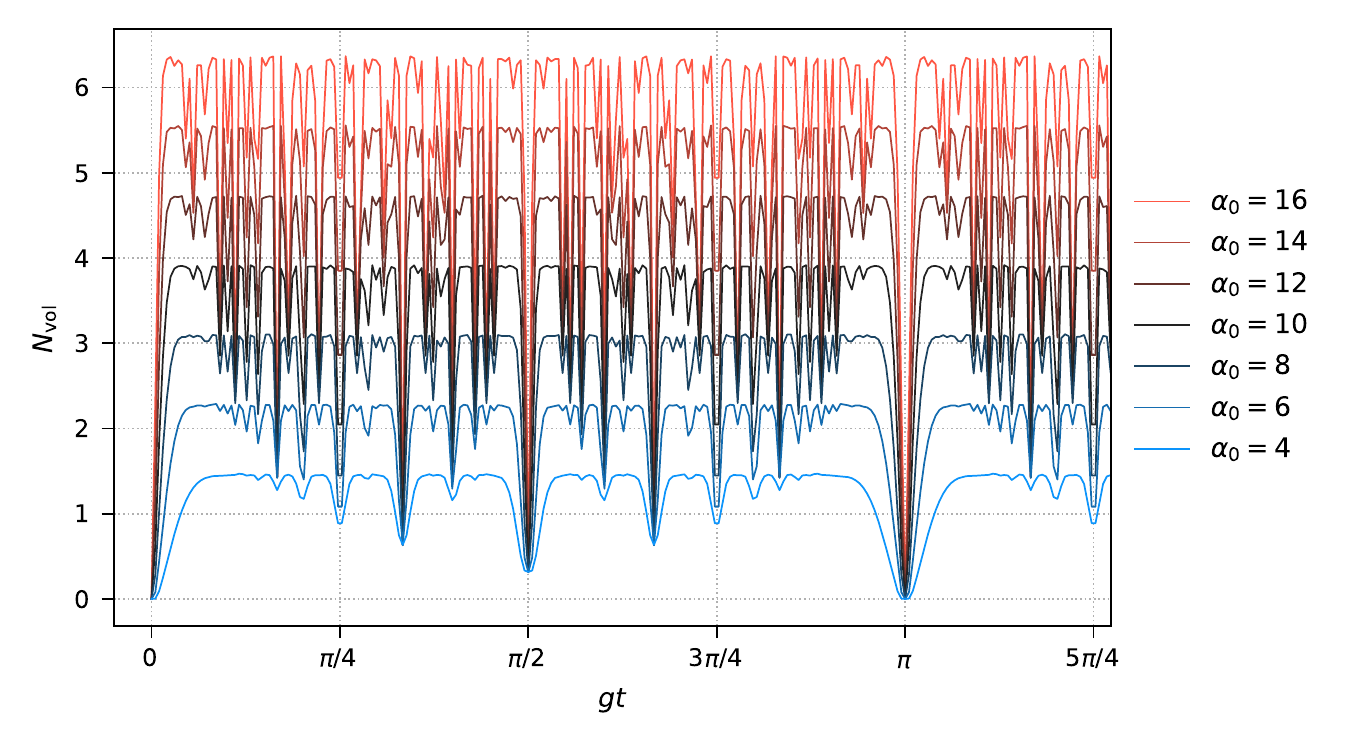}}}

\subfloat[\label{fig:negpeak-period-1}Negative peak]{\noindent \centering{}\captionsetup{position=top}\centerline{\includegraphics{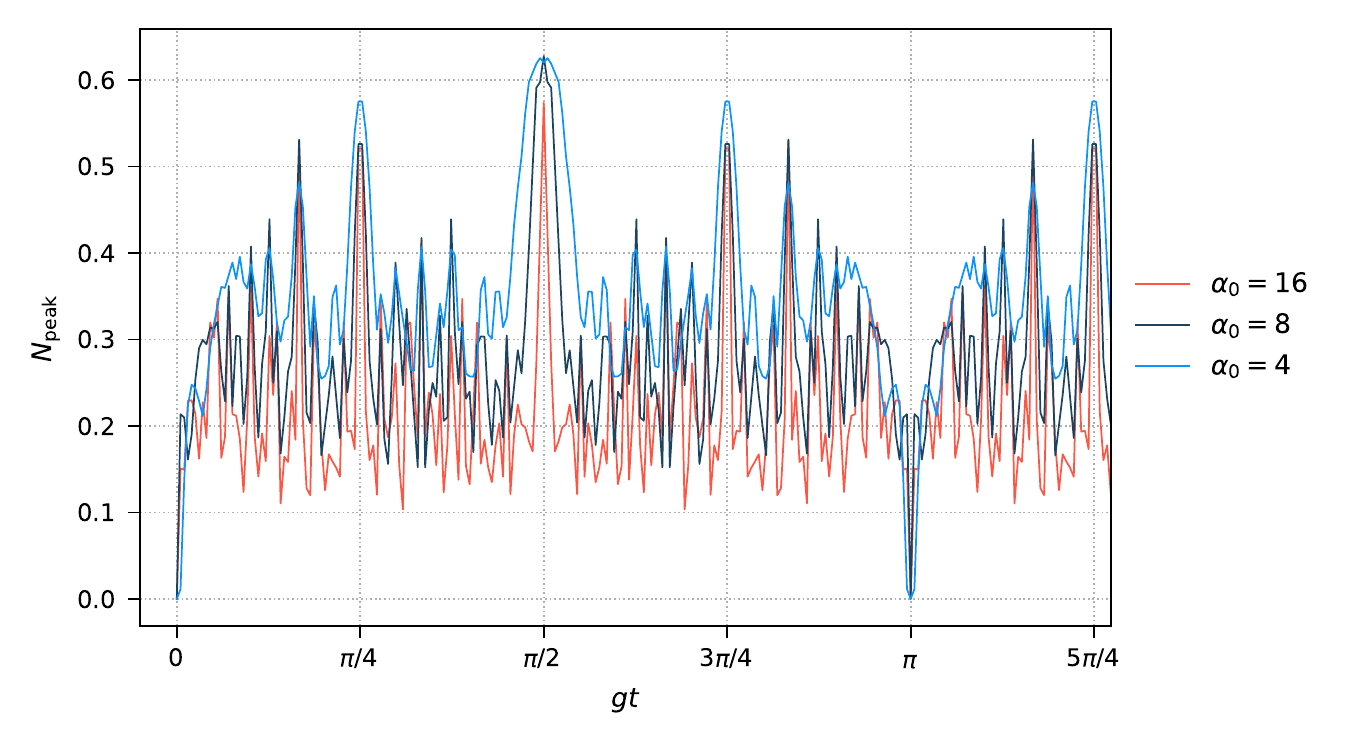}}}\caption[Negativity during periodic evolution of coherent state]{\label{fig:Negativity-during-periodic}\textbf{Negativity during
periodic evolution of coherent state.} The evolution of negativity
for coherent states $|\alpha_{0}\rangle$ over a full period $gt=\pi$.
The states shown in Figure \ref{fig:evo-gallery-coherent-2} are visible
as drops in negative volume $N_{\mathrm{vol}}$. The height of the
plateau scales linearly with $\alpha_{0}$ (see Figure \ref{fig:negpeak-period-1-1}).
Unlike the fractional revival states for the squeezed vacuum state
(compare with Figure \ref{fig:}), they show up as peaks in $N_{\mathrm{peak}}$.
The periodic evolution is discussed in Section \ref{subsec:coherent-state-periodic-evolution}.}
\end{figure}

\begin{figure}
\noindent \centering{}\centerline{\includegraphics{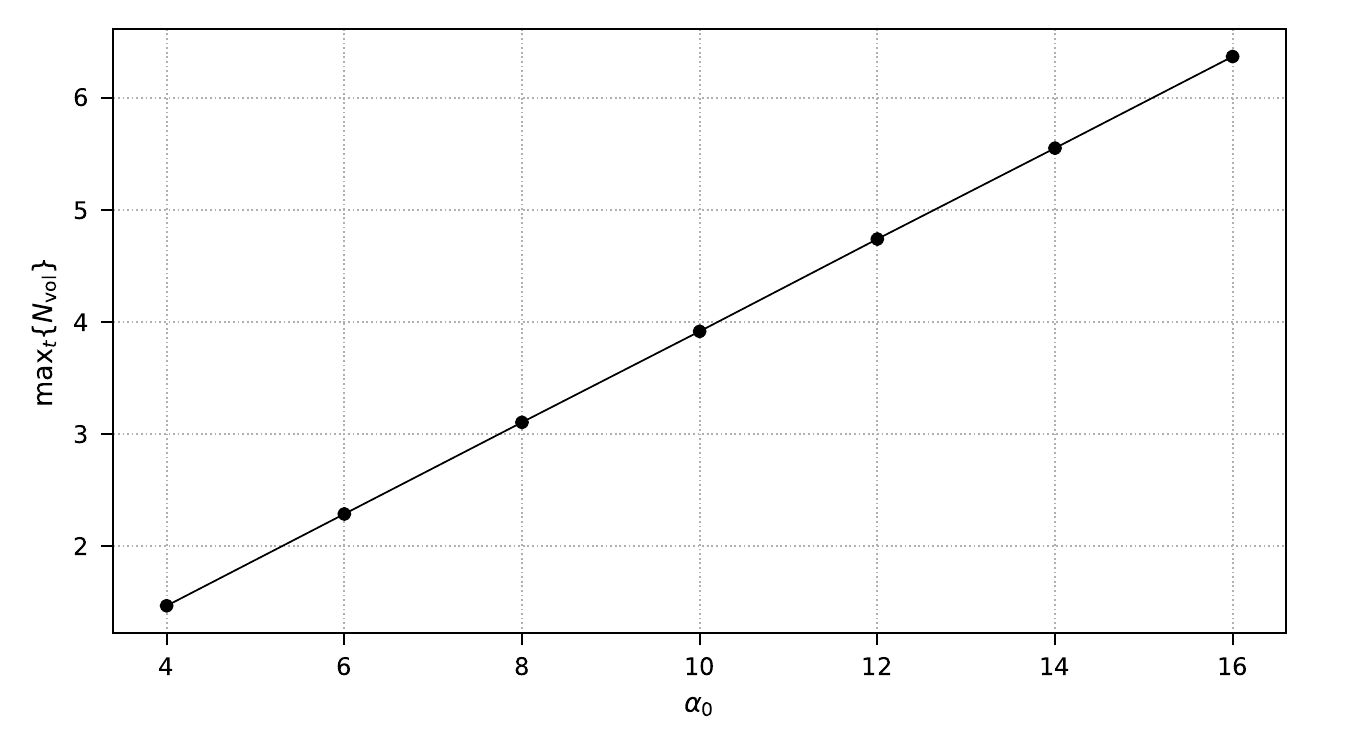}}\caption[Plateau height of negative volume for coherent state]{\label{fig:negpeak-period-1-1}\textbf{ Plateau height of negative
volume for coherent state. }Height of the plateau in negative volume
$N_{\mathrm{vol}}$ for the coherent state $|\alpha_{0}\rangle$ as
seen in Figure \ref{fig:negvol-period-1}. The linear relation is
described by $\max_{t}N_{\mathrm{vol}}=0.41\alpha_{0}-0.16$.}
\end{figure}

We briefly consider the periodic evolution in the negativity of the
state. $N_{\mathrm{vol}}$ and $N_{\mathrm{peak}}$ are plotted for
a full period in Figure \ref{fig:Negativity-during-periodic}. Consider
first $N_{\mathrm{vol}}$. The behavior of the negative volume of
the coherent state is qualitatively similar to that of the squeezed
vacuum state: $N_{\mathrm{vol}}$ increases monotonically until it
reaches a plateau-like region. The slope of the initial growth in
$N_{\mathrm{vol}}$ increases with squeezing. The states seen in Figure
\ref{fig:evo-gallery-coherent-2} are visible in Figure \ref{fig:negvol-period-1}
as drops in $N_{\mathrm{vol}}$. The negativity is mirrored around
the point halfway though the period $\pi/2$. The height of the plateau
appears to increase linearly with the initial state parameter $\alpha_{0}$
as seen in Figure \ref{fig:negpeak-period-1-1}. 

The quantity $N_{\mathrm{peak}}$ (shown in Figure \ref{fig:negpeak-period-1})
also increases however it appears to happen much faster than $N_{\mathrm{vol}}$.
For most of the period, $N_{\mathrm{peak}}$ fluctuates violently.
The states seen in Figure \ref{fig:evo-gallery-coherent-2} are visible
as peaks rather than drops as in the case of $N_{\mathrm{vol}}$.

\begin{figure}
\noindent \begin{centering}
\includegraphics{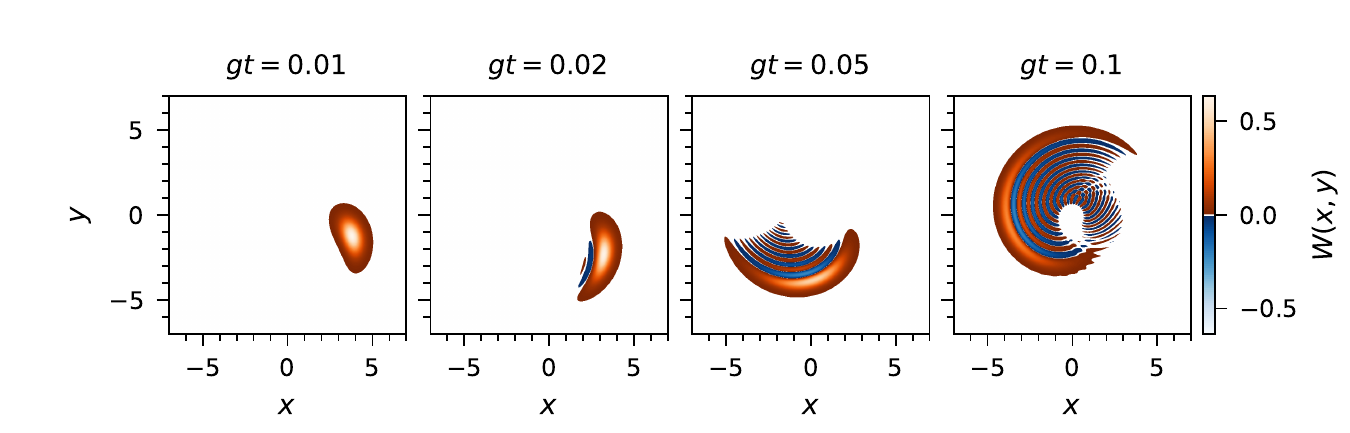}
\par\end{centering}
\caption[Short time unitary evolution of coherent state]{\label{fig:evo-gallery-coherent}\textbf{Short time unitary evolution
of coherent state.} The initial state (i.e. $gt=0$) is a coherent
state $|\alpha_{0}\rangle$ with $\alpha_{0}=4$ and may be seen in
Figure \ref{fig:evo-gallery-coherent-2}. As the coherent state evolves,
the parts farthest from the origin rotate with a relatively larger
angular frequency and the result is that the state is squeezed and
subsequently bent around the origin. This causes the appearance of
fringes with negativity. The short time evolution of the coherent
state is discussed in Section \ref{subsec:Evolution-over-Short-1}.}
\end{figure}

\subsection{Evolution over Short Time\label{subsec:Evolution-over-Short-1}}

\begin{figure}
\subfloat[Negative volume]{\noindent \begin{centering}
\captionsetup{position=top}\includegraphics{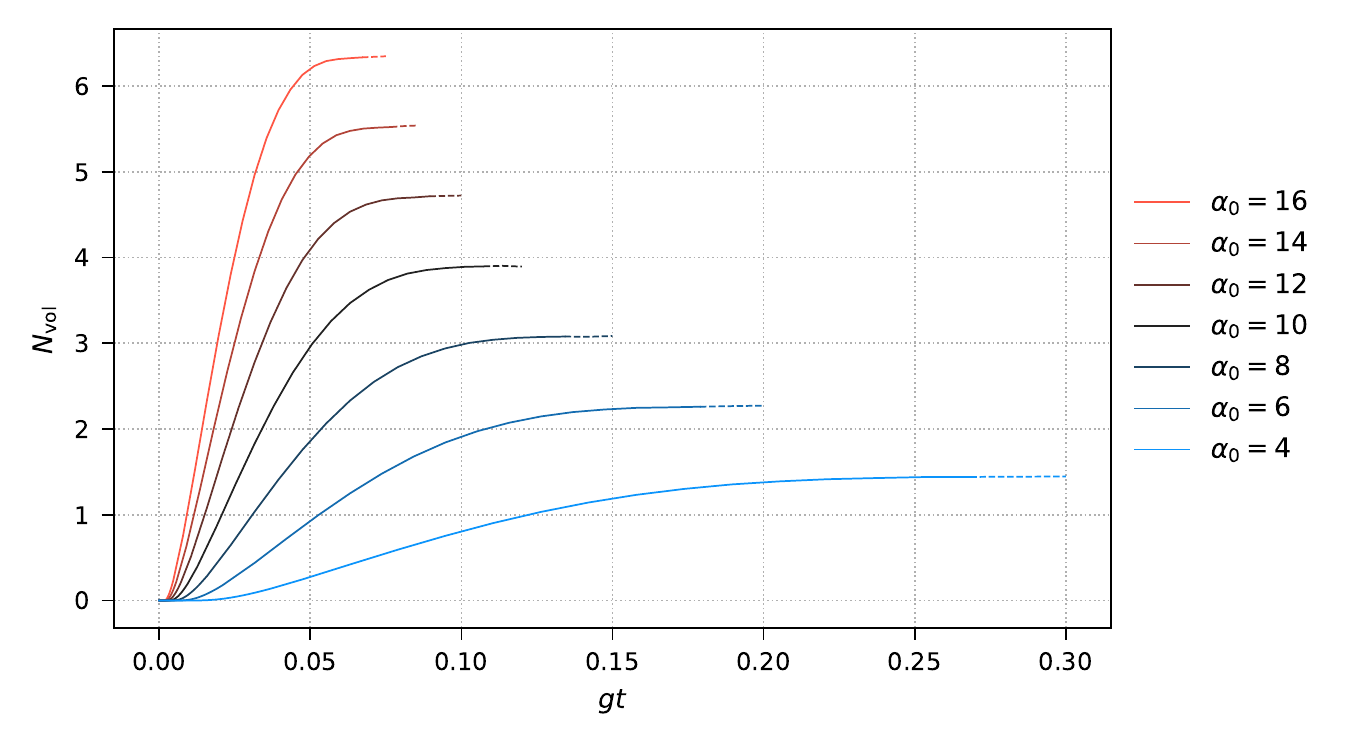}
\par\end{centering}
\noindent \centering{}}

\subfloat[Negative peak]{\noindent \begin{centering}
\captionsetup{position=top}\includegraphics{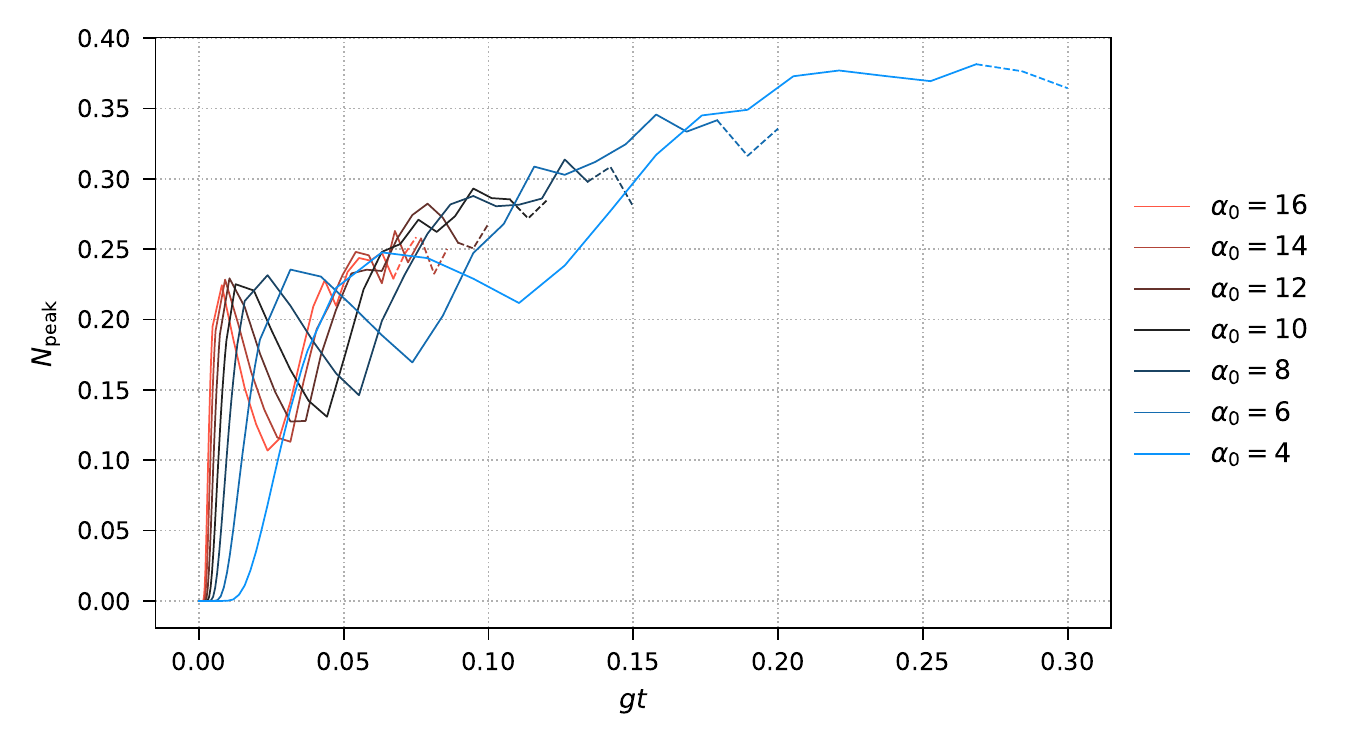}
\par\end{centering}
\noindent \centering{}}

\caption[Negativity during short time unitary evolution of coherent state]{\label{fig:No-scaling}\textbf{Negativity during short time unitary
evolution of coherent state.} The graphs show negativity the as a
function of the (unscaled) time $gt$ for the evolution of a coherent
state $|\alpha_{0}\rangle$. In the first stages of evolution, the
negativity grows monotonically with a rate that increases with $\alpha_{0}$.
However the initial monotone growth breaks down sooner for $N_{\mathrm{peak}}$
than $N_{\mathrm{vol}}$. See Section \ref{subsec:Evolution-over-Short-1}.}
\end{figure}

Like we did for the previous initial states, we consider the short
time evolution. The short time evolution of a coherent state with
$\alpha_{0}=4$ is depicted in Figure \ref{fig:evo-gallery-coherent}.
The unitary evolution of the Wigner function of an initially coherent
state under the Kerr Hamiltonian has been treated before \cite{Oliva_QuantumKerrOscillators_2019,Stobinska_WignerFunctionEvolution_2008,Yurke_GeneratingQuantumMechanical_1986}.
As the coherent state first evolves, due to the variation in the angular
frequency with amplitude, the Wigner density farthest from the origin
revolve around it faster than the density closer to it. The resulting
amplitude-dependent phase shift produces a squeezing effect on the
state \cite{White_KerrNoiseReduction_2000}. As the state evolves
further, the Wigner function is also bent around the origin. As the
bending increases fringes form and the Wigner function thus assumes
negative values. The initial evolution of the negativity is shown
in Figure \ref{fig:No-scaling}. We observe that $N_{\mathrm{peak}}$
departs from the initial monotone growth earlier than $N_{\mathrm{vol}}$.
By appropriately scaling the axes, some scalings may be determined
empirically. For very short times, the negativity (both $N_{\mathrm{vol}}$
and $N_{\mathrm{peak}}$) may be found to appear constant as a function
of $\alpha_{0}^{3/2}t$ for varying squeezing $r_{0}$. For $N_{\mathrm{vol}}$,
one may additionally find that the slope of the growth in the linear
region scales with $\alpha_{0}^{2}$. This is discussed further in
Appendix \ref{chap:Emperical-Scalings-for}.

\chapter{Coupling to an Environment \label{chap:coupling-to-the-environment}}

In this chapter, we apply the results of the previous chapter in a
more realistic setting by considering the evolution of a Kerr oscillator
described by the master equation (\ref{eq:general-master-equation}).
This combines the unitary dynamics explored in Chapter \ref{chap:nonlinear-oscillators}
with the decoherence effects of damping and dephasing. Before we apply
the general master equation (\ref{eq:general-master-equation}) however,
we consider each decoherence effect in an isolated setting.

Section \ref{sec:damping} considers damping. Section \ref{subsec:Fundamental-Solution}
describes the fundamental solution and uses it to derive a finite
negativity decay time general to any state. Section \ref{sec:kerr-state-damping}
considers the decay in negativity under damping. The consideration
of damping finishes in Section \ref{subsec:Damped-Kerr-Evolution},
where the evolution of a squeezed vacuum state of the damped Kerr
oscillator is studied. There, we compute the maximum negative volume
during evolution and demonstrate that it exhibits asymptotic scaling
in the limit of large squeezing.

Section \ref{sec:phase-decoherence} considers phase decoherence in
a similar way. Section \ref{sec:kerr-state-dephasing} considers decay
in negativity under dephasing, reusing the initial states of Section
\ref{sec:kerr-state-damping}. Section \ref{subsec:Kerr-Oscillator-with}
then considers the evolution of a squeezed vacuum state of the dephasing
Kerr oscillator and examines again the maximum negative volume.

Section \ref{sec:kerr-dynamics-combined-decoherence} concludes the
chapter by considering the combination of the previous decoherence
effects. The differences between damping and dephasing are discussed
in Section \ref{subsec:Dephasing-and-Damping}. Sections \ref{subsec:Equation-of-Motion}
and \ref{subsec:Rescaled-Coordinates-and} introduce appropriate equations
of motion and applies the large squeezing approximation. Finally,
Section \ref{subsec:Maximum-Negative-Volume} considers the maximum
negative volume and maximum negative peak given various strengths
of the decoherence effects.

\section{Damping\label{sec:damping}}

\begin{figure}
\noindent \begin{centering}
\includegraphics{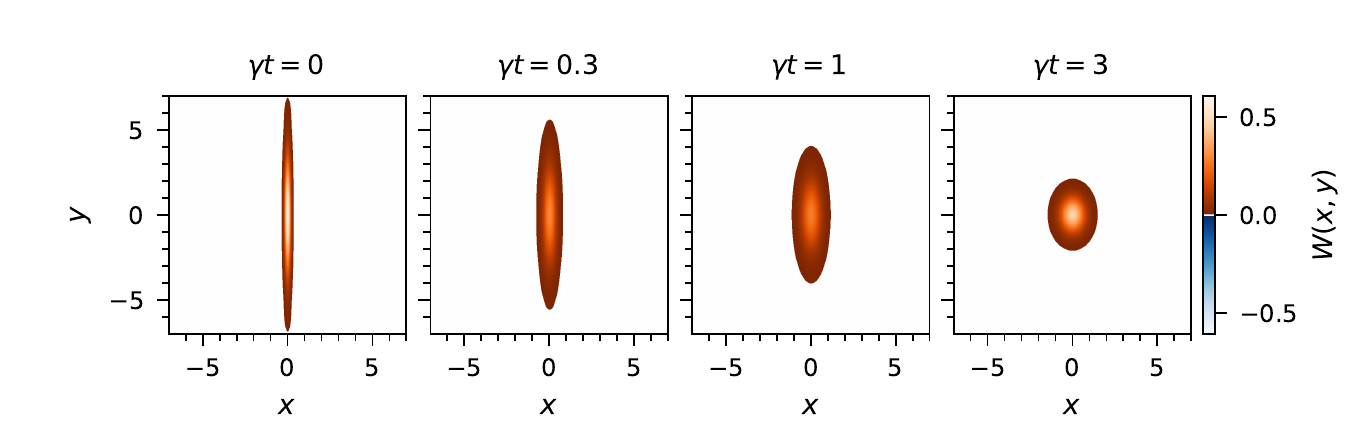}
\par\end{centering}
\caption[Damping of squeezed vacuum at zero temperature]{\label{fig:Damping-of-squeezed}\textbf{Damping of squeezed vacuum
at zero temperature.} The initial state is a squeezed vacuum state
(\ref{eq:-124}) with $\xi=r_{0}=1.5$. The effect of zero temperature
($\bar{n}=0$) damping is cooling towards the vacuum state. Thus the
steady state (which is approached by $\gamma t=3$) is the vacuum
state shown in Figure \ref{fig:vacuum-state-plot}. The state remains
Gaussian at all times as per Section \ref{sec:damping}.}
\end{figure}

We wish to initially study the isolated effects of damping and therefore
obtain the relevant master equation by removing the effects of unitary
evolution and dephasing from the general master equation (\ref{eq:general-master-equation}).
Setting $\hat{H}=0$ and $\gamma_{\phi}=0$ achieves this, leaving
only the terms shown in

\begin{equation}
\dot{\hat{\rho}}=\gamma\left(\bar{n}+1\right)\mathcal{D}[\hat{a}]\hat{\rho}+\gamma\bar{n}\mathcal{D}[\hat{a}^{\dagger}]\hat{\rho}.\label{eq:damped-me}
\end{equation}
Equation (\ref{eq:damped-me}) describes the coupling of the quantum
system to an environment in thermal equilibrium where $\gamma$ is
a frequency describing the coupling strength and $\bar{n}$ denotes
the mean occupancy of the oscillator when in thermal equilibrium with
the environment. One can think of the term $\gamma\left(\bar{n}+1\right)\mathcal{D}[\hat{a}]$
($\gamma\bar{n}\mathcal{D}[\hat{a}^{\dagger}]\hat{\rho}$) as cooling
(heating) since its effect is to decrease (increase) the expectation
value of the system energy $\langle\hat{H}\rangle$.

Using the techniques from Section \ref{sec:wigner-pde-derivation},
the partial differential equation for $W(x,y,t)$ corresponding to
(\ref{eq:damped-me}) is derived. In Cartesian coordinates this equation
is expressed as
\begin{equation}
\partial_{t}W(x,y,t)=\frac{\gamma}{4}\left(\bar{n}+\frac{1}{2}\right)\nabla^{2}W(x,y,t)+\frac{\gamma}{2}\partial_{x}\left(xW(x,y,t)\right)+\frac{\gamma}{2}\partial_{y}\left(yW(x,y,t)\right).\label{eq:damping-pde-xy}
\end{equation}
Let us also note that (\ref{eq:damping-pde-xy}) clearly separates
the effects of damping into a temperature-dependent part and a temperature-invariant
part. The term proportional to $\nabla^{2}W(x,y,t)$ describes a diffusive
effect in phase space. The strength of this effect increases with
temperature. The other terms are independent of temperature. These
other terms, proportional to $\partial_{x}\left(xW\right)$ or $\partial_{y}\left(yW\right)$,
causes a flow of Wigner density toward the origin.\footnote{Of course, (\ref{eq:damped-me}) can also be written in the form $\dot{\hat{\rho}}=\gamma\left(\bar{n}+\frac{1}{2}\right)\left(\mathcal{D}[\hat{a}]+\mathcal{D}[\hat{a}^{\dagger}]\right)\hat{\rho}+\frac{\gamma}{2}\left(\mathcal{D}[\hat{a}]-\mathcal{D}[\hat{a}^{\dagger}]\right)\hat{\rho}$,
separating it into a temperature dependent and a temperature invariant
part. In that form it is however less apparent effects of the superoperators
$(\mathcal{D}[\hat{a}]+\mathcal{D}[\hat{a}^{\dagger}])$ and $(\mathcal{D}[\hat{a}]-\mathcal{D}[\hat{a}^{\dagger}])$
are diffusion and flow toward the origin.} The evolution of the squeezed vacuum state $|\xi=1.5\rangle$ for
$\bar{n}=0$ is shown in Figure \ref{fig:Damping-of-squeezed}.

\begin{figure}
\noindent \begin{centering}
\includegraphics{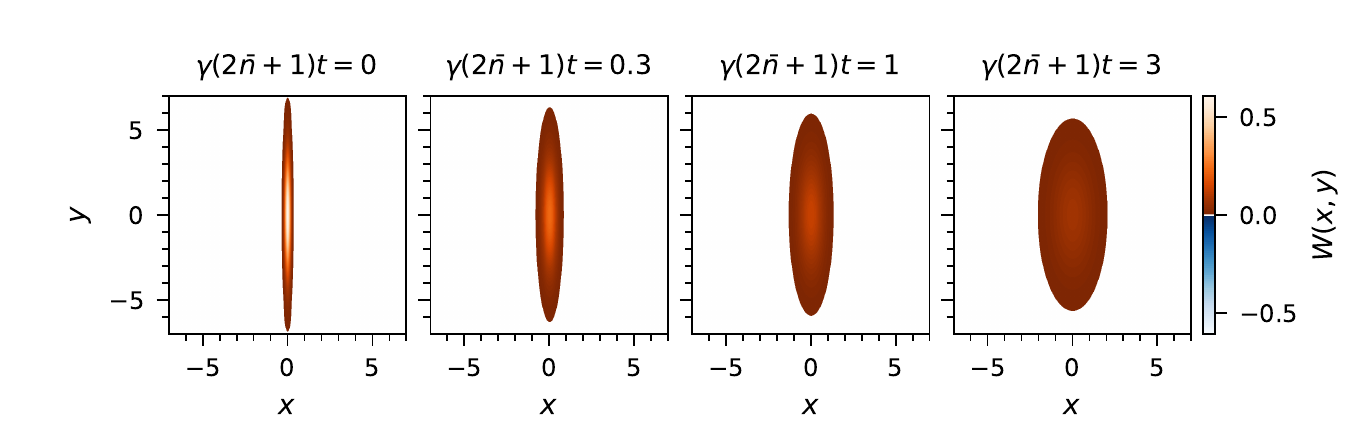}
\par\end{centering}
\caption[Damping of squeezed vacuum at high temperature]{\label{fig:Damping-of-squeezed-1}\textbf{Damping of squeezed vacuum
at high temperature.} The initial state is a squeezed vacuum state
(\ref{eq:-124}) with $\xi=r_{0}=1.5$. The temperature of the environment
is set by $\bar{n}=1000$. At large temperatures the diffusive effect
of (\ref{eq:damped-me}) dominates and the evolution is thus approximately
described by (\ref{eq:damping-pde-high-temp}). As per Section \ref{subsec:Fundamental-Solution}
the state remains Gaussian at all times.}
\end{figure}

The current chapter is mainly motivated by physical systems in the
limit of large temperature. We shall therefore focus on $\bar{n}\gg1$.
In this limit, (\ref{eq:damping-pde-xy}) reduces to the form
\begin{equation}
\partial_{t}W(x,y,t)=\frac{\gamma\bar{n}}{4}\nabla^{2}W(x,y,t).\label{eq:damping-pde-high-temp}
\end{equation}
This is simply the two-dimensional heat equation. This reduces the
number of parameters by one such that we need now only consider a
single parameter proportional to the product $\gamma(2\bar{n}+1)=2\gamma\bar{n}$.
The simple physical interpretation of (\ref{eq:damping-pde-high-temp})
is the coupling of the system to a bath of very large temperature.
We however take the limit where $\bar{n}$ tends to infinity while
the product $\gamma(2\bar{n}+1)$ is held constant. In this limit
$\gamma$ tends to zero. We understand this as examination of the
short time evolution before the system has had significant time to
cool. In this limit, expect the quantities $N_{\mathrm{vol}}$ and
$N_{\mathrm{peak}}$ to decay quickly compared to the time $1/\gamma$.
The evolution of the squeezed vacuum state $|\xi=1.5\rangle$ for
$\bar{n}=1000$ is shown in Figure \ref{fig:Damping-of-squeezed-1}.

\subsection{Fundamental Solution\label{subsec:Fundamental-Solution}}

Before we move on to consider specific initial state, we first note
that (\ref{eq:damping-pde-xy}) allows for the solution of an arbitrary
initial state through the use of a fundamental solution. Written in
the form (\ref{eq:damping-pde-xy}) the equation may be recognized
as the Fokker-Planck equation of an Ornstein--Uhlenbeck process in
two spatial dimensions \cite{Gardiner_HandbookStochasticProcesses_1985}.
The process has in this case as parameters a diffusion coefficient
$\frac{\gamma}{4}\left(\bar{n}+\frac{1}{2}\right)$ and a drift coefficient
$\frac{\gamma}{2}$. For an initial state given by a Gaussian function
the solution remains Gaussian at all times \cite{Walls_QuantumOptics_2008,Wang_TheoryBrownianMotion_1945}.
Taking the Fourier transform of the equation (\ref{eq:damping-pde-xy})
it may be shown that the problem with an initial function given by
the delta function (though this function cannot be thought of as the
Wigner function of a valid quantum state\footnote{$W_{\delta}$ does not represent the Wigner function of any physical
state. One way to see this is that the variances of $W_{\delta}$
(equations (\ref{eq:-195}) and (\ref{eq:-196})) violate the Heisenberg
uncertainty relation $\langle(\Delta x)^{2}\rangle_{t}\langle(\Delta y)^{2}\rangle_{t}\geq1/16$.
$W_{\delta}$ is simply a mathematical tool with which to express
the solution to (\ref{eq:damping-pde-xy}) given an arbitrary initial
state.})
\begin{equation}
W_{\delta}(x,y,0)=\delta(x-x_{0})\delta(y-y_{0})
\end{equation}
is solved by a Gaussian \cite{Wang_TheoryBrownianMotion_1945} with
expectation values
\begin{subequations}
\label{eq:-171}
\begin{align}
\langle x\rangle_{t} & =\int dx\,dy\,xW_{\delta}(x,y,t)=x_{0}e^{-\gamma t/2}\qquad\text{with \ensuremath{x_{0}=\langle x\rangle_{t=0}},}\\
\langle y\rangle_{t} & =\int dx\,dy\,yW_{\delta}(x,y,t)=y_{0}e^{-\gamma t/2}\qquad\text{with \ensuremath{y_{0}=\langle y\rangle_{t=0}},}
\end{align}
and (co)variances
\begin{align}
\langle(\Delta x)^{2}\rangle_{t} & =\int dx\,dy\,\left(x^{2}-x\langle x\rangle_{t}\right)W_{\delta}(x,y,t)=\frac{2\bar{n}+1}{4}\left(1-e^{-\gamma t}\right),\label{eq:-195}\\
\langle(\Delta y)^{2}\rangle_{t} & =\int dx\,dy\,\left(y^{2}-y\langle y\rangle_{t}\right)W_{\delta}(x,y,t)=\frac{2\bar{n}+1}{4}\left(1-e^{-\gamma t}\right),\label{eq:-196}\\
\left\langle \left(x-\langle x\rangle_{t}\right)\left(y-\langle y\rangle_{t}\right)\right\rangle _{t} & =\int dx\,dy\,\left(x-\langle x\rangle_{t}\right)\left(y-\langle y\rangle_{t}\right)W_{\delta}(x,y,t)=0.
\end{align}
\end{subequations}
The equations (\ref{eq:-171}) describe the fundamental solution to
(\ref{eq:damping-pde-xy}) and can be exploited to write the solution
of the system for an arbitrary initial state by convolution \cite{Ferraro_GaussianStatesQuantum_2005a}.

Even though $W_{\delta}$ does not represent a physical state, the
steady state solution coincides with the proper steady state quantum
state Wigner function of (\ref{eq:damping-pde-xy}). In the steady
state, the drift and diffusive effects balance such that the Wigner
function remains constant in time. Taking the limit of $t\to+\infty$
in (\ref{eq:-171}) and identifying $\bar{n}$ with $\bar{n}_{0}$
allows one to recover the thermal state Wigner function (\ref{eq:-21}).
It can be intuitively understood that an increase in temperature leads
to the steady state of the system assuming the form of a wider Gaussian
function.

\subsubsection{Decay of Negativity in Finite Time\label{subsec:Decay-of-Negativity}}

By relating the solution (\ref{eq:-171}) to the definition of the
Husimi Q function (hereafter Q function) we can establish a finite
bound for the time evolved under (\ref{eq:damping-pde-xy}) after
which the Wigner function is completely non-negative. Of course, this
bound only applies to Wigner functions evolved under damping alone.
Additional effects, such as unitary evolution terms, render the bound
void. The Q function is defined as \cite{Gerry_IntroductoryQuantumOptics_2004}
\begin{equation}
Q(\alpha,\alpha^{*})=\langle\alpha|\hat{\rho}|\alpha\rangle
\end{equation}
from which it is seen that\footnote{Note that the Q function is however not strictly positive. In fact,
the zeros of the Q function are related to the negative regions of
the Wigner function \cite{Korsch_ZerosHusimiDistribution_1997}.}
\begin{equation}
Q(\alpha,\alpha^{*})\geq0\qquad\text{for all \ensuremath{\alpha}.}\label{eq:-187}
\end{equation}
Additionally, the Q function is related to the Wigner function through
the convolution\footnote{This relation between the various quasiprobability distributions has
been used to define the non-classical depth \cite{Lee_MeasureNonclassicalityNonclassical_1991,Lee_MomentsFunctionsNonclassical_1992,Marchiolli_NonclassicalDepthPhase_2001}.
This is measure of non-classicality complementary to $N_{\mathrm{vol}}$
and $N_{\mathrm{peak}}$ (e.g. it is nonzero for a squeezed vacuum
state even though $N_{\mathrm{vol}}=N_{\mathrm{peak}}=0$) \cite{Kenfack_NegativityWignerFunction_2004}.} \cite{Cahill_DensityOperatorsQuasiprobability_1969a}
\begin{equation}
Q(\beta,\beta^{*})=\int d\alpha\,d\alpha^{*}\,W(\alpha,\alpha^{*})e^{-2|\alpha-\beta|^{2}}.\label{eq:-188}
\end{equation}
We can combine (\ref{eq:-171}), (\ref{eq:-187}) and (\ref{eq:-188})
to establish a finite time after which the negativity will have completely
vanished. Using (\ref{eq:-171}), we may write the solution of (\ref{eq:damping-pde-xy})
given an arbitrary initial state $W(x,y,0)$ as (note the rescaling
of the arguments)
\begin{equation}
W(xe^{-\gamma t/2},ye^{-\gamma t/2},t)=\int dx'\,dy'\,W(x',y',0)\exp\left(\frac{-2(x-x')^{2}-2(y-y')^{2}}{\left(2\bar{n}+1\right)\left(1-e^{-\gamma t}\right)e^{\gamma t}}\right).\label{eq:-189}
\end{equation}
Comparing (\ref{eq:-188}) and (\ref{eq:-189}), we see that choosing
$t_{\mathrm{decay}}$ such that 
\begin{equation}
\left(2\bar{n}+1\right)\left(\exp(\gamma t_{\mathrm{decay}})-1\right)=1
\end{equation}
we have
\begin{equation}
W(x\exp(-\gamma t_{\mathrm{decay}}/2),y\exp(-\gamma t_{\mathrm{decay}}/2),t_{\mathrm{decay}})=Q(x,y,0).
\end{equation}
Since the Q function is manifestly non-negative (\ref{eq:-187}) for
all states, 
\begin{equation}
t_{\mathrm{decay}}=\gamma^{-1}\log\left(1+\frac{1}{2\bar{n}+1}\right)\label{eq:-194}
\end{equation}
denotes a time at which the Wigner function is non-negative. Since
the evolution of a non-negative Wigner function under (\ref{eq:damping-pde-xy})
can never lead to negativity, the Wigner function remains non-negative
after $t_{\mathrm{decay}}$. We also note that $t_{\mathrm{decay}}$
is finite for finite $\gamma$. Thus the Wigner function loses all
negativity after a finite time under damping. As such, we can regard
$t_{\mathrm{decay}}$ as a characteristic time scale for damping.

In the high temperature limit, letting $\bar{n}\to\infty$ and $\gamma\to0$
such that the quantity $(2\bar{n}+1)\gamma$ is kept constant, we
expand the logarithm in (\ref{eq:-194}) to find
\begin{equation}
t_{\mathrm{decay}}=\frac{1}{(2\bar{n}+1)\gamma}=\frac{1}{2\bar{n}\gamma}\qquad\text{for large \ensuremath{\bar{n}}.}\label{eq:-226}
\end{equation}
Since the introduction of damping allows for the evolution of a pure
state into a mixed state (e.g. the steady state of (\ref{eq:damped-me})
is the thermal state (\ref{eq:-141})), the statement that $N_{\mathrm{vol}}\neq0$
and $N_{\mathrm{peak}}\neq0$ for all non-Gaussian pure states thus
no longer applies. In anticipation of Section \ref{sec:kerr-state-damping},
Table \ref{tab:Scaled-negativity-decay} shows $t_{\mathrm{decay}}$
expressed with the later derived effective damping rate for squeezed
states.

\begin{table}
\captionsetup{width=0.6\textwidth}

\renewcommand{\arraystretch}{1.5}
\noindent \centering{}%
\begin{tabular}{|c|c|}
\hline 
$r_{0}$ & $\gamma(2\bar{n}+1)s^{2}t_{\mathrm{decay}}$\tabularnewline
\hline 
\hline 
$0.5$ & $2.72$\tabularnewline
\hline 
$0.75$ & $4.48$\tabularnewline
\hline 
$1$ & $7.39$\tabularnewline
\hline 
$1.25$ & $12.18$\tabularnewline
\hline 
$1.5$ & $20.09$\tabularnewline
\hline 
$1.75$ & $33.12$\tabularnewline
\hline 
$2$ & $54.60$\tabularnewline
\hline 
\end{tabular}\caption[Scaled negativity decay times]{\label{tab:Scaled-negativity-decay}\textbf{Scaled negativity decay
times.} The time $t_{\mathrm{decay}}$describes the finite time after
which all negativity has vanished. It is computed in the high temperature
limit using (\ref{eq:-226}). The quantity $\gamma(2\bar{n}+1)e^{2r_{0}}=\gamma(2\bar{n}+1)s^{2}$
is found as the effective damping rate in Section \ref{sec:kerr-state-damping}.
The values of this table may be applied to Figure \ref{fig:Decay-of-squeezed}
where the decay caused by damping is shown for a particular initial
state.}
\end{table}

\subsection{Damping of Squeezed Kerr State \label{sec:kerr-state-damping}}

We continue our analysis of energy damping by considering the evolution
of a specific initial state under (\ref{eq:damped-me}). This will
give us some insight in how the quantities $N_{\mathrm{vol}}$ and
$N_{\mathrm{peak}}$ of relevant states decay under damping. Reusing
the initial states of the previous chapter, which were all Gaussian,
in the analysis of negativity however would yield trivial results:
Gaussian states evolved by (\ref{eq:damping-pde-xy}) remain Gaussian
\cite{Ferraro_GaussianStatesQuantum_2005a}. Hence if a Gaussian initial
state is chosen, $N_{\mathrm{vol}}$ and $N_{\mathrm{peak}}$ are
$0$ for all time $t$. Inspired by the results of Chapter \ref{chap:nonlinear-oscillators}
we instead introduce the squeezed Kerr state
\begin{equation}
|r_{0},\tilde{\tau}_{0}\rangle=\hat{U}_{K}(\tilde{\tau}_{0}e^{-4r_{0}})\hat{S}(r_{0})|0\rangle,\label{eq:squeezed-kerr-state-def}
\end{equation}
where $\hat{S}(r_{0})$ is the squeezing operator as defined in (\ref{eq:squeezing-operator})
and $\hat{U}_{K}(\tilde{\tau}_{0}e^{-4r_{0}})$ is the unitary transformation
\begin{equation}
\hat{U}_{K}(\tilde{\tau}_{0})=\exp\left(-i\hat{a}^{\dagger}\hat{a}^{\dagger}\hat{a}\hat{a}\tilde{\tau}e^{-4r_{0}}\right).
\end{equation}
This corresponds to the Kerr oscillator evolution of a squeezed state
for a time 
\begin{equation}
t=\tilde{\tau}_{0}/gs^{4}.\label{eq:-209}
\end{equation}
In the limit of large squeezing $s$, it is known from Chapter \ref{chap:nonlinear-oscillators}
that $N_{\mathrm{vol}}$ and $N_{\mathrm{peak}}$ both grow as functions
of the scaled time $gts^{4}$ in a way invariant of the squeezing
$s=e^{r_{0}}$. This is seen in Figures \ref{fig:sctime-negvol} and
\ref{fig:sctime-negpeak}. In the same limit, we furthermore know
that the Wigner function $\tilde{W}$ expressed in scaled coordinates
$(\tilde{x},\tilde{y})=(sx,y/s)$ also evolves as a function of the
scaled time $gts^{4}$ in a way invariant of $s$. This can be seen
by inserting (\ref{eq:-209}) into the expression for $\tilde{W}$
found in (\ref{eq:-51}) and is also demonstrated by Figure \ref{fig:scaled-squeezed-gallery-1}.
Hence we state that $\tilde{W}_{|r_{0},\tilde{\tau}_{0}\rangle}(\tilde{x},\tilde{y})$
is approximately independent of $r_{0}$ in the limit of large squeezing
(note that the state $|r_{0},\tilde{\tau}_{0}\rangle$ is not independent
of $r_{0}$, e.g. for $\tilde{\tau}_{0}=0$ is it the squeezed state
$|\xi{=}r_{0}\rangle$ which is manifestly dependent on $r_{0}$).

\subsubsection{Rescaled Coordinates}

We wish to find a scaled time for the quantities $N_{\mathrm{vol}}$
and $N_{\mathrm{peak}}$ when the system evolves under damping. We
reuse (\ref{eq:-19}) as the definition of $\tilde{W}$. To find the
appropriate scaling, we repeat now the steps of Sections \ref{subsec:negativity-mechanism}
to rescale the phase space damping dynamics and discover the equation
of motion for $\tilde{W}$. This again transfers the parameter $s$
from the initial state to the equation of motion. Using coordinates
$\tilde{x}$ and $\tilde{y}$ of (\ref{eq:-30-2}) the rescaled coordinate
equation of motion derived from (\ref{eq:damping-pde-xy}) takes the
form

\begin{align}
\partial_{t}\tilde{W}(\tilde{x},\tilde{y},t) & =\frac{\gamma s^{2}}{4}\left(\bar{n}+\frac{1}{2}\right)\partial_{\tilde{x}}^{2}\tilde{W}(\tilde{x},\tilde{y},t)+\frac{\gamma}{4s^{2}}\left(\bar{n}+\frac{1}{2}\right)\partial_{\tilde{y}}^{2}\tilde{W}(\tilde{x},\tilde{y},t)\label{eq:-90}\\
 & \qquad+\frac{\gamma}{2}\partial_{\tilde{x}}\left(\tilde{x}\tilde{W}(\tilde{x},\tilde{y},t)\right)+\frac{\gamma}{2}\partial_{\tilde{y}}\left(\tilde{y}\tilde{W}(\tilde{x},\tilde{y},t)\right).\nonumber 
\end{align}
We have included terms independent of $\bar{n}$ simply to demonstrate
that squeezing only applies to the diffusive terms while leaving the
drift term unchanged. As noted below (\ref{eq:damping-pde-xy}) we
consider the system for large $\bar{n}$ and as such the terms in
the second line are disregarded independently of their contained power
of $s$.

\subsubsection{Large Squeezing Approximation}

Keeping only the single term of (\ref{eq:-90}) which is proportional
to $s^{2}$, we are left with the equation
\begin{equation}
\partial_{t}\tilde{W}(\tilde{x},\tilde{y},t)=\frac{\gamma s^{2}}{4}\left(\bar{n}+\frac{1}{2}\right)\partial_{\tilde{x}}^{2}\tilde{W}(\tilde{x},\tilde{y},t).\label{eq:-103}
\end{equation}
We have no convenient analytical expression for the initial state\footnote{(\ref{eq:-102}) does give an approximate form of the initial state
$\tilde{W}_{|r_{0},\tau_{0}\rangle}(\tilde{x},\tilde{y})$ in the
form of a Fourier transform. This is trivially evolved further under
(\ref{eq:-103}) in the Fourier domain: $\tilde{W}(\tilde{x},\tilde{y},t)=\left(2\pi\right)^{-1/2}\int_{-\infty}^{\infty}dk\,h(k)\,e^{i(k\tilde{x}-2k\tau_{0}s^{4}\tilde{y}^{3}-\tau_{0}s^{4}k^{3}/8)}e^{-\gamma s^{2}t(2\bar{n}+1)/8}$
with $h(k)=\left(2\pi\right)^{-1/2}\frac{2}{\pi}\int_{-\infty}^{\infty}d\tilde{x}\,e^{-2\tilde{x}^{2}-2\tilde{y}^{2}}e^{-ik\tilde{x}}.$
We will not need this for the arguments in the text however.} 
\[
\tilde{W}(\tilde{x},\tilde{y},0)=\tilde{W}_{|r_{0},\tilde{\tau}_{0}\rangle}(\tilde{x},\tilde{y}).
\]
We can instead extract the required information directly from (\ref{eq:-102})
without requiring an expression $\tilde{W}_{|r_{0},\tilde{\tau}_{0}\rangle}(\tilde{x},\tilde{y})$.
We still require operating in the regime of large $s$ however. Simply
introduce a new scaled time coordinate 
\begin{equation}
\tilde{\tau}_{\gamma}=\gamma ts^{2}(2\bar{n}+1)/8\label{eq:-172}
\end{equation}
 and a function $\tilde{v}(\tilde{x},\tilde{y},\tilde{\tau}_{\gamma})$
such that
\begin{equation}
\tilde{W}(\tilde{x},\tilde{y},t)=\tilde{v}(\tilde{x},\tilde{y},\tilde{\tau}_{\gamma}{=}\gamma ts^{2}(2\bar{n}+1)/8)\label{eq:-104}
\end{equation}
with the equation of motion for $\tilde{v}(\tilde{x},\tilde{y},\tilde{\tau}_{\gamma})$
derived from (\ref{eq:-103}):
\begin{equation}
\partial_{\tilde{\tau}_{\gamma}}\tilde{v}(\tilde{x},\tilde{y},\tilde{t})=\partial_{\tilde{x}}^{2}v(\tilde{x},\tilde{y},\tilde{\tau}).\label{eq:-105}
\end{equation}
(\ref{eq:-105}) contains no reference to $s$. We also note that
the initial state
\begin{equation}
\tilde{v}(\tilde{x},\tilde{y},0)=\tilde{W}_{|r_{0},\tilde{\tau}_{0}\rangle}(\tilde{x},\tilde{y})
\end{equation}
is independent of $s$ in the limit of large squeezing as well. We
therefore expect the function $\tilde{v}(\tilde{x},\tilde{y},\tilde{\tau}_{\gamma})$
to exhibit asymptotic behavior in the limit of large squeezing. We
write the decaying Wigner function as
\begin{equation}
\tilde{W}(\tilde{x},\tilde{y},t)=\tilde{v}(\tilde{x},\tilde{y},8t/\gamma s^{2}(2\bar{n}+1)).\label{eq:-208}
\end{equation}
Given the previous arguments, we expect the full dependence on $s$
to be expressed in the rescaling of time as the third argument of
$\tilde{v}$ in (\ref{eq:-208}). This indicates that an increase
in squeezing also increases the effective strength of damping. In
other words, a state which is more squeezed is also damped more quickly.

\subsubsection{Decay of Negativity}

To examine the change in damping with varying squeezing, we numerically
investigate the decay of the quantities $N_{\mathrm{vol}}$ and $N_{\mathrm{peak}}$
for the states $|r_{0},\tilde{\tau}_{0}\rangle$. We can physically
think of this as a two-stage process wherein the system is first evolved
under the unitary dynamics arising from the Kerr Hamiltonian (\ref{eq:kerr-hamiltonian})
(forming the state $|r_{0},\tilde{\tau}_{0}\rangle$) and then subsequently
decays as described by (\ref{eq:damped-me}), i.e. with no unitary
evolution terms.

\begin{figure}
\subfloat[\label{fig:squeezed-kerrdecay-negvol}Negative volume]{\noindent \begin{centering}
\includegraphics{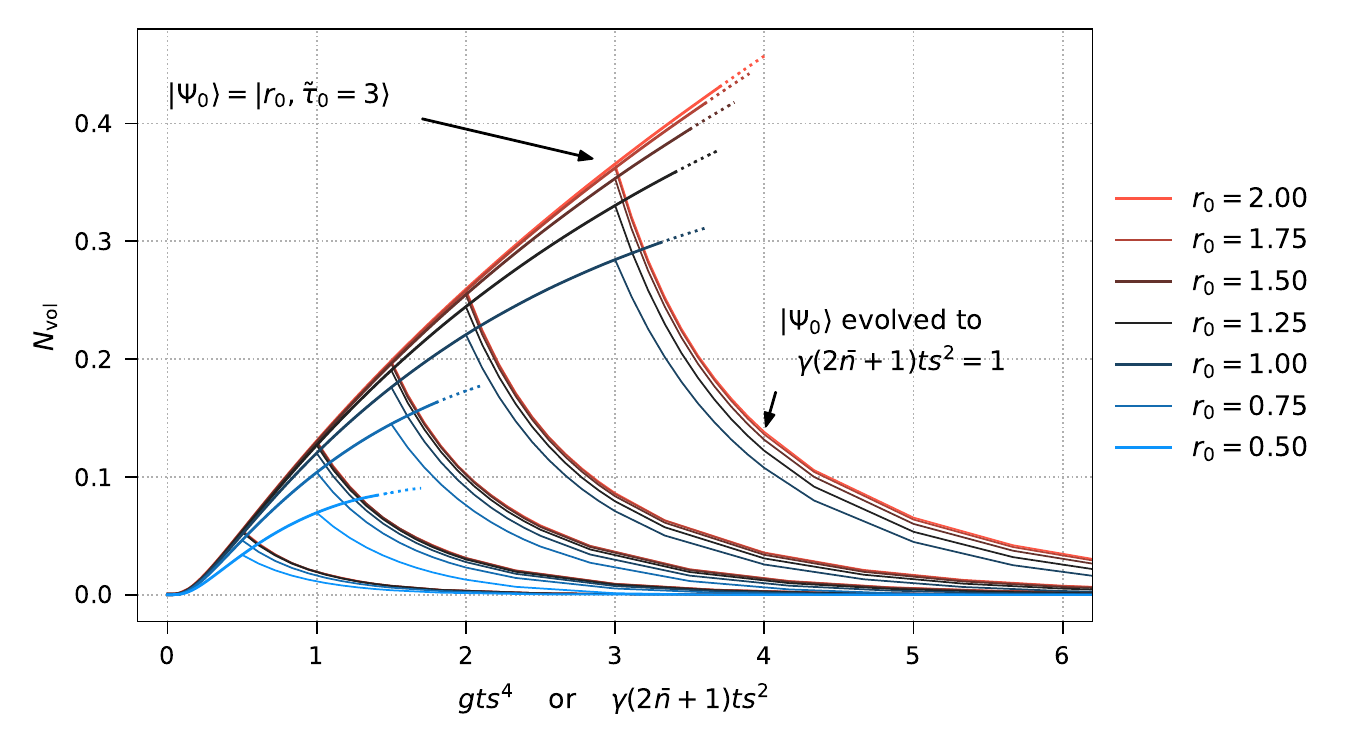}
\par\end{centering}
}

\subfloat[\label{fig:squeezed-kerrdecay-negpeak}Negative peak]{\noindent \begin{centering}
\includegraphics{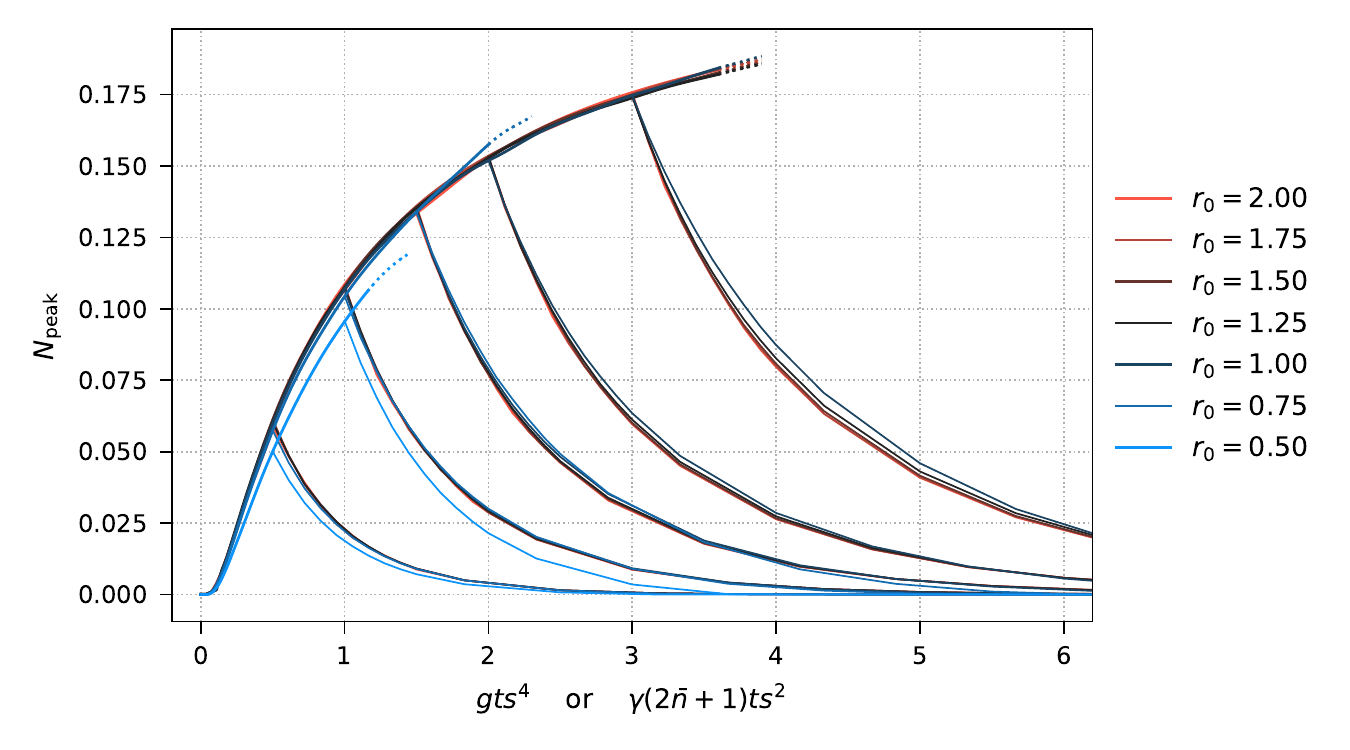}
\par\end{centering}
}

\caption[Decay of squeezed Kerr state negativity under damping]{\label{fig:Decay-of-squeezed}\textbf{Decay of squeezed Kerr state
negativity under damping. }The increasing graphs show the negativity
under unitary evolution as a function of the scaled time $gts^{4}$
(this mirrors Figure \ref{fig:-2}). At select points in time, the
instantaneous state is evolved under the damping master equation (\ref{eq:damped-me})
which causes a decay in negativity. This decay is plotted as a function
of the scaled time $\gamma(2\bar{n}+1)ts^{2}$ which is seen to describe
the decay well. The dimensions of the horizontal axes are described
with equation (\ref{eq:-210}). The decay complies with the bound
of $t_{\mathrm{decay}}$ found in Section \ref{subsec:Fundamental-Solution}
(see Table \ref{tab:Scaled-negativity-decay}). This bound can thus
enter into the deliberation of the validity of the approximation as
$r_{0}$ decreases. For large squeezing, $t_{\mathrm{decay}}$ is
however seen to be a bad indicator of the characteristic time scale
of the system. }
\end{figure}

When plotting $N_{\mathrm{vol}}$ and $N_{\mathrm{peak}}$, we wish
to scale the time axes of the graphs to demonstrate the asymptotic
behavior as was done in Figure \ref{fig:-2}. From these, we know
that the unitary evolution should by graphed as a function of $gts^{4}$.
Note that the points on a vertical line $gts^{4}=k_{1}$ will all
share the value of $\tilde{\tau}_{0}=k_{1}$. We similarly graph the
decay lines as functions of $\gamma t_{\gamma}(2\bar{n}+1)s^{2}$
with $t_{\gamma}$ denoting the time under decay. The graphs for unitary
evolution and decay are scaled in relation to each other such that\footnote{The rescaling done in (\ref{eq:-210}) may also be though as an adjustment
of the frequencies $\gamma$ and $g$ in relation to each other. With
this view, the symbol $t_{\gamma}$ is no longer required to distinguish
the time under damping from the time under unitary evolution.}
\begin{equation}
gts^{4}=\gamma t_{\gamma}(2\bar{n}+1)s^{2}\label{eq:-210}
\end{equation}
by which (\ref{eq:-210}) may be used as a neutral quantity to describe
any time interval of heterogeneous evolution (e.g. unitary evolution
followed by damping) and also to compare time intervals of damping
and unitary evolution. We note that all points on a vertical decay
line $\gamma t_{\gamma}(2\bar{n}+1)s^{2}=k_{2}$ will share $\tilde{\tau}_{\gamma}=k_{2}/8$.
With this established, Figure \ref{fig:Decay-of-squeezed} shows the
quantities $N_{\mathrm{vol}}$ and $N_{\mathrm{peak}}$ plotted as
functions of the neutral quantity (\ref{eq:-210}) (which matches
the $x$-axes in Figure \ref{fig:-2}). The initial unitary evolution
manifests itself as a monotonic growth of $N_{\mathrm{vol}}$ and
$N_{\mathrm{peak}}$ (as analyzed in Section \ref{sec:squeezed-vacuum-initial-state}).
At specific points in time, the resulting state is then evolved further
using (\ref{eq:damped-me}) (and vanishing unitary dynamics $\hat{H}=0$).
We see that both growth and decay of the negativity appear to have
a specific asymptotic behavior as $r_{0}$ increases. In addition
to the effect of the Kerr nonlinearity, the decay of negativity in
the limit of large squeezing is seen to be well described by the rescaled
time $\tilde{\tau}_{\gamma}$ of (\ref{eq:-172}).

\subsection{Damped Kerr Evolution of Squeezed Vacuum\label{subsec:Damped-Kerr-Evolution}}

As the next step, we combine the effects of the damping and Kerr dynamics.
Summing the right hand sides of the damping master equation (\ref{eq:damped-me})
and the von Neumann equation (\ref{eq:kerr-vonneumann-eq}) for the
unitary evolution of the Kerr oscillator, we arrive at the master
equation
\begin{equation}
\dot{\hat{\rho}}=-ig\left[\hat{a}^{\dagger}\hat{a}^{\dagger}\hat{a}\hat{a},\hat{\rho}\right]+\gamma\left(\bar{n}+1\right)\mathcal{D}[\hat{a}]\hat{\rho}+\gamma\bar{n}\mathcal{D}[\hat{a}^{\dagger}]\hat{\rho}.\label{eq:damped-kerr-me-1}
\end{equation}
The procedure for obtaining the equation of motion for the Wigner
function from a master equation treats each right hand side term separately.
Since all right hand side terms of (\ref{eq:damped-kerr-me-1}) have
been considered previously, the right hand side in the equation of
motion for $W$ is simply obtained as the sum of the right hand sides
of (\ref{eq:damping-pde-high-temp}) and (\ref{eq:-18}). We write
it here in Cartesian coordinates:
\begin{equation}
\partial_{t}W(x,y,t)=\begin{aligned}[t] & 2g\left(x^{2}+y^{2}-1\right)\left(-y\partial_{x}+x\partial_{y}\right)W(x,y,t)\\
 & -\frac{g}{8}\left(-y\partial_{x}+x\partial_{y}\right)\left(\partial_{x}^{2}+\partial_{y}^{2}\right)W(x,y,t)\\
 & +\frac{\gamma}{4}\left(\bar{n}+\frac{1}{2}\right)\left(\partial_{x}^{2}+\partial_{y}^{2}\right)W(x,y,t)\\
 & +\frac{\gamma}{2}\partial_{x}\left(xW(x,y,t)\right)+\frac{\gamma}{2}\partial_{y}\left(yW(x,y,t)\right).
\end{aligned}
\label{eq:-137}
\end{equation}
The discussion of individual terms in Sections \ref{subsec:Wigner-Function-Equation}
and \ref{sec:damping} apply to the terms of (\ref{eq:-137}) as well:
The terms on the first line persist in the classical limit and create
a rotation in phase space with a radially dependent angular frequency.
The terms on the second line of (\ref{eq:-137}) vanish in the classical
limit. These are the terms containing third-order derivatives which
give rise to the negative values of $W$. The damping produces a diffusive
effect proportional to $\gamma(2\bar{n}+1)$ and a drift toward the
origin proportional to $\gamma$.

\subsubsection{Evolution of a Squeezed Vacuum State}

We return now to the evolution of the squeezed vacuum state $|\xi\rangle$.
The dynamics (\ref{eq:damped-kerr-me-1}) and (\ref{eq:-137}) are
rotationally invariant. We can see this by briefly recasting (\ref{eq:-137})
in polar coordinates, yielding
\begin{align}
\partial_{t}W(r,\phi,t) & =\begin{aligned}[t] & 2g(r^{2}-1)\partial_{\phi}W(r,\phi,t)-\frac{g}{8}\nabla^{2}\partial_{\phi}W(r,\phi,t)\\
 & \frac{\gamma}{4}\left(\bar{n}+\frac{1}{2}\right)\nabla^{2}W(r,\phi,t)+\frac{\gamma}{2}r\partial_{r}W(r,\phi,t)+\gamma W(r,\phi,t).
\end{aligned}
\end{align}
Rotational invariance is seen from the lack of dependence on the angular
coordinate $\phi$. Even with the inclusion of damping we can therefore
continue to set $\theta_{0}=0$ in $|\xi\rangle$. The Wigner function
for the initial state is given in equation (\ref{eq:squeezed-vacuum-initial-state}).

\begin{figure}[p]
\noindent \begin{centering}
\centerline{\includegraphics[viewport=0bp 40bp 428bp 500bp,clip]{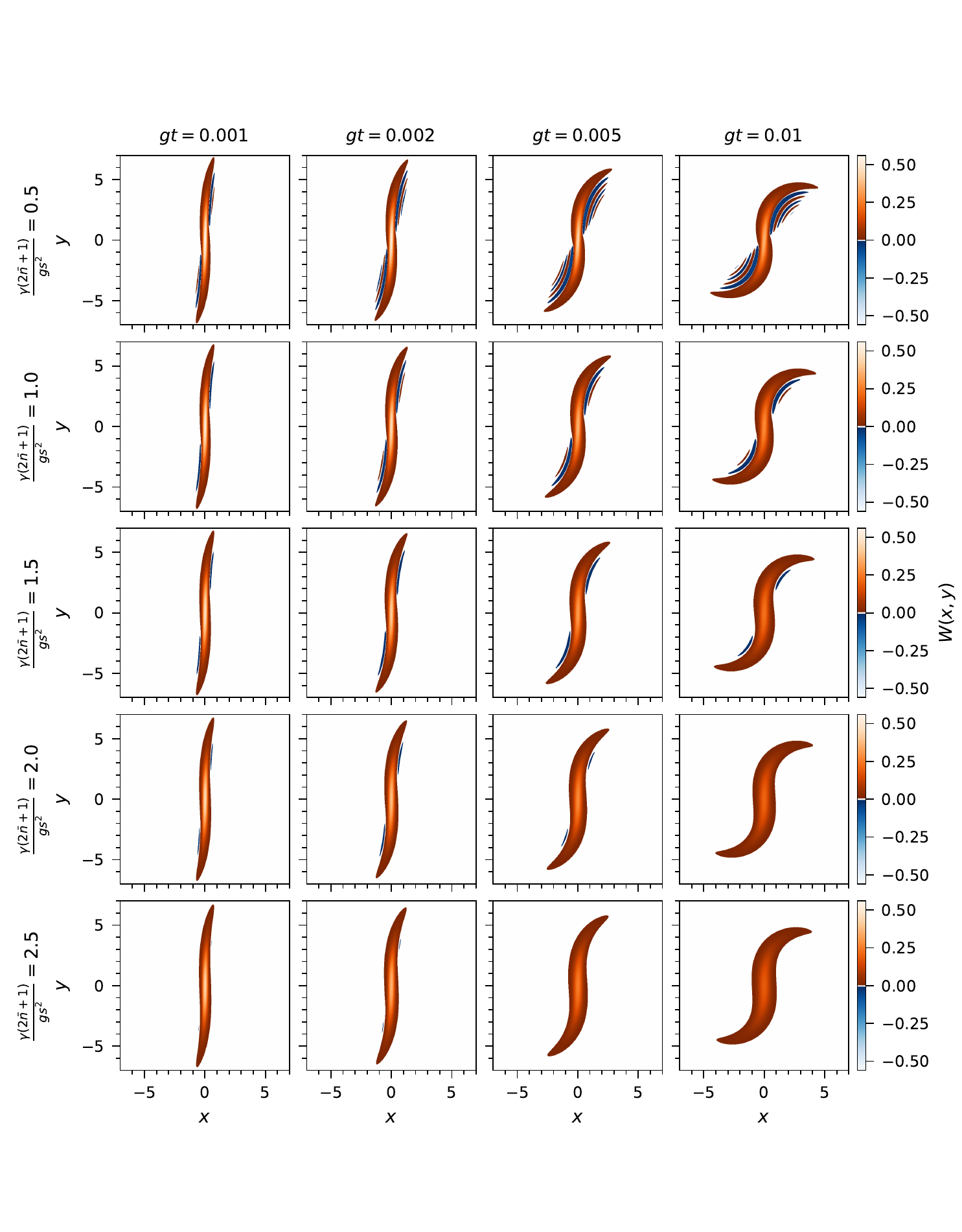}}
\par\end{centering}
\caption[Evolution for varying damping rates]{\label{fig:scaled-squeezed-gallery-1-3}\textbf{Evolution for varying
damping rates.} The squeezed vacuum state (\ref{eq:-73}) with $\xi=r_{0}=1.5$
is evolved under the damping master equation (\ref{eq:damped-kerr-me-1})
with varying effective damping rates $\gamma(2\bar{n}+1)/gs^{2}$
(see Section \ref{subsec:Damped-Kerr-Evolution}). The temperature
is kept fixed at $\bar{n}=1000$ and the damping is therefore well
described by the high temperature equation (\ref{eq:-103}) as a homogeneous
diffusive effect throughout phase space. The damped Kerr oscillator
is discussed in Section \ref{subsec:Kerr-Oscillator-with}. This initial
state is the same as the one used for Figure \ref{fig:asymp-medium-evo}
showing unitary evolution.}
\end{figure}

\begin{figure}
\captionsetup{position=top}\subfloat[\label{fig:squeezed-kerrdecay-negvol-1}Negative volume]{\noindent \begin{centering}
\captionsetup{position=top}\includegraphics{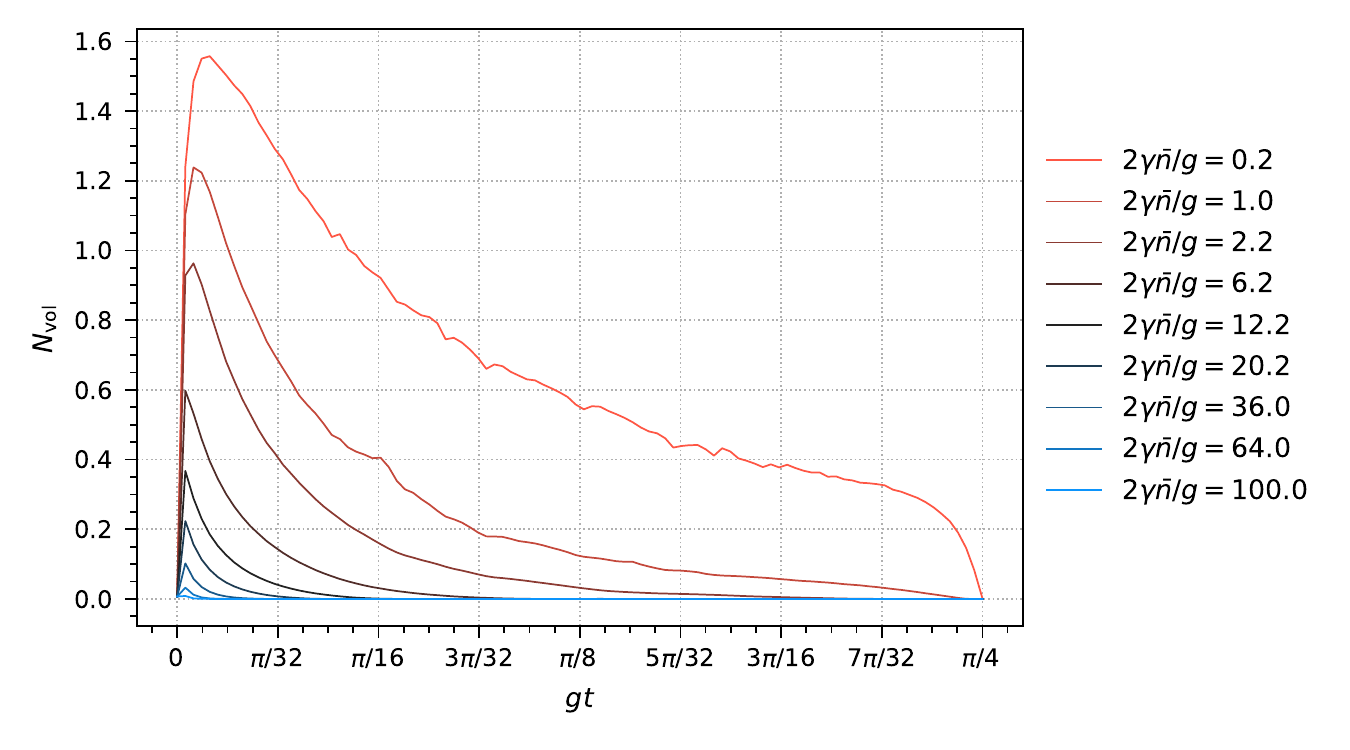}
\par\end{centering}
}

\subfloat[\label{fig:squeezed-kerrdecay-negpeak-1}Negative peak]{\noindent \begin{centering}
\captionsetup{position=top}\includegraphics{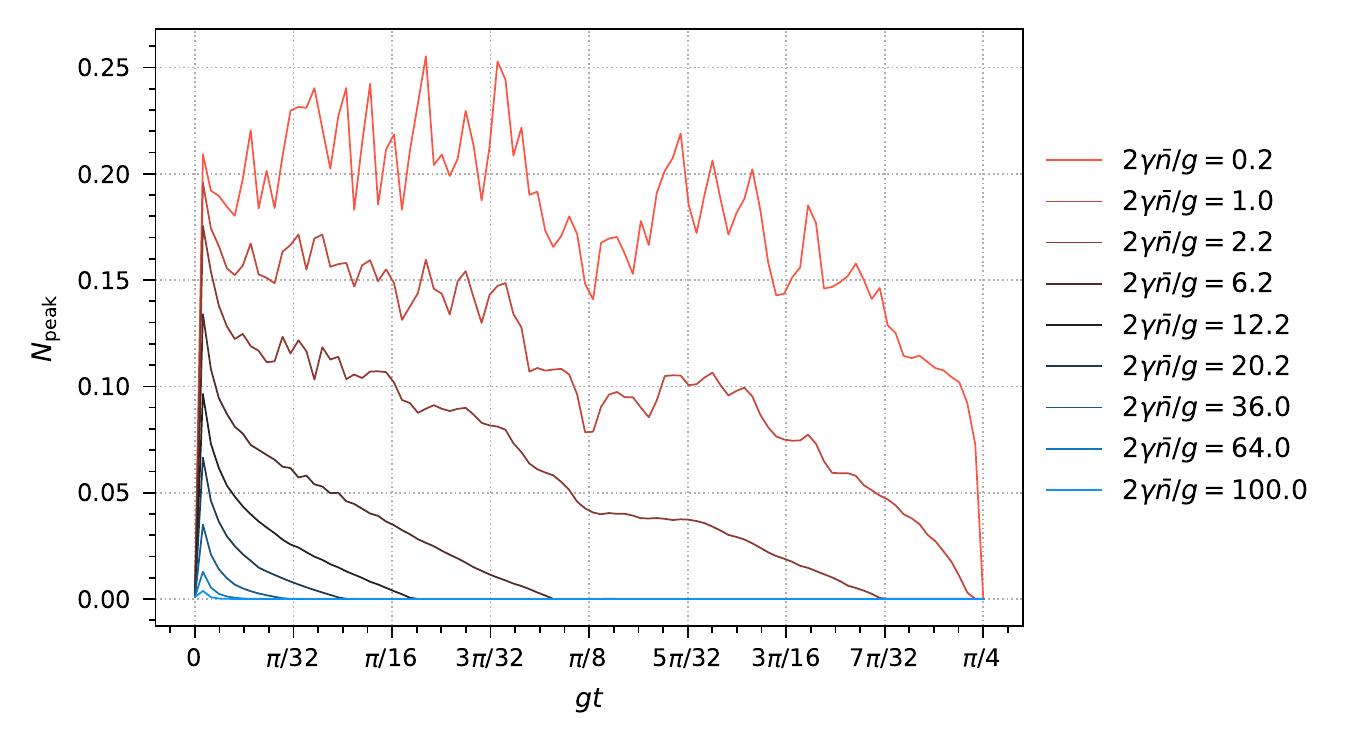}
\par\end{centering}
}

\caption[Negativity during long time damped evolution of squeezed vacuum]{\label{fig:Note-that-the}\textbf{Negativity during long time damped
evolution of squeezed vacuum.} Evolution of the squeezed vacuum state
(\ref{eq:-73}) with $\xi=r_{0}=2.0$. under the master equation (\ref{eq:damped-kerr-me-1}).
A time interval corresponding to a full period of unitary evolution
is shown. Increasing the damping compared with $g$ causes the negativity
to decrease. The evolution of a squeezed vacuum state of the damped
Kerr oscillator is discussed in Section \ref{subsec:Damped-Kerr-Evolution}.
Note that the number of points shown fail to express all details of
the evolution (a more accurate account of the frequency of fluctuations
is provided by Figure \ref{fig:}). }
\end{figure}

The evolution of the Wigner function for the particular initial state
$|\xi=1.5\rangle$ is shown in Figure \ref{fig:scaled-squeezed-gallery-1-3}
for various values of $2\gamma\bar{n}\approx\gamma(2\bar{n}+1)$ and
$\bar{n}=1000$. This initial state is the same as the one used for
Figure \ref{fig:asymp-medium-evo} showing unitary evolution. Increasing
$\gamma(2\bar{n}+1)$ (downward in Figure \ref{fig:asymp-medium-evo})
generally results in a softening of the Wigner function as it evolves.
The peak value of the Wigner function decreases while the variance
increases. The finer details of the Wigner function are reduced in
magnitude. This lessens the amplitude of the fringes forming in the
concave regions of the S-shape causing a reduction in $N_{\mathrm{vol}}$
and $N_{\mathrm{peak}}$. Figure \ref{fig:Note-that-the} show the
negativity over an entire period for the initial state $|\xi=2\rangle$.
It is seen that increased damping in all cases leads to a decrease
in negativity. It also causes the smaller details seen in the time
evolution to vanish. Hence the graphs of $N_{\mathrm{vol}}$ and $N_{\mathrm{peak}}$
appear smoother for larger damping. This is especially evident in
the case of $N_{\mathrm{peak}}$.

\subsubsection{Rescaled Coordinates and Large Squeezing Approximation \label{subsec:kerrdecay-approx}}

Retracing the steps of Section \ref{sec:kerr-state-damping}, we wish
to rescale the initial state and equation of motion. The terms for
unitary evolution and decoherence have already been scaled separately
in equations (\ref{eq:-31-2-1}) and (\ref{eq:-90}). The combined
right hand side is simply found by summing the right hand sides of
those two equations. We then arrive at
\begin{equation}
\partial_{t}\tilde{W}(\tilde{x},\tilde{y},t)=\begin{aligned}[t] & 2g\left(-\tilde{x}^{2}\tilde{y}\partial_{\tilde{x}}-s^{4}\tilde{y}^{3}\partial_{\tilde{x}}+\frac{1}{s^{4}}\tilde{x}^{3}\partial_{\tilde{y}}+\tilde{x}\tilde{y}^{2}\partial_{\tilde{y}}\right)\tilde{W}(\tilde{x},\tilde{y},t)\\
 & -2g\left(-s^{2}\tilde{y}\partial_{\tilde{x}}+\frac{1}{s^{2}}\tilde{x}\partial_{\tilde{y}}\right)\tilde{W}(\tilde{x},\tilde{y},t)\\
 & -\frac{g}{8}\left(-s^{4}\tilde{y}\partial_{\tilde{x}}^{3}+\frac{1}{s^{4}}\tilde{x}\partial_{\tilde{y}}^{3}+\tilde{x}\partial_{\tilde{y}}\partial_{\tilde{x}}^{2}-\tilde{y}\partial_{\tilde{x}}\partial_{\tilde{y}}^{2}\right)\tilde{W}(\tilde{x},\tilde{y},t)\\
 & +\frac{\gamma s^{2}}{4}\left(\bar{n}+\frac{1}{2}\right)\partial_{\tilde{x}}^{2}\tilde{W}(\tilde{x},\tilde{y},t)+\frac{\gamma}{4s^{2}}\left(\bar{n}+\frac{1}{2}\right)\partial_{\tilde{y}}^{2}\tilde{W}(\tilde{x},\tilde{y},t)\\
 & +\frac{\gamma}{2}\partial_{\tilde{x}}\left(\tilde{x}\tilde{W}(\tilde{x},\tilde{y},t)\right)+\frac{\gamma}{2}\partial_{\tilde{y}}\left(\tilde{y}\tilde{W}(\tilde{x},\tilde{y},t)\right).
\end{aligned}
\label{eq:-173}
\end{equation}
We see from this that the effect of diffusion changes in inverse proportion
to the variance in that axis, e.g. the squeezed axis variance $1/4s^{2}$
causes the corresponding diffusion to increase by a factor of $s^{2}$.We
are interested here in the specific regime of large squeezing $s$
where the effects of the Kerr nonlinearity and damping are both significant.
We therefore keep separately the terms from (\ref{eq:-173}) which
contains the highest power of $s$ in combination with $g$ and $\gamma$.
These are the terms proportional to $gs^{4}$ or $\gamma s^{2}\left(\bar{n}+\frac{1}{2}\right)$.
This leaves us with the equation

\begin{equation}
\partial_{t}\tilde{W}(\tilde{x},\tilde{y},t)=\begin{aligned}[t] & -2gs^{4}\tilde{y}^{3}\partial_{\tilde{x}}\tilde{W}(\tilde{x},\tilde{y},t)+\frac{g}{8}s^{4}\tilde{y}\partial_{\tilde{x}}^{3}\tilde{W}(\tilde{x},\tilde{y},t)\\
 & +\frac{\gamma s^{2}}{4}\left(\bar{n}+\frac{1}{2}\right)\partial_{\tilde{x}}^{2}\tilde{W}(\tilde{x},\tilde{y},t).
\end{aligned}
\label{eq:-136}
\end{equation}
This equation allows one to compare the effects of squeezing and damping.
We see from (\ref{eq:-136}), that the nonlinearity and the damping
effect enters into (\ref{eq:-136}) as terms containing different
powers of $s$. We therefore expect the Kerr effect to scale with
$s^{4}$ (as for the unitary evolution, see Section \ref{subsec:large-squeezing-approximation})
and the damping to scale with $s^{2}$ (as with the isolated damping,
see Section \ref{sec:kerr-state-damping}). To formalize this expectation,
we can extend the large squeezing Fourier space solution of Section
\ref{subsec:large-squeezing-approximation} to include damping. Define
again $\tilde{u}(\mu,\tilde{\tau})$ by
\begin{equation}
\tilde{W}(\tilde{x},\tilde{y},t)=\tilde{u}_{\tilde{y}}(\tilde{x}-2gs^{4}\tilde{y}^{3}t,gs^{4}\tilde{y}t/8).\label{eq:-158-1}
\end{equation}
From (\ref{eq:-136}), the equation of motion for $\tilde{u}(\mu,\tilde{\tau})$
is found as
\begin{subequations}
\label{eq:-178}

\begin{equation}
\partial_{\tilde{\tau}}\tilde{u}_{\tilde{y}}(\mu,\tilde{\tau})=\partial_{\mu}^{3}\tilde{u}_{\tilde{y}}(\mu,\tilde{\tau})+\beta_{\tilde{y}}\partial_{\mu}^{2}\tilde{u}_{\tilde{y}}(\mu,\tilde{\tau})\label{eq:-176}
\end{equation}
with
\begin{equation}
\beta_{\tilde{y}}=\frac{\gamma(2\bar{n}+1)}{gs^{2}\tilde{y}}.\label{eq:-177}
\end{equation}
\end{subequations}
The solution analogous to (\ref{eq:-174}) is
\begin{equation}
\tilde{u}_{\tilde{y}}(\mu,\tilde{\tau})=\frac{1}{\sqrt{2\pi}}\int_{-\infty}^{\infty}dk\,h_{\tilde{y}}(k)\,e^{i(k\mu-k^{3}\tilde{\tau})}e^{-\beta_{\tilde{y}}\tilde{\tau}}.\label{eq:-28-1}
\end{equation}
$h_{\tilde{y}}(k)$ is given by (\ref{eq:-80}). We see from (\ref{eq:-178})
that all problem parameters enter into (\ref{eq:-28-1}) only in the
form $\gamma(2\bar{n}+1)/gs^{2}$ and implicitly in $\tilde{\tau}$.
Thus, if thinking of rescaled time $\tilde{\tau}$ as the fundamental
time of the problem, we can think of $\gamma(2\bar{n}+1)/gs^{2}$
as an effective ratio between damping and nonlinearity. 

Of course, recovering through (\ref{eq:-158-1}) the solution for
$\tilde{W}(\tilde{x},\tilde{y},t)$ yields
\begin{equation}
\tilde{W}(\tilde{x},\tilde{y},t)=\frac{1}{\sqrt{2\pi}}\int_{-\infty}^{\infty}dk\,h_{\tilde{y}}(k)\,e^{i(k\tilde{x}-2kgs^{4}\tilde{y}^{3}t-k^{3}gs^{4}\tilde{y}t/8)}e^{-(2\bar{n}+1)\gamma s^{2}t/8},\label{eq:-252}
\end{equation}
and hence the absolute values of $gs^{4}$ and $(2\bar{n}+1)\gamma s^{2}$
must be kept separate if the time $t$ has relevance (this is unsurprising
since (\ref{eq:-28-1}) contains no explicit frequencies and thus
way to express a time scale measured in seconds).

\subsubsection{Maximum Negative Volume}

\begin{figure}
\noindent \begin{centering}
\centerline{\includegraphics{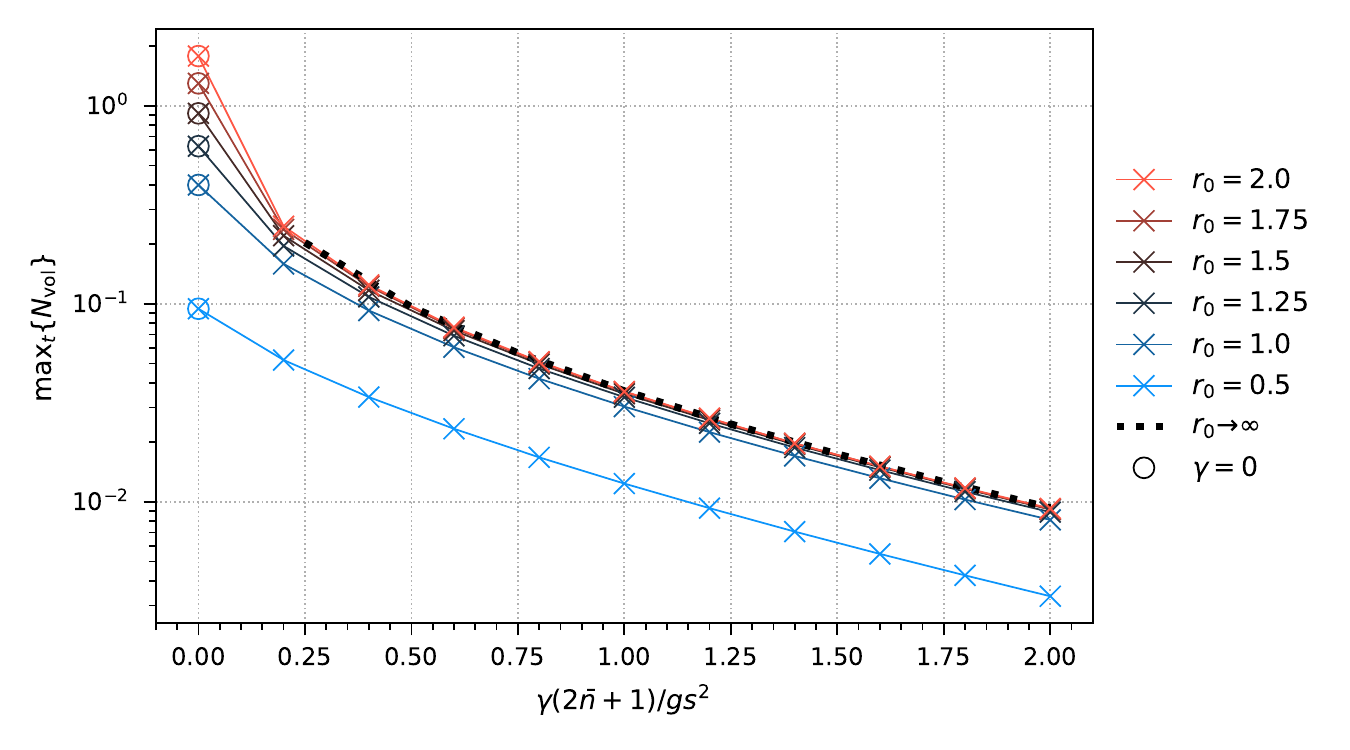}}
\par\end{centering}
\caption[Maximum negative volume versus damping rate]{\label{fig:-Maximum-negative}\textbf{ Maximum negative volume versus
damping rate.} Demonstration of the maximum negative volume $\max_{t}\left\{ N_{\mathrm{vol}}(t)\right\} $
as a function of the scaled decoherence rate. This quantity is defined
in (\ref{eq:-179}) as a measure of negativity that is independent
of the squeezing. The initial state is a squeezed vacuum state (\ref{eq:-73})
with squeezing parameters $\theta_{0}=0$ and $r_{0}$. For $\gamma\protect\neq0$,
master equation (\ref{eq:damped-kerr-me-1}) is used to evolve the
state. The values for unitary evolution ($\gamma=0$) describe the
plateau height as seen in Figure \ref{fig:maxvol-period}. The maximum
negative volume generally decreases with increasing $\gamma(2\bar{n}+1)/gs^{2}$.
For larger $\gamma(2\bar{n}+1)/gs^{2}$ however, the graphed quantity
is seen to tend asymptotically to a fixed value as $r_{0}$ increases.
The thick dotted line shows the asymptotic behavior obtained from
(\ref{eq:-252}) for $\gamma(2\bar{n}+1)/gs^{2}\protect\geq0.25$.}
\end{figure}

We wish to now construct a measure which summarizes the effects of
squeezing and damping. We therefore define
\begin{align}
(\text{maximum negative volume}) & =\max_{t}\left\{ N_{\mathrm{vol}}(t)\right\} .\label{eq:-179}
\end{align}
Graphically, the quantity $\max_{t}\left\{ N_{\mathrm{vol}}(t)\right\} $
measures the maximum of the graph of negative volume versus time,
examples of which may be found in Figure \ref{fig:squeezed-kerrdecay-negvol-1}.
Expressing it using the solution (\ref{eq:-28-1}), we have (see (\ref{eq:-162}))
\begin{subequations}
\label{eq:-175}
\begin{align}
\max_{t}\left\{ N_{\mathrm{vol}}(t)\right\}  & =-\min_{t}\int d\tilde{x}d\tilde{y}\,\min\{0,\tilde{u}_{\tilde{y}}(\tilde{x}-2gts^{4}\tilde{y}^{3},gts^{4}\tilde{y}/8)\}\\
 & =-\min_{\tilde{\tau}}\int d\mu d\tilde{y}\,\min\{0,\tilde{u}_{\tilde{y}}(\mu,\tilde{\tau})\}.
\end{align}
\end{subequations}
This quantity is insensitive to the characteristic time scale of the
problem $1/g$, i.e. it has no consequence if the outer minimum in
(\ref{eq:-175}) is taken with respect to $t\in[0,\infty)$ or $\tilde{\tau}\in[0,\infty)$.
Furthermore, the function $\tilde{u}_{\tilde{y}}$ depends only on
the effective ratio $\gamma(2\bar{n}+1)/s^{2}g$ since this the only
parameter present in its equation of motion (\ref{eq:-176}). We therefore
also expect $\gamma(2\bar{n}+1)/s^{2}g$ to be the relevant quantity
for $\max_{t}\left\{ N_{\mathrm{vol}}(t)\right\} $. We evolve the
squeezed initial state under (\ref{eq:damped-kerr-me-1}) and consider
the maximum negative volume as a function of time. Figure \ref{fig:-Maximum-negative}
shows $\max_{t}\left\{ N_{\mathrm{vol}}(t)\right\} $ as a function
of $\gamma(2\bar{n}+1)/gs^{2}$ for various squeezing parameters $r_{0}$.
The values appear to behave asymptotically as $r_{0}$is increased.
We expect this behavior to break down for smaller values of $r_{0}$
(e.g. $r_{0}=0$ leads to no negativity) which is clearly visible
for $r_{0}=0.5$ and less so for $r_{0}=1.0$. We can interpret this
as the collapse of the approximation of large squeezing which was
used to obtain (\ref{eq:-136}). The points of Figure \ref{fig:-Maximum-negative}
were obtained through successive refinement of the time resolution
in the around $\max_{t}\left\{ N_{\mathrm{vol}}(t)\right\} $ to estimate
the quantity more accurately than possible from the data shown in
Figure \ref{fig:squeezed-kerrdecay-negvol-1} alone. 

We also note that in the special case of no damping ($\gamma=0$),
the value of $\max_{t}\left\{ N_{\mathrm{vol}}(t)\right\} $ is the
plateau height discussed in Section \ref{subsec:Negativity-during-a}
and plotted in Figure \ref{fig:maxvol-period}. With little or no
damping there is not enough time for the negative regions of the Wigner
function to decay to zero before the system evolution transitions
away from its initial character, also rendering the large squeezing
approximation invalid. This transition can be seen in Figure \ref{fig:evo-gallery-medium}
in the case of $\gamma=0$.

\section{Phase Decoherence \label{sec:phase-decoherence}}

\begin{figure}
\noindent \begin{centering}
\includegraphics{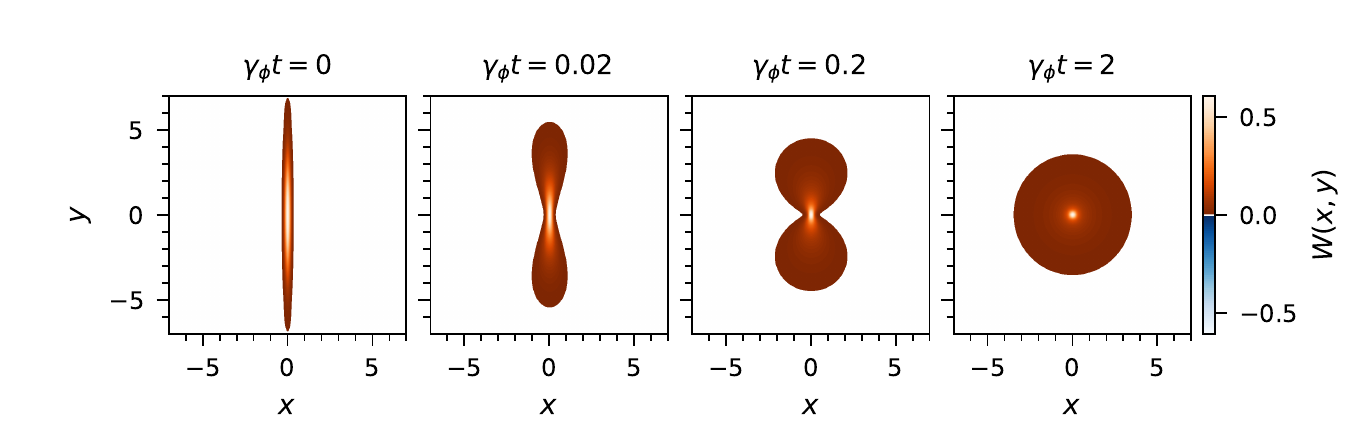}
\par\end{centering}
\caption[Dephasing of squeezed vacuum]{\label{fig:Damping-of-squeezed-1-1}\textbf{Dephasing of squeezed
vacuum. }The squeezed vacuum state (\ref{eq:-73}) with squeezing
parameter $r_{0}=1.5$ is evolved under the dephasing master equation
(\ref{eq:dephasing-me}). Dephasing causes a diffusive effect along
the angular coordinate as described in Section \ref{sec:phase-decoherence}.
This also increases the variance in the $\hat{X}$ quadrature. At
$\gamma_{\phi}t=2$, the state appears indistinguishable from the
steady state solution which is a rotationally symmetric state.}
\end{figure}

Outside of damping, many experimental systems are limited by phase
noise. We describe this phenomenon quantum mechanically by introducing
a dephasing term into the master equation. In the general master equation
(\ref{eq:general-master-equation}) such a term is included and written
proportional to the dephasing coefficient $\gamma_{\phi}$ which is
a frequency describing the strength of the effect. We construct for
now a master equation describing this effect alone. Setting all but
the dephasing term to zero in (\ref{eq:general-master-equation})
renders the dephasing master equation
\begin{equation}
\dot{\hat{\rho}}=\gamma_{\phi}\mathcal{D}[\hat{n}]\hat{\rho}.\label{eq:dephasing-me}
\end{equation}
We note here, that the equation is easily solved element-wise in the
number state basis (see Appendix \ref{app:operator-picture-dephasing}).
We apply to (\ref{eq:dephasing-me}) the procedure of Section \ref{sec:wigner-pde-derivation}
to discover the corresponding equation of motion for the Wigner function
$W$. The results can be stated in polar coordinates as
\begin{equation}
\partial_{t}W(r,\phi,t)=\frac{\gamma_{\phi}}{2}\partial_{\phi}^{2}W(r,\phi,t).\label{eq:dephasing-pde-polar}
\end{equation}
It is seen that (\ref{eq:dephasing-pde-polar}) describes a diffusion
process in the angular coordinate $\phi\in[0;2\pi)$. The requirement
that the Wigner function has a unique value imposes a periodic boundary
condition:
\begin{equation}
W(r,2\pi,t)=W(r,0,t).
\end{equation}
For a given initial state, we can therefore express the solution to
(\ref{eq:dephasing-pde-polar}) exactly by decomposing (\ref{eq:dephasing-pde-polar})
into eigenfunctions of $\partial_{\phi}^{2}$ resulting in a Fourier
series in the coordinate $\phi$. Appendix \ref{app:operator-picture-dephasing}
details the derivation of a conceptually similar solution formulated
in terms of density matrices. 

\subsection{Dephasing of Squeezed Kerr State \label{sec:kerr-state-dephasing}}

For now, we investigate dephasing of a specific initial state. The
deliberations of Section \ref{sec:kerr-state-damping} still apply:
Simply applying dephasing to a Gaussian state will hold $N_{\mathrm{vol}}=N_{\mathrm{peak}}=0$
(although the state will in most cases cease to be Gaussian). We reuse
the squeezed Kerr state $|r_{0},\tilde{\tau}_{0}\rangle$ as given
in (\ref{eq:squeezed-kerr-state-def}) and examine the decay of $N_{\mathrm{vol}}$
and $N_{\mathrm{peak}}$ under dephasing instead. As in Section \ref{sec:kerr-state-damping}
we will now rescale the initial state and the equation of motion for
the Wigner function. Since the initial state is reused from Section
\ref{sec:kerr-state-damping}, arguments for why this rescaling is
meaningful may be found there.

\subsubsection{Rescaled Coordinates}

To introduce the rescaled coordinates, we require the equation of
motion expressed in Cartesian coordinates. Recasting (\ref{eq:dephasing-pde-polar})
in Cartesian coordinates yields
\begin{equation}
\partial_{t}W(x,y,t)=\frac{\gamma_{\phi}}{2}\left(y^{2}\partial_{x}^{2}+x^{2}\partial_{y}^{2}-2xy\partial_{x}\partial_{y}-x\partial_{x}-y\partial_{y}\right)W(x,y,t).\label{eq:dephasing-pde-xy}
\end{equation}
We then introduce the rescaled coordinates $(\tilde{x},\tilde{y})$
as given in (\ref{eq:-30-2}) with corresponding differential operators
as given in (\ref{eq:-27}). In these coordinates, the initial state
takes again the simple form of (\ref{eq:-30}) while (\ref{eq:dephasing-pde-xy})
is transformed to 
\begin{equation}
\partial_{t}\tilde{W}(\tilde{x},\tilde{y},t)=\frac{\gamma_{\phi}}{2}\left(s^{4}\tilde{y}^{2}\partial_{\tilde{x}}^{2}+s^{-4}\tilde{x}^{2}\partial_{\tilde{y}}^{2}-2\tilde{x}\tilde{y}\partial_{\tilde{x}}\partial_{\tilde{y}}-\tilde{x}\partial_{\tilde{x}}-\tilde{y}\partial_{\tilde{y}}\right)\tilde{W}(\tilde{x},\tilde{y},t).\label{eq:dephasing-pde-scaled}
\end{equation}
As previously, this form makes the dependence on $s$ explicit in
the equation of motion. The initial state 
\[
\tilde{W}(\tilde{x},\tilde{y},0)=\tilde{W}_{|r_{0},\tilde{\tau}_{0}\rangle}(\tilde{x},\tilde{y})
\]
is approximately independent of $r_{0}$ in the limit of large squeezing
and small $\tilde{\tau}_{0}$.

\subsubsection{Large Squeezing Approximation}

Next, we discard all but the leading order terms of (\ref{eq:dephasing-pde-scaled}).
In this case, retain from (\ref{eq:dephasing-pde-scaled}) only the
single term containing $s^{4}$. This yields the equation

\begin{equation}
\partial_{t}\tilde{W}(\tilde{x},\tilde{y},t)=\frac{\gamma_{\phi}}{2}s^{4}\tilde{y}^{2}\partial_{\tilde{x}}^{2}\tilde{W}(\tilde{x},\tilde{y},t).\label{eq:dephasing-pde-approx}
\end{equation}
Moving from (\ref{eq:dephasing-pde-polar}) to (\ref{eq:dephasing-pde-approx})
changes from a diffusive process in the angular coordinate to a diffusive
process in the $\tilde{x}$-coordinate instead. One could think of
this as the linearization of the angular diffusion described by (\ref{eq:dephasing-pde-polar})
around the $\tilde{y}$-axis. Indeed, we are mainly interested in
the behavior of the Wigner function in proximity to the $\tilde{y}$-axis
since most of the Wigner density is concentrated here for the relevant
squeezed states (as determined by our choice of $\theta_{0}=0$).
The angular derivative operator $\partial_{\phi}$ contains an implicit
scaling factor of $r$ as can be seen from the equivalent differential
operator in Cartesian coordinates (see (\ref{eq:-14})). Due to this,
the strength of the diffusive effect described by (\ref{eq:dephasing-pde-polar})
actually scales as $r^{2}$ (the radial coordinate $r$ measures the
distance to the origin). This spatial dependence has been made explicit
in \ref{eq:dephasing-pde-approx} as the factor $\tilde{y}^{2}$.
Notice finally that the dephasing rate $\gamma_{\phi}$ appears in
the subexpression $s^{4}\gamma_{\phi}$. From this, we expect the
dephasing to scale with $s^{4}$. We investigate this in the following
section by looking at the decay of $N_{\mathrm{vol}}$ and $N_{\mathrm{peak}}$.
Prior to that, we repeat the arguments of Section \ref{sec:kerr-state-damping},
to express $\tilde{W}(\tilde{x},\tilde{y},t)$ with all dependence
on squeezing explicit in the expression.

Like in (\ref{eq:-172}), we introduce a rescaled time coordinate
\begin{equation}
\tilde{\tau}_{\phi}=\frac{\gamma_{\phi}}{2}s^{4}\tilde{y}^{2}t.
\end{equation}
Unlike (\ref{eq:-172}) (but similar to the rescaled time in (\ref{eq:-62}))
$\tilde{\tau}_{\phi}$ has an implicit dependence on $\tilde{y}$.
We also introduce the function $\tilde{v}(\tilde{x},\tilde{y},\tilde{\tau}_{\phi})$
such that
\begin{equation}
\tilde{W}(\tilde{x},\tilde{y},t)=\tilde{v}(\tilde{x},\tilde{y},\tilde{\tau}_{\phi}{=}\gamma_{\phi}s^{4}\tilde{y}^{2}t/2).\label{eq:-104-1}
\end{equation}
The equation of motion and initial state for $\tilde{v}$ are again
given by
\begin{equation}
\partial_{\tilde{\tau}_{\gamma}}\tilde{v}(\tilde{x},\tilde{y},\tilde{t})=\partial_{\tilde{x}}^{2}v(\tilde{x},\tilde{y},\tilde{\tau})\label{eq:-105-1}
\end{equation}
and
\begin{equation}
\tilde{v}(\tilde{x},\tilde{y},0)=\tilde{W}_{|r_{0},\tilde{\tau}_{0}\rangle}(\tilde{x},\tilde{y}).
\end{equation}
The function $\tilde{v}$ is seen to be independent of squeezing in
both its initial state and equation of motion. Hence, we expect all
dependence on squeezing of $\tilde{W}$ to be explicit in (\ref{eq:-104-1}).

\subsubsection{Decay of Negativity}

We expect from (\ref{eq:dephasing-pde-approx}), that the dephasing
scales with $s^{4}$. This may be seen from (\ref{eq:-104-1}). Notably,
this scaling is shared with the Kerr effect (see (\ref{eq:-31-2-1})).
In analogy with Section \ref{sec:kerr-state-damping}, conclude our
analysis of the squeezed Kerr state under the dephasing.

\begin{figure}
\subfloat[\label{fig:squeezed-kerrdephas-negvol}Negative volume]{\noindent \begin{centering}
\includegraphics{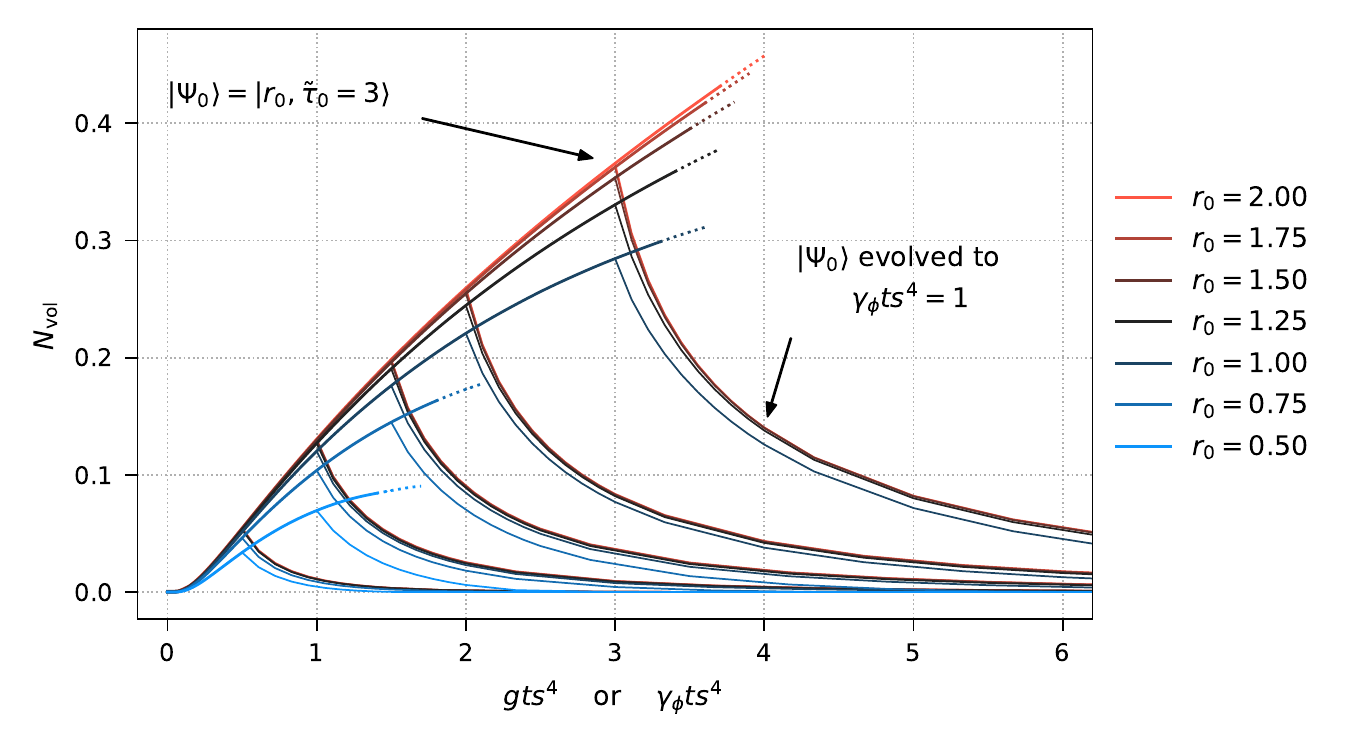}
\par\end{centering}
}

\subfloat[\label{fig:squeezed-kerrdephas-negpeak}Negative peak]{\noindent \begin{centering}
\includegraphics{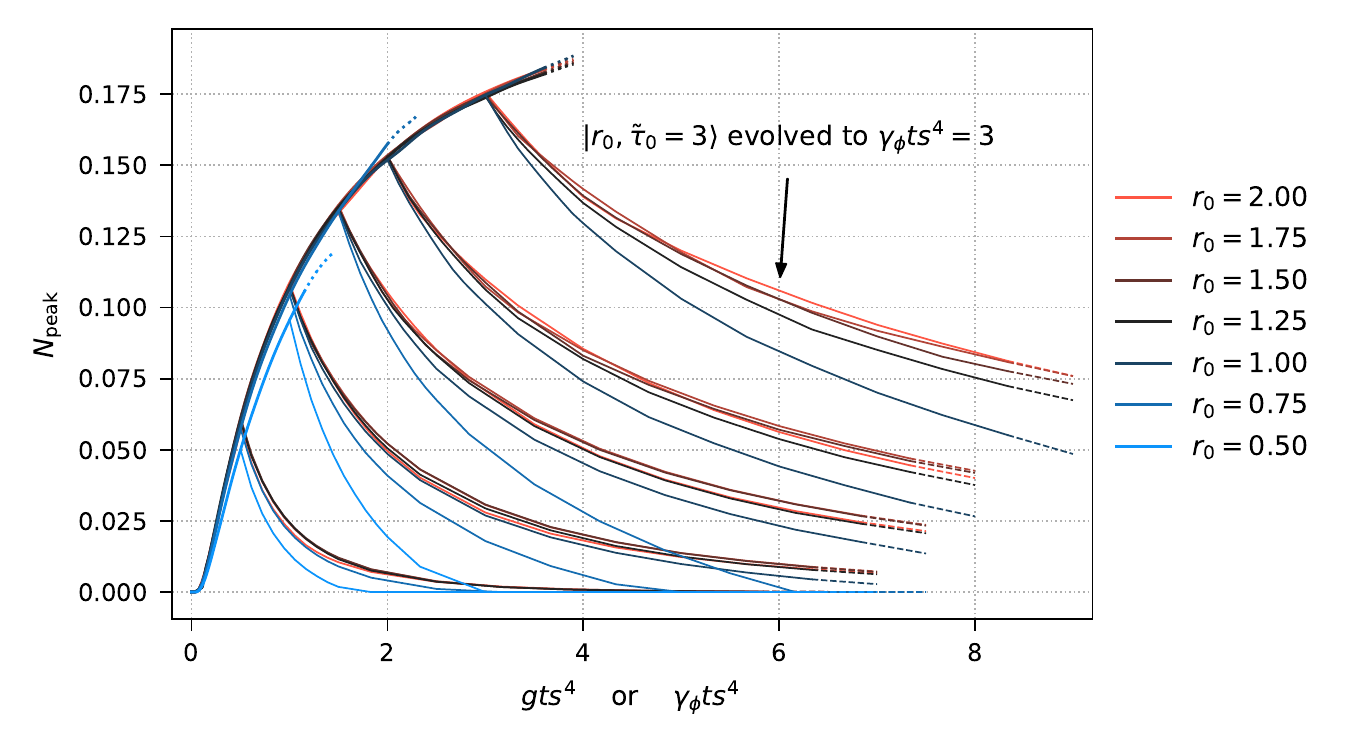}
\par\end{centering}
}

(\ref{eq:damped-me})\caption[Decay of squeezed Kerr state negativity under dephasing]{\label{fig:Decay-of-squeezed-1}\textbf{Decay of squeezed Kerr state
negativity under dephasing. }As in Figure \ref{fig:Decay-of-squeezed},
growing graphs show the unitary evolution in negativity of a squeezed
vacuum state (\ref{eq:-73}) with squeezing parameter $r_{0}$. At
select points in time, the instantaneous state is evolved under the
dephasing master equation \ref{eq:dephasing-me} which causes a decay
in negativity. This decay is plotted as a function of the scaled time
$\gamma_{\phi}ts^{4}$ which is seen to describe the decay well. The
dimensions of the horizontal axes are explained with equation (\ref{eq:-211}). }
\end{figure}

We evolve the squeezed state for a time $t_{0}=\tilde{\tau}_{0}/gs^{4}$
under the unitary dynamics of the Kerr Hamiltonian. The result is
the state $|r_{0},\tilde{\tau}_{0}\rangle$. This state is then evolved
under dephasing for a time $t_{\phi}$. We set
\begin{equation}
gts^{4}=\gamma_{\phi}ts^{4}\label{eq:-211}
\end{equation}
such that we may compare and sum times by scaling them as $\gamma_{\phi}ts^{4}$
or $gts^{4}$. Figures \ref{fig:squeezed-kerrdephas-negvol} and \ref{fig:squeezed-kerrdephas-negpeak}
show the negativity of the two staged process of Kerr evolution to
the state $|r_{0},\tilde{\tau}_{0}\rangle$ followed by dephasing
computed with (\ref{eq:dephasing-me}). When looking at $N_{\mathrm{vol}}$
(Figure \ref{fig:squeezed-kerrdephas-negvol}), these scalings seem
to fit well. The decay of the negative volume under dephasing is very
similar to the decay under damping (Figure \ref{fig:squeezed-kerrdecay-negvol}).

The negative peak (Figures \ref{fig:squeezed-kerrdephas-negvol})
departs from the asymptotic behavior more quickly. Compare with the
negativity decay from damping (Figure \ref{fig:squeezed-kerrdecay-negpeak}),
the asymptotic behavior is less pronounced. For weaker squeezing $r_{0}=0.5$
or $r_{0}=0.75$ the negativity decays more quickly to zero. This
could be an indicator that the negative peak lies close to the origin
compared to the overall negativity measured by $N_{\mathrm{vol}}$. 

\subsection{Kerr Oscillator with Dephasing \label{subsec:Kerr-Oscillator-with}}

Like it was done for damping to obtain (\ref{eq:damped-kerr-me-1}),
we wish to combine the effects of dephasing and unitary Kerr evolution.
Combining the relevant equations, (\ref{eq:dephasing-me}) and (\ref{eq:kerr-vonneumann-eq}),
we have
\begin{equation}
\dot{\hat{\rho}}=-ig\mathcal{C}[\hat{a}^{\dagger}\hat{a}^{\dagger}\hat{a}\hat{a}]\hat{\rho}+\gamma_{\phi}\mathcal{D}[\hat{n}]\hat{\rho}.\label{eq:dephased-kerr-me}
\end{equation}
$\gamma_{\phi}\mathcal{D}[\hat{n}]$ is the superoperator describing
dephasing. The unitary term is written using the superoperator $\mathcal{C}[\hat{a}^{\dagger}\hat{a}^{\dagger}\hat{a}\hat{a}]$
as defined in (\ref{eq:-212}). Because the superoperators $\gamma_{\phi}\mathcal{D}[\hat{n}]$
and $-ig\mathcal{C}[\hat{a}^{\dagger}\hat{a}^{\dagger}\hat{a}\hat{a}]$
are both diagonal in the number state basis, they commute (shown in
Appendix \ref{app:superoperator-commutation-relations}). The two
stage evolution of Kerr evolution followed by dephasing, explored
in the previous sections, is therefore identical to the simultaneous
dephasing and unitary Kerr-evolution. Formally, we can write\footnote{Alternately, the conclusion formalized in (\ref{eq:-91}) may be reached
by transforming to the interaction picture using $\hat{H}=\hbar g\hat{a}^{\dagger}\hat{a}\hat{a}^{\dagger}\hat{a}$
as the base Hamiltonian, solving for dephasing along and then finally
transforming back to the Schrödinger picture. This procedure is demonstrated
in Appendix \ref{app:operator-picture-dephasing} where it is used
to obtain an alternate form for the solution to a slightly generalized
version of (\ref{eq:dephased-kerr-me}).}
\begin{equation}
\hat{\rho}(t)=e^{-igt\mathcal{C}[\hat{a}^{\dagger}\hat{a}^{\dagger}\hat{a}\hat{a}]+\gamma_{\phi}t\mathcal{D}[\hat{n}]}\hat{\rho}(0)=e^{\gamma_{\phi}t\mathcal{D}[\hat{n}]}e^{-igt\mathcal{C}[\hat{a}^{\dagger}\hat{a}^{\dagger}\hat{a}\hat{a}]}\hat{\rho}(0).\label{eq:-91}
\end{equation}

\begin{figure}[p]
\noindent \begin{centering}
\centerline{\includegraphics[viewport=0bp 40bp 428bp 500bp,clip]{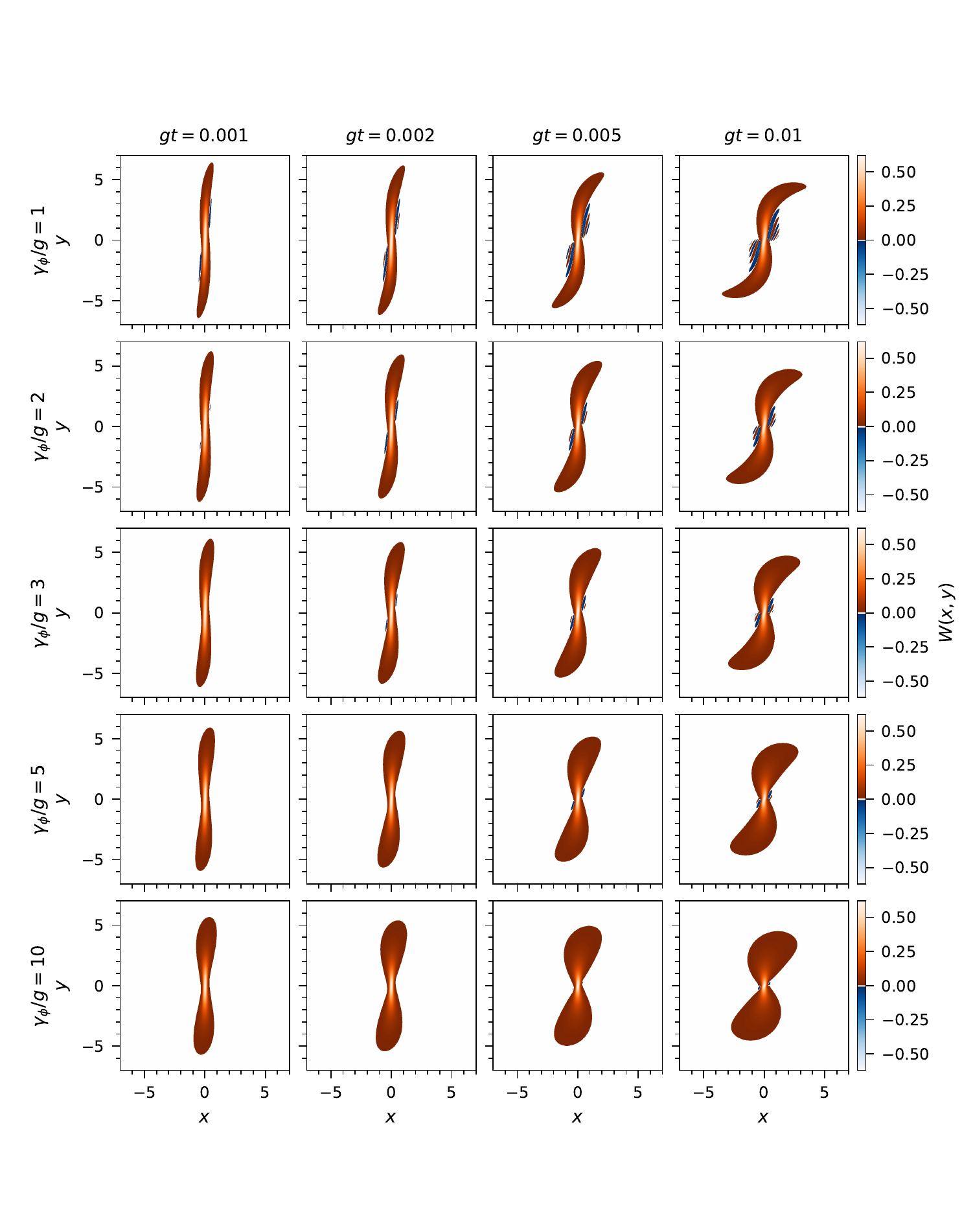}}
\par\end{centering}
\caption[Dephasing Kerr oscillator Wigner function evolution]{\label{fig:scaled-squeezed-gallery-1-3-1}\textbf{Dephasing Kerr
oscillator Wigner function evolution.} The squeezed vacuum state (\ref{eq:-73})
with $\xi=r_{0}=1.5$ is evolved under the dephasing master equation
(\ref{eq:dephased-kerr-me}) with varying dephasing rates $\gamma_{\phi}$.
Dephasing quickly diffuses the parts of the Wigner function far from
the origin while leaving the parts, including negativity, relatively
closer unaffected (compare with the effects of damping in Figure \ref{fig:scaled-squeezed-gallery-1-3}.
The dephasing Kerr oscillator is discussed in Section \ref{subsec:Kerr-Oscillator-with}.}
\end{figure}

Figure \ref{fig:scaled-squeezed-gallery-1-3-1} displays the evolution
of the squeezed state $|\xi=1.5\rangle$ for various values of $\gamma_{\phi}/g$.
The radial dependence of the diffusive effect is clearly visible.
The parts of the fringes far from the origin are quickly washed out
whereas a small region of negativity remains toward the origin, even
given substantial amount of dephasing. Increased dephasing decreases
the size of the negative region but does not, for the parameters shown,
clearly cause the negativity to vanish completely. A small amount
of negativity is seen close to the origin for all states.

\subsubsection{Maximum Negative Volume}

\begin{figure}
\noindent \begin{centering}
\centerline{\includegraphics{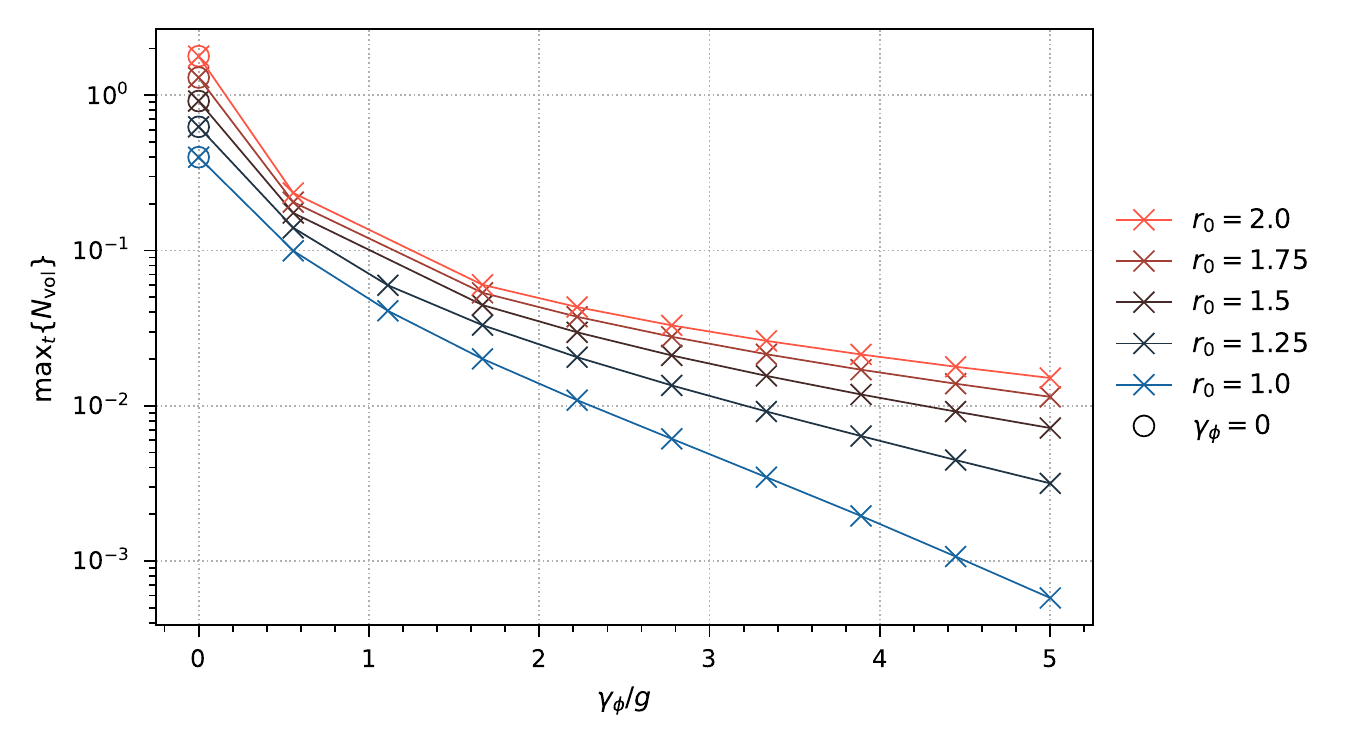}}
\par\end{centering}
\caption[Maximum negative volume as a function of dephasing rate]{\label{fig:maxnegvol-damping-1}\textbf{ Maximum negative volume
as a function of dephasing rate.} Increasing the squeezing parameter
$r_{0}$ leads to an increase in maximum negative volume. Meanwhile,
increasing the dephasing rate $\gamma_{\phi}$ decreases the maximum
negative volume. As $\gamma_{\phi}$ increases, the graphs appear
to diverge suggesting that an increase in squeezing does not leave
the maximum negative volume more vulnerable to dephasing. The evolution
of the Kerr oscillator with dephasing is discussed in Section \ref{subsec:Kerr-Oscillator-with}.}
\end{figure}

We now compute the maximum negative volume as defined in (\ref{eq:-179}).
This is shown in Figure \ref{fig:maxnegvol-damping-1}. For any given
dephasing rate $\gamma_{\phi}$, the maximum negative volume increases
monotonically as a function of the squeezing parameter $r_{0}$. This
shows that an increase in squeezing does not cause the negativity
to be more vulnerable to the effects of dephasing.

In the investigated parameter regime ($r_{0}\in[1,2]$ and $\gamma_{\phi}\in[0,5]$),
the graphs furthermore appear to diverge as $\gamma_{\phi}$ grows.
Hence an increase in dephasing rate appear to affect the states of
lower $r_{0}$ in the strongest way. This suggests that one may increase
the squeezing of the initial state without increasing the vulnerability
of the maximum negative volume toward dephasing. This point is relevant
if one wishes to compensate for a small $g$ by increasing $s$ since
it shows that dephasing is not worsened by this increase in $s$. 

In the investigated parameter regime, no conclusive statement can
be made about the potential existence of an asymptotic behavior with
respect to damping as the squeezing is increased. We can say that
the asymptotic behavior, if it exists, requires a greater amount of
squeezing to be visible than is the case for damping (compare Figures
\ref{fig:-Maximum-negative} and \ref{fig:maxnegvol-damping-1}).

\section{Decoherence Effects in Combination \label{sec:kerr-dynamics-combined-decoherence}}

We conclude this chapter by considering the Kerr oscillator with a
combination of energy damping and dephasing. This is the most general
system to be considered in this thesis. 

\subsection{Dephasing and Damping Compared\label{subsec:Dephasing-and-Damping}}

\begin{figure}
\noindent \begin{centering}
\makebox[0pt][c]{\mbox{\includegraphics{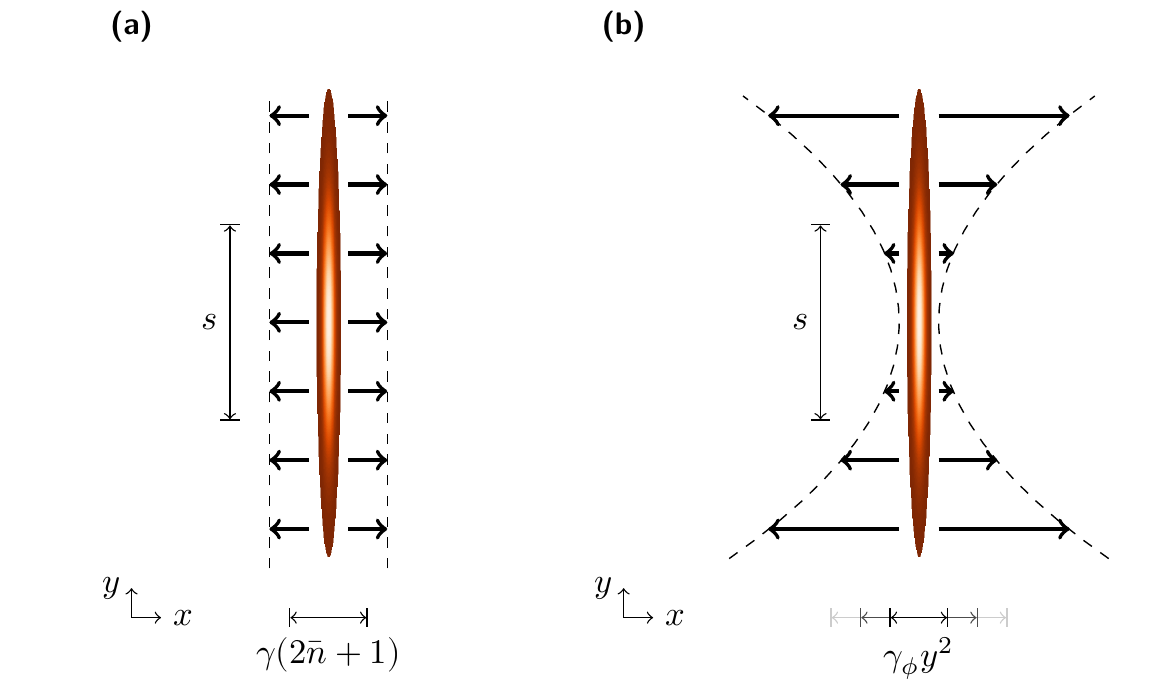}\hskip2em}}
\par\end{centering}
\noindent \centering{}\caption[Illustration of differences between damping and dephasing]{\label{fig:Illustration-of-differences}\textbf{Illustration of differences
between damping and dephasing.} The arrows are proportional to the
diffusion in the direction of the $x$-coordinate. (a) shows damping
and (b) shows dephasing. As discussed in Section \ref{subsec:Dephasing-and-Damping},
the diffusive effect of damping shown in (a) is constant and proportional
to $\gamma(2\bar{n}+1$). The diffusive effect of dephasing scales
with $\gamma_{\phi}y^{2}$. Conceptually, this means that the effect
of dephasing is large far from the origin and vanishes near it.}
\end{figure}

Before considering the combined effects of damping and dephasing however,
we remark on the differences between these effects. They are most
easily compared if the relevant equations of motion for the Wigner
function are expressed in the same coordinate systems. We therefore
rewrite the equation for damping (\ref{eq:damping-pde-xy}) in polar
coordinates, yielding (see also Appendix \ref{app:derivations-of-c-eqs})
\begin{equation}
\begin{aligned}\partial_{t}W(r,\phi,t)= & \frac{\gamma}{4}\left(\bar{n}+\frac{1}{2}\right)\underbrace{\left(\partial_{r}^{2}+\frac{1}{r}\partial_{r}+\frac{1}{r^{2}}\partial_{\phi}^{2}\right)}_{\nabla^{2}}W(r,\phi,t)\\
 & +\frac{\gamma}{2}r\partial_{r}W(r,\phi,t)+\gamma W(r,\phi,t).
\end{aligned}
\end{equation}
Dephasing and damping both describe a diffusive processes which can
be seen from the presence of second order derivatives on the right
hand side. The dephasing equation however contains no derivatives
with respect to the radial coordinate $r$. This means that dephasing,
in contrast to damping, only causes a flow of the Wigner density in
the angular direction. For this reason, there exist states for which
the Wigner function remains negative under any amount of evolution
under dephasing.\footnote{The number states $|n\rangle$ are a trivial example since they remain
constant under dephasing. This is easily concluded either from the
operator formalism (see (\ref{eq:dephasing-matrix-element})) or by
substitution of (\ref{eq:-83}) into (\ref{eq:dephasing-pde-polar}).
No such states exist for damping, since its effect is always decay
toward the steady state solution thermal state (see Section \ref{subsec:Decay-of-Negativity}).} 

It is also fruitful to note the difference between the terms causing
angular diffusion. Angular diffusion is caused by the differential
operator $\partial_{\phi}^{2}$. In the dephasing equation (\ref{eq:dephasing-pde-polar})
it has the constant coefficient $\gamma_{\phi}/2$. In the damping
equation the $\partial_{\phi}^{2}$-term has the coefficient $\gamma\left(2\bar{n}+1\right)/8r^{2}$.
As noted in Section \ref{sec:kerr-state-dephasing}, the operator
$\partial_{\phi}^{2}$ contains an implicit scaling of $r^{2}$. This
means that the diffusive effect of damping is constant everywhere
in phase space (the drift towards the origin is not however). For
the same reason, the diffusive effect of dephasing is proportional
to $y^{2}$ and therefore vanishes toward the origin. This is illustrated
in Figure \ref{fig:Illustration-of-differences}. In the context of
negativity, this means that negative regions closer to the origin
are relatively more vulnerable to damping\footnote{This view is consistent with the main result of Section \ref{sec:damping},
where it was shown that an increase in squeezing (which generally
moves features of the Wigner function farther from the origin) reduces
the adverse effect of damping.} whereas negative regions far from the origin are relatively more
vulnerable to dephasing. 

This complementary character of dephasing and damping is clearly visible
when comparing the evolution of the Wigner function. Under damping
(Figure \ref{fig:scaled-squeezed-gallery-1-3}), the diffusive effect
appears homogeneous throughout phase space. It appears to be the case
that the regions which would have been at the center of negativity
for the undamped oscillator maintain their negativity the longest
under damping. In contrast, the diffusive effect of dephasing (see
Figure \ref{fig:scaled-squeezed-gallery-1-3-1}) is clearly stronger
further from the origin. The result is that the negative regions become
concentrated close to the origin.

This may offer an explanation for why the large squeezing approximation
is more effective for damping than dephasing (as observed in Section
\ref{subsec:Kerr-Oscillator-with}): The decay of negativity under
dephasing happens only through angular diffusion. For positive and
negative regions close to the origin to mix thus eliminating the negativity
requires diffusion over an angle which is a appreciable fraction of
$\pi$. In the corresponding time, the diffusion farther from the
origin, which is described by the same angle, will have caused a significant
flow of Wigner density away from the $y$-axis and toward the $x$-axis
causing a deterioration in the large squeezing approximation from
considerations similar to those of Section \ref{subsec:Validity-of-Large}.

\subsection{Equation of Motion\label{subsec:Equation-of-Motion}}

We are now ready to construct the general equation of motion for the
combined effects of dephasing and damping. The relevant master equation
is obtained by including all terms of (\ref{eq:general-master-equation})
and inserting the Kerr Hamiltonian (\ref{eq:kerr-hamiltonian}) as
$\hat{H}$:

\begin{equation}
\dot{\hat{\rho}}=-ig\left[\hat{a}^{\dagger}\hat{a}^{\dagger}\hat{a}\hat{a},\hat{\rho}\right]+\gamma\left(\bar{n}+1\right)\mathcal{D}[\hat{a}]\hat{\rho}+\gamma\bar{n}\mathcal{D}[\hat{a}^{\dagger}]\hat{\rho}+\gamma_{\phi}\mathcal{D}[\hat{n}]\hat{\rho}.\label{eq:decoh-kerr-me}
\end{equation}
We can derive the corresponding equation of motion for the Wigner
function simply by combining (\ref{eq:-137}) and (\ref{eq:dephasing-pde-xy})
to obtain
\begin{equation}
\partial_{t}W(x,y,t)=\begin{aligned}[t] & 2g\left(x^{2}+y^{2}-1\right)\left(-y\partial_{x}+x\partial_{y}\right)W(x,y,t)\\
 & -\frac{g}{8}\left(-y\partial_{x}+x\partial_{y}\right)\left(\partial_{x}^{2}+\partial_{y}^{2}\right)W(x,y,t)\\
 & +\frac{\gamma}{4}\left(\bar{n}+\frac{1}{2}\right)\left(\partial_{x}^{2}+\partial_{y}^{2}\right)W(x,y,t)\\
 & +\frac{\gamma}{2}\partial_{x}\left(xW(x,y,t)\right)+\frac{\gamma}{2}\partial_{y}\left(yW(x,y,t)\right)\\
 & +\frac{\gamma_{\phi}}{2}\left(-y\partial_{x}+x\partial_{y}\right)\left(-y\partial_{x}+x\partial_{y}\right)W(x,y,t).
\end{aligned}
\label{eq:decoh-pde}
\end{equation}
The effects of the individual terms have already been described; see
the previous section as well as Sections \ref{subsec:Wigner-Function-Equation},
\ref{sec:damping} and \ref{sec:phase-decoherence}.

\subsection{Rescaled Coordinates and Large Squeezing Approximation\label{subsec:Rescaled-Coordinates-and}}

We introduce once again rescaled coordinates $(\tilde{x},\tilde{y})$
as given in (\ref{eq:-30-2}) with corresponding differential operators
as given in (\ref{eq:-27}). The initial state is now given by (\ref{eq:-30})
while (\ref{eq:decoh-pde}) is transformed to 
\begin{equation}
\partial_{t}\tilde{W}(\tilde{x},\tilde{y},t)=\begin{aligned}[t] & 2g\left(-\tilde{x}^{2}\tilde{y}\partial_{\tilde{x}}-s^{4}\tilde{y}^{3}\partial_{\tilde{x}}+\frac{1}{s^{4}}\tilde{x}^{3}\partial_{\tilde{y}}+\tilde{x}\tilde{y}^{2}\partial_{\tilde{y}}\right)\tilde{W}(\tilde{x},\tilde{y},t)\\
 & -2g\left(-s^{2}\tilde{y}\partial_{\tilde{x}}+\frac{1}{s^{2}}\tilde{x}\partial_{\tilde{y}}\right)\tilde{W}(\tilde{x},\tilde{y},t)\\
 & -\frac{g}{8}\left(-s^{4}\tilde{y}\partial_{\tilde{x}}^{3}+\frac{1}{s^{4}}\tilde{x}\partial_{\tilde{y}}^{3}+\tilde{x}\partial_{\tilde{y}}\partial_{\tilde{x}}^{2}-\tilde{y}\partial_{\tilde{x}}\partial_{\tilde{y}}^{2}\right)\tilde{W}(\tilde{x},\tilde{y},t)\\
 & +\frac{\gamma s^{2}}{4}\left(\bar{n}+\frac{1}{2}\right)\partial_{\tilde{x}}^{2}\tilde{W}(\tilde{x},\tilde{y},t)+\frac{\gamma}{4s^{2}}\left(\bar{n}+\frac{1}{2}\right)\partial_{\tilde{y}}^{2}\tilde{W}(\tilde{x},\tilde{y},t)\\
 & +\frac{\gamma}{2}\partial_{\tilde{x}}\left(\tilde{x}\tilde{W}(\tilde{x},\tilde{y},t)\right)+\frac{\gamma}{2}\partial_{\tilde{y}}\left(\tilde{y}\tilde{W}(\tilde{x},\tilde{y},t)\right).\\
 & +\frac{\gamma_{\phi}}{2}\left(s^{4}\tilde{y}^{2}\partial_{\tilde{x}}^{2}+s^{-4}\tilde{x}^{2}\partial_{\tilde{y}}^{2}-2\tilde{x}\tilde{y}\partial_{\tilde{x}}\partial_{\tilde{y}}-\tilde{x}\partial_{\tilde{x}}-\tilde{y}\partial_{\tilde{y}}\right)\tilde{W}(\tilde{x},\tilde{y},t).
\end{aligned}
\end{equation}
We extract the most significant terms in the limit of large squeezing:
\begin{equation}
\partial_{t}\tilde{W}(\tilde{x},\tilde{y},t)=\begin{aligned}[t] & -2gs^{4}\tilde{y}^{3}\partial_{\tilde{x}}\tilde{W}(\tilde{x},\tilde{y},t)+\frac{g}{8}s^{4}\tilde{y}\partial_{\tilde{x}}^{3}\tilde{W}(\tilde{x},\tilde{y},t)\\
 & +\frac{\gamma s^{2}}{4}\left(\bar{n}+\frac{1}{2}\right)\partial_{\tilde{x}}^{2}\tilde{W}(\tilde{x},\tilde{y},t)+\frac{\gamma_{\phi}}{2}s^{4}\tilde{y}^{2}\partial_{\tilde{x}}^{2}\tilde{W}(\tilde{x},\tilde{y},t).
\end{aligned}
\label{eq:-136-1}
\end{equation}
The right hand side above is the sum of the right hand sides of (\ref{eq:-136})
and (\ref{eq:dephasing-pde-approx}). This again reduces the problem
to a two-dimensional one. Thus define the function $\tilde{u}_{\tilde{y}}(\mu,\tilde{\tau})$
by
\begin{equation}
\tilde{W}(\tilde{x},\tilde{y},t)=\tilde{u}_{\tilde{y}}(\tilde{x}-2gs^{4}\tilde{y}^{3}t,gs^{4}\tilde{y}t/8).\label{eq:-64-1}
\end{equation}
From (\ref{eq:-136-1}), the equation of motion for $u_{\tilde{y}}(\mu,\tilde{\tau})$
is given by
\begin{subequations}
\label{eq:-138}
\begin{equation}
\partial_{\tilde{\tau}}\tilde{u}_{\tilde{y}}(\mu,\tilde{\tau})=\partial_{\mu}^{3}\tilde{u}_{\tilde{y}}(\mu,\tilde{\tau})+\beta_{\tilde{y}}\partial_{\mu}^{2}\tilde{u}_{\tilde{y}}(\mu,\tilde{\tau})\label{eq:-139}
\end{equation}
with
\begin{equation}
\beta_{\tilde{y}}=\left(\frac{\gamma(2\bar{n}+1)}{gs^{2}\tilde{y}}+\frac{4\gamma_{\phi}\tilde{y}}{g}\right).\label{eq:-218}
\end{equation}
\end{subequations}
(\ref{eq:-138}) describes a third-order dispersive process with diffusion.
The strength of the diffusion varies with $\tilde{y}$ and is described
by $\beta_{\tilde{y}}$. $\beta_{\tilde{y}}$ is a clear expression
of the differences between damping and dephasing discussed in Section
\ref{subsec:Dephasing-and-Damping}.\footnote{On the $\tilde{y}$-axis, the diffusion arising from damping is homogeneous
throughout phase space whereas the diffusion from dephasing increases
with $|\tilde{y}|^{2}$. The damping contribution to $\beta_{\tilde{y}}$
in (\ref{eq:-218}) appears to vary in proportion to $\tilde{y}^{-1}$,
however this is because (\ref{eq:-139}) is expressed in terms of
the rescaled time $\tilde{\tau}$ which carries an implicit factor
of $\tilde{y}$ (see (\ref{eq:-64-1})).} Additionally, increasing squeezing reduces the effects of damping
due to the factor of $s^{2}$ in the denominator. This is in contrast
to the dephasing term of $\beta_{\tilde{y}}$ which is invariant with
respect to the squeezing $s$.

One may solve for $\tilde{u}_{\tilde{y}}$ using a Fourier series
as done previously. The solution for $\tilde{W}(\tilde{x},\tilde{y},t)$
is then obtained from this solution and (\ref{eq:-64-1}) as
\begin{subequations}
\label{eq:-160-1}
\begin{align}
\tilde{W}(\tilde{x},\tilde{y},t) & =\frac{1}{\sqrt{2\pi}}\int_{-\infty}^{\infty}dk\,h_{\tilde{y}}(k)\,e^{i(k\tilde{x}-2kgts^{4}\tilde{y}^{3}-gts^{4}k^{3}\tilde{y}/8)}e^{-\gamma(2\bar{n}+1)s^{2}t/8}e^{-\gamma_{\phi}s^{4}\tilde{y}^{2}t/2}\label{eq:-102-1}
\end{align}
where the Fourier transform of the initial state is given by
\begin{equation}
h_{\tilde{y}}(k)=\frac{1}{\sqrt{2\pi}}\int_{-\infty}^{\infty}d\tilde{x}\,\tilde{W}(\tilde{x},\tilde{y},0)\,e^{-ik\tilde{x}}.
\end{equation}
\end{subequations}
This reveals the scaling of diffusion as $\tilde{y}^{2}$ and unit
for dephasing and damping respectively.

\subsection{Maximum Negative Volume and Peak\label{subsec:Maximum-Negative-Volume}}

\begin{figure}[p]
\noindent \begin{centering}
\subfloat[Maximum negative volume $\max_{t}N_{\mathrm{vol}}$]{\noindent \begin{centering}
\includegraphics{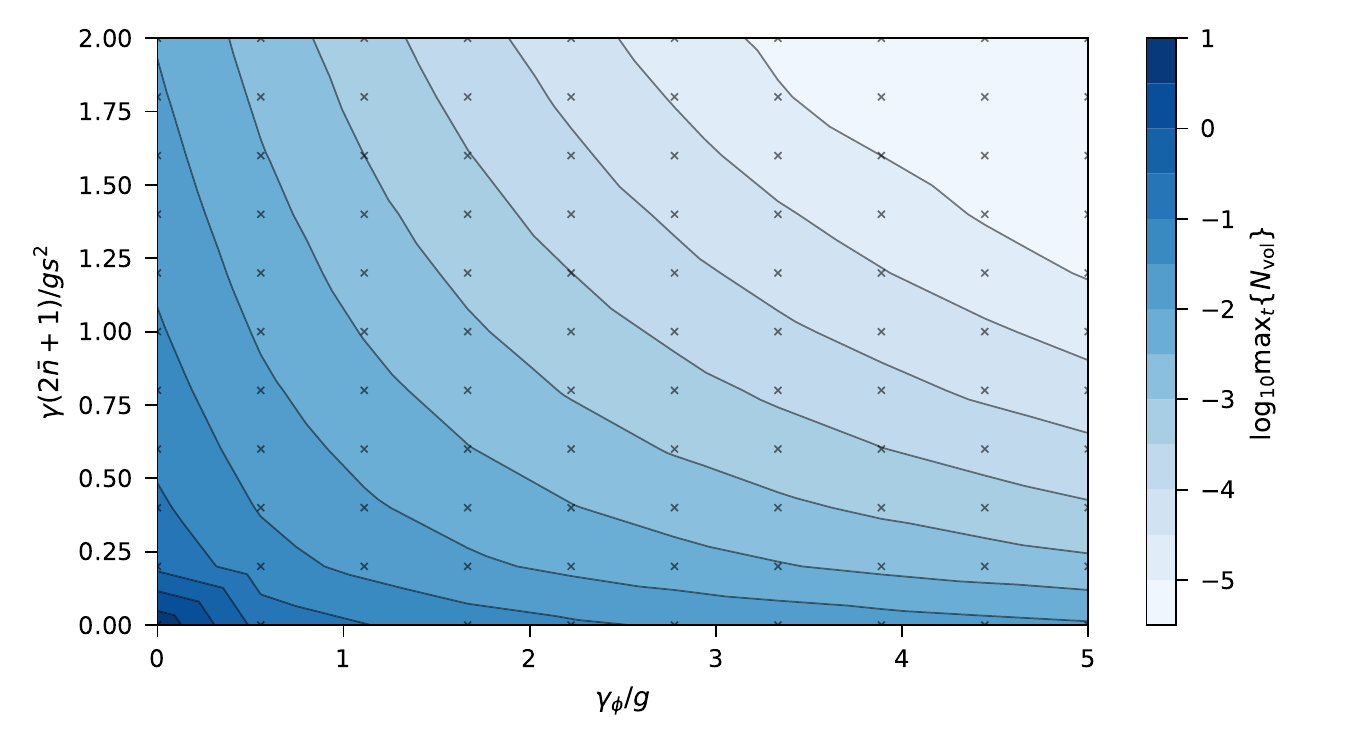}
\par\end{centering}
}
\par\end{centering}
\noindent \begin{centering}
\subfloat[Maximum negative peak $\max_{t}N_{\mathrm{peak}}$]{\noindent \begin{centering}
\includegraphics{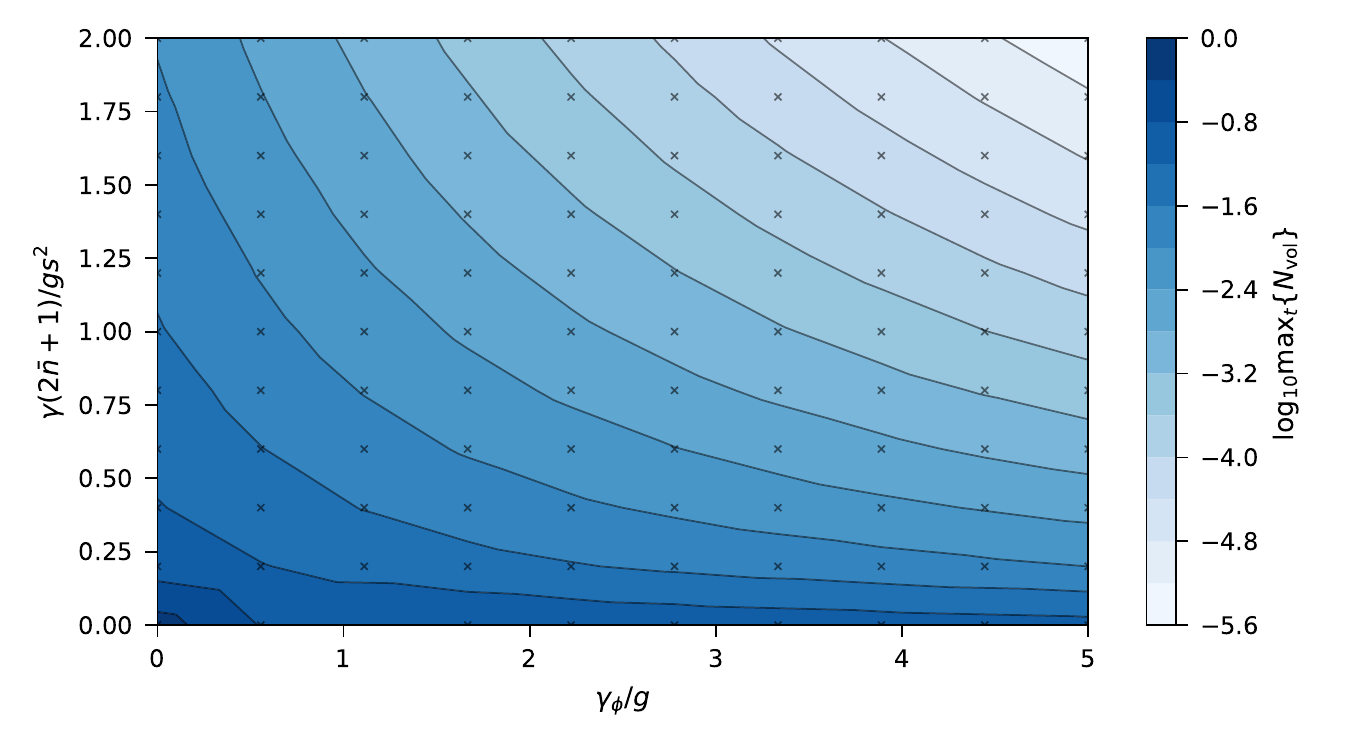}
\par\end{centering}
}
\par\end{centering}
\caption[Maximum negativity for combined decoherence]{\label{fig:contour-plot}\textbf{Maximum negativity for combined
decoherence.} Logarithmic plot of the maximum negative volume $\max_{t}N_{\mathrm{vol}}$
and maximum negative peak $\max_{t}N_{\mathrm{peak}}$. The plotted
quantities were obtained from the evolution of the squeezed vacuum
state (\ref{eq:-225}) with $r_{0}=1.75\Leftrightarrow s=5.75$. Each
simulation is marked by “$\times$”. The effective damping rate
$\gamma(2\bar{n}+1)s^{2}$ was set with $\bar{n}=1000$. }
\end{figure}
\begin{figure}[p]
\noindent \begin{centering}
\subfloat[Maximum negative volume $\max_{t}N_{\mathrm{vol}}$]{\noindent \begin{centering}
\includegraphics{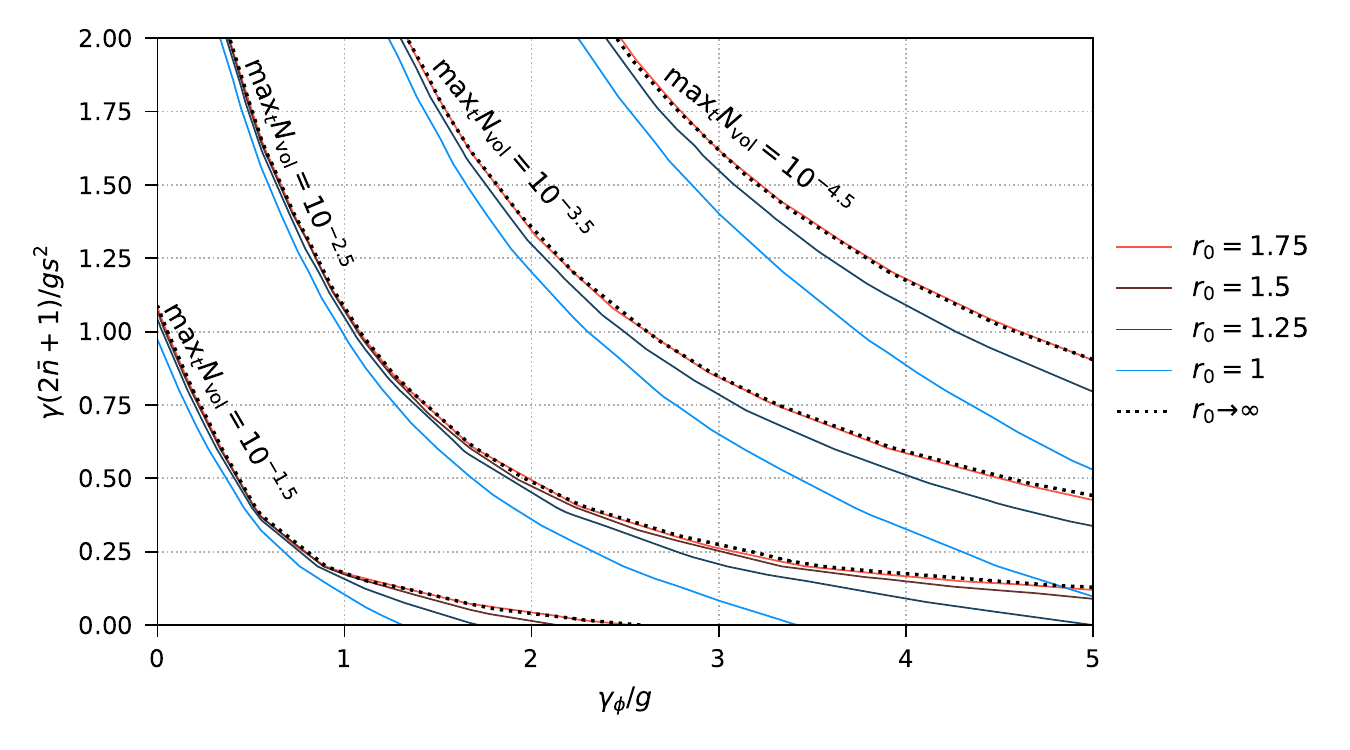}
\par\end{centering}
}
\par\end{centering}
\noindent \begin{centering}
\subfloat[Maximum negative peak $\max_{t}N_{\mathrm{peak}}$]{\noindent \begin{centering}
\includegraphics{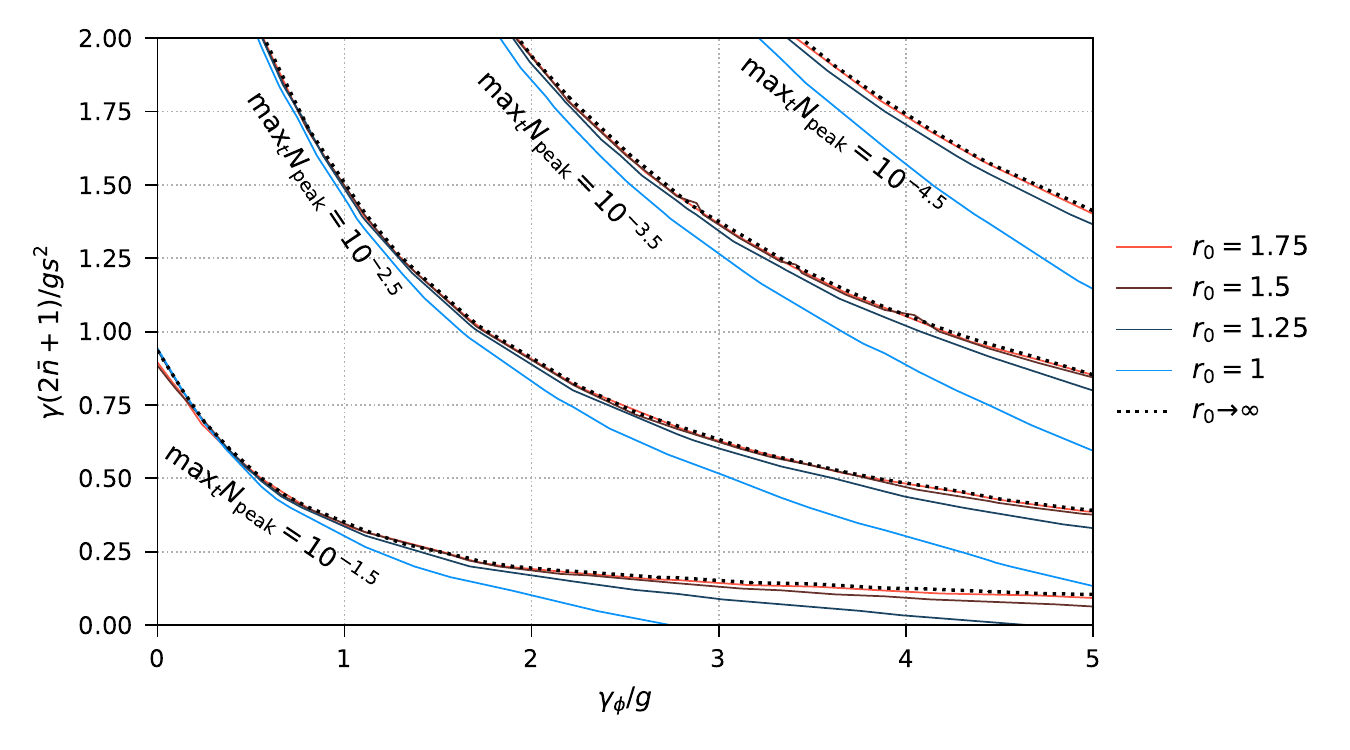}
\par\end{centering}
}
\par\end{centering}
\caption[Maximum negativity contours for combined decoherence]{\label{fig:contour-plot-1}\textbf{Maximum negativity contours for
combined decoherence. }Contours of maximum negative volume $\max_{t}N_{\mathrm{vol}}$
and maximum negative peak $\max_{t}N_{\mathrm{peak}}$ as a function
of decoherence rates. The effective damping rate $\gamma(2\bar{n}+1)s^{2}$
was set with $\bar{n}=1000$. Figure \ref{fig:contour-plot} shows
the source of the contour for $r_{0}=1.75$. For low dephasing, the
maximum negativity as a function of effective damping rate appear
independent of the squeezing $s$. For larger dephasing rates, the
negativity increases with $s$. The data is discussed in Section \ref{subsec:Maximum-Negative-Volume}.
Thick dotted lines ($r_{0}\to\infty$) show the contours obtained
from (\ref{eq:-160-1}). The contour at $10^{-4.5}$ is not shown
for $r_{0}=1.5$.}
\end{figure}

To quantify the interplay between damping and dephasing, we study
the maximum negative volume and peak for various values of the parameters
$r_{0}$, $\gamma_{\phi}$ and $\gamma(2\bar{n}+1)$. Figure \ref{fig:contour-plot}
shows the negativity in the case $r_{0}=1.75$. The parameters that
take values of frequency are normalized to the value of $g$. It is
seen that an increase in either $\gamma_{\phi}/g$ or $\gamma(2\bar{n}+1)/gs^{2}$
always cause a decrease in the maximum negativity irrespective of
the value of the other decoherence parameter. We also note that the
adverse effects on the maximum negativity combine in a super-linear
way and that one can therefore not describe the maximum negative volume
as the sum of the two effects, i.e. one cannot define functions $f_{\gamma}$
and $f_{\gamma_{\phi}}$ to make the following an equality:
\begin{equation}
\max_{t}N_{\mathrm{vol}}\neq f_{\gamma}\left(\gamma(2\bar{n}+1)\right)+f_{\gamma_{\phi}}(\gamma_{\phi}),\label{eq:-227}
\end{equation}
as equality in (\ref{eq:-227}) would have manifested itself as straight
line contours in Figure \ref{fig:contour-plot}. The same conclusion
is reached for $\max_{t}N_{\mathrm{peak}}$.

To investigate the dependence on the squeezing $s$, we superimpose
the contours of the maximum negativity with $r_{0}$ taking the values
$1$, $1.25$, $1.5$ and $1.75$ (of which Figure \ref{fig:contour-plot}
shows the case $r_{0}=1.75$) in Figure \ref{fig:contour-plot-1}.
This indicates the regions where the large squeezing approximation
of Section \ref{subsec:Rescaled-Coordinates-and} is valid. We see
that the contours assume common values for larger values of $r_{0}$
and effective damping $\gamma(2\bar{n}+1)/gs^{2}$ as expected. We
also show the contours obtained from the Fourier transformed solution
(\ref{eq:-160-1}) to the equation after the large squeezing approximation.
These match well with the contours of $r_{0}=1.75$ suggesting that
it may be possible to find an asymptotic behavior for the entire graph
given sufficient squeezing.

We also note that the contours appear to diverge as $\gamma_{\phi}/g$
is increased and $\gamma(2\bar{n}+1)/gs^{2}$ decreased. This means
that a large squeezing is required for the large squeezing approximation
to work with dephasing than with damping. This is a similar observation
to that made for the maximum negative volume as a function of $\gamma_{\phi}/g$
in Figure \ref{fig:maxnegvol-damping-1}. Assuming that increased
$\gamma_{\phi}/g$ can never lead to an increase in negativity (which
has not been proven here, but seems like a reasonable assumption given
that dephasing can be described as angular diffusion and commutes
with both effects of the unitary term and damping, see Appendix \ref{app:superoperator-commutation-relations}),
the conclusion is that an increase in squeezing can compensate for
a small ratio $\gamma(2\bar{n}+1)/g$ when measuring the maximum negativity
without any adverse effects even for nonzero $\gamma_{\phi}/g$. 

\section{Summary of Scalings for the Squeezed Vacuum State}

We end the chapter by summarizing the discovered scalings for the
evolution of negativity for the squeezed vacuum state $|\xi=r_{0}\rangle$.
These are listed in Table \ref{tab:Summary-of-scaling}. In Chapter
\ref{chap:nonlinear-oscillators} we considered the unitary dynamics
of the Kerr oscillator. For large squeezing only the most significant
terms were kept, rendering the approximated equation of motion in
a form in which $s$ and the Kerr coefficient $g$ entered only in
the combination $gs^{4}$ (displayed in Table \ref{tab:Summary-of-scaling}).
Since the frequency $gs^{4}$ is now the only parameter, the time
$t$ was be rescaled to $gts^{4}$ (also displayed in Table \ref{tab:Summary-of-scaling})
such that both the initial state and the equation of motion are free
of parameters. In this chapter, the appropriate scaling of time was
then established for the cases of isolated damping or dephasing (significant
terms in the equation of motion as well as appropriate scalings of
time listed in Table \ref{tab:Summary-of-scaling}). The respective
scaled times were found to describe the decay of negativity as a result
of both isolated damping and dephasing. The case of simultaneous Kerr
evolution and high temperature damping of effective rate $\gamma(2\bar{n}+1)$
were investigated and the maximum negativity was found to be well
described by the ratio of the scaled times $gs^{2}/\gamma(2\bar{n}+1)$.
In case of the dephasing however Kerr oscillator, the scaled time
did not by itself account for the relation between maximum negativity,
squeezing and dephasing rate.

\begin{table}
\centering\makebox[0pt][c]{\renewcommand{\arraystretch}{1.5}%
\begin{minipage}[t]{1.2\textwidth}%
\noindent \begin{center}
\begin{longtable}[c]{|c|c|c|c|}
\hline 
Effect & Scaled time & Significant terms & Relevant sections and figures\tabularnewline
\hline 
\hline 
\multirow{2}{*}{Kerr oscillator} & $tgs^{4}$ & $gs^{4}\left(-2\tilde{y}^{3}\partial_{\tilde{x}}+\frac{1}{8}\tilde{y}\partial_{\tilde{x}}^{3}\right)\tilde{W}$ & Sections \ref{subsec:Evolution-over-Short}--\ref{subsec:Evolution-of-Negativity-1}.\hfill{}\tabularnewline
 & \footnotesize from (\ref{eq:-64}) & \footnotesize from (\ref{eq:-31-2-1}) & Figures \ref{fig:scaled-squeezed-gallery-1} and \ref{fig:-2}.\hfill{}\tabularnewline
\hline 
\multirow{2}{*}{Damping} & $t\gamma(2\bar{n}+1)s^{2}$ & $\frac{1}{8}\gamma s^{2}\left(2\bar{n}+1\right)\partial_{\tilde{x}}^{2}\tilde{W}$ & Sections \ref{sec:kerr-state-damping} and \ref{subsec:Damped-Kerr-Evolution}.\hfill{}\tabularnewline
 & \footnotesize from (\ref{eq:-104}) & \footnotesize from (\ref{eq:-90}) & Figures \ref{fig:Decay-of-squeezed} and \ref{fig:-Maximum-negative}.\hfill{}\tabularnewline
\hline 
\multirow{2}{*}{Dephasing} & $t\gamma_{\phi}s^{4}$ & $\frac{1}{2}\gamma_{\phi}s^{4}\tilde{y}^{2}\partial_{\tilde{x}}^{2}\tilde{W}$ & Sections \ref{sec:kerr-state-dephasing} and \ref{subsec:Kerr-Oscillator-with}.\hfill{}\tabularnewline
 & \footnotesize from (\ref{eq:-104-1}) & \footnotesize from (\ref{eq:dephasing-pde-scaled}) & Figures \ref{fig:Decay-of-squeezed-1} and \ref{fig:maxnegvol-damping-1}.\hfill{}\tabularnewline
\hline 
\end{longtable}
\par\end{center}%
\end{minipage}}\setcounter{table}{1}

\caption[Summary of scaling relations for squeezed vacuum]{\label{tab:Summary-of-scaling}\textbf{Summary of scaling relations
for squeezed vacuum. }Table summarizing the main results for the short
time evolution of the squeezed vacuum state. The unitary Kerr evolution
as well as the decoherence effects of damping and dephasing are listed.
When plotting the negativity for short times as a function of the
scaled time, the graphs exhibit asymptotic behavior as $r_{0}$ is
increased. We can understand this behavior by rescaling the Wigner
function as described in Section \ref{subsec:intro-rescaled-coords}
such that the most significant terms in its equation of motion (also
tabulated) will appear with the squeezing explicit. The final column
references passages of particular relevance from this thesis.}
\end{table}

\cleardoublepage\phantomsection\addcontentsline{toc}{chapter}{Conclusion}

\chapter*{Conclusion}

In this thesis, we studied the evolution of a squeezed vacuum state
of the Kerr oscillator. We focus specifically on the formation of
negativity of the Wigner function. To gain a better understanding
of the system, we studied first the unitary dynamics. We explored
the analytical operator solution which describes a periodic evolution
in the state. Motivated by the damping of current experimental systems,
we then studied the evolution of the squeezed vacuum state over short
time intervals. Various tools were employed to gain a geometrical
understanding of the state evolution for short times. We then used
the Fokker-Planck-like equation for the evolution of the Wigner function
to develop a solution in the limit of large negativity. This solution
was then used to show that the nonlinearity $g$ scales as $ge^{4r_{0}}$
where $r_{0}$ denotes the squeezing parameter. We also considered
the open system of a damped and dephasing Kerr oscillator. Using the
large-squeezing solution for the unitary case, it was found that the
effective ratio between nonlinearity and damping was improved with
squeezing by a factor of $e^{2r_{0}}$. Combining the effects of dephasing
and damping, it was shown that the enhancement of the effective ratio
between nonlinearity and damping through squeezing does not have an
exacerbating effect on the decoherence arising from dephasing. Table
\ref{tab:Summary-of-scaling} lists relevant results for scaling.

Several avenues of further exploration have presented themselves during
this work. Building directly upon the results here, it appears plausible
from Chapter \ref{chap:nonlinear-oscillators} that the results of
Chapter \ref{chap:coupling-to-the-environment} can be straightforwardly
extended to include squeezed thermal states. On a more technical level,
work could be done to quantify the accuracy of the large squeezing
approximation for any Gaussian state beyond what has been done here.
Of course, the ability to strongly squeeze the vacuum state is essential
to the direct applicability of the results presented here and one
should thus survey the experimental results for this before proceeding.
Optical squeezing of up to $15$ dB has been observed experimentally
\cite{Vahlbruch_Detection15DB_2016}.

It has also been suggested to apply Kerr evolution to a coherent initial
state to form negativity in the Wigner function \cite{Stobinska_WignerFunctionEvolution_2008}.
This is experimentally attractive since coherent states are far easier
to prepare experimentally. Since the negativity for the coherent state
only forms near the coherent displacement amplitude in phase space
however (cf. Figure \ref{fig:evo-gallery-coherent}), the coherent
state may be more vulnerable to the effects of dephasing. This suggests
a comparison between the evolution in negativity for coherent and
squeezed states. The methods demonstrated for the squeezed state in
this thesis may find use in developing a similar understanding for
the formation of negativity in the evolution of a coherent state.
It may also be possible to show that limited squeezing of a coherent
state will amplify the nonlinear effects of the Kerr oscillator.

Experimental measurement of Wigner function negativity requires the
development of mechanical state tomography \cite{Vanner_OptomechanicalQuantumState_2015}.
It may be shown that squeezing a state does not hamper the ability
to reconstruct the state Wigner function from measurements of its
marginal distributions \cite{Fresta_ElementaryTestNonclassicality_2015}.
As part of this, the quadrature squeezing could also be introduced
as simultaneous with the Kerr evolution rather than as a prior step.
One could also explore alternate effects such as the use of feedback
to enhance the nonlinear effects or stabilize a state with negative
Wigner function \cite{Koppenhofer_HeraldedDissipativePreparation_2019}.

In summary, we have shown that one may compensate for a weak nonlinearity
through squeezing. This opens the door to witness a negative Wigner
function in a macroscopic mechanical system even if nonlinearity is
a scarce resource, provided that the ability to squeeze the system
state significantly exists. While the work in this thesis has been
motivated by nanomechanical systems, most of it is general to any
quantum system described by a Kerr oscillator where damping dominates
and which allows for quadrature squeezing.

\noindent \cleardoublepage\phantomsection\addcontentsline{toc}{chapter}{Bibliography}

\noindent \printbibliography
\cleardoublepage\phantomsection\addcontentsline{toc}{chapter}{List of Figures}

\noindent \listoffigures
\cleardoublepage\phantomsection\addcontentsline{toc}{chapter}{List of Tables}\listoftables
\newgeometry{left=3cm,right=3cm,top=3cm,bottom=3cm}

\noindent \cleardoublepage\phantomsection
\addcontentsline{toc}{chapter}{Appendices}

\appendix

\chapter{Interaction Picture and Rotating Wave Approximation\label{chap:Interaction-Picture-and}}

Consider the Hamiltonian
\begin{equation}
\hat{H}=\frac{\hat{p}^{2}}{2m}+\frac{1}{2}m\omega^{2}\hat{q}^{2}+\frac{\beta}{4}\hat{q}^{4}.\label{eq:-36}
\end{equation}
This describes the quantum mechanical Duffing oscillator \cite{Babourina-Brooks_QuantumNoiseNanomechanical_2008}
with mass $m$ and base frequency $\omega$. $\beta$ is the Duffing
parameter and has dimensions of mass times the square of ratio of
angular frequency to length, i.e.
\begin{equation}
[\beta]=\frac{(\text{energy})}{(\text{length})^{4}}=(\text{mass})\cdot\frac{(\text{angular frequency})^{2}}{(\text{length})^{2}}.
\end{equation}
The Duffing oscillator describes an oscillator where the (angular)
frequency is dependent on the displacement. $\sqrt{\beta}$ relates
the change in displacement to the corresponding change in (angular)
frequency. With the conventions used here, $\beta$ is additionally
scaled with the mass of the oscillator.

From (\ref{eq:-36}), we may derive Langevin equations for the system
operators $\hat{q}$ and $\hat{p}$ to be
\begin{equation}
\dot{\hat{q}}=\frac{\hat{p}}{m}\label{eq:-68}
\end{equation}
and
\begin{equation}
\dot{\hat{p}}=-m\omega^{2}\hat{q}-m\gamma\dot{\hat{q}}-\beta\hat{q}^{3}.\label{eq:-69}
\end{equation}
Equations (\ref{eq:-68}) and (\ref{eq:-69}), or equivalently
\begin{equation}
\ddot{\hat{q}}=-\omega^{2}\hat{q}-\gamma\dot{\hat{q}}-\frac{\beta}{m}\hat{q}^{3},
\end{equation}
are often used to introduce the dynamics of the Duffing oscillator
in place of (\ref{eq:-36}).

(\ref{eq:-36}) also allows us to write the Hamiltonian as
\begin{equation}
\hat{H}=\hbar\omega\left(\hat{a}^{\dagger}\hat{a}+\frac{1}{2}\right)+\frac{\hbar^{2}\beta}{16m^{2}\omega^{2}}\left(\hat{a}+\hat{a}^{\dagger}\right)^{4}.
\end{equation}
We wish to express $\hat{H}$ in the interaction picture with 
\begin{equation}
\hat{H}_{0}=\hbar\omega_{0}\hat{a}^{\dagger}\hat{a}\label{eq:-67}
\end{equation}
 as the base Hamiltonian. We can then transition to the interaction
picture by simply substituting \cite{Sakurai_ModernQuantumMechanics_2011a}:
\begin{align}
\hat{a} & \to\hat{a}_{I}=e^{i\hat{H}_{0}t}\hat{a}e^{-i\hat{H}_{0}t}=\hat{a}e^{-i\omega_{0}t}\,,\\
\hat{a}^{\dagger} & \to\hat{a}_{I}^{\dagger}=e^{i\hat{H}_{0}t}\hat{a}^{\dagger}e^{-i\hat{H}_{0}t}=\hat{a}^{\dagger}e^{i\omega_{0}t}\,.
\end{align}
Substitution yields the interaction Hamiltonian
\begin{align}
\hat{H}_{I} & =\begin{aligned}[t] & \frac{3\hbar^{2}\beta}{8m^{2}\omega_{0}^{2}}\left(\hat{a}^{\dagger}\hat{a}^{\dagger}\hat{a}\hat{a}+2\hat{a}^{\dagger}\hat{a}+\frac{1}{2}\right)+\frac{\hbar^{2}\beta}{8m^{2}\omega_{0}^{2}}\left[\left(2\hat{a}^{\dagger}\hat{a}^{3}+3\hat{a}^{2}\right)e^{-2i\omega_{0}t}+\mathrm{h.c.}\right]\\
 & +\frac{\hbar^{2}\beta}{16m^{2}\omega_{0}^{2}}\left(\hat{a}^{4}e^{-4i\omega_{0}t}+\mathrm{h.c.}\right)+\hbar(\omega-\omega_{0})\left(\hat{a}^{\dagger}\hat{a}\right).
\end{aligned}
\label{eq:-13}
\end{align}
”h.c.” is a placeholder for the Hermitian conjugate of the other
terms within its innermost containing parentheses. Setting $\omega_{0}=\omega$,
the base oscillator frequency is removed. Under the assumption of
a relatively large base oscillator frequency $\omega$,
\[
\omega\gg g,\tag{{\ref{eq:-214}}}
\]
 the oscillating terms can be neglected and we are left with\footnote{The coefficient of the term $\hat{a}^{\dagger}\hat{a}$ in (\ref{eq:-66})
is arbitrary in the sense that it can be made to have any value by
appropriate choice of the interaction picture frequency $\omega_{0}$
in (\ref{eq:-67}). Choosing $\omega_{0}$ as the real root of the
third degree polynomial $4m^{2}\omega_{0}^{2}(\omega-\omega_{0})+3\hbar^{2}\beta$
would cause the term to vanish. With the assumption (\ref{eq:-214})
however, we may consider $\omega_{0}=\omega$ to be a root of the
polynomial.}
\begin{equation}
\hat{H}_{\mathrm{RWA}}=\frac{3\hbar^{2}\beta}{8m^{2}\omega^{2}}\left(\hat{a}^{\dagger}\hat{a}^{\dagger}\hat{a}\hat{a}+2\hat{a}^{\dagger}\hat{a}+\frac{1}{2}\right).\label{eq:-66}
\end{equation}
This is the rotating wave approximation (hence the change of subscript).
To consolidate with the Hamiltonian central to this thesis:
\[
\hat{H}=\hbar g\hat{a}^{\dagger}\hat{a}^{\dagger}\hat{a}\hat{a},\tag{{\ref{eq:kerr-hamiltonian}}}
\]
we identify
\[
g=\frac{3\hbar\beta}{8m^{2}\omega^{2}}.\tag{{\ref{eq:-215}}}
\]

\chapter{Equations of Motion for the Wigner Function \label{app:derivations-of-c-eqs}}

A procedure for deriving equivalent c-number equations from the master
equation is described in Section \ref{sec:wigner-pde-derivation}.
We start from the master equation (\ref{eq:general-master-equation})
and insert a general Hamiltonian containing a simple harmonic oscillator
term, a Kerr nonlinearity term and parametric squeezing terms proportional
to the respective coefficients $\omega$, $g$ and $|\eta|$:
\begin{equation}
\hat{H}=\hbar\omega\hat{a}^{\dagger}\hat{a}+\hbar g\hat{a}^{\dagger}\hat{a}^{\dagger}\hat{a}\hat{a}+i\hbar\left(\eta^{*}\hat{a}\hat{a}-\eta\hat{a}^{\dagger}\hat{a}^{\dagger}\right).\label{eq:-25}
\end{equation}
(\ref{eq:-25}) is inserted into the general master equation, (\ref{eq:general-master-equation}),
to arrive at the master equation for the density matrix $\hat{\rho}$:
\begin{subequations}
\label{eq:-37-1}
\begin{align}
\dot{\hat{\rho}} & =-i\omega\left[\hat{a}^{\dagger}\hat{a},\hat{\rho}\right]-ig\left[\hat{a}^{\dagger}\hat{a}^{\dagger}\hat{a}\hat{a},\hat{\rho}\right]+\left[\eta^{*}\hat{a}\hat{a}-\eta\hat{a}^{\dagger}\hat{a}^{\dagger},\hat{\rho}\right]\label{eq:-37}\\
 & \qquad+\gamma(\bar{n}+1)\mathcal{D}\left[\hat{a}\right]\hat{\rho}+\gamma\bar{n}\mathcal{D}\left[\hat{a}^{\dagger}\right]\hat{\rho}+\gamma_{\phi}\mathcal{D}\left[\hat{n}\right]\hat{\rho}.\nonumber 
\end{align}
\end{subequations}
Equation (\ref{eq:-37-1}) is the most general master equation we
will need to consider in this thesis. Below, corresponding terms in
the partial differential equation for the Wigner function are separately
derived for each term in (\ref{eq:-37-1}). Symbolic manipulations
were performed with the help of the Python library SymPy \cite{Meurer_SymPySymbolicComputing_2017}.

The steps to move from the master equation to the corresponding partial
differential equation for the Wigner function are described in Section
\ref{sec:wigner-pde-derivation}. The procedure is briefly outlined
here as well: First, the time derivative of the symmetrically ordered
characteristic function\footnote{The derivations in the following sections take us by the function
$\chi(\lambda,\lambda^{*},t)=\left\langle \hat{D}(\lambda)\right\rangle $.
For consistency in expressions where it and expectation values of
other operator quantities appear however, we shall keep it expressed
simply as the expectation value of $\hat{D}(\lambda)$.} is written $\partial_{t}\chi(\lambda,\lambda^{*},t)=\Tr\left[\dot{\hat{\rho}}\hat{D}(\lambda)\right]$.
Replacing $\dot{\hat{\rho}}$ by its right hand side from (\ref{eq:-37}),
one obtains the trace of a sum of several operators. The linearity
of the trace can be used to consider the trace of each term separately
instead. The argument of the trace in each term consists of the factors
$\hat{\rho}$ and $\hat{D}(\lambda)$ with some number of interspersed
ladder operators $\hat{a}$ and $\hat{a}^{\dagger}$. Next, the cyclic
property of the trace is employed to in turn apply each ladder operator
to $\hat{D}(\lambda)$ from either left or right. The exhaustive list
of possibilities is found in (\ref{eq:-5}). Neither $\partial_{\lambda}$,
$\partial_{\lambda^{*}}$, $\lambda$ or $\lambda^{*}$ are operator
quantities, so moving them outside the trace allows each term to take
the general form $\text{\ensuremath{\lambda^{m}\left(\lambda^{*}\right)^{n}\partial_{\lambda}^{p}\partial_{\lambda^{*}}^{q}\left\langle \hat{D}(\lambda)\right\rangle }}$.
The resulting expression is a partial differential equation in $\lambda$,
$\lambda^{*}$ and $t$ for $\left\langle \hat{D}(\lambda)\right\rangle $.
Finally, for each term in this differential equation, derive the corresponding
term in the partial differential equation for the Wigner function
by considering the complex Fourier transform from the symmetrically
ordered characteristic function to the Wigner function. This correspondence
is computed for the general term as given by (\ref{eq:-79-1}).

\setcounter{section}{-1}

\section{Coordinate Systems \label{app:phase-space-coordinate-coords}}

Before we catalogue the equations of motion arising from the effects
studied in this thesis, we state the identities necessary for conversion
from the coordinate $\alpha$ to Cartesian and polar coordinates.

Cartesian coordinates $(x,y)$ were defined by

\begin{align}
\alpha & =x+iy,\tag{{\ref{eq:-17}}}\nonumber \\
\alpha^{*} & =x+iy,
\end{align}
with the inverse relations
\begin{align*}
x & =\Re\alpha=\frac{1}{2}(\alpha+\alpha^{*}),\tag{{\ref{eq:-242}}}\\
y & =\Im\alpha=\frac{1}{2i}(\alpha-\alpha^{*}).\tag{{\ref{eq:-243}}}
\end{align*}
The differential operators $\partial_{\alpha}$ and $\partial_{\alpha^{*}}$
are then
\begin{subequations}
\label{eq:-241}
\begin{align}
\partial_{\alpha} & =\left(\partial_{\alpha}x\right)\partial_{x}+\left(\partial_{\alpha}y\right)\partial_{y}=\frac{1}{2}\partial_{x}+\frac{1}{2i}\partial_{y},\\
\partial_{\alpha^{*}} & =\left(\partial_{\alpha^{*}}x\right)\partial_{x}+\left(\partial_{\alpha^{*}}y\right)\partial_{y}=\frac{1}{2}\partial_{x}-\frac{1}{2i}\partial_{y}.
\end{align}
\end{subequations}

Polar coordinates were defined by
\begin{align}
\alpha & =re^{i\phi}\tag{{\ref{eq:-16}}}\nonumber \\
\alpha^{*} & =re^{-i\phi}
\end{align}
with the inverse relations
\begin{align}
r & =|\alpha|,\\
\phi & =\arg\alpha.
\end{align}
The differential operators $\partial_{\alpha}$ and $\partial_{\alpha^{*}}$
are then
\begin{subequations}
\label{eq:-241-1}
\begin{align}
\partial_{\alpha} & =\left(\partial_{\alpha}r\right)\partial_{r}+\left(\partial_{\alpha}\phi\right)\partial_{\phi}=\frac{1}{2}e^{-i\phi}\partial_{r}+\frac{1}{2ir}e^{-i\phi}\partial_{\phi},\\
\partial_{\alpha^{*}} & =\left(\partial_{\alpha^{*}}r\right)\partial_{r}+\left(\partial_{\alpha^{*}}\phi\right)\partial_{\phi}=\frac{1}{2}e^{i\phi}\partial_{r}-\frac{1}{2ir}e^{i\phi}\partial_{\phi}.
\end{align}
\end{subequations}

We also note the following direct relations between the Cartesian
and polar coordinates. In particular, the Laplacian $\nabla^{2}$
is used in both Cartesian and polar coordinate systems where appropriate.
\begin{subequations}
\label{eq:36} 
\begin{align}
r\partial_{r} & =x\partial_{x}+y\partial_{y},\label{eq:7}\\
\partial_{\phi} & =-y\partial_{x}+x\partial_{y},\label{eq:-14}\\
\nabla^{2} & =\partial_{x}^{2}+\partial_{y}^{2}=\left(\partial_{r}^{2}+r^{-2}\partial_{\phi}^{2}+r^{-1}\partial_{r}\right),\label{eq:-1-2}\\
\partial_{\phi}^{2} & =y^{2}\partial_{x}^{2}+x^{2}\partial_{y}^{2}-2xy\partial_{x}\partial_{y}-x\partial_{x}-y\partial_{y}.
\end{align}
\end{subequations}

For reference, we also note that the divergence of a vector 
\begin{equation}
\mathbf{v}=v_{x}\hat{\mathbf{x}}+v_{y}\hat{\mathbf{y}}=v_{r}\hat{\mathbf{r}}+v_{\phi}\bm{{\phi}}
\end{equation}
 is given by
\begin{equation}
\nabla\cdot\mathbf{v}=\partial_{x}v_{x}+\partial_{y}v_{y}=\frac{1}{r}\partial_{r}\left(rv_{r}\right)+\frac{1}{r}\partial_{\phi}v_{\phi}.
\end{equation}

\section{Harmonic Oscillator}

The von Neumann equation for the simple harmonic oscillator takes
the form
\begin{equation}
\dot{\hat{\rho}}=-i\omega\left[\hat{a}^{\dagger}\hat{a},\hat{\rho}\right].
\end{equation}
This dynamics of the term are notable for commuting with all other
dynamics of (\ref{eq:-37-1}) except for parametric squeezing. Translating
each term into an equation for the displacement operator yields the
terms
\begin{subequations}
\label{eq:-244}
\begin{align}
-i\omega\left\langle \hat{D}(\lambda)\hat{a}^{\dagger}\hat{a}\right\rangle  & =-i\omega\left(-\frac{\lambda^{*}}{2}+\partial_{\lambda}\right)\left(-\frac{\lambda}{2}-\partial_{\lambda^{*}}\right)\left\langle \hat{D}(\lambda)\right\rangle ,\\
i\omega\left\langle \hat{a}^{\dagger}\hat{a}\hat{D}(\lambda)\right\rangle  & =i\omega\left(\frac{\lambda}{2}-\partial_{\lambda^{*}}\right)\left(\frac{\lambda^{*}}{2}+\partial_{\lambda}\right)\left\langle \hat{D}(\lambda)\right\rangle 
\end{align}
\end{subequations}
which add to

\begin{equation}
\partial_{t}\left\langle \hat{D}(\lambda)\right\rangle =i\omega\left(\lambda\partial_{\lambda}-\lambda^{*}\partial_{\lambda^{*}}\right)\left\langle \hat{D}(\lambda)\right\rangle .\label{eq:-79}
\end{equation}
Taking the Fourier transform and applying (\ref{eq:-79-1}), (\ref{eq:-79})
is rewritten to the equation for $W(\alpha,\alpha^{*})$:
\begin{equation}
\partial_{t}W(\alpha,\alpha)=i\omega\left(\alpha\partial_{\alpha}-\alpha^{*}\partial_{\alpha^{*}}\right)W(\alpha,\alpha^{*}).\label{eq:-94}
\end{equation}
In Cartesian coordinates (\ref{eq:-94}) takes the form
\begin{equation}
\partial_{t}W(x,y)=\omega\left(-y\partial_{x}+x\partial_{y}\right)W(x,y).
\end{equation}
In polar coordinates (\ref{eq:-94}) takes the form
\begin{equation}
\partial_{t}W(r,\phi)=\omega\partial_{\phi}W(r,\phi).
\end{equation}

\section{Kerr Oscillator}

The von Neumann equation for the Kerr oscillator takes the form
\begin{equation}
\dot{\hat{\rho}}=-ig\left[\hat{a}^{\dagger}\hat{a}^{\dagger}\hat{a}\hat{a},\hat{\rho}\right].
\end{equation}
Translating each term into an equation for the displacement operator
yields
\begin{subequations}
\label{eq:-245}
\begin{align}
\partial_{t}\left\langle \hat{D}(\lambda)\right\rangle  & =2ig\left(-\lambda\partial_{\lambda^{*}}\partial_{\lambda}^{2}+\lambda^{*}\partial_{\lambda^{*}}^{2}\partial_{\lambda}\right)\left\langle \hat{D}(\lambda)\right\rangle \\
 & \qquad+2ig\left(-\lambda\partial_{\lambda}+\lambda^{*}\partial_{\lambda^{*}}\right)\left\langle \hat{D}(\lambda)\right\rangle \\
 & \qquad-\frac{ig}{2}\left(-\lambda^{2}\lambda^{*}\partial_{\lambda}+\lambda\left(\lambda^{*}\right)^{2}\partial_{\lambda^{*}}\right)\left\langle \hat{D}(\lambda)\right\rangle .
\end{align}
\end{subequations}
Taking the Fourier transform and applying (\ref{eq:-79-1}), (\ref{eq:-245})
is rewritten to the equation for $W(\alpha,\alpha^{*})$:
\begin{subequations}
\label{eq:-246}
\begin{align}
\partial_{t}W(\alpha,\alpha^{*}) & =2ig\left(\alpha^{2}\alpha^{*}\partial_{\alpha}-\alpha\left(\alpha^{*}\right)^{2}\partial_{\alpha^{*}}\right)W(\alpha,\alpha^{*})\\
 & \qquad-2ig\left(\alpha\partial_{\alpha}-\alpha^{*}\partial_{\alpha^{*}}\right)W(\alpha,\alpha^{*})\\
 & \qquad-\frac{ig}{2}\left(\alpha\partial_{\alpha^{*}}\partial_{\alpha}^{2}-\alpha^{*}\partial_{\alpha}\partial_{\alpha^{*}}^{2}\right)W(\alpha,\alpha^{*}).
\end{align}
\end{subequations}
In Cartesian coordinates (\ref{eq:-246}) takes the form
\begin{align}
\partial_{t}W(x,y) & =2g\left(x^{2}+y^{2}-1\right)\left(-y\partial_{x}+x\partial_{y}\right)W(x,y)-\frac{g}{8}\left(-y\partial_{x}+x\partial_{y}\right)\left(\partial_{x}^{2}+\partial_{y}^{2}\right)W(x,y)
\end{align}
In polar coordinates (\ref{eq:-246}) takes the form (cf. \textcite{Stobinska_WignerFunctionEvolution_2008})
\begin{equation}
\partial_{t}W(r,\phi)=2g(r^{2}-1)\partial_{\phi}W(r,\phi)-\frac{g}{8}\nabla^{2}\partial_{\phi}W(r,\phi).
\end{equation}

\section{Damping}

The master equation for a system under the influence of damping only
is given by
\begin{equation}
\dot{\hat{\rho}}=\gamma(\bar{n}+1)\left(\hat{a}\hat{\rho}\hat{a}^{\dagger}-\frac{1}{2}\hat{a}^{\dagger}\hat{a}\hat{\rho}-\frac{1}{2}\hat{\rho}\hat{a}^{\dagger}\hat{a}\right)+\gamma\bar{n}\left(\hat{a}^{\dagger}\rho\hat{a}-\frac{1}{2}\hat{a}\hat{a}^{\dagger}\hat{\rho}-\frac{1}{2}\hat{\rho}\hat{a}\hat{a}^{\dagger}\right).
\end{equation}
Translating each term into an equation for the displacement operator
yields
\begin{equation}
\partial_{t}\left\langle \hat{D}(\lambda)\right\rangle =-\gamma\left(\bar{n}+\frac{1}{2}\right)\lambda\lambda^{*}\left\langle \hat{D}(\lambda)\right\rangle -\gamma\lambda\partial_{\lambda}\left\langle \hat{D}(\lambda)\right\rangle -\gamma\lambda^{*}\partial_{\lambda^{*}}\left\langle \hat{D}(\lambda)\right\rangle \label{eq:-251}
\end{equation}
Taking the Fourier transform and applying (\ref{eq:-79-1}), (\ref{eq:-251})
is rewritten to the equation for $W(\alpha,\alpha^{*})$ (cf. \textcite{Walls_QuantumOptics_2008}):
\begin{equation}
\partial_{t}W(\alpha,\alpha^{*},t)=\gamma\left(\bar{n}+\frac{1}{2}\right)\partial_{\alpha}\partial_{\alpha^{*}}W(\alpha,\alpha^{*},t)+\frac{\gamma}{2}\partial_{\alpha}\left(\alpha W(\alpha,\alpha^{*},t)\right)+\frac{\gamma}{2}\partial_{\alpha^{*}}\left(\alpha^{*}W(\alpha,\alpha^{*},t)\right)\label{eq:-89}
\end{equation}
In Cartesian coordinates (\ref{eq:-89}) takes the form
\begin{equation}
\partial_{t}W(x,y,t)=\frac{\gamma}{4}\left(\bar{n}+\frac{1}{2}\right)\left(\partial_{x}^{2}+\partial_{y}^{2}\right)W(x,y,t)+\frac{\gamma}{2}\partial_{x}\left(xW(x,y,t)\right)+\frac{\gamma}{2}\partial_{y}\left(yW(x,y,t)\right)\label{eq:-92}
\end{equation}
In polar coordinates (\ref{eq:-89}) takes the form (cf. \textcite{Stobinska_WignerFunctionEvolution_2008})
\begin{equation}
\partial_{t}W(r,\phi,t)=\frac{\gamma}{4}\left(\bar{n}+\frac{1}{2}\right)\underbrace{\left(\partial_{r}^{2}+\frac{1}{r}\partial_{r}+\frac{1}{r^{2}}\partial_{\phi}^{2}\right)}_{\nabla^{2}}W(r,\phi,t)+\frac{\gamma}{2}r\partial_{r}W(r,\phi,t)+\gamma W(r,\phi,t)\label{eq:-93}
\end{equation}

\section{Dephasing}

The master equation for a system under the influence of dephasing
only is given by
\begin{equation}
\dot{\hat{\rho}}_{\gamma_{\phi}}=\gamma_{\phi}\left(\hat{n}\hat{\rho}\hat{n}-\frac{1}{2}\hat{n}^{2}\hat{\rho}-\frac{1}{2}\hat{\rho}\hat{n}^{2}\right)
\end{equation}
Translating each term into an equation for the displacement operator
yields
\begin{equation}
\partial_{t}\left\langle \hat{D}(\lambda)\right\rangle =-\frac{\gamma_{\phi}}{2}\lambda^{2}\partial_{\lambda}^{2}\left\langle \hat{D}(\lambda)\right\rangle -\frac{\gamma_{\phi}}{2}\left(\lambda^{*}\right)^{2}\partial_{\lambda^{*}}^{2}\left\langle \hat{D}(\lambda)\right\rangle +\gamma_{\phi}\lambda\lambda^{*}\partial_{\lambda}\partial_{\lambda^{*}}\left\langle \hat{D}(\lambda)\right\rangle .\label{eq:-250}
\end{equation}
Taking the Fourier transform and applying (\ref{eq:-79-1}), (\ref{eq:-250})
is rewritten to the equation for $W(\alpha,\alpha^{*})$:
\begin{subequations}
\label{eq:-247}
\begin{align}
\partial_{t}W(\alpha,\alpha^{*},t) & =-\frac{\gamma_{\phi}}{2}\alpha^{2}\partial_{\alpha}^{2}W(\alpha,\alpha^{*},t)-\frac{\gamma_{\phi}}{2}\left(\alpha^{*}\right)^{2}\partial_{\alpha^{*}}^{2}W(\alpha,\alpha^{*},t)+\gamma_{\phi}\alpha\alpha^{*}\partial_{\alpha}\partial_{\alpha^{*}}W(\alpha,\alpha^{*},t)\\
 & \qquad-\frac{\gamma_{\phi}}{2}\alpha^{*}\partial_{\alpha^{*}}W(\alpha,\alpha^{*},t)-\frac{\gamma_{\phi}}{2}\alpha\partial_{\alpha}W(\alpha,\alpha^{*},t)\\
 & =-\frac{\gamma_{\phi}}{2}\left(\alpha\partial_{\alpha}-\alpha^{*}\partial_{\alpha^{*}}\right)^{2}W(\alpha,\alpha^{*},t).
\end{align}
\end{subequations}
In Cartesian coordinates (\ref{eq:-247}) takes the form
\begin{equation}
\partial_{t}W(x,y,t)=\left(-y\partial_{x}+x\partial_{y}\right)^{2}W(x,y,t).
\end{equation}
In polar coordinates (\ref{eq:-247}) takes the form
\begin{equation}
\partial_{t}W(r,\phi,t)=\frac{\gamma_{\phi}}{2}\partial_{\phi}^{2}W(r,\phi,t).
\end{equation}

\section{Parametric Squeezing}

Parametric squeezing was introduced in Section \ref{sec:quadrature-squeezing}
and further treated in Section \ref{sec:return-to-parametric-squeezing}.
The von Neumann equation takes the form
\begin{equation}
\dot{\hat{\rho}}_{\eta}=\left[\eta^{*}\hat{a}\hat{a}-\eta\hat{a}^{\dagger}\hat{a}^{\dagger},\hat{\rho}\right].
\end{equation}
Translating each term into an equation for the displacement operator
yields the terms
\begin{equation}
\partial_{t}\left\langle \hat{D}(\lambda)\right\rangle =2\left(\eta^{*}\lambda\partial_{\lambda^{*}}+\eta\lambda^{*}\partial_{\lambda}\right)\left\langle \hat{D}(\lambda)\right\rangle .\label{eq:-248}
\end{equation}
Taking the Fourier transform and applying (\ref{eq:-79-1}), (\ref{eq:-248})
is rewritten to the equation for $W(\alpha,\alpha^{*})$:
\begin{equation}
\partial_{t}W(\alpha,\alpha^{*},t)=2\eta^{*}\alpha\partial_{\alpha^{*}}W(\alpha,\alpha^{*},t)+2\eta\alpha^{*}\partial_{\alpha}W(\alpha,\alpha^{*},t).\label{eq:-249}
\end{equation}
In Cartesian coordinates (\ref{eq:-249}) takes the form
\begin{equation}
\partial_{t}W(x,y,t)=2\left(x\,\Re\eta+y\,\Im\eta\right)\frac{\partial}{\partial x}W(x,y,t)+2\left(x\,\Im\eta-y\,\Re\eta\right)\frac{\partial}{\partial y}W(x,y,t).
\end{equation}
In polar coordinates (\ref{eq:-249}) takes the form
\begin{equation}
\partial_{t}W(r,\phi,t)=2r\Re\left(\eta e^{-2i\phi}\right)\frac{\partial}{\partial r}W\left(r,\phi,t\right)+2\Im\left(\eta e^{-2i\phi}\right)\frac{\partial}{\partial\phi}W\left(r,\phi,t\right).
\end{equation}

\chapter{Superoperator Commutation Relations \label{app:superoperator-commutation-relations}}

Consider the master equation (\ref{eq:general-master-equation}) with
$\hat{H}=\hbar g\hat{a}^{\dagger}\hat{a}^{\dagger}\hat{a}\hat{a}$.
It reads

\begin{equation}
\dot{\hat{\rho}}=\mathcal{L}_{g}\hat{\rho}+\mathcal{L}_{\gamma}\hat{\rho}+\mathcal{L}_{\gamma_{\phi}}\hat{\rho}.
\end{equation}
The Kerr oscillator unitary dynamics are described by the superoperator
$\mathcal{L}_{g}\hat{\rho}=-ig\left[\hat{a}^{\dagger}\hat{a}^{\dagger}\hat{a}\hat{a},\hat{\rho}\right]$.
Dephasing and damping effects have been summarized as the superoperators
$\mathcal{L}_{\gamma_{\phi}}\hat{\rho}=\gamma_{\phi}\mathcal{D}[\hat{n}]\hat{\rho}$
and $\mathcal{L}_{\gamma}\hat{\rho}=\gamma\left(\bar{n}+1\right)\mathcal{D}[\hat{a}]\hat{\rho}+\gamma\bar{n}\mathcal{D}[\hat{a}^{\dagger}]\hat{\rho}$. 

The operators describing unitary evolution and dephasing are both
diagonal in the number state basis, from which it follows that they
commute:

\begin{equation}
\mathcal{L}_{g}\mathcal{L}_{\gamma_{\phi}}\hat{\rho}=\mathcal{L}_{\gamma_{\phi}}\mathcal{L}_{g}\hat{\rho}\label{eq:-49-1}
\end{equation}
for any density matrix $\hat{\rho}$. It may also be shown that damping
and dephasing commute:
\begin{equation}
\mathcal{L}_{\gamma_{\phi}}\mathcal{L}_{\gamma}\hat{\rho}=\mathcal{L}_{\gamma}\mathcal{L}_{\gamma_{\phi}}\hat{\rho}.\label{eq:-49}
\end{equation}
$\mathcal{L}_{g}$ and $\mathcal{L}_{\gamma}$ do not commute however,
and one has that
\begin{equation}
\mathcal{L}_{g}\mathcal{L}_{\gamma}\hat{\rho}=\mathcal{L}_{\gamma}\mathcal{L}_{g}\hat{\rho}-2ig\gamma\left(\bar{n}+1\right)\left[\hat{a}^{\dagger}\hat{a},\hat{a}\hat{\rho}\hat{a}^{\dagger}\right]+2ig\gamma\bar{n}\left[\hat{a}^{\dagger}\hat{a},\hat{a}^{\dagger}\hat{\rho}\hat{a}\right].
\end{equation}

\paragraph{Damping and dephasing.}

Equation (\ref{eq:-49}) follows from the fact that the superoperator
$\mathcal{D}[\hat{n}]$ commutes with both $\mathcal{D}[\hat{a}]$
and $\mathcal{D}[\hat{a}^{\dagger}]$. This is shown below by explicitly
computation of the commutator expression for both cases. Define for
notational convenience 
\begin{equation}
\hat{a}_{-}=\hat{a},\qquad\qquad\hat{a}_{+}=\hat{a}^{\dagger},\qquad\text{and}\qquad\hat{\xi}_{\pm}=\hat{a}_{\mp}\hat{\rho}\hat{a}_{\pm},\label{eq:-60}
\end{equation}
noting that $\hat{a}_{\pm}\hat{n}=(\hat{n}\mp1)\hat{a}_{\pm}$. The
two parts of the relevant superoperator commutator can be written

\begin{equation}
\mathcal{D}[\hat{n}]\mathcal{D}[\hat{a}_{\mp}]\hat{\rho}=\begin{aligned}[t] & \hat{n}\hat{\xi}_{\pm}\hat{n}-\frac{1}{2}\hat{n}^{2}\hat{\xi}_{\pm}-\frac{1}{2}\hat{\xi}_{\pm}\hat{n}^{2}\\
 & -\frac{1}{2}\hat{n}\hat{n}\hat{\rho}\hat{n}+\frac{1}{4}\hat{n}^{3}\hat{\rho}+\frac{1}{4}\hat{n}\hat{\rho}\hat{n}^{2}\\
 & -\frac{1}{2}\hat{n}\hat{\rho}\hat{n}\hat{n}+\frac{1}{4}\hat{n}^{2}\hat{\rho}\hat{n}+\frac{1}{4}\hat{\rho}\hat{n}^{3}
\end{aligned}
\label{eq:-32}
\end{equation}
and
\begin{subequations}
\label{eq:}

\begin{align}
\mathcal{D}[\hat{a}_{\mp}]\mathcal{D}[\hat{n}]\hat{\rho} & =\begin{aligned}[t] & \hat{a}_{\mp}\hat{n}\hat{\rho}\hat{n}\hat{a}_{\pm}-\frac{1}{2}\hat{n}\hat{n}\hat{\rho}\hat{n}-\frac{1}{2}\hat{n}\hat{\rho}\hat{n}\hat{n}\\
 & -\frac{1}{2}\hat{a}_{\mp}\hat{n}^{2}\hat{\rho}\hat{a}_{\pm}+\frac{1}{4}\hat{n}\hat{n}^{2}\hat{\rho}+\frac{1}{4}\hat{n}^{2}\hat{\rho}\hat{n}\\
 & -\frac{1}{2}\hat{a}_{\mp}\hat{\rho}\hat{n}^{2}\hat{a}_{\pm}+\frac{1}{4}\hat{n}\hat{\rho}\hat{n}^{2}+\frac{1}{4}\hat{\rho}\hat{n}^{2}\hat{n}
\end{aligned}
\\
 & =\begin{aligned}[t] & \left(\hat{n}\pm1\right)\hat{\xi}_{\pm}\left(\hat{n}\pm1\right)-\frac{1}{2}\hat{n}\hat{n}\hat{\rho}\hat{n}-\frac{1}{2}\hat{n}\hat{\rho}\hat{n}\hat{n}\\
 & -\frac{1}{2}\left(\hat{n}\pm1\right)^{2}\hat{\xi}_{\pm}+\frac{1}{4}\hat{n}\hat{n}^{2}\hat{\rho}+\frac{1}{4}\hat{n}^{2}\hat{\rho}\hat{n}\\
 & -\frac{1}{2}\hat{\xi}_{\pm}\left(\hat{n}\pm1\right)^{2}+\frac{1}{4}\hat{n}\hat{\rho}\hat{n}^{2}+\frac{1}{4}\hat{\rho}\hat{n}^{2}\hat{n}.
\end{aligned}
\label{eq:-33}
\end{align}
\end{subequations}
Subtracting (\ref{eq:-32}) and (\ref{eq:-33}), one may neglect all
terms in which only operators diagonal in the number state basis,
are applied to $\hat{\rho}$. Hence 
\begin{align}
\mathcal{D}[\hat{a}_{\mp}]\mathcal{D}[\hat{n}]\hat{\rho}-\mathcal{D}[\hat{n}]\mathcal{D}[\hat{a}_{\mp}]\hat{\rho} & =\begin{aligned}[t] & \left(\hat{n}\pm1\right)\hat{\xi}_{\pm}\left(\hat{n}\pm1\right)-\hat{n}\hat{\xi}_{\pm}\hat{n}\\
 & -\frac{1}{2}\left(\hat{n}\pm1\right)^{2}\hat{\xi}_{\pm}+\frac{1}{2}\hat{n}^{2}\hat{\xi}_{\pm}-\frac{1}{2}\hat{\xi}_{\pm}\left(\hat{n}\pm1\right)^{2}+\frac{1}{2}\hat{\xi}_{\pm}\hat{n}^{2}
\end{aligned}
\\
 & =\pm\hat{n}\hat{\xi}_{\pm}\pm\hat{\xi}_{\pm}\hat{n}\mp\hat{n}\hat{\xi}_{\pm}\mp\hat{\xi}_{\pm}\hat{n}=0,
\end{align}
demonstrating (\ref{eq:-49}).

\paragraph{Kerr nonlinearity and damping.}

The definitions (\ref{eq:-60}) are reused here and an extra factor
of $\delta$ is inserted into the Hamiltonian, generalizing the derivation
to a frame rotating at an arbitrary frequency. The two parts of the
relevant superoperator commutator can be written

\begin{equation}
\left[\hat{n}^{2}-\delta\hat{n},\mathcal{D}[\hat{a}_{\mp}]\hat{\rho}\right]=\begin{aligned}[t] & \left(\hat{n}^{2}-\delta\hat{n}\right)\hat{\xi}_{\pm}-\hat{\xi}_{\pm}\left(\hat{n}^{2}-\delta\hat{n}\right)\\
 & -\frac{1}{2}\left(\hat{n}^{2}-\delta\hat{n}\right)\hat{n}\hat{\rho}+\frac{1}{2}\hat{n}\hat{\rho}\left(\hat{n}^{2}-\delta\hat{n}\right)\\
 & -\frac{1}{2}\left(\hat{n}^{2}-\delta\hat{n}\right)\hat{\rho}\hat{n}+\frac{1}{2}\hat{\rho}\hat{n}\left(\hat{n}^{2}-\delta\hat{n}\right)
\end{aligned}
\label{eq:-32-1}
\end{equation}
and
\begin{subequations}
\label{eq:-55}

\begin{align}
\mathcal{D}[\hat{a}_{\mp}]\left(\left[\hat{n}^{2}-\delta\hat{n},\hat{\rho}\right]\right) & =\begin{aligned}[t] & \hat{a}_{\mp}\left(\hat{n}^{2}-\delta\hat{n}\right)\hat{\rho}\hat{a}_{\pm}-\frac{1}{2}\hat{n}\left(\hat{n}^{2}-\delta\hat{n}\right)\hat{\rho}-\frac{1}{2}\left(\hat{n}^{2}-\delta\hat{n}\right)\hat{\rho}\hat{n}\\
 & -\hat{a}_{\mp}\hat{\rho}\left(\hat{n}^{2}-\delta\hat{n}\right)\hat{a}_{\pm}-\frac{1}{2}\hat{n}\hat{\rho}\left(\hat{n}^{2}-\delta\hat{n}\right)-\frac{1}{2}\hat{\rho}\left(\hat{n}^{2}-\delta\hat{n}\right)\hat{n}
\end{aligned}
\label{eq:-56}\\
 & =\begin{aligned}[t] & \left(\left(\hat{n}\pm1\right)^{2}-\delta\left(\hat{n}\pm1\right)\right)\hat{\xi}_{\pm}-\frac{1}{2}\hat{n}\left(\hat{n}^{2}-\delta\hat{n}\right)\hat{\rho}-\frac{1}{2}\left(\hat{n}^{2}-\delta\hat{n}\right)\hat{\rho}\hat{n}\\
 & -\hat{\xi}_{\pm}\left(\left(\hat{n}\pm1\right)^{2}-\delta\left(\hat{n}\pm1\right)\right)-\frac{1}{2}\hat{n}\hat{\rho}\left(\hat{n}^{2}-\delta\hat{n}\right)-\frac{1}{2}\hat{\rho}\left(\hat{n}^{2}-\delta\hat{n}\right)\hat{n}
\end{aligned}
\end{align}
\end{subequations}
Subtracting (\ref{eq:-32-1}) and (\ref{eq:-56}), one may neglect
all terms in which only operators diagonal in the number state basis,
are applied to $\hat{\rho}$. Hence 
\begin{subequations}
\label{eq:-57}
\begin{align}
\left[\hat{n}^{2}-\delta\hat{n},\mathcal{D}[\hat{a}_{\mp}]\hat{\rho}\right]-\mathcal{D}[\hat{a}_{\mp}]\left(\left[\hat{n}^{2}-\delta\hat{n},\hat{\rho}\right]\right) & =\begin{aligned}[t] & \left(\left(\hat{n}\pm1\right)^{2}-\delta\left(\hat{n}\pm1\right)-\hat{n}^{2}+\delta\hat{n}\right)\hat{\xi}_{\pm}\\
 & -\hat{\xi}_{\pm}\left(\left(\hat{n}\pm1\right)^{2}-\delta\left(\hat{n}\pm1\right)-\hat{n}^{2}+\delta\hat{n}\right)
\end{aligned}
\label{eq:-58}\\
 & =\pm2\hat{\xi}_{\pm}\hat{n}\mp2\hat{n}\hat{\xi}_{\pm}.\label{eq:-59}
\end{align}
\end{subequations}
The absence of $\delta$ in (\ref{eq:-59}) shows that $\mathcal{D}[\hat{a}_{\mp}]$
and $\mathcal{C}[\hat{n}]$ (which constitutes the operator part of
the right hand side of the von Neumann equation for the simple harmonic
oscillator) do commute.

\chapter{Wigner Function for a Displaced Squeezed Thermal State}

To demonstrate the various parameters used for the initial state,
we compute the Wigner function for the displaced quadrature-squeezed
thermal state 

\begin{equation}
\hat{\rho}_{0}=\hat{D}(\alpha_{0})\hat{S}(\xi_{0})\hat{\rho}_{\bar{n}_{0}}\hat{S}^{\dagger}(\xi_{0})\hat{D}^{\dagger}(\alpha_{0}).
\end{equation}
The state is parametrized by $\bar{n}_{0}$, $\xi_{0}$ and $\alpha_{0}$.
The state is obtained as follows: $\bar{n}_{0}$ is a real number
describing the mean occupancy of the starting thermal state. This
state is then squeezed to a degree given by the complex squeezing
parameter $\xi_{0}$. The resulting squeezed thermal state is then
finally displaced as described by the complex displacement parameter
$\alpha_{0}$ to obtain $\hat{\rho}_{0}$. Computational steps to
obtain the Wigner function for this state in various coordinate systems
are listed below.

\section{Thermal State Wigner Function\label{app:thermal-state-wig}}

A thermal state for a harmonic oscillator with base frequency $\omega$
is given by \cite{Gerry_IntroductoryQuantumOptics_2004} 
\begin{equation}
\hat{\rho}=\frac{e^{-\hbar\omega\hat{n}^{2}/k_{B}T}}{\Tr\left[e^{-\hbar\omega\hat{n}^{2}/k_{B}T}\right]}.
\end{equation}
Let $\bar{n}_{0}=\left(e^{-\hbar\omega/k_{B}T}-1\right)^{-1}$ denote
the expectation value $\Tr[\hat{n}\hat{\rho}]$ (the index is to distinguish
it from the temperature of the environment if open systems are considered).
This allows one to write the Q-function as \cite{Gerry_IntroductoryQuantumOptics_2004}
\begin{align}
Q_{\hat{\rho}}(\alpha,\alpha^{*}) & =\frac{1}{\pi}\langle\alpha|\hat{\rho}|\alpha\rangle=\frac{1}{\pi(\bar{n}_{0}+1)}\exp\left(-\frac{|\alpha|^{2}}{\bar{n}_{0}+1}\right).\label{eq:-50}
\end{align}
The inverse Fourier transform allows one to find the anti-normal ordered
characteristic function $\chi_{A}$ \cite{Gerry_IntroductoryQuantumOptics_2004}:
\begin{equation}
\chi_{A}(\lambda,\lambda^{*})=\frac{1}{\pi(\bar{n}_{0}+1)}\int d\alpha\,e^{\alpha^{*}\lambda-\alpha\lambda^{*}}\,\exp\left(-\frac{|\alpha|^{2}}{\bar{n}_{0}+1}\right)=\exp\left[-\left(\bar{n}_{0}+1\right)|\lambda|^{2}\right].
\end{equation}
Using the disentangling theorem (\ref{eq:disentangling-theorem}),
the symmetrized characteristic function is expressed from the anti-normal
ordered one as 
\begin{equation}
\chi(\lambda,\lambda^{*})=\chi_{A}(\lambda,\lambda^{*})e^{|\lambda|^{2}/2}=\exp\left[-\left(\bar{n}_{0}+\frac{1}{2}\right)|\lambda|^{2}\right].
\end{equation}
Finally, $W(\alpha,\alpha^{*})$ is found from $\chi(\lambda,\lambda^{*})$
by means of the Fourier transform, whereby\footnote{Comparing (\ref{eq:-50}) and (\ref{eq:-21}), it is seen that the
parameter $\bar{n}_{0}$ appears in a sum with $1$ in the Q-function
whereas it appears in a sum with $\frac{1}{2}$ in the Wigner function.
The same relationship is also seen with the parameter $\bar{n}$ in
the equations of motion for $W$ (see (\ref{eq:-89}--\ref{eq:-93}))
and $Q$ (see e.g. \textcite{DAriano_TimeEvolutionAnharmonic_1995}).
This can also be extended to the P-function where the addend is $0$
in both the thermal state \cite{Gerry_IntroductoryQuantumOptics_2004}
and the equation of motion \cite{Walls_QuantumOptics_2008}. This
is unsurprising since we expect the parameter in the steady state
(which should appear somewhere in the equation of motion) to match
the parameter of the thermal state.}
\begin{equation}
W_{\hat{\rho}}(\alpha,\alpha^{*})=\int e^{\alpha\lambda^{*}-\alpha^{*}\lambda}\exp\left[-\left(\bar{n}_{0}+\frac{1}{2}\right)|\lambda|^{2}\right]=\frac{1}{\pi\left(\bar{n}_{0}+\frac{1}{2}\right)}\exp\left(-\frac{|\alpha|^{2}}{\bar{n}_{0}+\frac{1}{2}}\right).\label{eq:-21}
\end{equation}
The parameter $\sigma=\sqrt{2\bar{n}_{0}+1}$, proportional to the
standard deviation of the Gaussian, will also sometimes be used instead
of $\bar{n}_{0}$. With $\sigma$, $W$ takes the form
\begin{equation}
W_{\hat{\rho}}(\alpha,\alpha^{*})=\frac{2}{\pi\sigma^{2}}\exp\left(-\frac{2|\alpha|^{2}}{\sigma^{2}}\right).\label{eq:-21-1}
\end{equation}
At vanishing temperature, i.e. $\bar{n}_{0}=0$ and $\sigma=1$, the
vacuum state Wigner function, equation (\ref{eq:vacuum-state-wigner-function}),
is recovered.

\section{Squeezed Thermal State Wigner Function \label{app:squeezed-thermal-state-wig}}

Recall the squeezing operator 
\[
\hat{S}(\xi)=e^{\frac{1}{2}\left(\xi^{*}\hat{a}\hat{a}-\xi\hat{a}^{\dagger}\hat{a}^{\dagger}\right)}.\tag{{\ref{eq:squeezing-operator}}}
\]
and its action on $\hat{a}$ and $\hat{a}^{\dagger}$:
\begin{align*}
\hat{S}^{\dagger}(\xi)\hat{a}\hat{S}(\xi) & =\hat{a}\cosh r-\hat{a}^{\dagger}e^{i\theta}\sinh r,\tag{{\ref{eq:-71}}}\\
\hat{S}^{\dagger}(\xi)\hat{a}^{\dagger}\hat{S}(\xi) & =\hat{a}^{\dagger}\cosh r-\hat{a}e^{-i\theta}\sinh r.\tag{{\ref{eq:-72}}}
\end{align*}
We may use (\ref{eq:-114-1-1}) to apply $\hat{S}$ to $W_{\hat{\rho}}(\alpha,\alpha^{*})$
to arrive at the Wigner function for the squeezed thermal state $\hat{S}(\xi)\hat{\rho}\hat{S}^{\dagger}(\xi).$
Prior to that though, we expand on the steps of (\ref{eq:-114-1-1}).
We wish to perform an integral substitution to write 
\begin{equation}
W_{\hat{S}(\xi_{0})\hat{\rho}\hat{S}^{\dagger}(\xi_{0})}(\alpha,\alpha^{*})=\int d\lambda\,d\lambda^{*}\,e^{\alpha\lambda^{*}-\alpha^{*}\lambda}\Tr\left[\hat{\rho}e^{\left(\lambda\cosh r_{0}-\lambda^{*}e^{i\theta_{0}}\sinh r_{0}\right)\hat{a}^{\dagger}-\left(\lambda^{*}\cosh r_{0}-\lambda e^{-i\theta_{0}}\sinh r_{0}\right)\hat{a}}\right]
\end{equation}
Define the new complex coordinate $\mu$ by 
\begin{equation}
\left(\begin{array}{c}
\mu\\
\mu^{*}
\end{array}\right)=\left(\begin{array}{cc}
\cosh r & -e^{i\theta}\sinh r\\
-e^{-i\theta}\sinh r & \cosh r
\end{array}\right)\left(\begin{array}{c}
\lambda\\
\lambda^{*}
\end{array}\right)
\end{equation}
with the inverse transform given by

\begin{equation}
\left(\begin{array}{c}
\lambda\\
\lambda^{*}
\end{array}\right)=\left(\begin{array}{cc}
\cosh r & e^{i\theta}\sinh r\\
e^{-i\theta}\sinh r & \cosh r
\end{array}\right)\left(\begin{array}{c}
\mu\\
\mu^{*}
\end{array}\right).
\end{equation}
$W_{\hat{S}(\xi_{0})\hat{\rho}\hat{S}^{\dagger}(\xi_{0})}(\alpha,\alpha^{*})$
can now be written
\begin{subequations}
\label{eq:-219}
\begin{align}
W_{\hat{S}(\xi_{0})\hat{\rho}\hat{S}^{\dagger}(\xi_{0})}(\alpha,\alpha^{*}) & =\begin{aligned}[t]\int d\mu d\mu^{*}\, & \left|\begin{array}{cc}
\cosh r & e^{i\theta}\sinh r\\
e^{-i\theta}\sinh r & \cosh r
\end{array}\right|\\
 & \cdot e^{\alpha\left(\mu^{*}\cosh r+\mu e^{i\theta}\sinh r\right)-\alpha^{*}\left(\mu\cosh r-\mu^{*}e^{-i\theta}\sinh r\right)}\,\Tr\left[\hat{\rho}e^{\mu\hat{a}^{\dagger}-\mu^{*}\hat{a}}\right]
\end{aligned}
\\
 & =\int d\mu d\mu^{*}\,e^{\left(\alpha\cosh r+\alpha^{*}e^{i\theta}\sinh r\right)\mu^{*}-\left(\alpha^{*}\cosh r+\alpha e^{-i\theta}\sinh r\right)\mu}\,\chi(\mu).\label{eq:-22}
\end{align}
\end{subequations}
Hence, we can write 
\begin{equation}
W_{\hat{S}(\xi)\hat{\rho}\hat{S}^{\dagger}(\xi)}(\alpha,\alpha^{*},t)=W_{\rho}(\alpha\cosh r+\alpha^{*}e^{i\theta}\sinh r,\alpha^{*}\cosh r+\alpha e^{-i\theta}\sinh r)\,.\label{eq:-23}
\end{equation}
This is applied to (\ref{eq:-21}) to write

\begin{equation}
W_{\hat{S}(\xi)\hat{\rho}\hat{S}^{\dagger}(\xi)}(\alpha,\alpha^{*})=\frac{1}{\pi\left(\bar{n}_{0}+\frac{1}{2}\right)}\exp\left[-\frac{|\alpha\cosh r_{0}+\alpha^{*}e^{i\theta_{0}}\sinh r_{0}|^{2}}{\bar{n}_{0}+\frac{1}{2}}\right].\label{eq:general-ics-complex-amplitude-1}
\end{equation}

\section{Displaced Squeezed Thermal State Wigner Function \label{app:displaced-squeezed-thermal-state-wig}}

Combining (\ref{eq:-21}), (\ref{eq:-23}) and (\ref{eq:-114}), the
Wigner function for the displaced squeezed thermal state $\hat{D}(\alpha_{0})\hat{S}(\xi)\hat{\rho}_{\bar{n}_{0}}\hat{S}^{\dagger}(\xi)\hat{D}^{\dagger}(\alpha_{0})$
may be written
\begin{equation}
W(\alpha,\alpha^{*})=\frac{1}{\pi\left(\bar{n}_{0}+\frac{1}{2}\right)}\exp\left[-\frac{|\left(\alpha-\alpha_{0}\right)\cosh r_{0}+\left(\alpha-\alpha_{0}\right)^{*}e^{i\theta_{0}}\sinh r_{0}|^{2}}{\bar{n}_{0}+\frac{1}{2}}\right].\label{eq:general-ics-complex-amplitude}
\end{equation}
The parameter $\bar{n}_{0}$ is the mean occupancy of the thermal
state, $\xi=r_{0}e^{i\theta_{0}}$ describes the squeezing and $\alpha_{0}$
describes the displacement. This can be expressed in Cartesian coordinates
as
\begin{equation}
W(x,y)=\frac{1}{\pi\left(\bar{n}_{0}+\frac{1}{2}\right)}\begin{aligned}[t]\exp & \left[-\frac{\left(\left(x-x_{0}\right)e^{r_{0}}\cos\frac{\theta_{0}}{2}+\left(y-y_{0}\right)e^{r_{0}}\sin\frac{\theta_{0}}{2}\right)^{2}}{\bar{n}_{0}+\frac{1}{2}}\right.\\
 & \left.\qquad-\frac{\left(\left(x-x_{0}\right)e^{-r_{0}}\sin\frac{\theta_{0}}{2}-\left(y-y_{0}\right)e^{-r_{0}}\cos\frac{\theta_{0}}{2}\right)^{2}}{\bar{n}_{0}+\frac{1}{2}}\right]
\end{aligned}
\label{eq:general-ics-cartesian}
\end{equation}
where $x_{0}=\Re\alpha_{0}$ and $y_{0}=\Im\alpha_{0}$. In polar
coordinates it reads
\begin{equation}
W(r,\phi)=\frac{1}{\pi\left(\bar{n}_{0}+\frac{1}{2}\right)}\exp\left[-\frac{|\left(re^{i\phi}-\alpha_{0}\right)\cosh r_{0}+\left(re^{-i\phi}-\alpha_{0}^{*}\right)e^{i\theta_{0}}\sinh r_{0}|^{2}}{\bar{n}_{0}+\frac{1}{2}}\right].\label{eq:general-ics-polar}
\end{equation}
The state of (\ref{eq:general-ics-complex-amplitude}), (\ref{eq:general-ics-cartesian})
and (\ref{eq:general-ics-polar}) includes as special cases a coherent
state (where $\alpha_{0}$ is the coherent amplitude with other parameters
set to zero), a squeezed vacuum state (with squeezing parameter $\xi=r_{0}e^{i\theta_{0}}$
and other parameters set to zero) and a thermal state (where $\bar{n}_{0}$
is the mean occupancy of the thermal state and the other parameters
are set to zero).

\chapter{Effect of Dephasing in the Operator Picture \label{app:operator-picture-dephasing}}

A simple description of the effect of dephasing is found when unitary
system dynamics are diagonal in the number state basis. Consider the
dynamics of a system governed by the master equation
\begin{equation}
\dot{\hat{\rho}}(t)=-\frac{i}{\hbar}[H(\hat{n}),\hat{\rho}(t)]+\gamma_{\phi}\mathcal{D}[\hat{n}]\hat{\rho}(t)\label{eq:1-1}
\end{equation}
where setting $\hat{H}=H(\hat{n})$ indicates that the Hamiltonian
is diagonal in the number state basis:
\begin{equation}
[H(\hat{n}),\hat{n}]=0.\label{eq:-38}
\end{equation}
In particular, (\ref{eq:-38}) is obeyed for the Kerr oscillator (see
(\ref{eq:kerr-hamiltonian})). The dephasing has strength $\gamma_{\phi}$
with the Lindblad superoperator given by $\mathcal{D}[\hat{n}]\hat{\rho}=\hat{n}\hat{\rho}\hat{n}^{\dagger}-\tfrac{1}{2}\{\hat{n}^{\dagger}\hat{n},\hat{\rho}\}$.

Here it is shown that the solution to (\ref{eq:1-1}) can also be
written as

\begin{equation}
\hat{\rho}(t)=e^{-iH(\hat{n})t/\hbar}\left[\frac{1}{\sqrt{4\pi\gamma_{\phi}t}}\int_{-\infty}^{\infty}d\phi\,e^{-\phi^{2}\big/4\gamma_{\phi}t}e^{i\hat{n}\phi}\hat{\rho}(0)e^{-i\hat{n}\phi}\right]e^{iH(\hat{n})t/\hbar}\label{eq:-41}
\end{equation}
or, commuting the exponentials, equivalently
\begin{equation}
\hat{\rho}(t)=\frac{1}{\sqrt{4\pi\gamma_{\phi}t}}\int_{-\infty}^{\infty}d\phi\,e^{-\phi^{2}\big/4\gamma_{\phi}t}e^{i\hat{n}\phi}\left[e^{-iH(\hat{n})t/\hbar}\hat{\rho}(0)e^{iH(\hat{n})t/\hbar}\right]e^{-i\hat{n}\phi}.\label{eq:-42}
\end{equation}
Formally, one may write
\begin{equation}
e^{(-it/\hbar)[H(\hat{n}),\cdot]}e^{\gamma_{\phi}t\mathcal{D}[\hat{n}]}=e^{(-it/\hbar)[H(\hat{n}),\cdot]+t\gamma_{\phi}\mathcal{D}[\hat{n}]}=e^{\gamma_{\phi}\mathcal{D}[\hat{n}]}e^{(-it/\hbar)[H(\hat{n}),\cdot]}.
\end{equation}
The symbolic notation $[H,\cdot]$ denotes the superoperator that
applies the commutator with the Hamiltonian to an operator, i.e. $[H(\hat{n}),\cdot]\hat{\rho}=[H(\hat{n}),\hat{\rho}]$. 

Equations (\ref{eq:-41}) and (\ref{eq:-42}) also suggest a way to
compute the effect of dephasing under the condition (\ref{eq:-38}):
Computing the unitary evolution and the integral over $\phi$ separately.
Since the operations commute, the order of these steps is insignificant.

Before we continue to the derivation, let us note that the solution
to (\ref{eq:1-1}) is  expressed readily in componentwise form. By
computing
\begin{equation}
\frac{d}{dt}\langle m|\hat{\rho}(t)|n\rangle=-\frac{i}{\hbar}\langle m|[H(\hat{n}),\hat{\rho}(t)]|n\rangle+\langle m|\gamma_{\phi}\mathcal{D}[\hat{n}]\hat{\rho}(t)|n\rangle
\end{equation}
one obtains

\begin{equation}
\frac{d}{dt}\langle m|\hat{\rho}(t)|n\rangle=\left[-i\left(H(m)-H(n)\right)-\frac{\gamma_{\phi}}{2}(m-n)^{2}\right]\langle m|\hat{\rho}(t)|n\rangle
\end{equation}
whose solution is simply
\begin{equation}
\langle m|\hat{\rho}(t)|n\rangle=e^{-i\left(H(m)-H(n)\right)t-\frac{\gamma_{\phi}}{2}(m-n)^{2}t}\langle m|\hat{\rho}(0)|n\rangle.
\end{equation}
In the above expressions, $H(m)$ represents the expression for $H(\hat{n})$
where $\hat{n}$ has been substituted with the c-number $\hat{n}$.

\paragraph{Transforming to the interaction picture.}

Since the entire right hand side of (\ref{eq:1-1}) is diagonal in
the number state basis, transforming to the interaction picture leaves
the dephasing terms of (\ref{eq:1-1}) invariant. This allows for
a simple description of the effect of dephasing. Following the steps
in Section \ref{sec:interaction-picture}, transform now to the interaction
picture as described by the unitary transformation
\begin{equation}
\hat{U}_{I}(t)=e^{-iH(\hat{n})t}.\label{eq:-34}
\end{equation}
This is equivalent to choosing $\hat{H}_{0}=H(\hat{n})$ and $V=0$.
(\ref{eq:-34}) defines interaction picture quantities
\begin{equation}
\hat{\rho}_{I}(t)=\hat{U}^{\dagger}(t)\hat{\rho}(t)\hat{U}(t)
\end{equation}
and 
\begin{equation}
\hat{H}_{I}=\hat{U}^{\dagger}(t)H(\hat{n})\hat{U}(t)=H(\hat{n}).
\end{equation}
From (\ref{eq:-34}), the equation of motion for $\rho_{I}(t)$ may
be derived as
\begin{equation}
\dot{\hat{\rho}}_{I}(t)=-\frac{i}{\hbar}\left[H(\hat{n}),\hat{\rho}_{I}(t)\right]-\frac{i}{\hbar}\left[\hat{H}_{I},\hat{\rho}_{I}(t)\right]+\gamma_{\phi}\mathcal{D}[\hat{n}]\hat{\rho}_{I}=\gamma_{\phi}\mathcal{D}[\hat{n}]\hat{\rho}_{I}\label{eq:-35}
\end{equation}
where it is noted that
\begin{equation}
\hat{U}^{\dagger}(t)\left(\mathcal{D}[\hat{n}]\hat{\rho}_{I}(t)\right)\hat{U}(t)=\mathcal{D}[\hat{n}]\left(\hat{U}^{\dagger}(t)\hat{\rho}(t)\hat{U}(t)\right),
\end{equation}
cf. (\ref{eq:lindblad-superoperator}).

\paragraph{Expression of problem equivalent to the master equation.}

Recall from Section \ref{sec:operator-transformations}, the rotation
operator
\[
\hat{R}(\phi)=e^{i\hat{n}\phi}\,.\tag{\ref{eq:rotation-operator}}
\]
As an ansatz, assume that one may choose some density matrix $\hat{\rho}_{0}$
such that the solution to (\ref{eq:-35}), $\hat{\rho}_{I}(t)$, can
be written 
\begin{equation}
\hat{\rho}_{I}(t)=\int_{-\infty}^{\infty}d\phi\,p(\phi;t)\hat{R}(\phi)\hat{\rho}_{0}\hat{R}^{\dagger}(\phi).\label{eq:3-1}
\end{equation}
It is seen from this that $p(\phi;t)$ should be normalized as if
it was a probability density in $\phi$:\footnote{Note that the integral is over the interval $(-\infty,\infty)$ and
not the interval $[0,2\pi)$. We can not really think of $p(\phi;t)$
as a probability density function of an angle in the normal sense,
since the rotation operator wraps around at $2\pi$: $\hat{R}(\phi+2\pi)=\hat{R}(\phi)$.
Choosing instead the interval $[0,2\pi)$ will impose a periodic boundary
condition on (\ref{eq:15}). It can still be solved using a Fourier
series, but the concise expression (\ref{eq:-43}) is lost.}
\begin{equation}
1=\Tr\left[\hat{\rho}_{I}(t)\right]=\int_{-\infty}^{\infty}d\phi\,p(\phi;t)\,\Tr\left[\hat{R}(\phi)\hat{\rho}_{0}\hat{R}^{\dagger}(\phi)\right]=\int_{-\infty}^{\infty}d\phi\,p(\phi;t).\label{eq:-46}
\end{equation}
Here, the cyclic property of the trace has been used, as has the normalization
condition $\Tr\hat{\rho}_{0}=1$. Inserting the ansatz into the right
and left hand sides of the master equation (\ref{eq:-35}), the resulting
expressions should be equal if (\ref{eq:3-1}) is to be a solution.
The left hand side yields
\begin{subequations}
\label{eq:6} 
\begin{align}
\dot{\hat{\rho}}_{I}(t) & =\frac{d}{dt}\int_{-\infty}^{\infty}d\phi\,p(\phi;t)\hat{R}(\phi)\hat{\rho}_{0}\hat{R}^{\dagger}(\phi)\label{eq:4}\\
 & =\int_{-\infty}^{\infty}d\phi\,\frac{\partial p(\phi;t)}{\partial t}\hat{R}(\phi)\hat{\rho}_{0}\hat{R}^{\dagger}(\phi)\,,\label{eq:5}
\end{align}
\end{subequations}
while the right hand side yields
\begin{subequations}
\label{eq:7-1} 
\begin{align}
\mathcal{D}[\hat{n}]\hat{\rho} & =\int_{-\infty}^{\infty}d\phi\,p(\phi;t)\left[\hat{n}\hat{R}(\phi)\hat{\rho}_{0}\hat{R}^{\dagger}(\phi)\hat{n}-\tfrac{1}{2}\hat{n}^{2}\hat{R}(\phi)\hat{\rho}_{0}\hat{R}^{\dagger}(\phi)-\tfrac{1}{2}\hat{R}(\phi)\hat{\rho}_{0}\hat{R}^{\dagger}(\phi)\hat{n}^{2}\right]\label{eq:8}\\
 & =\begin{aligned}[t]\int_{-\infty}^{\infty}d\phi\,p(\phi;t)\bigg[ & -i\frac{\partial}{\partial\phi}\hat{R}(\phi)\rho_{0}i\frac{\partial}{\partial\phi}\hat{R}^{\dagger}(\phi)\\
 & -(-i)^{2}\frac{1}{2}\frac{\partial^{2}}{\partial\phi^{2}}\hat{R}(\phi)\hat{\rho}_{0}\hat{R}^{\dagger}(\phi)-i^{2}\frac{1}{2}\hat{R}(\phi)\hat{\rho}_{0}\frac{\partial^{2}}{\partial\phi^{2}}\hat{R}^{\dagger}(\phi)\bigg]
\end{aligned}
\label{eq:9}\\
 & =\int_{-\infty}^{\infty}d\phi\,p(\phi;t)\frac{\partial^{2}}{\partial\phi^{2}}\left[\hat{R}(\phi)\hat{\rho}_{0}\hat{R}^{\dagger}(\phi)\right]\label{eq:10}\\
 & =\begin{aligned}[t] & \int_{-\infty}^{\infty}d\phi\,\frac{\partial^{2}p(\phi;t)}{\partial\phi^{2}}\left[\hat{R}(\phi)\hat{\rho}_{0}\hat{R}^{\dagger}(\phi)\right]\\
 & +\left(p(\phi;t)\frac{\partial}{\partial\phi}\left[\hat{R}(\phi)\hat{\rho}_{0}\hat{R}^{\dagger}(\phi)\right]-\frac{\partial p(\phi;t)}{\partial\phi}\left[\hat{R}(\phi)\hat{\rho}_{0}\hat{R}^{\dagger}(\phi)\right]\right)\Bigg|_{\phi=-\infty}^{\infty}
\end{aligned}
\label{eq:11}\\
 & =\int_{-\infty}^{\infty}d\phi\,\frac{\partial^{2}p(\phi;t)}{\partial\phi^{2}}\left[\hat{R}(\phi)\hat{\rho}_{0}\hat{R}^{\dagger}(\phi)\right]\,.\label{eq:12}
\end{align}
\end{subequations}
$p(\phi;t)$ and $\frac{\partial p(\phi;t)}{\partial\phi}$ are both
assumed to vanish as $\phi$ tends to $\pm\infty$, allowing for the
removal of the boundary terms in (\ref{eq:11}). Using (\ref{eq:6})
and (\ref{eq:7-1}), the dephasing master equation (\ref{eq:-35})
requires that 
\begin{equation}
\int_{-\infty}^{\infty}d\phi\,\frac{\partial p(\phi;t)}{\partial t}\left[\hat{R}(\phi)\hat{\rho}_{0}\hat{R}^{\dagger}(\phi)\right]=\gamma_{\phi}\int_{-\infty}^{\infty}d\phi\,\frac{\partial^{2}p(\phi;t)}{\partial\phi^{2}}\left[\hat{R}(\phi)\hat{\rho}_{0}\hat{R}^{\dagger}(\phi)\right]\,.\label{eq:14}
\end{equation}
One way to satisfy (\ref{eq:14}), is to choose $p(\phi;t)$ to be
the solution to the partial differential equation
\begin{equation}
\frac{\partial p(\phi;t)}{\partial t}=\gamma_{\phi}\frac{\partial^{2}p(\phi;t)}{\partial\phi^{2}}.\label{eq:15}
\end{equation}
Such a solution may in turn be used to express a solution to the dephasing
master equation. Given initial conditions for the master equation,
one needs to first construct corresponding $\hat{\rho}_{0}$ and $p(\phi;0)$
satisfying (\ref{eq:3-1}) for $t=0$, then solve (\ref{eq:15}) with
this $p(\phi;0)$ as initial condition.

\paragraph{Construction of master equation solution.}

Given an initial condition in the interaction picture, $\hat{\rho}_{I}(0)$,
choosing 
\begin{equation}
\hat{\rho}_{0}=\hat{\rho}_{I}(0)\label{eq:-39}
\end{equation}
 and 
\begin{equation}
p(\phi;0)=\delta(\phi),\label{eq:-40}
\end{equation}
it is seen that (\ref{eq:3-1}) is satisfied. The solution to (\ref{eq:15})
with (\ref{eq:-40}) is given by \cite{Arfken_MathematicalMethodsPhysicists_2012}
\begin{equation}
p(\phi;t)=\frac{1}{\sqrt{4\pi\gamma_{\phi}t}}e^{-\phi^{2}\big/4\gamma_{\phi}t}\,.\label{eq:17}
\end{equation}
Together then, equations (\ref{eq:3-1}) and (\ref{eq:17}) describe
the evolution of an arbitrary density matrix under dephasing:
\begin{equation}
\hat{\rho}_{I}(t)=\frac{1}{\sqrt{4\pi\gamma_{\phi}t}}\int_{-\infty}^{\infty}d\phi\,e^{-\phi^{2}\big/4\gamma_{\phi}t}e^{i\hat{n}\phi}\hat{\rho}_{I}(0)e^{-i\hat{n}\phi}\,.\label{eq:-43}
\end{equation}

\paragraph{Return from the Schrödinger picture.}

The interaction picture initial state is trivially obtained as identical
to the Schrödinger picture initial state\footnote{This is the case for the choice of interaction picture given by (\ref{eq:-34})
and not generally true.}
\begin{equation}
\hat{\rho}_{I}(0)=\hat{U}^{\dagger}(0)\hat{\rho}(0)\hat{U}(0)=\hat{\rho}(0).\label{eq:-44}
\end{equation}
The Schrödinger picture density matrix is obtained using the inverse
transformation:
\begin{equation}
\hat{\rho}(t)=\hat{U}(t)\hat{\rho}_{I}(t)\hat{U}^{\dagger}(t).\label{eq:-45}
\end{equation}
Combining (\ref{eq:-34}), (\ref{eq:-43}), (\ref{eq:-44}) and (\ref{eq:-45}),
(\ref{eq:-41}) is obtained. Since $\hat{U}(t)$ and $\hat{R}(\phi)$
commute, this result may also be expressed as (\ref{eq:-42}).

\paragraph{Note on non-uniqueness.}

While (\ref{eq:-41}) or (\ref{eq:-42}) express a general solution
to (\ref{eq:1-1}), the particular choice of $\hat{\rho}_{0}$ and
$p(\phi;t)$ is not unique.

$p(\phi;t)$ can trivially be remapped to $p(\phi+2n;t)$ for integer
$n$ without changing (\ref{eq:3-1}). A simple example which may
be described by an infinite number of different choices for $p(\phi;t)$
is provided by the number state $|n\rangle\langle n|$. Using this
as the initial state in (\ref{eq:1-1}), it is seen to be constant
under both the unitary dynamics of $H(\hat{n})$ and dephasing. Equating
the trivial constant solution with (\ref{eq:3-1}) to get
\begin{equation}
|n\rangle\langle n|=\int_{-\infty}^{\infty}d\phi\,p(\phi;t)|n\rangle\langle n|\,,
\end{equation}
it is seen that any $p(\phi;t)$ satisfying the simple normalization
condition (\ref{eq:-46}) at all times $t$ yields a solution --
$p(\phi;t)$ may even be negative for certain arguments $\phi$.

As an alternate example, consider an arbitrary initial interaction
picture state $\hat{\rho}_{I}(0)$ evolved to a time $t_{1}$. The
resulting state is then given by
\begin{equation}
\hat{\rho}_{I}(t_{1})=\frac{1}{\sqrt{4\pi\gamma_{\phi}t_{1}}}\int_{-\infty}^{\infty}d\phi\,e^{-\phi^{2}\big/4\gamma_{\phi}t_{1}}e^{i\hat{n}\phi}\hat{\rho}_{I}(0)e^{-i\hat{n}\phi}.\label{eq:-47}
\end{equation}
For a new time $t_{2}>t_{1}$, (\ref{eq:-47}) is still valid when
$t_{1}$ is replaced by $t_{2}$. However, one may also choose $\hat{\rho}(t_{1})$
as a new initial state and evolve $\hat{\rho}(t_{1})$ to a new time
$t_{2}$ with the formula
\begin{equation}
\hat{\rho}(t_{2})=\frac{1}{\sqrt{4\pi\gamma_{\phi}\left(t_{2}-t_{1}\right)}}\int_{-\infty}^{\infty}d\phi\,e^{-\phi^{2}\big/4\gamma_{\phi}(t_{2}-t_{1})}e^{i\hat{n}\phi}\hat{\rho}_{I}(t_{1})e^{-i\hat{n}\phi}.\label{eq:-48}
\end{equation}
Comparing (\ref{eq:-47}) and (\ref{eq:-48}), the form of the solutions
differ in both the initial state ($\hat{\rho}_{I}(0)$ versus $\hat{\rho}_{I}(t_{1})$)
and the weight function (without $\phi$-normalization, $e^{-\phi^{2}\big/4\gamma_{\phi}t_{2}}$
versus $e^{-\phi^{2}\big/4\gamma_{\phi}(t_{2}-t_{1})}$). Even so,
the resulting density matrix, $\hat{\rho}(t_{2})$, is the same assuming
uniqueness of the solution to the master equation (\ref{eq:1-1}).

\chapter{Periodic Evolution of Scaled Wigner Function \label{app:supporting-figures}}

\noindent 
\begin{figure}[H]
\noindent \begin{centering}
\centerline{\includegraphics[viewport=0bp 120bp 428bp 500bp,clip]{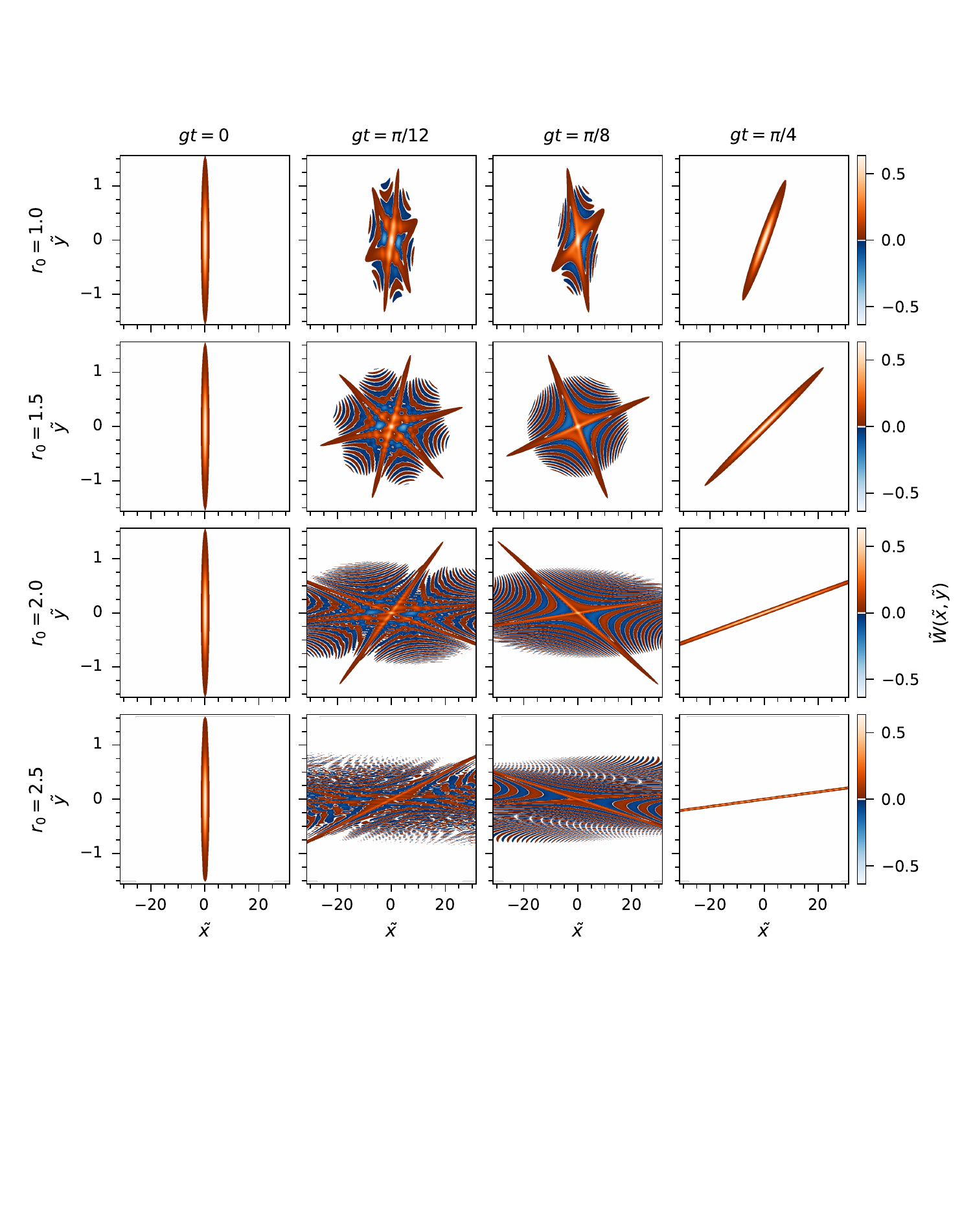}}
\par\end{centering}
\caption[Periodic evolution of scaled Wigner function]{\label{fig:scaled-squeezed-gallery-1-1}\textbf{Periodic evolution
of scaled Wigner function.} Scaled Wigner function (\ref{eq:-19})
evolved to fractional multiples of the period. This demonstrates that
the scaled Wigner function is useful only in considering the evolution
over short times. Note that the maximum value of the squeezed Wigner
function (which occurs when $gt$ is an integer multiple of $\pi/4$)
is given the same for all (namely $W(0,0,t)=4/\pi$, cf. (\ref{eq:-27})).
The axes are shared with Figure \ref{fig:scaled-squeezed-gallery-1}
but the depicted times are shared with Figure \ref{fig:evo-gallery-long}. }
\end{figure}

\chapter{Emperical Scalings for Coherent State Negativity\label{chap:Emperical-Scalings-for}}

\begin{figure}
\captionsetup{position=top}\subfloat[\label{fig:-4}Plot of $N_{\mathrm{vol}}/\alpha^{1/2}$ as a function
of $gt\alpha_{0}^{3/2}$. ]{\noindent \begin{centering}
\includegraphics{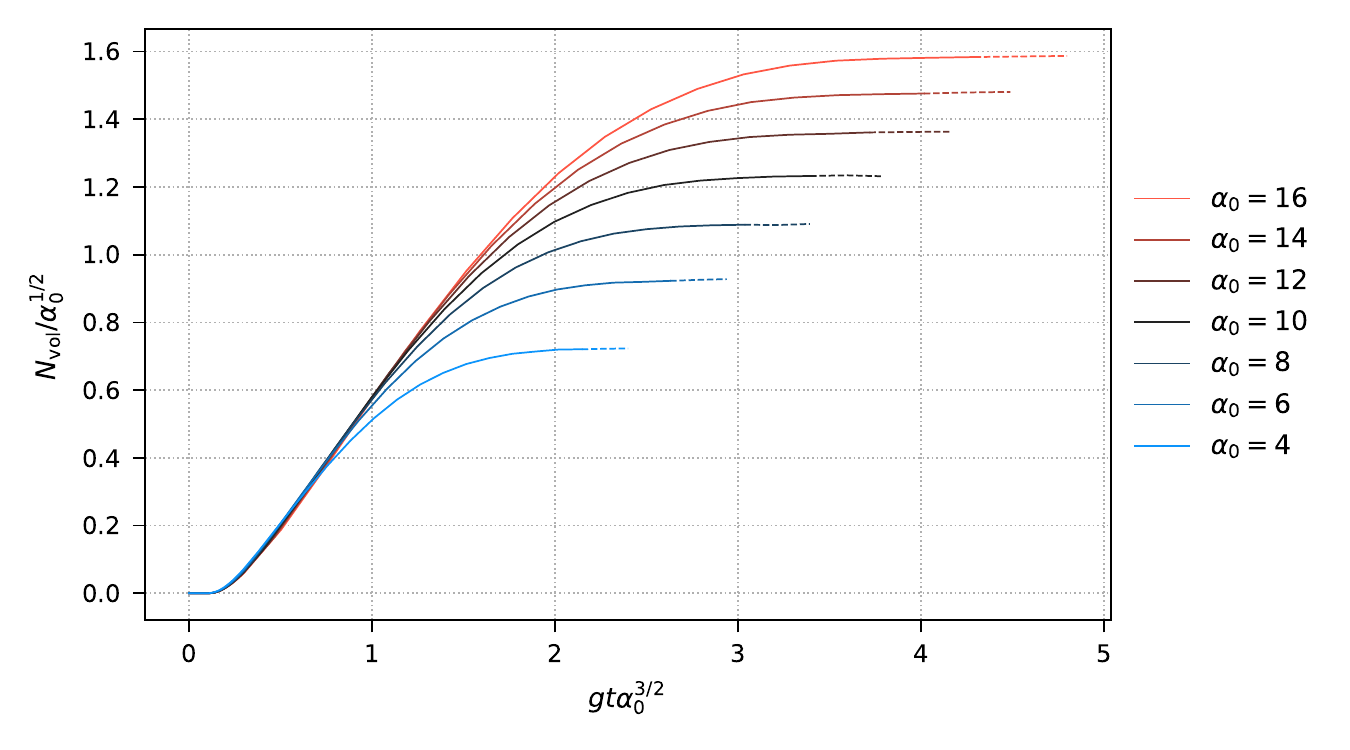}
\par\end{centering}
\noindent \centering{}}

\subfloat[Plot of $N_{\mathrm{vol}}$ as a function of $gt\alpha_{0}^{3/2}$.]{\noindent \begin{centering}
\includegraphics{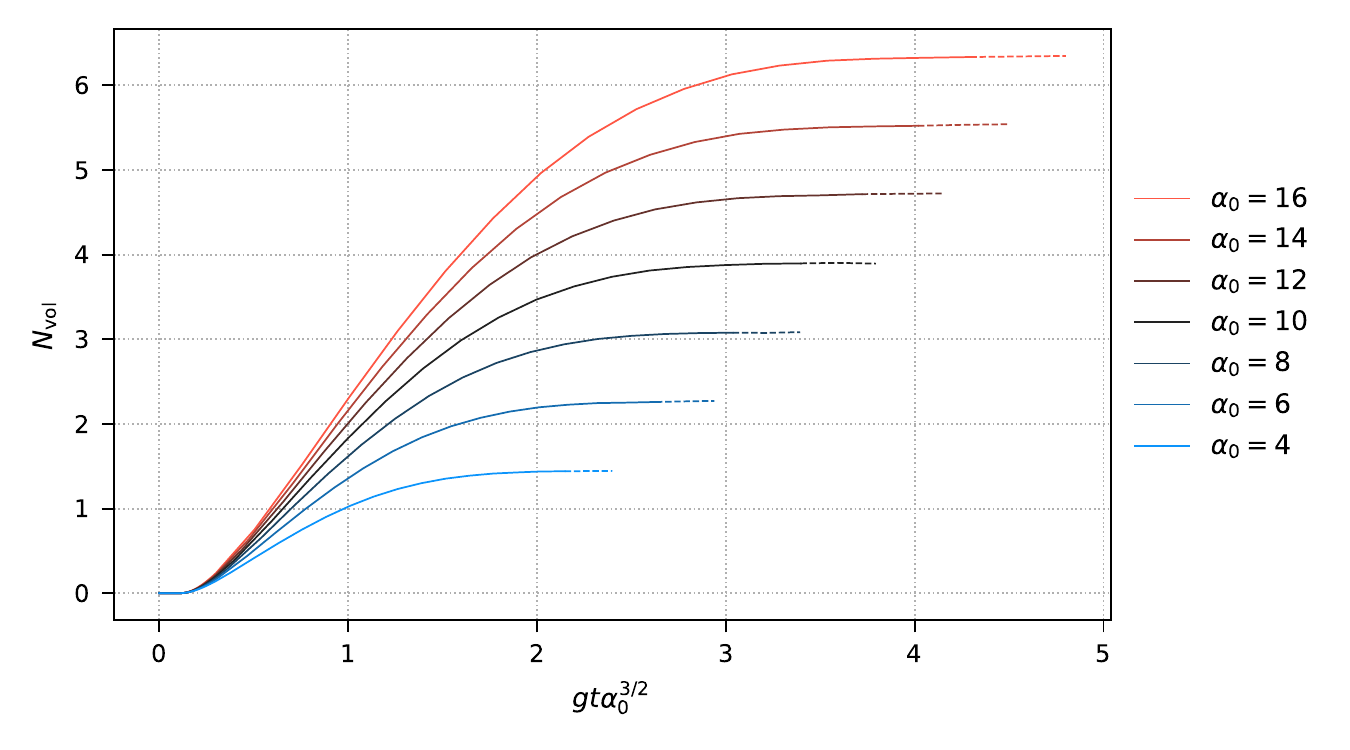}
\par\end{centering}
\noindent \centering{}}

\caption[Graphs for the negative volume for the coherent state.]{\textbf{Graphs for the negative volume $N_{\mathrm{vol}}$ for the
coherent state.} The short time evolution of the coherent state is
discussed in Section \ref{subsec:Evolution-over-Short-1}.}
\end{figure}

\begin{figure}
\captionsetup{position=top}\subfloat[\label{fig:-4-1}Plot of $N_{\mathrm{peak}}/\alpha^{1/2}$ as a function
of $gt\alpha_{0}^{3/2}$.]{\noindent \begin{centering}
\includegraphics{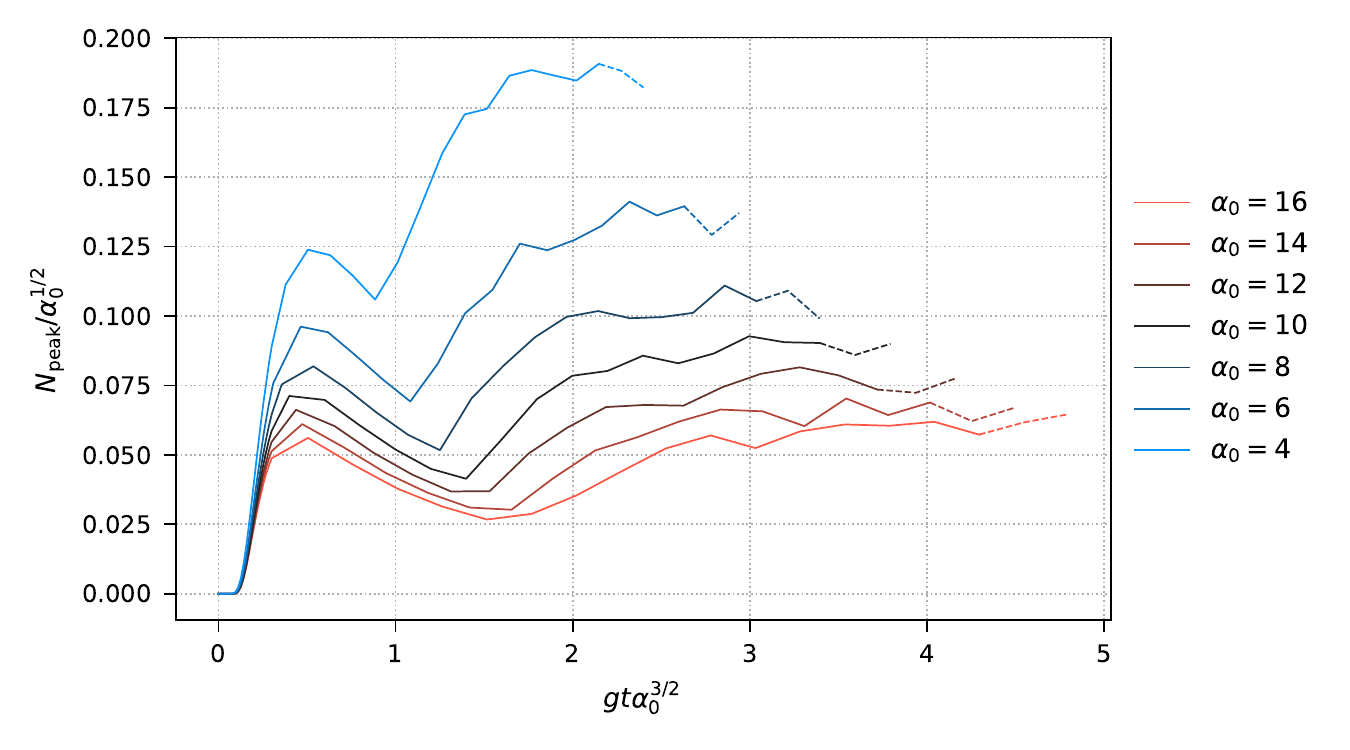}
\par\end{centering}
\noindent \centering{}}

\subfloat[\label{fig:-7}Plot of $N_{\mathrm{peak}}$ as a function of $gt\alpha_{0}^{3/2}$.]{\noindent \begin{centering}
\includegraphics{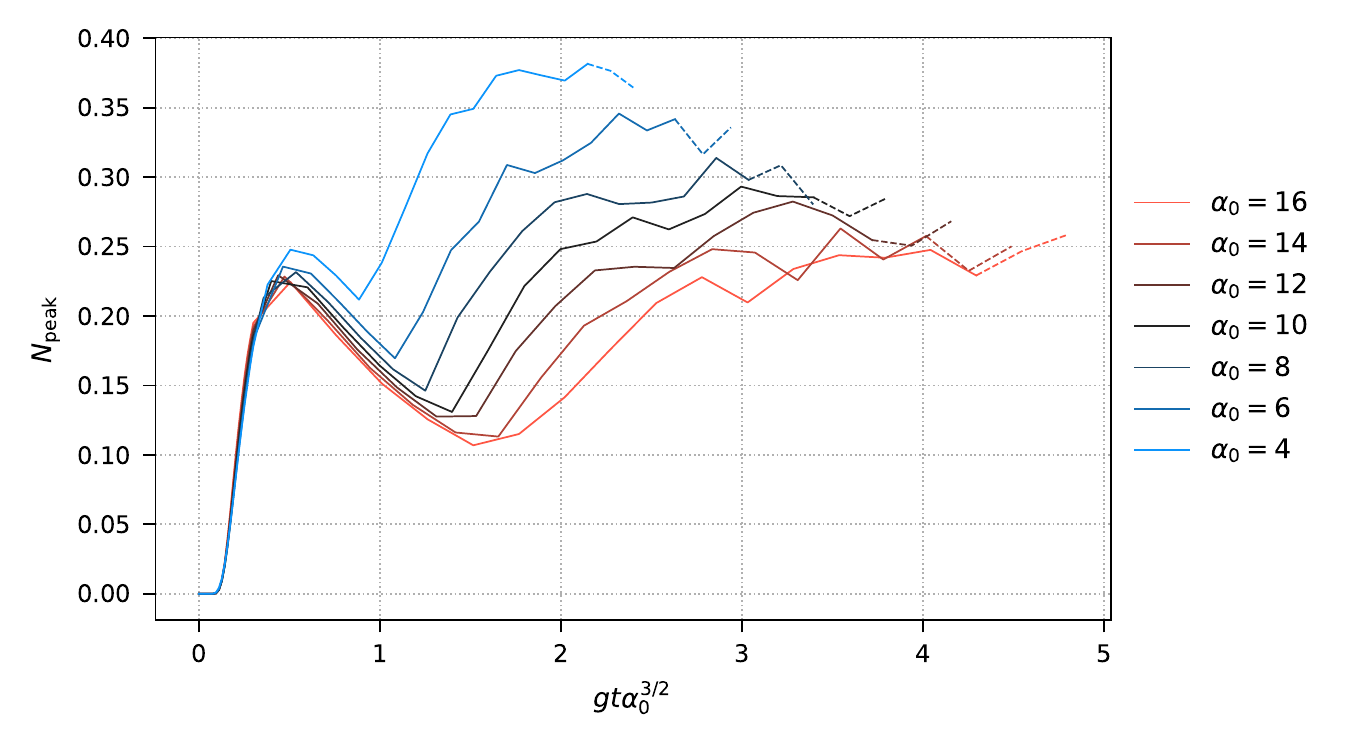}
\par\end{centering}
\noindent \centering{}}

\caption[Graphs for the negative peak for the coherent state]{\textbf{Graphs for the negative peak $N_{\mathrm{peak}}$ for the
coherent state.} The short time evolution of the coherent state is
discussed in Section \ref{subsec:Evolution-over-Short-1}.}
\end{figure}

\begin{figure}
\captionsetup{position=top}\subfloat[Plot of $N_{\mathrm{vol}}$ as a function of $gt\alpha_{0}^{3/2}$.]{\noindent \begin{centering}
\includegraphics{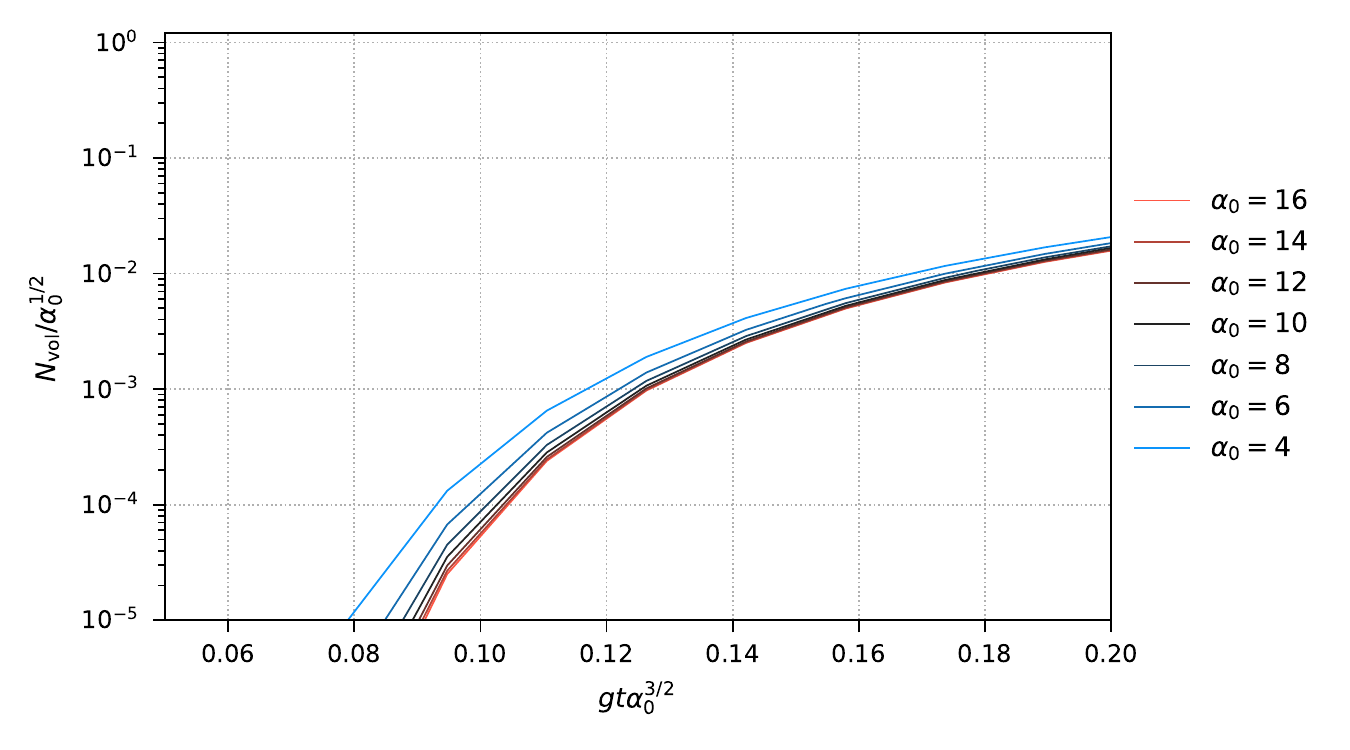}
\par\end{centering}
\noindent \centering{}}

\subfloat[\label{fig:-5}Plot of $N_{\mathrm{vol}}$ as a function of $gt\alpha_{0}^{3/2}$.]{\noindent \begin{centering}
\includegraphics{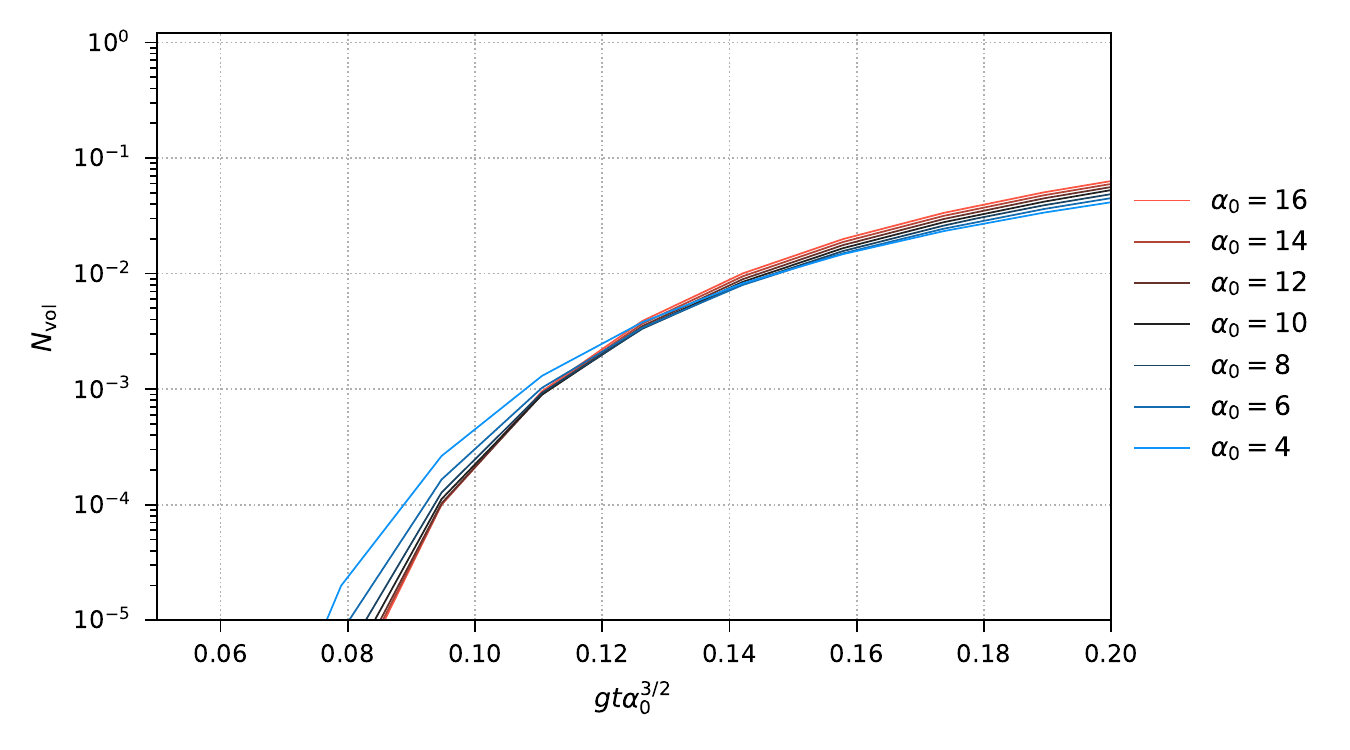}
\par\end{centering}
\noindent \centering{}}

\caption[Graphs for the negative volume for the coherent state]{\textbf{Graphs for the negative volume $N_{\mathrm{vol}}$ for the
coherent state}. The short time evolution of the coherent state is
discussed in Section \ref{subsec:Evolution-over-Short-1}.}
\end{figure}

\begin{figure}
\captionsetup{position=top}\subfloat[Plot of $N_{\mathrm{peak}}/\alpha^{1/2}$ as a function of $gt\alpha_{0}^{3/2}$.]{\noindent \begin{centering}
\includegraphics{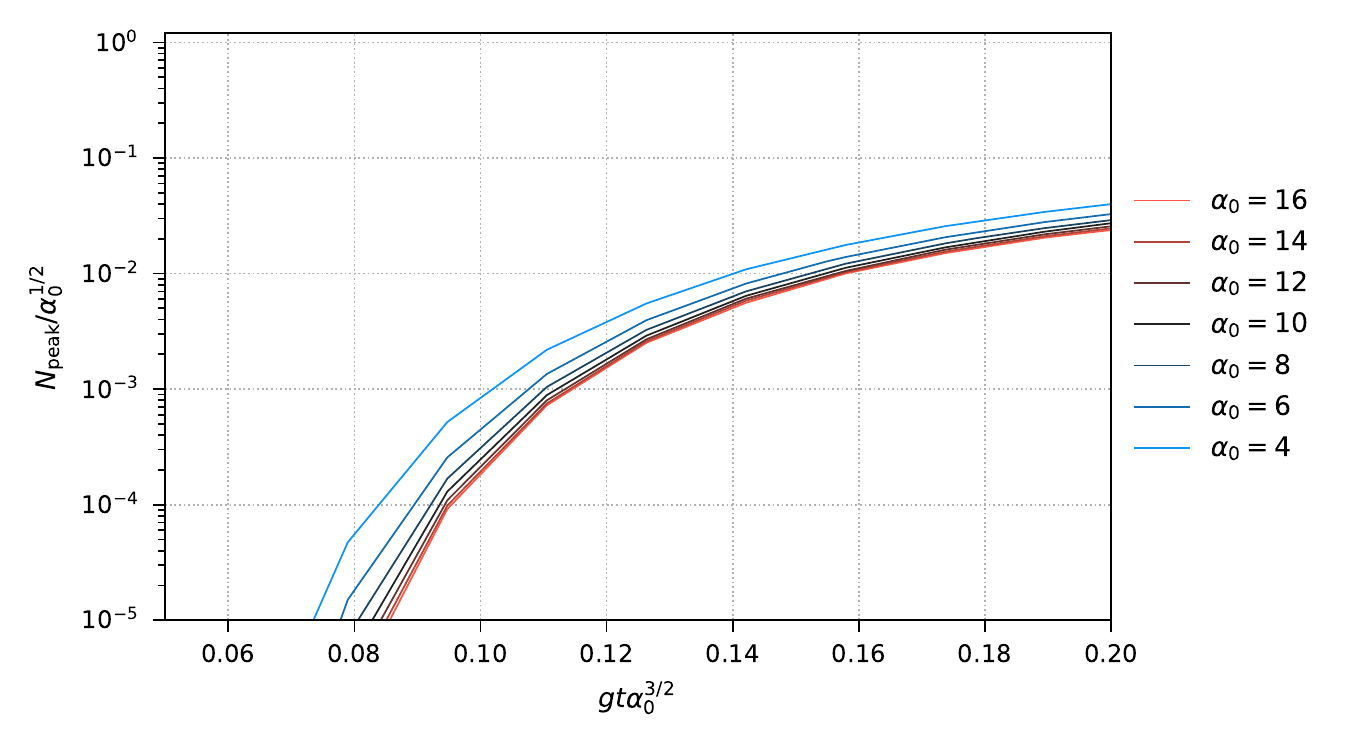}
\par\end{centering}
\noindent \centering{}}

\subfloat[\label{fig:-6}Plot of $N_{\mathrm{peak}}$ as a function of $gt\alpha_{0}^{3/2}$.]{\noindent \begin{centering}
\includegraphics{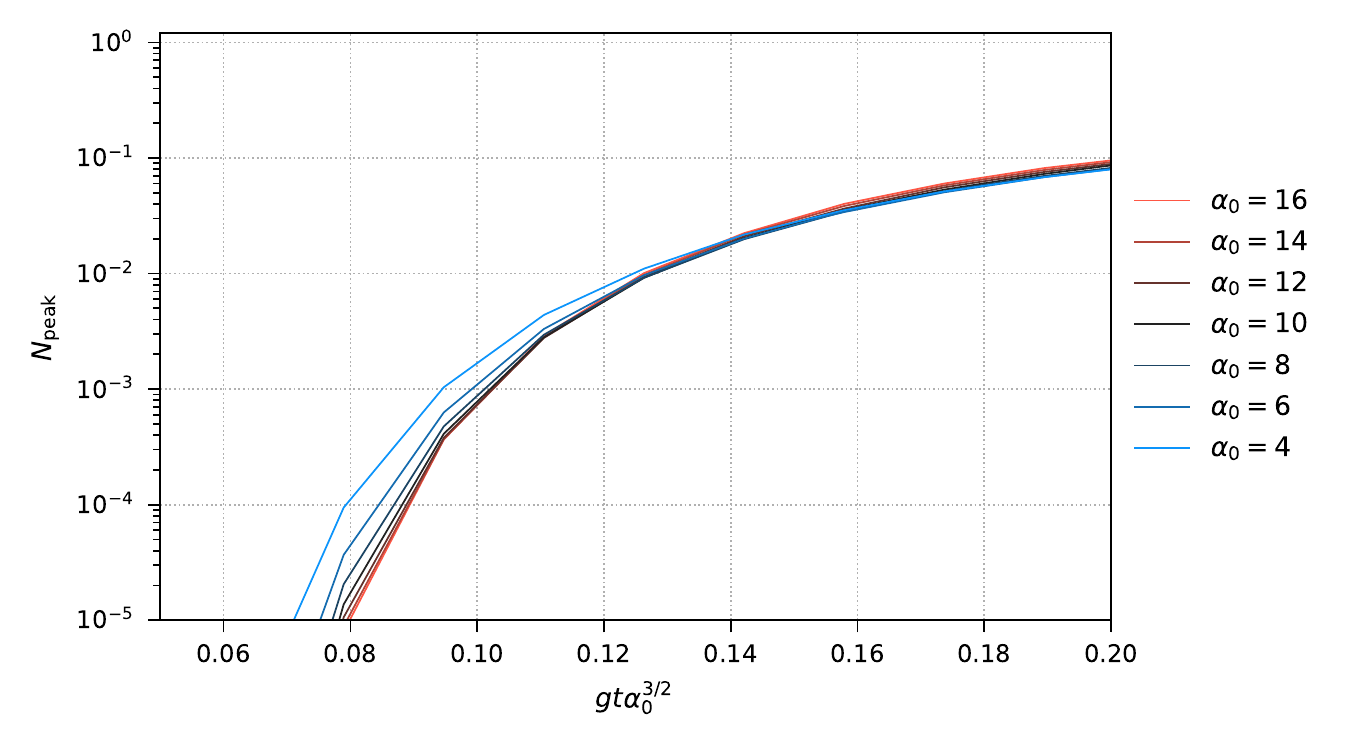}
\par\end{centering}
\noindent \centering{}}

\caption[Graphs for the negative peak for the coherent state]{\textbf{Graphs for the negative peak $N_{\mathrm{peak}}$ for the
coherent state.} The short time evolution of the coherent state is
discussed in Section \ref{subsec:Evolution-over-Short-1}.}
\end{figure}

The following figures show the short time evolution in negativity
for a coherent state as discussed in Section \ref{subsec:Evolution-over-Short-1}.
Graphing $\alpha_{0}^{-1/2}N_{\mathrm{vol}}$ as a function of $gt\alpha_{0}^{3/2}$
leaves the graphs overlapping in the region of constant growth in
negativity (Figure \ref{fig:-4}). This shows that the slope of $N_{\mathrm{vol}}$
in this region is approximately $\alpha_{0}^{2}$. For even shorter
times, the graphs appear to match when plotting just $N_{\mathrm{vol}}$
as a function of $gt\alpha_{0}^{3/2}$ (Figure \ref{fig:-5}). With
respect to $N_{\mathrm{peak}}$, the graphs appear to match well when
graphing $N_{\mathrm{peak}}$ as a function of $gt\alpha_{0}^{3/2}$
for both short and intermediate times (Figures \ref{fig:-7} and \ref{fig:-6}).

\end{document}